\documentclass{scrreprt}

\author{Thomas Wolkanowski-Gans}
\title{Dissertation}
\date{06. Juni 2016}

\usepackage[german,english]{babel}
\usepackage{graphics, graphicx}
\usepackage{latexsym}
\usepackage[margin=10pt,font=small,labelfont=bf]{caption}
\usepackage{amsmath}
\usepackage{cite}
\usepackage{booktabs}
\usepackage{hyperref}
\usepackage{amsfonts}
\usepackage{bbm}
\usepackage{epstopdf}
\usepackage{romannum}
\usepackage{amssymb}
\usepackage{setspace}
\usepackage{color}
\usepackage{simplewick}
\usepackage{cancel}
\usepackage{array,blindtext}
\usepackage[square,comma,sort&compress,numbers]{natbib}
\usepackage{doi}
\usepackage{notoccite}
\usepackage{enumitem}
\usepackage{rotating}
\usepackage{rotfloat}
\usepackage{wrapfig}

\DeclareMathOperator{\disc}{Disc}

\DeclareMathOperator{\diag}{diag}
\DeclareMathOperator{\abs}{Abs}

\DeclareMathOperator{\trace}{Tr}

\addtokomafont{disposition}{\boldmath}

\newcommand{\tagarray}{\mbox{}\refstepcounter{equation}(\theequation)}
\newcounter{bibpage}

\begin{document}

\onehalfspacing

\begin{titlepage}
	\begin{center}
		{\vspace{0.3cm}\hspace{0.4cm}\includegraphics[scale=0.8]{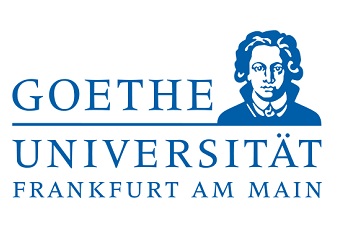}\\[1.7cm]}
		{\LARGE Dynamical generation of hadronic resonances in effective models with derivative interactions\\[2.3cm]}
		{\large \emph{Dissertation zur Erlangung des Doktorgrades\\
		der Naturwissenschaften}\\[0.5cm]}
		{\large \emph{vorgelegt beim Fachbereich Physik\\
		der Johann Wolfgang Goethe-Universit\"at\\
		in Frankfurt am Main}\\[2.3cm]}
		{\large von\\
		Thomas Wolkanowski-Gans\\
		aus Oppeln (Polen)}\\[0.5cm]
		Frankfurt am Main (2016)\\[0.5cm]
		(D 30)\\[2.9cm]
	\end{center}
\end{titlepage}

\thispagestyle{empty}
\

\newpage
\clearpage

\thispagestyle{empty}
\

\fontsize{11}{1.5}

\thispagestyle{empty}
\normalsize
\vspace{7.0cm}\noindent vom Fachbereich Physik der\\
Johann Wolfgang Goethe-Universit\"at als Dissertation angenommen

\vspace{6.0cm}

\noindent\small\textbf{Dekan:}\\[0.3cm]
\normalsize Prof.\ Dr.\ Ren\'{e} Reifarth\\[0.5cm]

\noindent\small\textbf{Gutachter:}\\[0.3cm]
\normalsize PD Dr.\ Francesco Giacosa\\
\normalsize Prof.\ Dr.\ Dirk H.\ Rischke\\[0.5cm]

\noindent Datum der Disputation: 10.08.2016

\newpage
\clearpage

\thispagestyle{empty}
\

\newpage
\clearpage

\normalsize
\thispagestyle{empty}
\ \\[7.0cm]

\hspace{0.0cm}``When I was young, I observed that nine out of ten things I did were failures.

\noindent\hspace{0.5cm} So I did ten times more work.''\\[-0.5cm]

\noindent\hspace{8.2cm}\footnotesize{-- George Bernard Shaw}

\newpage
\clearpage

\thispagestyle{empty}
\

\newpage
\clearpage

\normalsize
\thispagestyle{empty}
\ \\[7.0cm]

\hspace{2.2cm}Dies ist, um mein Versprechen einzuhalten...

\newpage
\clearpage

\thispagestyle{empty}
\

\normalsize
\thispagestyle{empty}
\

\thispagestyle{empty}
\

\newpage
\clearpage

\thispagestyle{empty}

\section*{Abstract}

\medskip

Light scalar mesons can be understood as dynamically generated resonances. They
arise as 'companion poles\grq\ in the propagators of quark-antiquark seed states 
when accounting for hadronic loop contributions to the self-energies of the latter. Such a mechanism may explain 
the overpopulation in the scalar sector -- there exist more resonances with total spin $J=0$ than can be described within a quark model.

Along this line, we study an effective Lagrangian approach where the isovector state $a_{0}(1450)$ 
couples via both non-derivative and derivative interactions to pseudoscalar mesons. It is demonstrated that the propagator has two poles: 
a companion pole corresponding to $a_{0}(980)$ and a pole of the seed state $a_{0}(1450)$. The positions of these
poles are in quantitative agreement with experimental data. Besides that, we investigate similar models for the isodoublet state $K_{0}^{\ast}(1430)$ by performing a fit to 
$\pi K$ phase shift data in the $I=1/2,$ $J=0$ channel.\ We show that, in order to fit the data accurately, a companion 
pole for the $K_{0}^{\ast}(800)$, that is, the light $\kappa$, is required.\ A large-$N_{c}$ study confirms that both resonances below $1$ GeV are predominantly four-quark states, while the heavy states are quarkonia.
\\
\vspace{-0.05cm}
\\
This thesis is based on the following publications:
\begin{itemize}[leftmargin=*]
\item T.\ Wolkanowski, M.\ So\l tysiak, and F.\ Giacosa, \emph{$K_{0}^{\ast}(800)$ as a companion pole of $K_{0}^{\ast}(1430)$}, Nucl.\ Phys.\ B \textbf{909}, 418 (2016) \href{http://arxiv.org/abs/1512.01071}{arXiv:1512.01071 [hep-ph]}
\item T.\ Wolkanowski and F.\ Giacosa, \emph{$a_{0}(980)$ as a companion pole of $a_{0}(1450)$}, PoS \textbf{CD15}, 131 (2016) \href{http://arxiv.org/abs/1510.05148}{arXiv:1510.05148 [hep-ph]}
\item T.\ Wolkanowski, F.\ Giacosa, and D.\ H.\ Rischke, \emph{$a_{0}(980)$ revisited}, Phys.\ Rev.\ D \textbf{93}, 014002 (2016) \href{http://arxiv.org/abs/arXiv:1508.00372}{arXiv:1508.00372 [hep-ph]}
\item T.\ Wolkanowski, \emph{Dynamical generation of hadronic resonances}, Acta Phys.\ Polon.\ B Proceed.\ Suppl.\ \textbf{8}, 273 (2015) \href{http://arxiv.org/abs/arXiv:1410.7022}{arXiv:1410.7022 [hep-ph]}
\item J.\ Schneitzer, T.\ Wolkanowski, and F.\ Giacosa, \emph{The role of the next-to-leading order triangle-shaped diagram in two-body hadronic decays}, Nucl.\ Phys.\ B \textbf{888}, 287 (2014) \href{http://arxiv.org/abs/1407.7414}{arXiv:1407.7414 [hep-ph]}
\item T. Wolkanowski and F. Giacosa, \emph{The scalar-isovector sector in the extended Linear Sigma Model}, Acta Phys.\ Polon.\ B Proceed.\ Suppl.\ \textbf{7}, 469 (2014)\\ \href{http://arxiv.org/abs/1404.5758}{arXiv:1404.5758 [hep-ph]}
\end{itemize}

\newpage
\clearpage

\thispagestyle{empty}

\

\newpage
\clearpage

\pagenumbering{gobble}

\begin{otherlanguage}{german}
\section*{Deutsche Zusammenfassung}
%\label{chap:german}

\medskip

Die QCD ist die Theorie der starken Wechselwirkung. Sie beschreibt im Allgemeinen die Kraft zwischen farbgeladenen Quarks als einen Austausch von ebenfalls farbigen Gluonen. Wegen des Ph\"anomens des Confinements kommen keine farbgeladenen Teilchen isoliert in der Natur vor, so dass Quarks und Gluonen in Form von Hadronen gebunden vorliegen m\"ussen. Im Speziellen m\"usste man aus der QCD diese gebundenen Zust\"ande als L\"osungen erhalten, es ist allerdings bis heute nicht m\"oglich, die Theorie ohne Gebrauch von N\"aherungsmethoden vollst\"andig zu l\"osen. Erschwerend kommt hinzu, dass die meisten Hadronen instabil sind und somit relativ schnell zerfallen.\ In der Regel k\"onnen sie anhand ihrer Zerfallscharakteristika identifiziert und wahlweise direkt oder indirekt vermessen werden; Quarkmodelle waren imstande, viele der bekannten Hadronen qualitativ wie quantitativ zu erkl\"aren. Weitere Experimente haben in der Vergangenheit aber gezeigt, dass es mehr Teilchen mit gleichen Quantenzahlen zu geben scheint, als mit einfachen Quarkmodellen konstruierbar sind. Insbesondere im skalaren Sektor ($J=0$) spricht man hierbei von \"Uberbev\"olkerung.

\vspace{0.15cm}
\noindent Ein Beispiel mag dies verdeutlichen: Es ist weitl\"aufig akzeptiert, dass es zwei Mesonen mit Isospin $I=1$ gibt, das schwere $a_{0}(1450)$ und das leichte $a_{0}(980)$. Da beide jeweils ein Isotriplet bilden, existieren je drei Zust\"ande mit unterschiedlicher elektrischer Ladung. Sofern das positive $a_{0}^{+}$ ein Quark-Antiquark-Paar darstellt, weist das Quarkmodell ihm die Zusammensetzung $u\bar{d}$ zu. Damit sind die M\"oglichkeiten ausgesch\"opft, die geforderten Quantenzahlen korrekt wiederzugeben, es gibt also keine Freiheit mehr. Unklar bleibt nun aber, in welchem der beiden Isotriplets der obige Zustand vorzufinden ist. Und selbst wenn man diese Frage beantworten k\"onnte, verblieben drei geladene Teilchen einer Sorte ohne Erkl\"arung.

\vspace{0.15cm}
\noindent In den letzten Jahren hat sich zunehmend gezeigt, dass die leichten Skalare, also auch das $a_{0}(980)$, durch hadronische Schleifen-Beitr\"age erzeugt werden k\"onnten. Letztere sind quantenfeldtheoretische Korrekturen, die in effektiven Modellen der QCD ber\"ucksichtigt werden k\"onnen. W\"ahrend die schweren Skalare als Quark-Antiquark-Paare angenommen werden, werden die leichten Partner nicht explizit ber\"ucksichtigt, sondern stattdessen als Mischzust\"ande gedeutet, die durch die Schleifen-Beitr\"age generiert werden. Wie kann man das verstehen?

\vspace{0.15cm}
\noindent Strenggenommen setzt sich der Zustandsvektor eines Mesons aus mehreren Beitr\"agen zusammen, weil das Teilchen die M\"oglichkeit hat zu zerfallen. Wenn es zum Beispiel in zwei andere Mesonen zerfallen kann, dann beinhaltet der Zustandsvektor neben dem $q\bar{q}$-Anteil unter anderem Vier-Quark- bzw.\ Zwei-Meson-Beitr\"age. Bei Vektormesonen sind die erstgenannten dominant und bestimmen haupts\"achlich ihre Eigenschaften; die anderen Beitr\"age entstammen aus den Schleifen und verschieben den jeweiligen Propagatorpol nur geringf\"ugig weg von der reellen Achse. Im Gegensatz dazu glaubt man, dass bei Skalaren der Sachverhalt anders ist. Zum einen sind die zus\"atzlichen Anteile oft relativ gr\"o{\ss}er als im Fall der Vektoren und ihr Einfluss auf Masse und Zerfallsbreite (also auf den Resonanzpol) ist somit nicht zu vernachl\"assigen. Daneben ist die Idee aber, dass sie weitere Pole auf die komplexe Ebene f\"uhren, die (manchmal) als neue Teilchen identifiziert werden k\"onnen. Es ist genau dieses Bild, mit dem man versucht, die \"Uberbev\"olkerung im skalaren Sektor zu erkl\"aren.
\\
\\
Die Idee einer solchen dynamischen Erzeugung von skalaren Resonanzen ist in der Literatur auf unterschiedliche Weise im Rahmen von effektiven Modellen verfolgt worden. In der vorliegenden Arbeit haben wir uns diesen Bem\"uhungen angeschlossen. Hierzu wurden durch das sogenannte ``erweiterte Lineare Sigma Modell'' (eLSM)\ \cite{eLSM1-2,denisphd,eLSM2} inspirierte effektive Theorien mit derivativen Kopplungen eingesetzt, um den eingangs erw\"ahnten Isovektor $a_{0}(980)$ und  das Isodoublet $K_{0}^{\ast}(800)$ (auch $\kappa$ genannt) zu beschreiben.
\\
\\
Zun\"achst wurde in Kapitel\ \ref{chap:chapter3} der Formalismus zur Berechnung von Schleifen-Beitr\"agen insbesondere mit derivativen Kopplungen erarbeitet. Dabei konnte gezeigt werden, dass der Zugang \"uber Dispersionsrelationen nicht identisch ist mit den \"ublichen Feynman-Regeln. W\"ahrend n\"amlich im zweiten Fall die Quantisierung eines Wechselwirkungsterms mit Ableitungen vor den Zerfallsprodukten zu Kaulquappen-Diagrammen f\"uhrt, die nach korrekter Behandlung der Ableitungen im weiteren Verlauf sich gegen gleiche Beitr\"age mit umgekehrtem Vorzeichen wegheben, verbleiben im ersten Fall diese \"uberz\"ahligen Terme.

\vspace{0.15cm}
\noindent Ein gravierender Effekt tritt auf, wenn zus\"atzlich eine Ableitung vor dem zerfallenden Teilchen vorhanden ist. Auch hier werden Kaulquappen-Diagramme erzeugt, die jetzt au{\ss}erdem energieabh\"angig sind; bei korrekter Behandlung werden sie \"ahnlich wie gerade beschrieben unwirksam gemacht. Dar\"uber hinaus wird aber die Normierung der Spektralfunktion zerst\"ort und muss durch eine Renormierung der Felder wiederhergestellt werden.
\\
\\
In Kapitel\ \ref{chap:chapter4} widmeten wir uns dann der Erweiterung einiger \"alterer Arbeiten von T\"ornqvist und Roos\ \cite{tornqvist,tornqvist2}, sowie Boglione und Pennington\ \cite{pennington} zum Thema der dynamischen Erzeugung im skalaren Sektor.\ Nach erfolgreicher Reproduktion der dortigen Ergebnisse erweiterten wir das zugrundeliegende Modell und brachten Licht in einige damals get\"atigte Aussagen. Boglione und Pennington argumentierten beispielsweise, sie h\"atten das schwere $a_{0}(1450)$ durch eine Analyse der Breit--Wigner-Massen gefunden. Tats\"achlich aber haben wir in ihrem Modell keinen entsprechenden Pol finden k\"onnen, sondern lediglich einen mit zu hoher Masse. Obwohl auch insgesamt die Polstruktur in quantitativer Hinsicht nur schlecht den experimentellen Befunden entsprach, erzeugte das Modell tats\"achlich zus\"atzliche Pole. Dies deuteten wir so, dass ein verbessertes Modell vielleicht imstande w\"are, auch quantitativ zu \"uberzeugen.

\vspace{0.15cm}
\noindent 
Deshalb untersuchten wir anschlie{\ss}end eine eigene effektive Theorie daraufhin, ob zwei Isotriplets gleichzeitig beschreibbar sind. Wir forderten, dass die beiden Resonanzen als Pole im Propagator existieren und die experimentellen Verzweigungsverh\"altnisse von $a_{0}(1450)$ richtig wiedergegeben wurden. Dadurch konnte der Parameterbereich unserer freien Modellparameter hinreichend gut eingeschr\"ankt werden; aus dem relevanten Fenster lie{\ss}en sich Werte entnehmen, die zum gew\"unschten Ergebnis f\"uhrten. Genauere Resultate sind nicht m\"oglich, da das Modell mehr Parameter besitzt, als Gleichungen seitens des Experiments zu l\"osen w\"aren bzw.\ es keine ad\"aquaten Daten f\"ur einen besseren Fit gibt (z.B. keine Daten zu Phasenverschiebungen).\ Bemerkenswert ist allerdings, dass unsere Absch\"atzung der relativen Kopplungsst\"arken von $a_{0}(980)$ zu seinen Zerfallskan\"alen darauf hindeutet, dass der Kaon-Kaon-Kanal dominant ist. Au{\ss}erdem machten wir Vorhersagen f\"ur die Phasenverschiebungen und Inelastizit\"at im Sektor mit Isospin $I=1$. Beide Untersuchungen best\"atigten andere fr\"uhere Arbeiten.

\vspace{0.15cm}
\noindent Zuletzt f\"uhrten wir eine Untersuchung unseres Modells f\"ur eine gro{\ss}e Anzahl an QCD-Farben durch: Wenn der Zustand $a_{0}(980)$ durch die hadronischen Wechselwirkungen zustande kommt, dann muss er verschwinden, sobald die Kopplungsst\"arke zu diesen Kan\"alen klein genug wird. Im Grenzfall einer gro{\ss}en Anzahl an QCD-Farben konnten wir genau das beobachten. Der Pol f\"ur $a_{0}(980)$ n\"aherte sich f\"ur kleiner werdende Kopplungen der reellen Achse an, verschwand aber f\"ur einen kritischen Wert. Der Pol f\"ur $a_{0}(1450)$ dagegen verschwand nicht, sondern wurde f\"ur kleiner werdende Kopplungen zum Pol eines stabilen Teilchens. All das best\"atigte unsere Annahme, dass die Resonanz unter $1$ GeV eine Form von Vier-Quark-Zustand ist, w\"ahrend wir f\"ur den schweren Partner genau das Gegenteil fanden.
\\
\\
In Kapitel\ \ref{chap:chapter5} verfolgten wir die gleiche Idee f\"ur den Sektor mit Isospin $I=1/2$, wendeten unser effektives Modell aber auf andere Weise an. Anstatt einen Satz von Parametern zu suchen, der zwei Pole f\"ur die beiden ben\"otigten Zust\"ande $K_{0}^{\ast}(1430)$ und $K_{0}^{\ast}(800)$ generiert, f\"uhrten wir einen Fit der experimentell ermittelten $\pi K$-Phasenverschiebung durch\ \cite{astonpion}.\ Dabei wurden vier verschiedene Varianten unseres Modells benutzt:
\begin{enumerate}[leftmargin=*]
\item \vspace{-0.17cm}Nicht-derivative und derivative Kopplungen: Dieser Fall entsprach der Situation wie bei Isospin $I=1$ und lieferte die beste Beschreibung der Daten. Wir konnten hieraus zwei Pole extrahieren, deren Position sehr gut mit den vorhandenen Ergebnissen aus dem PDG\ \cite{olive} \"ubereinstimmt -- unsere Fehler sind aber deutlich kleiner. Die Untersuchung in einer gro{\ss}en Anzahl an QCD-Farben zeigte das gleiche Bild wie f\"ur $I=1$, n\"amlich dass die leichte Resonanz $\kappa$ ein dynamisch generiertes Vier-Quark-Objekt ist, w\"ahrend $K_{0}^{\ast}(1430)$ einen gew\"ohnlichen Quark-Antiquark-Zustand darstellt.
\item \vspace{-0.28cm}Nur nicht-derivative Kopplungen: Es zeigte sich, dass diese Version des Modells nicht imstande ist, die Daten wiederzugeben. Hinzu kommt allerdings, dass es nicht m\"oglich war, \"uberhaupt einen Pol f\"ur das $\kappa$ zu generieren. Wir schlossen daraus, dass zumindest f\"ur unser Modell die Ableitungsterme im Allgemeinen sehr wichtig f\"ur die akkurate Beschreibung der experimentellen Befunde, speziell f\"ur die Anwesenheit des leichten skalaren Kaons aber essentiell notwenig sind. \"Ahnliche Aussagen lassen sich auch f\"ur den Fall des Isotriplets machen.
\item \vspace{-0.28cm}Nur derivative Kopplungen: Hierbei war der Fit zwar deutlich besser als vorher, wurde aber nach statistischer Auswertung als nicht ad\"aquat verworfen. Das Weglassen nicht-derivativer Kopplungen hatte im Vergleich zur Version des Modells unter $1.$ nur geringen Einfluss auf die Position des Pols von $K_{0}^{\ast}(1430)$. Ein dynamisch generierter zus\"atzlicher Pol f\"ur $K_{0}^{\ast}(800)$ wurde aber in der komplexen Ebene weiter nach rechts und n\"aher an die reelle Achse gef\"uhrt, was eine zu geringe Zerfallsbreite lieferte als erwartet. Insgesamt ist dadurch klar geworden, dass beide Arten von Kopplungen notwendig sind.
\item \vspace{-0.28cm}Wie unter $1.$, nun aber mit abgewandeltem Formfaktor: Auch dieser Fit stellte sich als nicht akzeptabel heraus. Des Weiteren lieferte der dynamisch generierte Pol f\"ur $\kappa$ eine verh\"altnism\"a{\ss}ig gro{\ss}e Masse. Das Verhalten des Pols war an sich auch deutlich anders als in den F\"allen zuvor: Der Pol startete im Grenzfall einer gro{\ss}en Anzahl an QCD-Farben tief in der komplexen Ebene und es konnte kein kritischer Wert der Kopplungskonstanten f\"ur sein Erscheinen bestimmt werden. Da der Formfaktor eine Auspr\"agung der Modellabh\"angigkeit unseres Ansatzes darstellt, konnten wir mit unserer Studie die These st\"arken, dass der Gau{\ss}'sche Formfaktor in der Tat eine hervorragende Wahl ist.
\end{enumerate}
%\vspace{-0.15cm}
\noindent Im Anschluss \"anderten wir unser Modell so ab, dass es nur Terme mit zus\"atzlich derivativen Kopplungen vor den zerfallenden Teilchen enthielt. Die Qualit\"at des Fits war vergleichbar mit dem unter $1.$\ Bemerkenswert ist der Umstand, dass die beiden darin existierenden Pole ziemlich genau die gleiche Position haben wie unter $1.$ Dies l\"asst den Schluss zu, dass im Rahmen unseres Ansatzes ein guter Fit nur dann m\"oglich ist, wenn ein akzeptabler Pol f\"ur das $\kappa$ erzeugt wird. Wie dem auch sei, als problematisch empfinden wir, dass -- entweder durch die Pr\"asenz der spezifischen Wechselwirkung oder wegen eines numerischen Problems -- eine Kopplungskonstante deutlich ausgepr\"agtere Fehler erh\"alt, als in allen anderen F\"allen zuvor.

\vspace{0.15cm}
\noindent Das gleiche Ph\"anomen beobachteten wir beim letzten untersuchten Modell, dem eLSM mit freien Parametern. Der Fit lieferte ein Ergebnis erneut vergleichbar mit dem Fall unter $1.$ Auch die Lage der Pole zeigte sich sehr \"ahnlich, jedoch kam es bei mindestens zwei Kopplungskonstanten zu deutlich gr\"o{\ss}eren Fehlern. Viel auff\"alliger war aber der Umstand, dass der Parameter $m_{0}$, welcher die nackte Quark-Antiquark-Paar Masse widerspiegelt, stark herabgesetzt wurde. Das ist deswegen sonderbar, weil in vergleichbaren Modellen die Hinzunahme eines Strange-Quarks (wie f\"ur den Sektor mit $I=1/2$ gegeben) diesen Wert normalerweise erh\"oht.
\\
\\
Die Hauptaussage der vorliegenden Arbeit ist, dass einige der leichten skalaren Mesonen im Rahmen von spezifischen hadronischen Modellen tat\"achlich als dynamisch generierte Resonanzen auftreten k\"onnen -- was eine m\"ogliche L\"osung f\"ur das eingangs beschriebene Problem der \"Uberbev\"olkerung w\"are. Es konnte sogar gezeigt werden, dass ein dem eLSM \"aquivalentes Modell dieses Ph\"anomen enthalten kann. Der wesentliche n\"achste Schritt w\"are deshalb, zun\"achst unseren Mechanismus an den Isoskalaren mit $I=0$ zu testen, um letztlich die elf freien Parameter des eLSM durch einen simultanen Fit einer gr\"o{\ss}eren Auswahl von experimentellen Daten zu fixieren. Ein positives Ergebnis w\"urde eine Antwort auf die Frage liefern, ob das eLSM seine Erfolgsgeschichte weiter schreiben kann. Die n\"otige Vorarbeit, um das zu \"uberpr\"ufen, wurde hier geleistet.

\end{otherlanguage}

\thispagestyle{empty}
\

\newpage
\clearpage

\pagenumbering{roman}
\setcounter{page}{1}

\renewcommand{\baselinestretch}{1.3}
\small\normalsize
\setcounter{tocdepth}{1}
\tableofcontents
\renewcommand{\baselinestretch}{1}
\small\normalsize
\clearpage

\renewcommand{\baselinestretch}{1.3}
\small\normalsize
\listoffigures
\addcontentsline{toc}{chapter}{List of figures}
\renewcommand{\baselinestretch}{1}
\small\normalsize

\renewcommand{\baselinestretch}{1.3}
\small\normalsize
\listoftables
\addcontentsline{toc}{chapter}{List of tables}
\renewcommand{\baselinestretch}{1}
\small\normalsize
\setcounter{bibpage}{\value{page}}
\stepcounter{bibpage}
\clearpage

\thispagestyle{empty}
\

\newpage
\clearpage

\thispagestyle{empty}
\

\newpage
\clearpage

\thispagestyle{empty}
\

\newpage
\clearpage

\pagenumbering{arabic}
\setcounter{page}{1}

\onehalfspacing
\fontsize{11}{1.5}

\chapter{Introduction}
\label{chap:chapter1}

\medskip

\section{Historical remarks}
The beginning of the $20$th century was a fascinating time of confusion.\ Physicists all around the world became puzzled by some unexpected experimental observations and new ideas concerning the microscopic structure of nature (later on incorporated in the theory\footnote{In various cases during this thesis, we will not strictly distinguish between terms like 'theory\grq, 'scientific theory\grq, 'model\grq, or 'theory limit\grq\ as actually proposed by philosophy of science.} of quantum mechanics). In retrospect, this time marks one of few crucial turning points not only in the thousands-year-long history of science, but also in the mere way of how human beings look at the world surrounding them. As a consequence, all coming generations have been left behind with a mixture of amusement and curiosity about the universe.\ While a huge number of our ancestors believed that they were close in obtaining a deep and conclusive understanding of the world, something very different seems nowadays to be apparent: this kind of search for knowledge may never reach a final end. This can be unsatisfying -- yet, some of us have arranged with it. Indeed, there are less people trying to reach for the answers to all things. Nevertheless, we started a new venture at the beginning of the $21$st century since it is up to us clarifying what our ancestors have left behind.

Besides philosophical and fundamental challenges after finding the appropriate mathematical formalism, (non-relativistic) quantum mechanics faced a huge problem in establishing a theory of nuclear forces. In $1935$, it was Yukawa who applied field-theoretical methods to derive the nucleon-nucleon force as an interaction through one-pion exchange\ \cite{yukawa}.\ Although this description finally turned out to be not the right path to follow, it was the motivation for a vast amount of new approaches in particle physics during the next decades.\ We will not try to review all those ideas, failures, and milestones.\ However, one principle can lead us to an understanding of this time: physicists usually believe that every description of nature should be made as simple as possible -- but no simpler.\footnote{This quote is often attributed to A.\ Einstein.} The basic first pages in some textbooks on particle physics for example start with this paradigm\ \cite{martin}.\ We therefore try to build up all matter from very few and hopefully simple blocks of matter, which are called elementary particles. This approach was first not successful; experimentalists discovered more and more heavy (unstable) particles known as {\em hadrons} in the early $60$s and their existence was not covered by the theoretical models constructed before.\ It was realized soon after that most of the new particles were very short-lived states, so called {\em resonances}.\ They did not hit the detectors directly but showed up as enhancements in process amplitudes during scattering reactions, and were identified mostly from their decay products.\ It became clear that they could not be taken as elementary.

After seminal works by Gell-Mann\ \cite{gell-mann}, Ne'eman, and Zweig\ \cite{zweig}, a classification scheme for the new and already known particles was established, as well as a unified theory for explaining hadrons and their interactions. Gell-Mann and Zweig proposed a solution using group-theoretical methods, namely, they treated all the different hadronic states as manifestations of multiplets within the $SU(3)$ (flavor) group.\ This required the existence of {\em quarks}, that is, elementary particles with spin $1/2$ as building blocks of hadrons, which interact via an octet of vector gauge bosons, the {\em gluons}.\ The fundamental theory of the interaction between quarks and gluons is {\em quantum chromodynamics (QCD)}\ \cite{foundQCD}.\ One main property of QCD, known as {\em confinement}, is the fact that the strong force between the particles does not decrease with distance.\ It is therefore believed that quarks and gluons can never be separated from hadrons. This is related to the technical problem that the whole theory is non-perturbative in the low-energy regime, which is relevant for describing hadrons and also atomic nuclei.

Despite huge efforts in recent years, it was up to now not possible to solve QCD analytically. In particular, {\em lattice QCD} is under continuous growth, where one tries to map the fundamental theory on a discretized space-time grid and performs specific calculations by using a large amount of computational power. Even the treatment of dynamical issues like the application of a coupled-channel scattering formalism seems to be coming within range, see {\em e.g.}\ Ref.\ \cite{dudek}.\ Besides many not yet solved problems, lattice QCD definitely has become a well-established non-perturbative approach for QCD. Other strategies have been found by using holographic models and the gauge/gravity correspondence, for instance to extract meson masses with good accuracy\ \cite{erlich,karch}.\ As will be discussed later, another very successful approach to QCD relies on the concept of {\em effective field theories (EFTs)}.\ There, one maps the fundamental theory onto a low-energy description by following a very general prescription -- as a consequence, the relevant degrees of freedom become hadrons and their interactions. Chiral perturbation theory (chPT)\ \cite{chpt,chpt2}\ as a prototype of this concept has been applied {\em e.g.}\ to meson-meson scattering.

\section{The quark model and QCD}
As already mentioned, one motivation for a new fundamental theory of hadronic particles was the lack of a classification scheme. Concerning dynamics there was another important question: why do most of the new unstable particles not decay into all other particles when their decays would be kinematically allowed? This suggested that there must be some 'rules\grq\ at work, restricting the amount of allowed decay channels. Strictly speaking, composite hadrons would possess some quantum numbers that are conserved under the strong interaction -- leading us to symmetries.

Today we know that one can interpret the lightest hadrons, the pion isotriplet, in terms of quark content as $\pi^{+}\sim u\bar{d}$, $\pi^{0}\sim 1/\sqrt{2}\hspace{0.05cm}(u\bar{u}-d\bar{d})$, and $\pi^{-}\sim d\bar{u}$.\ This is a natural consequence of isospin or $SU(2)$ flavor symmetry which is (nearly) exact in QCD, because the difference in mass between up and down quarks is very small compared to the hadronic scale. Consequently, all hadrons built from those quarks will be arranged within an $SU(2)$ multiplet, like in the case of the pion isotriplet, and have (nearly) the same mass. Adding a strange quark, slightly heavier than the up and down quark but still light enough, gives rise to $SU(3)$ multiplets like the pseudoscalar octet. The general mathematical formalism can be introduced by using the basis of strong isospin $T_{3}$ and hypercharge $Y$, yielding the state vectors of those three quarks:
\begin{equation}
|u\rangle = |T_{3},Y\rangle = |\frac{1}{2},\frac{1}{3}\rangle \ , \ \ \ |d\rangle = |\hspace{-0.08cm}-\hspace{-0.08cm}\frac{1}{2},\frac{1}{3}\rangle \ , \ \ \ |s\rangle = |0,-\frac{2}{3}\rangle \ .
\end{equation}
The multiplets are then constructed from this fundamental triplet and the antitriplet formed by the corresponding antiquarks. Here, the substructure of the resulting mesons obeys a $q\bar{q}$ pattern.\ The physical mesons form a singlet and an octet, while for the baryons ($qqq$ states) we find a singlet, two octets, and a decuplet. We also know today that, in addition to the three light quarks, there exist three heavy quarks: the charm, bottom, and top quark. This highly increases the number of physical particles\ \cite{olive}. The properties of all six quarks can be found in Table\ \ref{tab:quarks}.
\begin{table}[t]
\center
\begin{tabular}[c]{cccccccccc}
\toprule Flavor & Mass [GeV] & $Q$ [$e$] & $Y$ & $J$ & $B$ & $S$ & $C$ & $B^{\prime}$ & $T$ \\
\midrule\\[-0.25cm]
$u$ & $\left(2.3_{-0.5}^{+0.7}\right)\cdot10^{-3}$ & $2/3$ & $1/3$ & $1/2$ & $1/3$ & $0$ & $0$ & $0$ & $0$ \\[0.1cm]
$d$ & $\left(4.8_{-0.3}^{+0.5}\right)\cdot10^{-3}$ & $-1/3$ & $1/3$ & $1/2$ & $1/3$ & $0$ & $0$ & $0$ & $0$ \\[0.1cm]
$c$ & $1.275\pm0.025$ & $2/3$ & $4/3$ & $1/2$ & $1/3$ & $0$ & $1$ & $0$ & $0$ \\[0.1cm]
$s$ & $(95\pm5)\cdot10^{-3}$ & $-1/3$ & $-2/3$ & $1/2$ & $1/3$ & $-1$ & $0$ & $0$ & $0$ \\[0.1cm]
$t$ & $173.21\pm0.51\pm0.71$ & $2/3$ & $4/3$ & $1/2$ & $1/3$ & $0$ & $0$ & $0$ & $1$ \\[0.1cm]
$b$ & $4.66\pm0.03$ & $-1/3$ & $-2/3$ & $1/2$ & $1/3$ & $0$ & $0$ & $-1$ & $0$ \\
\\[-0.1cm]
\toprule
\end{tabular}
\caption{The quantum numbers listed are: electric charge ($Q$), hypercharge ($Y$), total spin ($J$), baryon number ($B$), strangeness ($S$), charmness ($C$), bottomness ($B^{\prime}$), and topness ($T$). See Ref.\ \cite{olive} for further discussion.}
\label{tab:quarks}
\end{table}

The upper classification scheme for hadrons in terms of their valence quarks is the famous {\em quark model}.\ After the discovery of the $\Delta^{++}$ baryon it was possible to assign the correct spin and flavor content to its state vector by using the quark model -- the only way to do so and obtain a charge $+2$ state is by having three up quarks. This leads to a symmetric flavor, spin, and spatial wave function, in particular $\Delta^{++}\sim u_{\uparrow}u_{\uparrow}u_{\uparrow}$. Therefore, the total (many-body) wave function is also symmetric. But this result is in contradiction to the fact that a fermionic many-body wave function has to be antisymmetric. In order to resolve this a new {\em color} degree of freedom for quarks was introduced: they carry either red, green, or blue color charge. Assuming that the $\Delta^{++}$ baryon is an antisymmetric superposition in color space, it is straightforward to construct its total antisymmetric wave function, which is also 'white\grq, {\em i.e.}, invariant under $SU(3)$ rotations in color space: $\Delta^{++}\sim \sum_{i,j,k=1}^{3}\varepsilon_{ijk}u_{\uparrow}^{i}u_{\uparrow}^{j}u_{\uparrow}^{k}$. Here, $\varepsilon_{ijk}$ is the Levi-Civita symbol and the summation runs over the three colors ($1\ \hat{=}$ red etc.). The corresponding expression for a $q\bar{q}$ meson like the pion would be $\pi^{+}\sim \sum_{i=1}^{3}u_{\uparrow}^{i}\bar{d}_{\downarrow}^{\hspace{0.03cm}i}$. Note that the number of colors can be determined from the experiment either from the neutral $\pi^{0}\rightarrow\gamma\gamma$ decay or the ratio of the cross sections for $e^{+}+e^{-}\rightarrow hadrons$ and $e^{+}+e^{-}\rightarrow \mu^{+}+\mu^{-}$.\ The best correspondence with experimental data is unambiguously obtained if the number of colors is $N_{c}=3$.

Now, the Lagrangian of QCD is constructed by starting from the Dirac version for massive spin-$1/2$ particles, where the quarks are incorporated as spinors with $N_{f}$ flavors, each in the fundamental representation of the $SU(3)_{c}$ (color) gauge group. The Lagrangian is then invariant under global $SU(3)_{c}$ transformations. For the same reason as in QED, we postulate the transformations to depend on the space-time coordinate, hence one requires the Lagrangian to be invariant under local transformations.\ This is only possible if one includes some further pieces transforming in such a way as to cancel the additional terms caused by the derivative in the Dirac operator.\ Since the latter brings in a Lorentz index, the required modification introduces eight new spin-$1$ fields, the gluons, living in the adjoint representation of the $SU(3)_{c}$ symmetry group, the octet. It is generated from the direct product of the color triplet and the antitriplet; thus gluons do carry color charge which is one main difference to QED. However, as the photon they are massless.

As mentioned, the QCD Lagrangian fulfills $SU(3)_{c}$ gauge invariance because a quark field $q_{f}$ in the fundamental representation transforms as
\begin{equation}
q_{f}\rightarrow q_{f}^{\prime} = \exp\left[-i\theta^{a}(x)t^{a}\right]q_{f} = U_{c}(x)q_{f} \ ,
\end{equation}
where the $t^{a}=\lambda^{a}/2$ denote the $SU(3)$ generators, $\lambda_{a}$ the Gell-Mann matrices, and $\theta^{a}(x)$ the group parameters (here, $a=1\dots N_{c}^{2}-1$). In analogy to the Dirac Lagrangian we therefore have
\begin{equation}
\mathcal{L}_{\text{QCD}} = \bar{q}_{f}(i\gamma^{\mu}D_{\mu}-m_{f})q_{f}-\frac{1}{4}G_{\mu\nu}^{a}G_{a}^{\mu\nu} \ ,
\label{eq:LagQCD}
\end{equation}
with implied summation over the flavor index $f$.\ The covariant derivative
\begin{equation}
D_{\mu} = \partial_{\mu}-ig\mathcal{A}_{\mu}
\end{equation}
contains the eight gluon gauge fields $\mathcal{A}_{\mu}=A_{\mu}^{a}t^{a}$.\ They transform under the gauge group according to
\begin{equation}
\mathcal{A}_{\mu}\rightarrow\mathcal{A}_{\mu}^{\prime} = U_{c}(x)\mathcal{A}_{\mu}U_{c}^{\dagger}(x)-\frac{i}{g}\big[\partial_{\mu}U_{c}(x)\big]U_{c}^{\dagger}(x) \ ,
\end{equation}
such that the covariant derivative transforms as
\begin{equation}
D_{\mu}\rightarrow D_{\mu}^{\prime} = U_{c}(x)D_{\mu}U_{c}^{\dagger}(x) \ ,
\end{equation}
making the first term in Eq.\ (\ref{eq:LagQCD}) invariant under $SU(3)_{c}$ transformations. The second part of the QCD Lagrangian represents the kinetic term for the gluons, given by the square of the field-strengths associated with the gauge fields.\footnote{Analogously to QED, the kinetic term is given by the square of the field-strength tensor, $\mathcal{G}_{\mu\nu}$, associated with the gauge fields. It is in general defined as the commutator of covariant derivatives:
\begin{equation}
\mathcal{G}_{\mu\nu} = \frac{i}{g}[D_{\mu},D_{\nu}] = \partial_{\mu}\mathcal{A}_{\nu}-\partial_{\nu}\mathcal{A}_{\mu}-ig[\mathcal{A}_{\mu},\mathcal{A}_{\nu}] \ . \nonumber
\end{equation}
One can therefore also write for the kinetic term $-\frac{1}{4}G_{\mu\nu}^{a}G_{a}^{\mu\nu}=-\frac{1}{2}\trace{(\mathcal{G}_{\mu\nu}\mathcal{G}^{\mu\nu})}$\hspace{0.02cm}.} The field-strengths are
\begin{equation}
G_{\mu\nu}^{a} = \partial_{\mu}A_{\nu}^{a}-\partial_{\nu}A_{\mu}^{a}+gf^{abc}A_{\mu}^{b}A_{\nu}^{c} \ ,
\end{equation}
with $f^{abc}$ the totally antisymmetric $SU(3)$ structure constants.

\begin{figure}[t]
\centering
\hspace{0.0cm}
\includegraphics[scale=0.35]{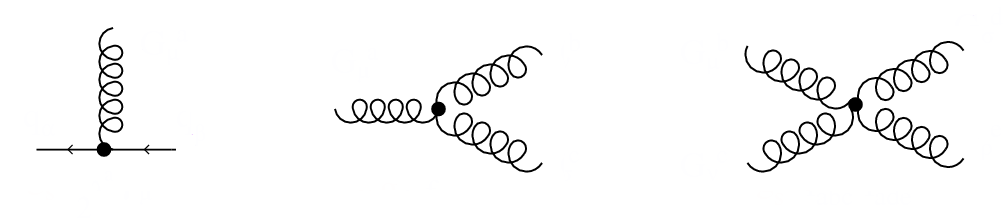}
\caption{Tree-level vertices of QCD: The solid straight lines represent quark and antiquark propagators, while the gluons are depicted as spiral lines.}
\label{fig:QCDvertices}
\end{figure}

It is worthwhile to look at the tree-level vertex structure of the QCD Lagrangian\footnote{Here, we ignore the ghost-gluon vertex. Note also that QCD obeys some additional symmetries and that some of them are broken. This topic will be further elaborated in the next chapter.}, see Figure\ {\ref{fig:QCDvertices}}.
The latter shows first the interaction vertex between quarks and gluons which is induced by the covariant derivative. In the second and third panel one recognizes three- and four-gluon interactions, where the former is momentum-dependent.\ These two vertices are a consequence of the non-abelian group structure of $SU(3)$.\ Note that the four-gluon vertex is of order $\mathcal{O}(g^{2})$ in the gauge coupling constant $g$.

\section{Aim of this work}
Intense research during the past decades has demonstrated that the majority of mesons can be understood as being predominantly $q\bar{q}$ states\ \cite{olive}.\ However, the quark model is not the end of the story. Most important for us in this thesis is the phenomenon of {\em overpopulation} in the scalar sector: it is not possible to assign all known mesons as quarkonia. For example, the state $a_{0}$ with $I=1$, $J=0$ lives in the isotriplet of the scalar meson octet, meaning that there exist three resonances with different electric charge. They have the same quark content as the pseudoscalar pions; since isospin symmetry is nearly exact in QCD, they are also nearly degenerated in mass. This isotriplet can now be identified with either the resonance $a_{0}(980)$ or the $a_{0}(1450)$.\ The quark model cannot explain which is the correct assignment, but can give however an interpretation of one isovector state (see also Table\ \ref{tab:mesons}).
\begin{table}[t]
\center
\begin{tabular}[c]{ccccc}
\toprule Particle & Quark content & $I$ & $J^{PC}$ & Mass [MeV] \\
\midrule\\[-0.25cm]
$\pi^{+}$, $\pi^{-}$, $\pi^{0}$ & $u\bar{d}$, $d\bar{u}$, $\frac{u\bar{u}-d\bar
{d}}{\sqrt{2}}$ & $1$ & $0^{-+}$ & $139.57$, $134.98$\\[0.15cm]
$K^{+}$, $K^{-}$, $K^{0}$, $\bar{K}^{0}$ & $u\bar{s}$, $s\bar{u}$, $d\bar{s}$, $s\bar{d}$
& $1/2$ & $0^{-+}$ & $493.68$, $497.61$\\[0.15cm]
$\eta$ & $\approx\frac{u\bar{u}+d\bar{d}-2s\bar
{s}}{\sqrt{6}}$ & $0$ & $0^{-+}$ & $547.86$\\[0.15cm]
$\eta^{\prime}$ & $\approx\frac{u\bar{u}+d\bar{d}+s\bar{s}}{\sqrt{3}}$ & $0$ & $0^{-+}$ & $957.78$\\[0.15cm]
$a_{0}^{+}$, $a_{0}^{-}$, $a_{0}^{0}$ & $u\bar{d}$, $d\bar{u}$, $\frac{u\bar{u}-d\bar{d}}{\sqrt{2}}$ & $1$ & $0^{++}$ & $1474\pm19$\\[0.15cm]
$K_{0}^{\ast+}$, $K_{0}^{\ast-}$, $K_{0}^{\ast0}$, $\bar{K}_{0}^{\ast0}$ &
$u\bar{s}$, $s\bar{u}$, $d\bar{s}$, $s\bar{d}$ & $1/2$ & $0^{++}$ & $1425\pm50$\\
\\[-0.1cm]
\toprule
\end{tabular}
\caption{Possible assignment of the most important physical mesons in this work concerning quark content. Here, $I$ is isospin, $P$ is parity, and $C$ is charge conjugation. No errors are given for the pseudoscalars. See Ref.\ \cite{olive} for further discussion.}
\label{tab:mesons}
\end{table}

In the literature, many suggestions have been discussed to solve this problem, such as the introduction of various unconventional mesonic states such as glueballs, hybrids, and four-quark states\ \cite{amslerrev}.\ Along this line, a specific concept of {\em dynamically generated} states was put forward {\em e.g.}\ in Refs.\ \cite{dullemond,morgan,tornqvist,pennington}.\ The main idea is that these states are not constructed, as in the quark model, from some building blocks and a confining potential, but rather arise from interactions between conventional $q\bar{q}$ mesons -- they appear as {\em companion poles} in the relevant process amplitude.\ We will present a more detailed explanation of this idea at the end of Chapter\ \ref{chap:chapter2}. Our aim will be to describe some of the physical mesons as dynamically generated states. This will be successfully performed for the isovector ($I=0$) and isodoublet ($I=1/2$) sectors, that is, we will show that for the heavy quarkonia states $a_{0}(1450)$ and $K_{0}^{\ast}(1430)$ the couplings to their decay channels are capable of dynamically generating the light states $a_{0}(980)$ (Chapter\ \ref{chap:chapter4}) and $K_{0}^{\ast}(800)$, also known as $\kappa$ (Chapter\ \ref{chap:chapter5}), respectively. To this end, we will apply a hadronic model that includes meson-meson interactions via derivative and non-derivative terms.\ In order to cope with these, Chapter\ \ref{chap:chapter3} is dedicated to work out the formalism of such interactions.
\\
\\
Organization of the thesis:
\begin{itemize}
\item Chapter\ \ref{chap:chapter2}: After a short introduction on resonances, we present the framework where we want to 
study scalar resonances: the extended Linear Sigma Model (eLSM) 
as an example for an effective model of QCD. We also illustrate the idea of dynamical generation via hadronic loop 
contributions in other effective theories.
\item Chapter\ \ref{chap:chapter3}: Since derivative and non-derivative interaction terms play an important role in our models, 
we present in detail how they are incorporated in order to calculate hadronic loop contributions.\ We also show that there is 
an apparent discrepancy between using ordinary Feynman rules and dispersion relations.
\item Chapters\ \ref{chap:chapter4} and\ \ref{chap:chapter5}: We apply the idea of dynamical generation introduced in the 
second chapter by discussing and extending previous calculations in the isovector sector with $I=1$. Then, we construct 
effective Lagrangians where $a_{0}(1450)$ and $K_{0}^{\ast}(1430)$ couple to pseudoscalar mesons by both non-derivative 
and derivative interactions. For both cases we look for companion poles that can be assigned to the corresponding resonances below $1$ GeV, {\em i.e.}, the $a_{0}(980)$ and the $K_{0}^{\ast}(800)$.
\end{itemize}

\newpage
\clearpage

\

\newpage
\clearpage

\chapter{Resonances}

\medskip

\label{chap:chapter2}
\section{Unstable particles and resonances}
The ideal quark model introduced in the previous chapter demonstrated that it is in principle capable of describing some of the most important aspects of nature, {\em i.e.}, the baryonic and mesonic ground states which are arranged as an octet and a decuplet, and a nonet.\ All of them can be considered as built from quarks and antiquarks, where the specific composition depends on some very few quantum numbers like spin $S$ and angular momentum $L$.\ States with higher total spin such as the vector mesons decay to pseudoscalars by the strong interaction, unveiling their constituent nature by decay patterns. One distinguishes hadrons between {\em particles} and {\em resonances}. In the framework of quantum field theory, the first term is assigned to quanta of some fields; they are able to propagate over sufficiently large time scales ({\em e.g.}\ from a creation reaction to a detector) and hence can be identified in experiments. In particular, they possess distinct measurable properties and consequently should satisfy the energy dispersion relation.

The further terminology can be fixed in the following way: a {\em stable} particle is able to propagate over an indefinite amount of time specific for the relevant interactions the particle obeys (for example, pions are stable for what concerns the strong interaction). This holds true until interactions with other particles occur. If the former does not hold true, we speak of {\em unstable} particles. For example, we know that charged pions as part of the particle shower in secondary cosmic rays have a mean life time $\tau$ of about $10^{-8}$ s. They can be described as nearly stable as long as the propagation and interaction time is much smaller than the mean life time (including relativistic time-dilation effects). Nevertheless, when considering time scales of some seconds those particles decay into other particles, namely muons and neutrinos.

The baryon decuplet with total spin $J=3/2$ contains the $\Delta$ baryon which possesses an extremely short mean life time on the order of $10^{-22}$ s, the time scale of the strong interaction.\ Though clearly an unstable particle, in this case it makes more sense to treat this particle like an excitation emerging when investigating nuclear matter and when performing high-energy collision experiments, respectively. The correct term would therefore be resonance. When traveling nearly at the speed of light those resonant states could only overcome distances of about $10^{-14}$ m before decaying. Yet, formally they can nevertheless be interpreted as fluctuations of some underlying field and so we may use the terms 'unstable particle\grq \ and 'resonance\grq \ interchangeably.

In general, by treating a particle decay as a Poisson process one usually defines
\begin{equation}
\tau = \Gamma^{-1} \ ,
\end{equation}
where $\Gamma$ is called the {\em decay width} of the resonance associated with a specific set of final states, namely its decay products. As a direct consequence, an exponential decay law for the survival probability $p(t)$ of the particle in its rest frame is obtained,
\begin{eqnarray}
p(t) = e^{-\Gamma t} \ .
\end{eqnarray}
One can show that this in fact is only a simplified picture, valid for narrow resonances with relatively large mean life times only\ \cite{brown}.\ For example, positively charged pions with dominant leptonic decay channel $\pi^{+}\rightarrow\mu^{+}\nu_{\mu}$ and a mass of about $M\simeq140$ MeV possess a mean life time of about $10^{-8}$ s. The ratio $\Gamma/M$ yields $\sim10^{-16}$, while for neutral pions with $\pi^{0}\rightarrow\gamma\gamma$ and a mean life time of about $10^{-17}$ s with $M\simeq135$ MeV one obtains $\Gamma/M\sim10^{-8}$.\ This measure can be implemented within a rule of thumb: whenever mass and decay width become comparable, a resonance leaves the realm of the exponential decay law.

Among such and other difficulties, very short-lived unstable particles in particular cannot be directly observed. Their existence is established from some {\em scattering processes}, like the inelastic reaction $A+B\rightarrow C+D$ of two incoming (stable) particles $A$ and $B$, and a set of outgoing particles $C$ and $D$, where the subset $C$ contains an intermediate resonance $R$ such that $R\rightarrow C$ without detection. Another possibility may be the elastic process $A+B\rightarrow R\rightarrow A+B$, where a resonance is created during the fusion of the incoming particles and finally decays without detection. A huge area of research is the extraction of resonance information from the corresponding scattering data.

\section{Parameterization of experimental data}
In the old days when QCD was not yet (fully) developed, a framework called $\hat S$-matrix theory was applied to interpret the experimental data.\ It was founded on the very basic understanding of quantum mechanics and some few postulates that mainly consist of unitarity, relativistic invariance, conservation of energy-momentum and angular momentum, and analyticity. The whole field, rich by its own history and methods, cannot be summarized appropriately in this work. For classic literature see for example Refs.\ \cite{chew,polkinghorne}, though the foundations were formulated much earlier by Wheeler\ \cite{wheeler} and Heisenberg\ \cite{heisenberg}.\ However, it may be possible to give a very interesting quote made by Chew and Frautschi\ \cite{chew2} in the context of this theory. While pointing out a definition for 'pure potential scattering\grq \, they stated, it is plausible ``[...]\ that {\em none} of the strongly interacting particles are {\em completely} independent but that each is a dynamical consequence of interactions between others.'' This remark shall guide us in some sense throughout the thesis at hand.

For the moment let us recall that, in context of scattering theory, the general expression for the decay width has nearly the same formal structure as the differential cross section $\text{d}\sigma$\ \cite{peskin,gross}.\ By performing a scattering experiment, {\em e.g.}\ of the type $A+B\rightarrow A+B$ with intermediate resonance $R$, and measuring the invariant mass distribution of the outgoing particles, one may find a peak in the differential cross section located around a value $\sqrt{s}\approx m_{R}$, the mass of the resonance $R$.\ This is because the elastic differential cross section is obtained as the squared scattering amplitude,
\begin{equation}
\left(\frac{d\sigma}{d\Omega}\right)_{\text{el}} = |F(\theta)|^{2} \ ,
\end{equation}
such that
\begin{eqnarray}
\sigma_{\text{el}} \ = \ \frac{4\pi}{k^{2}}\sum_{l=0}^{\infty}(2l+1)\sin^{2}\delta_{l} & = & \frac{4\pi}{k^{2}}\sum_{l=0}^{\infty}(2l+1)\left|\frac{e^{2i\delta_{l}}-1}{2i}\right|^{2} \nonumber \\[0.1cm]
& = & \frac{4\pi}{k^{2}}\sum_{l=0}^{\infty}(2l+1)\hspace{0.085cm}|f_{l}|^{2}\ .
\end{eqnarray}
Here, $k$ is the absolute value of three-momentum of one of the outgoing particles in the rest frame of the resonance $R$, while $l$ represents the angular momentum of the partial wave amplitude $f_{l}$, and $\delta_{l}$ is the corresponding phase shift. For a resonance with total spin $J$ the relevant partial wave has a maximum at $\delta_{l}=\pi/2$.\ One finds by Taylor expansion that near the resonance mass $\sqrt{s}\approx m_{R}$ the total elastic cross section is
\begin{equation}
\sigma_{\text{el}} \approx \frac{4\pi}{k^{2}}\frac{2J+1}{(2S_{A}+1)(2S_{B}+1)}\frac{s\hspace{0.02cm}\Gamma_{R}^{2}}{(s-m_{R}^{2})^{2}+m_{R}^{2}\Gamma_{R}^{2}} \ ,
\label{eq:crossBW}
\end{equation}
with $S_{A}$ and $S_{B}$ as the total spins of the incoming particles. The last factor is called the relativistic {\em Breit--Wigner distribution}.\ The above expression holds true for a single separated resonant state with only one decay channel $R\rightarrow AB$ and total decay width $\Gamma_{R}=\Gamma_{R\rightarrow AB}$.\ The obtained curve is a good approximation of the rate in the region of the resonance only; its mass simply corresponds to the maximum, while the physical width is the full width at half maximum.

One should note that this parameterization in principle introduces a pole on the complex energy plane according to $s_{\text{pole}}=m_{R}^{2}-i\hspace{0.02cm}m_{R}\Gamma_{R}$. However, the physical mass and width of a resonance are found from the position of the nearest pole on the appropriate unphysical Riemann sheet\footnote{If the reader is unfamiliar with Riemann sheets and multi-valued complex functions, see Appendix\ \ref{app:appendixD} for a short introduction.} of the relevant process amplitude (that is, the $\hat S$-matrix),
\begin{equation}
\sqrt{s_{\text{pole}}}=m_{\text{pole}}-i\hspace{0.02cm}\frac{\Gamma_{\text{pole}}}{2} \ ,
\label{eq:polecoor}
\end{equation}
a procedure going back to Peierls\ \cite{peierls}.\ The corresponding pole mass and width in general do not agree with the values a Breit--Wigner parameterization imposes on data, but they do for a narrow and well-separated resonance, in particular, far away from the opening of decay channels. The realization of this mere fact was crucial: compared to the vector and tensor mesonic states, the issue of scalar mesons has been the subject of a vivid debate among the physical community for a long time. Their identification and explanation in terms of quarks and gluons turned out to be very difficult and furthermore, some of those particles possess large decay widths, several decay channels, and a huge background.

Hence, one should remark that Eq.\ (\ref{eq:crossBW}) only describes a non-interfering production cross section of a single resonant state with two incoming (stable) particles, while usually background reactions and other multi-channel effects distort the pure contribution from the resonance, such that it is harder to observe if there is really something or not. For instance, one can be faced with very broad structures that cannot be separated from the background, the same as with line shapes partly deformed because of nearby decay opening channels. In such cases only the presence of a pole and its real part provides a good definition of a resonance mass. Furthermore, the existence and position of the pole is independent of the specific reaction studied. The general procedure of extracting the pole would then be to construct the $\hat S$-matrix and partial wave amplitude, respectively, which is then applied directly to fit experimental data or from which a suitable function can be derived to perform the fit (like the phase shift).

\section{The extended Linear Sigma Model in a nutshell}
A substantial progress in hadron physics was achieved when the concept of an {\em effective field theory (EFT)} was applied to the low-energy regime of QCD. Weinberg has pointed out the general ideas in Ref.\ \cite{weinberg}, {\em i.e.}, the key point is to identify the appropriate degrees of freedom and to write down the most general Lagrangian consistent with the assumed symmetries.\ As a consequence, it is not necessary anymore to solve the underlying fundamental theory due to the fact that within the new framework the degrees of freedom ('the basis\grq) are not quarks and gluons, but composite particles, namely hadrons.\footnote{Good introductions to the topic of EFT can be found in Refs.\ \cite{scherer,mosel,weise}.}  An effective Lagrangian for QCD will have the same symmetries as the latter -- and some of them will be broken. For instance, the QCD Lagrangian has an exact $SU(3)_{c}$ local gauge symmetry and is also approximately invariant under 
global $U(3)_{R}\times U(3)_{L}$ flavor rotations.\ The latter is of course the chiral symmetry for a number of $N_{f}=3$ quarks.\ Because of confinement, the low-energy regime is supposed to be mainly dominated by the chiral 
symmetry and its spontaneous, explicit, and anomalous breaking.

In this thesis, our Lagrangians will be inspired by an effective model called the {\em extended Linear Sigma Model (eLSM)}\ \cite{eLSM1-2,denisphd,eLSM2}, in which a linear representation of chiral symmetry is incorporated\ \cite{schwinger,levy,weinberg2} and where both the scalar and pseudoscalar degrees of freedom are present.\ This allows to introduce $G$-parity, conserved by the strong interaction, and corresponding eigenvectors for the pions.\ The chiral partner of the pion was found to be the $f_{0}(1370)$ state (and not the $f_{0}(500)$).\ The eLSM was formulated for $N_{f}=3$ quark flavors, vanishing temperatures and densities, and includes vector and axial-vector mesons, in some versions also candidates for the lowest lying scalar and pseudoscalar glueballs\ \cite{eLSM1-3,stani}.\ Further extensions can be found in Refs.\ \cite{eLSM1,achim,strb}.

The main ingredients of the eLSM are composite fields, all assigned as $q\bar{q}$ states. This can be proven by using large-$N_{c}$ arguments\ \cite{largeNc1,largeNc2}: the masses and decay widths obtained within the model scale as $N_{c}^{0}$ and $N_{c}^{-1}$, respectively. The assignment of the required meson matrices for all sectors is summarized by
\vspace{-0.4cm}
\begin{align}
\intertext{\ \ \ \ \ \textbullet \ \ (Pseudo-)Scalars \(\Phi_{ij} \sim(q_L \bar{q}_R)_{ij} \sim \frac{1}{\sqrt{2}} (q_i \bar q_j - q_i \gamma_5 \bar{q}_j)\):}
\Phi &=\frac{1}{\sqrt{2}}\begin{pmatrix} \frac{(\sigma_{N}+a_{0}^{0})}{\sqrt{2}}+\frac{i(\eta_{N}+\pi^{0})}{\sqrt{2}} & a_{0}^{+} +i\pi^{+} & K_{0}^{\ast+}+iK^{+}\\
a_{0}^{-}+i\pi^{-} & \frac{(\sigma_{N}-a_{0}^{0})}{\sqrt{2}}+\frac{i(\eta_{N}-\pi^{0})}{\sqrt{2}} & K_{0}^{\ast0}+iK^{0}\\
K_{0}^{\ast-}+iK^{-} & \bar{K}_{0}^{\ast0}+i\bar{K}^{0} & \sigma_{S}+i\eta_{S} \end{pmatrix} \ ,
\intertext{\ \ \ \ \ \textbullet \ \ Left-handed \(L_{ij}^\mu \sim(q_L \bar{q}_L)_{ij} \sim \frac{1}{\sqrt{2}} (q_i \gamma^\mu \bar{q}_j + q_i \gamma_5 \gamma^\mu \bar{q}_j)\):}
L^{\mu}& =\frac{1}{\sqrt{2}}\begin{pmatrix} \ \frac{\omega_{N}+\rho^{0}}{\sqrt{2}}+\frac{f_{1N}+a_{1}^{0}}{\sqrt{2}} & \rho^{+}+a_{1}^{+} & K^{\ast+}+K_{1}^{+} \\ 
\rho^{-}+a_{1}^{-} & \frac{\omega_{N}-\rho^{0}}{\sqrt{2}}+\frac{f_{1N} -a_{1}^{0}}{\sqrt{2}} & K^{\ast0}+K_{1}^{0}\\
K^{\ast-}+K_{1}^{-} & \bar{K}^{\ast0}+\bar{K}_{1}^{0} & \omega_{S}+f_{1S} \end{pmatrix}^\mu \ ,
\intertext{\ \ \ \ \ \textbullet \ \ Right-handed \(R_{ij}^\mu \sim(q_R \bar{q}_R)_{ij} \sim \frac{1}{\sqrt{2}} (q_i \gamma^\mu \bar{q}_j - q_i \gamma_5 \gamma^\mu \bar{q}_j)\):}
R^{\mu} & = \frac{1}{\sqrt{2}}\begin{pmatrix} \frac{\omega_{N}+\rho^{0}}{\sqrt{2}}-\frac{f_{1N}+a_{1}^{0}}{\sqrt{2}} & \rho^{+}-a_{1}^{+} & K^{\ast+}-K_{1}^{+}\\
\rho^{-}-a_{1}^{-} & \frac{\omega_{N}-\rho^{0}}{\sqrt{2}}-\frac{f_{1N} -a_{1}^{0}}{\sqrt{2}} & K^{\ast0}-K_{1}^{0}\\
K^{\ast-}-K_{1}^{-} & \bar{K}^{\ast0}-\bar{K}_{1}^{0} & \omega_{S}-f_{1S}\end{pmatrix}^{\mu} \ .
\end{align}
The first matrix represents the scalar and pseudoscalar mesons, the other two combine left- and right-handed vector and axial-vector mesons. Note that such an assignment restricts the number of possible quark-antiquark states, for instance there is only one $q\bar{q}$ state that forms the scalar isotriplet with $I=1$. Since it is known that two isotriplets exist\ \cite{olive}, it may be realized as either the $a_{0}(980)$ or the $a_{0}(1450)$.\ The eLSM in fact gives an answer which of the two it addresses (namely, the one above $1$ GeV), but the general problem of overpopulation (in the scalar sector) is not solved in the present form of the model. As mentioned at the end of the previous chapter, we will present a possible solution in this work.

For dimensional reasons, the meson matrices are not identical to the perturbative quark currents; the $\sim$ sign just states that both sides transform in the same way under global chiral transformations:
\begin{equation}
\Phi\rightarrow U_{L}\Phi U_{R}^{\dagger} \ \ , \ \ \ R^{\mu}\rightarrow U_{R}R^{\mu} U_{R}^{\dagger} \ \ , \ \ \ L^{\mu}\rightarrow U_{L}L^{\mu} U_{L}^{\dagger} \ \ ,
\label{eq:chiralTrafos}
\end{equation}
with the chiral rotations
\begin{eqnarray}
U_{L} & = & \exp\left(-\frac{i}{2}\theta_{L}^{a}\lambda^{a}\right) \ \approx \ 1-\frac{i}{2}\theta_{L}^{a}\lambda^{a}+\mathcal{O}(\theta_{L}^{2}) \ , \nonumber \\
U_{R}^{\dagger} & = & \exp\left(\frac{i}{2}\theta_{R}^{a}\lambda^{a}\right) \ \approx \ 1+\frac{i}{2}\theta_{R}^{a}\lambda^{a}+\mathcal{O}(\theta_{R}^{2}) \ .
\end{eqnarray}
The $\lambda^{a}$ are the ordinary Gell-Mann matrices (here, $a=0\dots N_{f}^{2}-1$). This brings us to a short discussion of symmetries: where are the QCD symmetries hidden and where are they broken in the eLSM? The mesonic part of the eLSM Lagrangian in its 'full glory\grq\ has the following form:
\begin{eqnarray}
\mathcal{L}_{\rm meson}^{\text{eLSM}} & = & \mathop{\mathrm{Tr}}[(D_{\mu}\Phi)^{\dagger}(D^{\mu}\Phi)] - m_{0}^{2}\mathop{\mathrm{Tr}}(\Phi^{\dagger}\Phi) - \lambda_{1}[\mathop{\mathrm{Tr}}(\Phi^{\dagger}\Phi)]^{2} - \lambda_{2}\mathop{\mathrm{Tr}}(\Phi^{\dagger}\Phi)^{2} \label{eq:eLSMLag} \\
&  & + \ c_{1}(\det\Phi - \det\Phi^{\dagger})^{2} + \mathop{\mathrm{Tr}}[H(\Phi + \Phi^{\dagger})] - \frac{1}{4}\mathop{\mathrm{Tr}}(L_{\mu\nu}^{2} + R_{\mu\nu}^{2}) \nonumber \\
&  & + \ \mathop{\mathrm{Tr}}\left[\left(\!\frac{m_{1}^{2}}{2}\!+\!\Delta\!\!\right)(L_{\mu}^{2} + R_{\mu}^{2})\right] + \frac{g_{2}}{2}(\mathop{\mathrm{Tr}}\{L_{\mu\nu}[L^{\mu},L^{\nu}]\} + \mathop{\mathrm{Tr}}\{R_{\mu\nu}[R^{\mu},R^{\nu}]\}) \nonumber \\
&  & + \ \frac{h_{1}}{2}\mathop{\mathrm{Tr}}(\Phi^{\dagger}\Phi)\mathop{\mathrm{Tr}}(L_{\mu}^{2} + R_{\mu}^{2}) + h_{2}\mathop{\mathrm{Tr}}[(L_{\mu}\Phi)^{2} + (\Phi R_{\mu} )^{2}] + 2h_{3}\mathop{\mathrm{Tr}}(L_{\mu}\Phi R^{\mu}\Phi^{\dagger}) \nonumber \\[0.15cm]
&  & + \ \text{chirally invariant vector and axial-vector four-point interaction vertices.} \nonumber
\end{eqnarray}
Here, the field-strength tensors 
\begin{equation}
R^{\mu\nu} = \partial^{\mu}R^{\nu}-\partial^{\nu}R^{\mu} \ \ , \ \ \ L^{\mu\nu} = \partial^{\mu}L^{\nu}-\partial^{\nu}L^{\mu}
\end{equation}
have been defined together with
\begin{equation}
D^{\mu}\Phi = \partial^{\mu}\Phi-ig_{1}(L^{\mu}\Phi-\Phi R^{\mu}) \ \ , \ \ \ H = \diag(h_{0}^{1},h_{0}^{2},h_{0}^{3}) \ \ , \ \ \ \Delta = \diag(\delta_{u},\delta_{d},\delta_{s}) \ \ .
\end{equation}
The constants $m_{0}$, $\lambda_{1}$, $\lambda_{2}$, $c_{1}$, $g_{1}$, $g_{2}$, $h_{1}$, $h_{2}$, and $h_{3}$ are model parameters with specific large-$N_{c}$ behavior.\ For instance, the bare mass $m_{0}$ is directly related to the shift of the gluonic field in the dilaton part of the model (not shown here), which goes like $m_{0}\propto N_{c}^{0}$, while $\lambda_{2}$ scales as $\lambda_{2}\propto N_{c}^{-1}$ because it is associated with quartic meson interaction vertices. For a detailed discussion see Ref.\ \cite{denisphd}.

Now, the Lagrangian\ (\ref{eq:eLSMLag}) must implement the QCD symmetries and their breaking:
\begin{itemize}
\item The $SU(3)_{c}$ gauge symmetry is exact in QCD. Since the degrees of freedom in the eLSM are colorless hadrons and confinement is trivially fulfilled, this symmetry is present from the very beginning by construction.
\item The $U(3)_{R}\times U(3)_{L}$ chiral symmetry is exact for vanishing bare quark masses in QCD and in fact realized there as a global one.\ Because of the transformation behavior\ (\ref{eq:chiralTrafos}) of our meson matrices, most of the terms shown in Eq.\ (\ref{eq:eLSMLag}) are invariant under chiral rotations. For example,
\begin{equation}
m_{0}^{2}\mathop{\mathrm{Tr}}(\Phi^{\dagger}\Phi) \rightarrow m_{0}^{2}\mathop{\mathrm{Tr}}(U_{R}\Phi^{\dagger}U_{L}^{\dagger}U_{L}\Phi U_{R}^{\dagger}) = m_{0}^{2}\mathop{\mathrm{Tr}}(\Phi^{\dagger}\Phi) \ ,
\end{equation}
where the unitarity property of the chiral rotations was used together with the fact that the trace in flavor space is invariant under cyclic permutations. Chiral symmetry breaking needs to be modeled separately for the different mesonic sectors; this is accounted for by the remaining non-invariant terms.
\item In the (pseudo)scalar sector, the term $\mathop{\mathrm{Tr}}[H(\Phi + \Phi^{\dagger})]$ generates explicit chiral symmetry breaking due to non-vanishing quark masses. The term contains the matrix $H$ with diagonal entries $h_{0}^{i}$, with flavor index $i=1\dots3$, where the entries are proportional to the $i$-th quark mass (with $h_{0}^{1}=h_{0}^{2}=h_{0}$ for exact isospin symmetry for up and down quarks).
\item In the (axial-)vector sector, the term containing the $\Delta$ matrix is responsible for explicit symmetry breaking since
\begin{equation}
\Delta\sim\diag(m_{u}^{2},m_{d}^{2},m_{s}^{2}) \ ,
\end{equation}
and hence introduces terms proportional to the squared quark masses as required.
\item Chiral symmetry is also spontaneously broken in QCD because of a non-vanishing expectation value of the quark condensate, $\langle q\bar{q}\rangle\neq0$.\ For $N_{f}=3$, this leads to the emergence of eight Nambu--Goldstone bosons which should be massles. However, they are not massless, because the symmetry is also explicitly broken by the $H$ term (for instance, it is $m_{\pi}^{2}\propto h_{0}$). This results in eight light pseudo-Nambu--Goldstone bosons, the inhabitants of the well-known octet of pseudoscalar mesons. The eLSM incorporates spontaneous chiral symmetry breaking due to the sign of $m_{0}^{2}<0$.
\item The chiral symmetry is broken by quantum effects, too; in QCD this is known as the chiral or $U(1)_{A}$ anomaly that induces a mass splitting between the pion and the $\eta$ meson, as well as the exceptional higher mass of the singlet state $\eta^{\prime}$ around $1$ GeV.\ This becomes evident via an extra term in the divergence of the axial-vector singlet current even when all quark masses vanish.\ The eLSM accounts for this by the term proportional to $c_{1}$\ \cite{chiralanomaly}, see also Refs.\ \cite{eLSM1-3,stani}.\ It is invariant under $SU(3)_{R}\times SU(3)_{L}$ but not under $U(1)_{A}$.
\item The gauge sector of QCD in the classical limit (strictly speaking, the classical action) is invariant under dilatation transformations, which is also true for the quark sector in the chiral limit. This symmetry is therefore explicitly broken for finite quark masses, but it is also anomalously broken when quantum corrections are considered: the trace of the energy-momentum tensor, which represents the conserved current, picks up a term proportional to the $\beta$-function of QCD. The running of the strong coupling constant then renders this term unequal to zero. The eLSM describes the trace anomaly by including a dilaton field with a convenient potential\ \cite{eLSM1}, such that dilatation symmetry is broken explicitly in the chiral limit. The corresponding part $\mathcal{L}_{\text{dil}}$ is not displayed in Eq.\ (\ref{eq:eLSMLag}).
\item All terms in the effective Lagrangian are $CPT$ invariant. This is evident from the construction of the meson matrices and the transformation behavior of the quark fields.\ When applying charge conjugation on the (pseudo)scalar meson matrix, $\Phi\rightarrow\Phi^{T}$, one finds for example
\begin{equation}
m_{0}^{2}\mathop{\mathrm{Tr}}(\Phi^{\dagger}\Phi) \rightarrow m_{0}^{2}\mathop{\mathrm{Tr}}({\Phi^{\dagger}}^{T}\Phi^{T}) = m_{0}^{2}\mathop{\mathrm{Tr}}([\Phi\Phi^{\dagger}]^{T}) = m_{0}^{2}\mathop{\mathrm{Tr}}(\Phi^{\dagger}\Phi) \ ,
\end{equation}
where we used that a matrix and its transpose have the same trace, together with the fact that the trace in flavor space is invariant under cyclic permutations.
\end{itemize}

Effective descriptions (of QCD) have their own issues. The eLSM Lagrangian contains only terms up to order four in dimension. This is not because one would like to preserve renormalizability, since an effective model can in principle not be valid up to arbitrarily large scales.\footnote{The validity of the eLSM is determined by the energy of the heaviest state present, thus up to $\sim1.8$ GeV.} In fact, once a dilaton field is included, this restricts possible terms to have just dimensionless couplings\footnote{With the exception of the explicitly dilatation symmetry breaking terms $\sim c_{1}$, $\sim H$, and $\sim\Delta$.} -- otherwise $(i)$ it is not possible to model the trace anomaly in the chiral limit in the same manner as in QCD and $(ii)$ one would allow terms of inverse order of the dilaton field, leading to singularities when it vanishes. Furthermore, vertices with derivative interactions are present. The spontaneous symmetry breaking mechanism requires to shift the $\sigma$ field by its vacuum expectation value, yielding mixing terms between the pseudoscalar and axial-vector sectors. They are removed from the Lagrangian by shifting the affected fields appropriately and hence introducing derivatively coupled pseudoscalars.

Although such new characteristics may complicate the handling of the model, it turns out that perturbative calculations can be applied in order to calculate tree-level masses and decay widths of resonances.\footnote{Relevant model results will be discussed in the respective sections, for all the details see Refs.\ \cite{denisphd,eLSM2}.} A pure two-body tree-level decay is the easiest non-trivial process in quantum field
theory.\ For example, an unstable bosonic particle $S$ may decay into two identical particles, denoted as
$\phi$.\ The decay amplitude is simply a constant in the case of scalar
particles and non-derivative interactions, $\mathcal{L}_{\text{int}}=gS\phi\phi$ (see also next chapter). 
Effective models can be studied by taking into account (hadronic) 
loop contributions in the relevant process amplitudes. The leading contribution to the self-energy 
would then be an ordinary one-loop diagram with circulating decay products. Both the mass and the width of the decaying 
particle are influenced by the quantum fluctuations due to the coupling to hadronic intermediate states. As wee shall see, 
this is in particular 
very important for scalar resonances. The optical theorem assures that the imaginary part of the one-loop diagram 
coincides with the formal expression of the tree-level decay width.

\begin{figure}[t]
\hspace*{-0.4cm}
\begin{minipage}[hbt]{8cm}
\centering
\includegraphics[scale=0.75]{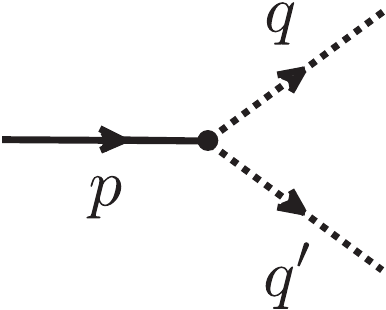}
\end{minipage}
\hspace*{-2.0cm}
\begin{minipage}[hbt]{8cm}
\centering
\vspace*{0.0cm}
\includegraphics[scale=0.75]{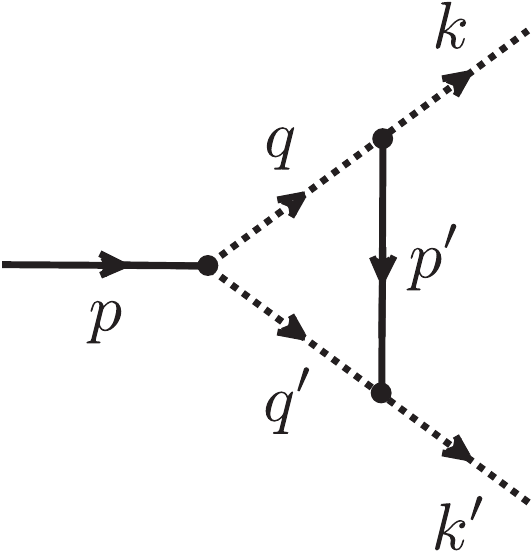}
\end{minipage}
\vspace*{0.2cm}
\caption{Left panel: LO diagram of the two-body decay $S\rightarrow\phi\phi$. Right panel: Triangle-shaped NLO diagram from Ref.\ \cite{jonas}, where the decaying particle $S$ is exchanged as a higher-order process.}
\label{fig:NLO}
\end{figure}

It was demonstrated in Ref.\ \cite{jonas} that the next-to-leading order (NLO) triangle diagram of a hadronic decay, depicted 
in Figure\ \ref{fig:NLO}, can be safely neglected in the case of a simple scalar theory without derivatives. This approximation has been used to study the well-known isoscalar resonances 
$f_{0}(500)$, $f_{0}(980)$, $f_{0}(1370)$, and $f_{0}(1500)$, the $\pi\pi$ and $K\bar{K}$ decay 
channels of $f_{0}(1710)$, and 
the $K\bar{K}$ decay of the isovector state $a_{0}(1450)$.\ Except for $f_{0}(500)$, one can therefore 
justify {\em a posteriori} all studies in which triangle diagram contributions were not taken into account. Since in the field of
hadron physics there are usually other (and even larger) sources of uncertainties due to various 
(and sometimes subtle) approximations and simplifications, the restriction to the leading-order tree-level diagram and 
to the (resummed) one-loop quantum corrections is reasonable and usually sufficient.\ In this work quantum corrections will therefore be only considered up to one-loop level.

The eLSM turned out to be quite successful. However, if hadronic resonances are constructed as $q\bar{q}$ states, then the problem of the overpopulation in the scalar sector remains unsolved.\ The scalar sector is described by the predominantly 
quark-antiquark states $f_{0}(1370)$,\ $f_{0}(1500)$,\ $K_{0}^{\ast}(1430)$, and the $a_{0}(1450)$, while the $f_{0}(1710)$ is predominantly gluonic.\ Then, the light resonances below $1$ GeV, namely $f_{0}(500)$, $f_{0}(980)$, $K_{0}^{\ast}(800)$, 
and $a_{0}(980)$, should not be part of the eLSM and form a nonet of predominantly some sort of four-quark objects. Some very different model approaches that try to include those particles are presented shortly in the following. The key message will be that at least some (scalar) mesons cannot just be predominantly quarkonia but, since they are highly influenced by the dynamics of their hadronic decay channels, they could be rather {\em dynamically generated} \ objects. This is based on the idea mediated by the prior quote of Chew and Frautschi\ \cite{chew2}.\ Please note that the following presentation can neither be sufficient nor complete; further details, as well as other approaches or general achievements, will only be mentioned via citations in suitable places within this thesis.

\section{Dynamical generation: Different approaches}
\subsection{Unitarized Quark Model (UQM)}
In quark models the quarks and antiquarks are assumed to be confined by the strong interaction. Then, the 
constituent quark masses are the result of absorbing the main interaction with the gluons -- what remains in dynamics is 
transformed into some confining potential which is used to form hadronic particles. Among others, T\"ornqvist and Roos\ 
\cite{tornqvistOld,tornqvist,tornqvist2} and later also Boglione and Pennington\ \cite{boglione,pennington} studied extensions 
by including meson-loop contributions to preexisting quark-antiquark states -- this method was used to unitarize the amplitudes which was called {\em unitarization}.\ They came to 
the conclusion that it could indeed be possible to generate more (scalar) states than actually feasible in a quark model, when 
starting from those preexisting mesons as bare seed states.

\begin{figure}[t]
\vspace{0.3cm}
\centering
\hspace{0.0cm}
\includegraphics[scale=0.4]{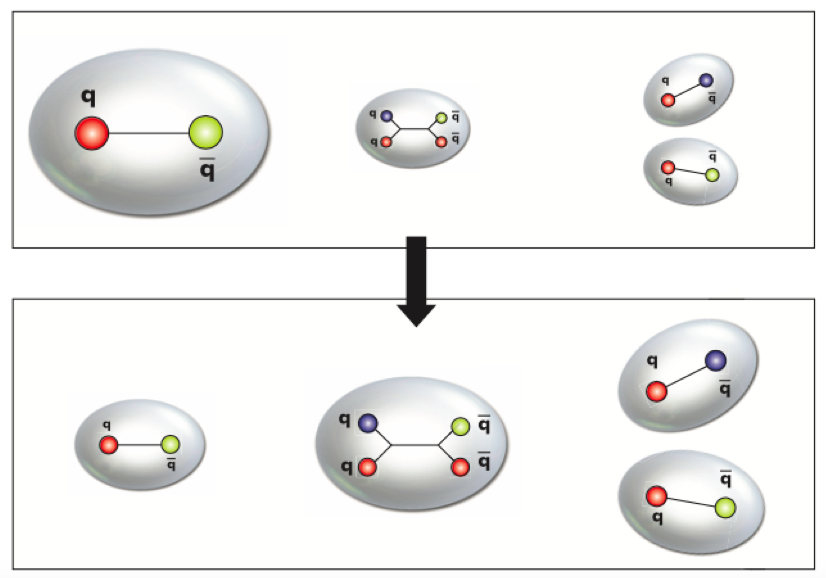}
\vspace{0.3cm}
\caption{Pictorial representation of an unstable meson's Fock space, taken from Ref.\ \cite{penningtonActa}.\ Since the initial quark-antiquark state has the possibility to decay, the Fock space must contain at least four-quark components that are also part of the particle's state vector. Consequently, the meson is better described as the combination of all its contributions, see first line. There, however, the $q\bar{q}$ component is dominant. In the second line the situation has changed: now, the four-quark components dominate because of the nature of the $S$-wave coupling.}
\label{fig:PennContri}
\end{figure}

How this can be understood? Let us consider as an intuitive example the $\phi$ meson which possesses the main decay 
channel $\phi\rightarrow K^{+}K^{-}$.\ It fits very well into the ideal octet of vector states with predominant flavor configuration 
of $s\bar{s}$, while the pseudoscalar kaons are $u\bar{s}$ and $\bar{u}s$, respectively. In terms of QCD, the $\phi$ decay is 
described by creating a $u\bar{u}$ pair out of the vacuum, while the decay into three pions is suppressed by the OZI rule. 
Since there {\em is} indeed the possibility to decay, {\em i.e.}, to end up with a configuration of four quarks, the Fock space of 
the initial particle must contain at least four-quark components\ \cite{penningtonActa}.\ Consequently, the meson is better described 
as the combination of all its contributions:
\begin{equation}
|\phi\rangle = a\hspace{0.025cm}|s\bar{s}\rangle+b\hspace{0.025cm}|K^{+}K^{-}\rangle+\dots \ ,
\end{equation}
where $a\propto N_{c}^{0}$ and $b\propto N_{c}^{-1/2}$\ \cite{giacosaLargeNC}.
For the vector mesons, however, the $q\bar{q}$ 
component is dominant, see also Figure\ \ref{fig:PennContri}. In contrast, for the scalar resonances below $1$ GeV one can imagine the situation where the four-quark 
components dominate -- which is the case because of the nature of the $S$-wave coupling -- and would bind. The latter 
need not necessarily be the actual microscopic picture: such states would not be pure molecules but contain some residue 
of their quarkonia seeds.

\begin{figure}[t]
\vspace{0.3cm}
\centering
\hspace{1.35cm}
\includegraphics[scale=0.5]{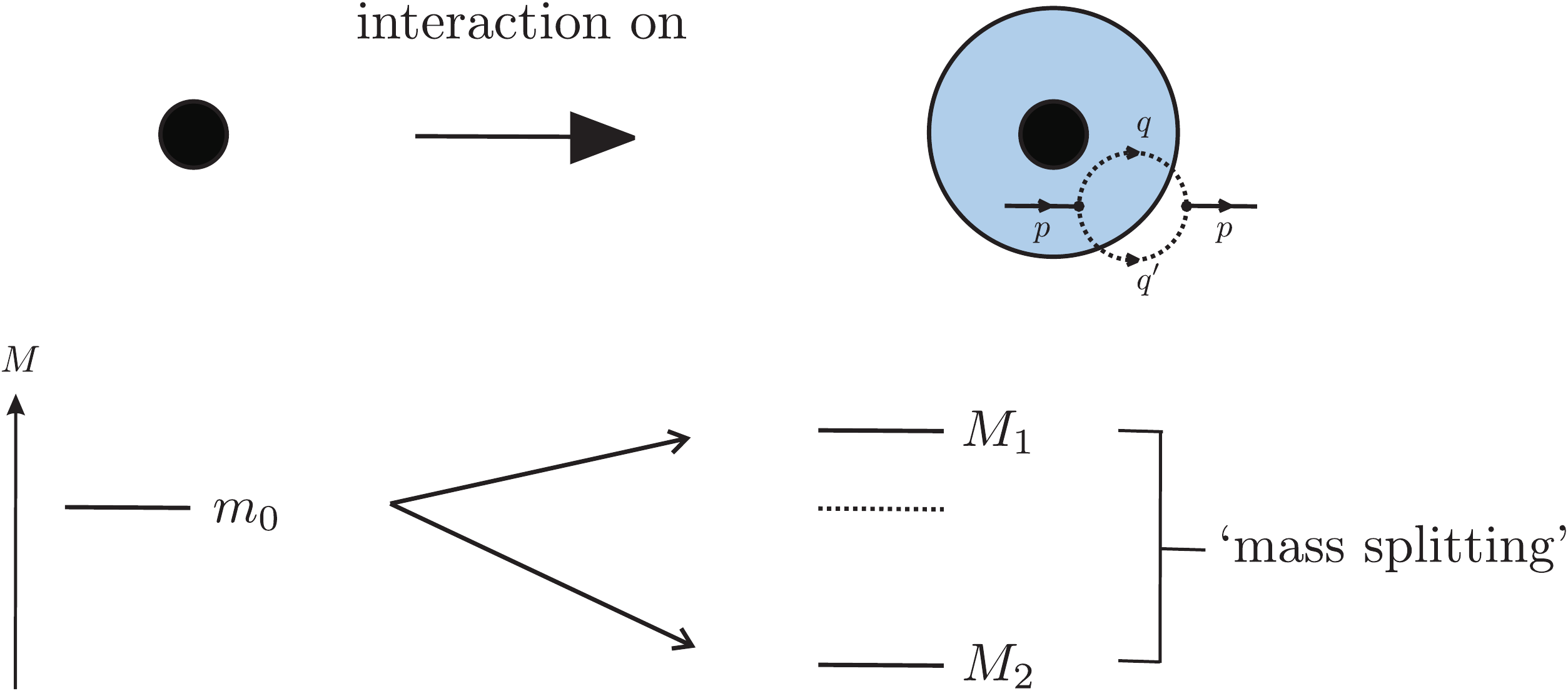}
\vspace{0.3cm}
\caption{Visualization of the simplest possible way to generate an additional resonance from a preexisting seed state: The latter (indicated as a black filled circle) starts with bare mass $M=m_{0}$, which is shifted when interactions are turned on and the state can communicate with intermediate (hadronic) loop contributions. This is usually called dressing, indicated as 
blue cloud around the seed. As an accompanying effect, an orthogonal state is obtained because the hadronic degrees of freedom tend to bind. This leads to a pair of resonances; one obtains the mass $M_{1}$ and another one $M_{2}$, both usually different from $m_{0}$. They appear as poles in the relevant process amplitude, where only one of them is present at the beginning (with no interaction).}
\label{fig:splitting}
\end{figure}

This reasoning is what leads to the mechanism called {\em dynamical generation}.\ Since the unstable particle's 
propagator represents the probability amplitude for propagating from one space-time point to another, all intermediate 
interactions in the form of hadronic loops can occur. This establishes access to the four-quark contributions. As usual, 
they shift the seed state pole from the real energy axis into the complex plane of an unphysical Riemann sheet, but in the 
case of the scalar sector it furthermore may create new poles. Those poles can be extracted from scattering data and some of them could be identified with physical resonances -- see Figure\ \ref{fig:splitting} for a visualization of this agument.

It became apparent that in order to obtain additional resonance poles, the UQM needs to include Adler zeros and a further 
$s$-dependence in the amplitudes, respectively. With these modifications it was possible to find at least some extra poles, 
in particular a putative pole for the states $f_{0}(500)$ and one for $a_{0}(980)$.\ This is one of the reasons why the corresponding publications\ 
\cite{tornqvist,tornqvist2} are very famous. On the other hand, no pole was found that could have been assigned to the 
$K_{0}^{\ast}(800)$, while two poles were obtained in the isovector sector.\footnote{An improved UQM was later 
applied in Ref.\ \cite{zhiyong}, where the $\kappa$ pole was indeed found.} The formal details of the UQM together with a 
critical analysis of its results concerning this last sector will be presented in Chapter\ \ref{chap:chapter4}.

\subsection{Resonance Spectrum Expansion model (RSE)}
\begin{figure}[h]
\vspace*{0.3cm}
\hspace*{-1.425cm}
\begin{minipage}[hbt]{8cm}
\centering
\includegraphics[scale=0.71]{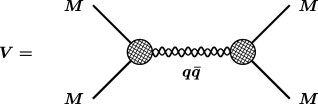}
\end{minipage}
\hspace*{-0.9cm}
\begin{minipage}[hbt]{8cm}
\centering
\vspace*{0.0cm}
\includegraphics[scale=0.725]{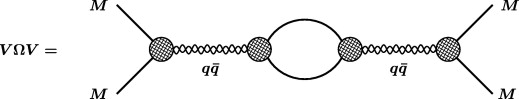}
\end{minipage}
\vspace*{0.3cm}
\caption{Born expansion of the RSE's transition operator, taken from Ref.\ \cite{beverenV}.\ Here, $V$ is the effective meson-meson potential and $\Omega$ is the meson-meson loop function. The wiggly lines represent the intermediate $s$-channel $q\bar{q}$ propagators between some vertex functions (shown as circles), modeled as spherical Bessel functions in momentum space.}
\label{fig:beverenV}
\end{figure}

This quark-meson model has quite a long history because the original version was published already in the $80$'s\ \cite{dullemond2,dullemond}, while further developments have been achieved during the past decades, see {\em e.g.}\ Refs.\ \cite{2006beveren,coitox,coitophd} and references therein. As in the case of the UQM, the RSE model is based on the unitarization of bare (scalar) states by their strong coupling to $S$-wave two-meson channels, in particular it is a coupled-channel model that describes elastic meson scattering of the form $A+B\rightarrow A+B$.\ The transition operator (entering the scattering formalism) contains an effective two-meson potential $V$ which is assumed to contain only intermediate $s$-channel exchanges of infinite towers of $q\bar{q}$ seed states, see also Figure\ \ref{fig:beverenV}.\ This corresponds to the spectrum of a confining potential which is chosen as a harmonic oscillator with constant frequency.\ The power of the formalism lies in the separable form of the interaction matrix elements, resulting in a closed form of the off-shell $\hat T$-matrix, and giving the possibility to study for example resonance poles.

Although in general there is no need for any approximation, for pedagogical reasons one can state that for low energies 
the infinity tower can be reduced to an effective constant; one is left with a contact term that dynamically generates exactly 
one pole in the case of scalar mesons\ \cite{beverenV}.\ The tower still can be approximated for somewhat higher energies 
by its leading term and an effective constant for the remaining sum. Then one obtains, apart from the dynamically 
generated resonance, another pole associated with the leading seed state, that is, the leading propagator mode. While 
the specific elaboration of the RSE model is very different from the UQM, both rely on the incorporation of 
hadronic loop contributions.

The RSE model was applied to different flavors, including charm and bottom, and needs only one elementary set of parameters. 
Surprising and most interesting for us is the fact that after the parameters are fixed by the vector and pseudoscalar spectra, all the low-lying scalar states are fully generated as resonance 
poles. It was therefore suggested to assign them to 
another distinct nonet of low-mass scalars purely obtained from dynamics.

\subsection{Coupled-channel unitarity approach by Oller, Oset, and Pel\'{a}ez}
Oller and Oset generated low-lying scalar mesons dynamically in the framework of a coupled-channel 
Lippmann--Schwinger (LS) approach\ \cite{oller,oller2}.\ The starting point here is the standard chiral Lagrangian in lowest 
$\mathcal{O}(p^{2})$ order of chPT\ \cite{chpt,chpt2}.\ It contains the most general low-energy interactions of the pseudoscalar mesons at 
this order.\ From this Lagrangian the tree-level amplitudes for scattering are obtained and consequently the meson-meson potential 
terms needed for the coupled-channel analysis. It turns out that it is possible to reduce the LS equation (where relativistic 
meson propagators are applied) to pure algebraic relations, yielding a simple form for the $\hat T$-matrix. Unitarizing this 
amplitude creates for instance the pole of the $a_{0}(980)$ in the isovector sector.

One advantage of this approach is that it requires the use of just one free parameter, namely a cutoff in the loop integrals 
coming from the LS equation, which is fixed to experimental data. In the further extension of Ref.\ \cite{oller3}, the 
next-order $\mathcal{O}(p^{4})$ Lagrangian of chPT is taken into account, where additional parameters entering by 
this procedure are also fitted to data. In this study also a pole for the $\kappa$ was obtained in collaboration with 
Pel\'{a}ez\ \cite{oller3} (the case $I=1/2$ was not investigated in the previous works). Later, in Ref.\ \cite{oller4}, it was 
demonstrated that for the scalar sector the unitarization of the $\mathcal{O}(p^{2})$ chPT amplitude is strong enough to 
dynamically generate the low-lying resonances including the $K_{0}^{\ast}(800)$.\ It was {\em a priori} not possible to say if the 
generated states in this approach are in fact quark-antiquark or four-quark resonances, and if they can be linked to the 
heavier mesons or not, see Ref.\ \cite{giacosaDynamical} for a detailed discussion of this issue. Yet, in Ref.\ \cite{oller4} a 
preexisting octet (and singlet) of bare resonances around $1.4$ GeV (and $1.0$ GeV) was included as a set of CDD poles\ 
\cite{cddpoles}.\ It was then found that {\em e.g.}\ the physical $a_{0}(1450)$ in fact originates from the octet, giving a clear 
statement about its nature.

Quite interestingly, in Ref.\ \cite{oller4} Oller {\em et al.} estimated the influence of the unphysical cuts for the 
elastic $\pi\pi$ and $\pi K$ $S$-waves with $I=0$ and $I=1/2$, respectively. Such cuts have been included in their 
previous works only in a perturbative sense (they were absorbed for example in one free parameter mentioned 
above) -- strictly speaking, no loop effects in the $t$- and $u$-channels were considered. It was argued that this kind of 
simplification is indeed justified due to the quality of the results: the {\em relevant} infinite series were summed up in the 
$s$-channel. In Ref.\ \cite{oller4}, however, the unphysical cuts were incorporated in terms of chPT up to $\mathcal{O}(p^{4})$, 
together with the exchange of resonances in the $t$- and $u$-channels. It was shown that the contributions from the former are 
rather small, because of cancellations with contributions coming from the latter, supporting the view of treating the unphysical 
cuts in a perturbative way.

\subsection{J\"ulich mesonic $t$-channel exchange model}
The J\"ulich meson-exchange model\ \cite{julich} was extended in Ref.\ \cite{janssen} to account for further meson-meson 
interactions. The approach was first based on a coupled-channel analysis of the $\pi\pi$- and $K\bar{K}$-channels, where it 
was found that the $f_{0}(980)$ can be generated by vector-meson exchanges in the $K\bar{K}$-channel, that are strong 
enough to produce a bound state pole in the $\hat T$-matrix.\ After extending this consistently to the $\eta\pi$ system, it was 
possible to demonstrate that the isovector sector is governed by the same dynamics, that is, by the coupling to the kaon-kaon channel.\ However, it was stressed that the $a_{0}(980)$ was rather a dynamically generated threshold effect with a 
relatively low-lying pole, because the important $\rho$ exchange between the two kaons becomes repulsive, not allowing to form a $K\bar{K}$ molecule.

One should note that a putative pole for the $f_{0}(500)$ was also obtained, generated from a strong $t$-channel $\rho$ 
exchange in the $\pi+\pi\rightarrow\pi+\pi$ potential. Nevertheless, Harada {\em et al.}\ observed that neglecting this 
contribution did not remove such a pole from the scattering amplitude of their own model, but only changed its position 
slightly\ \cite{harada}.\ It was also argued elsewhere that the pole generated by the J\"ulich group was considerably lighter 
and broader than generally accepted.

\section{Concluding remarks}
We have presented some very different models that try to generate additional (scalar) resonances. In simplified terms, 
except for the J\"ulich model, the generation mechanisms are exploiting hadronic loop contributions either by dressing 
$q\bar{q}$ seed states or by relying on meson-meson loops in the scattering amplitude only -- the overall dynamics are 
thus highly influenced by the decay channels. We therefore ask in the following if it is possible to generate low-lying scalar 
mesons by a similar mechanism within the eLSM.

\newpage
\clearpage

\

\newpage
\clearpage

\

\newpage
\clearpage

\chapter{Derivative interactions and dispersion relations}
\label{chap:chapter3}

\medskip

In this thesis we exploit the idea of dynamical generation in order to study two types of mesonic resonances: the scalar--isovector states $a_{0}(980)$ and $a_{0}(1450)$, and the isodoublet resonances $K_{0}^{\ast}(800)$ and $K_{0}^{\ast}(1430)$.\ It was mentioned at the end of the previous chapter that models of dynamical generation focus on the unitarization of bare scalar (seed) states via strong couplings to intermediate (hadronic) states. To this end, it becomes necessary to compute such loop contributions.\ If a 3d form factor (or regularization function) is applied, usually any one-loop diagram can be obtained from a {\em dispersion relation}. However, it will be demonstrated in the following that care is needed in the case of derivative interactions which naturally appear in our effective Lagrangians: there is an apparent discrepancy between ordinary Feynman rules and dispersion relations.

\section{Dispersion relations}
Multi-valued complex functions are typically manipulated by introducing branch cuts when performing contour integrations in the complex plane.\ Assuming the following properties of a function $f(z)$ defined on the complex plane,
\begin{itemize}
\item hermitian-analyticity,
\begin{equation}
f(z) = f^{*}(z^{*}) \ ,
\end{equation}
\item holomorphy except at the cut, where for real $s$ it is
\begin{eqnarray}
\lim_{\epsilon \to 0^{+}}\Big[f(s+i\epsilon)-f(s-i\epsilon)\Big] & = & 2i\lim_{\epsilon \to 0^{+}}\operatorname{Im}f(s+i\epsilon) \nonumber \\[0.1cm]
& = & \disc f(s+i\epsilon) \ ,
\end{eqnarray}
\item and vanishing faster than $\mathcal{O}(1/|z|^{\eta})$ (with $\eta>0$),
\end{itemize}
the function can be expressed due to Cauchy's integral formula in the limit $\epsilon\rightarrow0^{+}$ by using its imaginary part right above the cut\ \cite{pucker}:
\begin{equation}
f(z) = -\frac{1}{\pi}\int_{a}^{b}\text{d}s' \ \frac{\lim_{\epsilon \to 0^{+}}\operatorname{Im}f(s'+i\epsilon)}{z-s'} \ . \label{equation_dispersionrel}
\end{equation}
Here, $a$ and $b$ mark the branch points on the real axis. The full function is determined by the discontinuity only and can be calculated by evaluating the dispersion integral in Eq.\ (\ref{equation_dispersionrel}).

As an instructive example we take the complex root function $f(z)=\sqrt{z}$. Since it does not decrease for $|z|\rightarrow\infty$, we need to modify the dispersion integral by using a slightly modified function $g(z)$:
\begin{equation}
g(z) = \frac{\sqrt{z}}{z} \ .
\end{equation}
This new function has the same branch cut structure as $f(z)$ (note that there is no simple pole at $z=0$). The discontinuity of the pure root function at the cut is simply two times itself. For the new function this means
\begin{equation}
\disc g(-s) = -2\hspace{0.02cm}i\hspace{0.02cm}\frac{\sqrt{s}}{s} \ .
\end{equation}
The dispersion integral can then be computed by using a Hankel contour path of integration $\mathcal{C}$ with left open end:
\begin{eqnarray}
g(z) \ \ = \ \ \frac{1}{2\pi i}\oint_{\mathcal{C}}\text{d}\xi \ \frac{g(\xi)}{\xi-z} & = & \frac{1}{\pi}\int_{-\infty}^{0}\text{d}\rho' \ \frac{-\sqrt{-\rho'}}{\rho'(z-\rho')} \nonumber \\
& \stackrel{\rho'\rightarrow-s'}{=} & \frac{1}{\pi}\int_{0}^{\infty}\text{d}s' \ \frac{\sqrt{s'}}{s'(z+s')} \ .
\end{eqnarray}
Finally, the original function $f(z)$ can be denoted as
\vspace{0.15cm}
\begin{eqnarray}
f(z) & = & \frac{z}{\pi}\int_{0}^{\infty}\text{d}s' \ \frac{\sqrt{s'}}{s'(z+s')} \nonumber \\
& \stackrel{s'\rightarrow x^2}{=} & \frac{2z}{\pi}\int_{0}^{\infty}\text{d}x \ \frac{1}{z+x^{2}} \nonumber \\
& = & \frac{\sqrt{z}}{\pi}\int_{0}^{\infty}\text{d}x \ \left(\frac{1}{\sqrt{z}+ix}+\frac{1}{\sqrt{z}-ix}\right) \nonumber \\
& = & \frac{\sqrt{z}}{\pi}\pi \ , \ \ \text{for $\operatorname{Im}z\neq0\vee\operatorname{Re}z\geq0$} \nonumber \\
& = & \sqrt{z} \ .
\end{eqnarray}
We have used the identity
\begin{equation}
\int\text{d}x \ \left(\frac{1}{\sqrt{z}+ix}+\frac{1}{\sqrt{z}-ix}\right) = 2\arctan\left(\frac{x}{\sqrt{z}}\right) \ .
\end{equation}

The dispersion integral in Eq.\ (\ref{equation_dispersionrel}) is not valid on the real $z$-axis; this limit cannot be performed in a naive way since the integration contour could collide with a pole. We either use the Sokhotski--Plemelj theorem to identify the real part as the Cauchy principal value integral or simply obtain it from
\vspace{0.15cm}
\begin{eqnarray}
\operatorname{Re}f(s+i\epsilon) & = & \frac{1}{2}\hspace{0.07cm}\Big[f(s+i\epsilon)+f^{*}(s+i\epsilon)\Big] \nonumber \\
& = & \frac{1}{2\pi}\left[\int_{a}^{b}\text{d}s' \ \frac{\lim_{\epsilon \to 0^{+}}-\operatorname{Im}f(s'+i\epsilon)}{s-s'+i\epsilon}+\int_{a}^{b}\text{d}s' \ \frac{\lim_{\epsilon \to 0^{+}}-\operatorname{Im}f(s'+i\epsilon)}{s-s'-i\epsilon}\right] \nonumber \\
& = & \frac{1}{\pi} \ -\hspace{-0.435cm}\int_{a}^{b}\text{d}s^{\prime} \ \frac{\lim_{\epsilon \to 0^{+}}-\operatorname{Im}f(s'+i\epsilon)}{s-s'} \ . \label{eq:Redispersionrel} \\[-0.3cm]
\nonumber
\end{eqnarray}

\section{Non-derivative interaction of the form $\mathcal{L}_{\text{int}}=gS\phi\phi$}
\label{seq:PolFunc}
The simplest application of the dispersion integral in quantum field theory is the calculation of the one-loop contribution to the self-energy $\Pi(s)$ in a theory with two scalar fields, $S$ and $\phi$, containing the decay process $S\rightarrow\phi\phi$:
\begin{eqnarray}
\mathcal{L}_S & = & \frac{1}{2} \left( \partial_\mu S \partial^\mu S - m_{0}^2 S^2 \right) \ , \ \ \ \mathcal{L}_\phi \ = \ \frac{1}{2} \left( \partial_\mu \phi \partial^\mu \phi - m^2 \phi^2 \right) \ ,
\nonumber \\
\mathcal{L}_{\text{int}}& =& gS\phi\phi \ .
\label{eq:LagSphiphi}
\end{eqnarray}
Such a model was studied for example in Refs.\ \cite{veltman,giacosapagliara,thomasthesis}.\ The resulting one-loop diagram with circulating $\phi$-particles requires the presence of a form factor (or regularization function) $F(s)$, which depends on a UV cutoff scale $\Lambda$, to make the otherwise logarithmically divergent integral finite and the imaginary part of $\Pi(s)$ vanishing sufficiently fast at infinity, respectively. This property reflects the finite size of hadrons\ \cite{tornqvist,giacosapagliara}. The integral is then evaluated by assuming the form factor to depend only on the absolute value of the decay channel's c.m.\ three-momentum,\footnote{This actually violates Lorentz symmetry but we accept this drawback in order to arrive at simple analytic expressions. However, note that there is a trick to obtain the same form factor by using a covariant regularization function, see Ref.\ \cite{giacosaBo}.}
\begin{equation}
k(s) = \sqrt{\frac{s}{4}-m^{2}} \ .
\end{equation}
The analytic expression for the diagram is known\ \cite{thomasthesis}:
\begin{equation}
\label{eq:loopSolution}
\Pi(s+i\epsilon) = -\frac{g^{2}}{\pi^{2}}\int_{0}^{\infty}\text{d}u \ \frac{u^{2}F^{2}(u)}{\sqrt{u^{2}+m^{2}}\big[4(u^{2}+m^{2})-s-i\epsilon\big]} \ ,
\end{equation}
where $u$ is used to write the momenta of the $\phi$ particles inside the loop and $m$ is their mass. It is now straightforward to rewrite this expression into an imaginary part and a dispersion integral for the real part by using the Cauchy principal value:
\begin{eqnarray}
\Pi(s+i\epsilon) & = & \frac{g^{2}}{\pi^{2}}\int_{0}^{\infty}\text{d}u \ \frac{u^{2}F^{2}(u)}{\sqrt{u^{2}+m^{2}}\big[s-4(u^{2}+m^{2})+i\epsilon\big]} \\[0.2cm]
& = & \frac{1}{\pi}\int_{4m^{2}}^{\infty}\text{d}t \ \frac{\frac{\sqrt{\frac{t}{4}-m^{2}}}{4\pi\sqrt{t}}g^{2}F^{2}\left(\sqrt{\frac{t}{4}-m^{2}}\right)}{s-t+i\epsilon} \nonumber \\[0.2cm]
& = & \frac{1}{\pi} -\hspace{-0.435cm}\int_{4m^{2}}^{\infty}\text{d}t \ \frac{\frac{\sqrt{\frac{t}{4}-m^{2}}}{4\pi\sqrt{t}}g^{2}F^{2}\left(\sqrt{\frac{t}{4}-m^{2}}\right)}{s-t}-i\frac{\sqrt{\frac{s}{4}-m^{2}}}{4\pi\sqrt{s}}g^{2}F^{2}\left(\sqrt{\frac{s}{4}-m^{2}}\right) \ , \nonumber
\end{eqnarray}
where in the second step the variable transformation $u\rightarrow\sqrt{t/4-m^{2}}$ has been introduced.\ This calculation just proves that $\Pi(s)$ can in fact be expressed by a dispersion relation, if the optical theorem for Feynman diagrams is applicable in order to compute the imaginary part of the corresponding self-energy loop:
\begin{gather}
\label{eq:optical_simple}
\int\text{d}\Gamma \ |\hspace{0.03cm}\text{--}\hspace{0.05cm}i\mathcal{M}(s)|^{2} = \sqrt{s}\hspace{0.07cm}\Gamma^{\text{tree}}(s) = -\operatorname{Im}\Pi(s) \ , \\
\label{eq:tree_simple}
\Gamma^{\text{tree}}(s) = \frac{k(s)}{16\pi s}|\hspace{0.03cm}\text{--}\hspace{0.05cm}i\mathcal{M}(s)|^{2} \ .
\end{gather}
The ($s$-dependent) tree-level decay width $\Gamma^{\text{tree}}(s)$ is obtained by performing the phase space integral over the invariant amplitude,
\begin{equation}
-i\mathcal{M}(s=M^{2})=2igF\left(u=\sqrt{\frac{M^{2}}{4}-m^{2}}\right) \ .
\label{eq:SphiphiAmp}
\end{equation}
The infinitesimal expressions are
\begin{eqnarray}
\text{d}\Gamma & = & \frac{\mathcal{S}}{2M} \ \text{d}\varphi_{n} \ , \\
\text{d}\varphi_{n} & = & (2\pi)^{4}\delta^{(4)}\bigg(p-\sum_{i=1}^{n}p_{i}\bigg)\prod_{i=1}^{n}\frac{\text{d}^{3}p_{i}}{(2\pi)^{3}2E_{i}} \ .
\end{eqnarray}
Here, we evaluate all equations in the rest frame of the decaying particle $S$ with $p=(M,\textbf{0})$.\ $\mathcal{S}$ is a symmetry factor and $n$ is the number of particles created in the final state. For a general discussion of these formulas see for example Ref.\ \cite{gross}.\ In our case the symmetry factor is just one half -- since the directions of the outgoing momenta are determined by conservation laws, only a half sphere in position space needs to be taken into account. There is no angular dependence as a consequence of Lorentz invariance: decay products at rest of a spinless interaction have no preferred direction in which they are emitted.

The form factor $F(s)$ is not present at the Lagrangian level, hence it would not appear in the invariant amplitude\ (\ref{eq:SphiphiAmp}) of our simple theory. It is included by hand, however, this can be avoided when allowing non-local interactions in the Lagrangian:
\vspace{-0.1cm}
\begin{equation}
\mathcal{L}_{\text{int}}=gS(x)\phi(x)\phi(x)\rightarrow gS(x)\int\text{d}^{4}y \ \phi(x+y/2)\phi(x-y/2)\Phi(y) \ ,
\vspace{-0.1cm}
\end{equation}
where $\Phi(y)$ will be related to the form factor.\ 
When including such a non-local interaction term, the tree-level decay width is modified by the form factor only as a multiplicative factor.\ In order to see that, we start by writing down the lowest orders of the $\hat S$-matrix in the non-local case:\footnote{In this thesis we adopt the Feynman rules used by Peskin and Schroeder\ \cite{peskin}.}
\begin{eqnarray}
\hat{S} & = & \mathbbm{1}+ig\int\text{d}^{4}x \ \mathcal{T}\big{\{}\textbf{:}S(x)\int\text{d}^{4}y \ \phi(x+y/2)\phi(x-y/2)\Phi(y)\textbf{:}\big{\}} \nonumber \\
& = & \mathbbm{1}+\hat{S}^{(1)} \ ,
\end{eqnarray}
with the time-ordering operator $\mathcal{T}$ and where the dots mark the normal ordering prescription. The crossed-out terms in the resulting transition matrix element,
\vspace{-0.1cm}
\begin{eqnarray}
\langle\text{final}|\hat{S}|\text{initial}\rangle & = & \langle\textbf{p}_{1}\textbf{p}_{2}|\hat{S}^{(1)}|\textbf{p}\rangle \nonumber \\
& = & \langle\textbf{p}_{1}\textbf{p}_{2}| \ ig\int\text{d}^{4}x \ \textbf{:} \ \xcancel{S^{(+)}\int\text{d}^{4}y \ \phi^{(+)}\phi^{(+)}\Phi(y)} \nonumber \\
&  & + \ \xcancel{S^{(+)}\int\text{d}^{4}y \ \phi^{(+)}\phi^{(-)}\Phi(y)}+\xcancel{S^{(+)}\int\text{d}^{4}y \ \phi^{(-)}\phi^{(+)}\Phi(y)} \nonumber \\
&  & + \ \xcancel{S^{(+)}\int\text{d}^{4}y \ \phi^{(-)}\phi^{(-)}\Phi(y)}+S^{(-)}\int\text{d}^{4}y \ \phi^{(+)}\phi^{(+)}\Phi(y) \nonumber \\
&  & + \ \xcancel{S^{(-)}\int\text{d}^{4}y \ \phi^{(+)}\phi^{(-)}\Phi(y)}+\xcancel{S^{(-)}\int\text{d}^{4}y \ \phi^{(-)}\phi^{(+)}\Phi(y)} \nonumber \\
&  & + \ \xcancel{S^{(-)}\int\text{d}^{4}y \ \phi^{(-)}\phi^{(-)}\Phi(y)} \ \textbf{:} \ |\textbf{p}\rangle \ ,
\end{eqnarray}
give no contribution because the creation and annihilation operators combine in such a way that their scalar product vanishes.\ Note that the superscript at the $S$ and $\phi$ fields denotes those parts of the field that contain a creation or annihilation operator, for instance
\vspace{-0.3cm}
\begin{eqnarray}
\phi^{(+)} & \equiv & \phi^{(+)}(x+y/2) \nonumber \\
& = &  \int\frac{\text{d}^{3}p}{(2\pi)^{3}}\frac{1}{\sqrt{2E_{\textbf{p}}}} \ b_{\textbf{p}}^{\dagger}e^{ip\cdot \left(x+\frac{y}{2}\right)} \ .
\end{eqnarray}
\\[-0.3cm]
Writing out the Fourier expansion of the fields in full detail and using general commutation relations for the bosonic creation and annihilation operators, the remaining steps are lengthy but straightforward:
\begin{eqnarray}
& = & ig\int\text{d}^{4}x\int\text{d}^{3}p_{1}^{\prime}\int\text{d}^{3}p_{2}^{\prime}\int\text{d}^{3}p^{\prime} \ \frac{\sqrt{2E_{\textbf{p}_{1}}2E_{\textbf{p}_{2}}2E_{\textbf{p}}}}{\sqrt{2E_{\textbf{p}_{1}^{\prime}}2E_{\textbf{p}_{2}^{\prime}}2E_{\textbf{p}^{\prime}}}(2\pi)^{9}} \nonumber \\[0.15cm]
&  & \times \int\text{d}^{4}y \ e^{ip_{1}^{\prime}\cdot\left(x+\frac{y}{2}\right)}e^{ip_{2}^{\prime}\cdot\left(x-\frac{y}{2}\right)}e^{-ip^{\prime}\cdot x}\Phi(y)\langle0|b_{\textbf{p}_{2}}b_{\textbf{p}_{1}}b_{\textbf{p}_{1}^{\prime}}^{\dagger}b_{\textbf{p}_{2}^{\prime}}^{\dagger}a_{\textbf{p}^{\prime}}a_{\textbf{p}}^{\dagger}|0\rangle \nonumber \\[0.3cm]
& = & ig\int\text{d}^{4}x\int\text{d}^{3}p_{1}^{\prime}\int\text{d}^{3}p_{2}^{\prime}\int\text{d}^{3}p^{\prime} \ \frac{\sqrt{2E_{\textbf{p}_{1}}2E_{\textbf{p}_{2}}2E_{\textbf{p}}}}{\sqrt{2E_{\textbf{p}_{1}^{\prime}}2E_{\textbf{p}_{2}^{\prime}}2E_{\textbf{p}^{\prime}}}}\int\text{d}^{4}y \ e^{ip_{1}^{\prime}\cdot\left(x+\frac{y}{2}\right)}e^{ip_{2}^{\prime}\cdot\left(x-\frac{y}{2}\right)}e^{-ip^{\prime}\cdot x}\Phi(y) \nonumber \\[0.15cm]
&  & \times \ \Big[\delta^{(3)}(\textbf{p}^{\prime}-\textbf{p})\delta^{(3)}(\textbf{p}_{1}-\textbf{p}_{2}^{\prime})\delta^{(3)}(\textbf{p}_{2}-\textbf{p}_{1}^{\prime})+\delta^{(3)}(\textbf{p}^{\prime}-\textbf{p})\delta^{(3)}(\textbf{p}_{1}-\textbf{p}_{1}^{\prime})\delta^{(3)}(\textbf{p}_{2}-\textbf{p}_{2}^{\prime})\Big] \nonumber \\[0.3cm]
& = & ig\int\text{d}^{4}x\int\text{d}^{4}y\left[e^{ip_{2}\cdot\left(x+\frac{y}{2}\right)}e^{ip_{1}\cdot\left(x-\frac{y}{2}\right)}e^{-ip\cdot x}+e^{ip_{1}\cdot\left(x+\frac{y}{2}\right)}e^{ip_{2}\cdot\left(x-\frac{y}{2}\right)}e^{-ip\cdot x}\right]\Phi(y) \nonumber \\[0.3cm]
& = & ig\int\text{d}^{4}x \ e^{i(p_{1}+p_{2}-p)\cdot x}\int\text{d}^{4}y\left[e^{iy\cdot\left(\frac{p_{2}-p_{1}}{2}\right)}+e^{iy\cdot\left(\frac{p_{1}-p_{2}}{2}\right)}\right]\Phi(y) \ .
\end{eqnarray}
By identifying the form factor $F(u)$ as the Fourier transform of $\Phi(y)$ with dependence only on the magnitude of the corresponding three-momentum, the invariant amplitude is changed only by the former:
\begin{eqnarray}
\langle\text{final}|\hat{S}|\text{initial}\rangle & = & 2igF(|\textbf{p}_{1}|)(2\pi)^{4}\delta^{(4)}(p_{1}+p_{2}-p) \\
& \stackrel{!}{=} & -i\mathcal{M}(2\pi)^{4}\delta^{(4)}(p-p_{1}-p_{2}) \ \nonumber , \\[0.15cm]
\Rightarrow \ \ \ -i\mathcal{M} & = & 2igF(|\textbf{p}_{1}|) \ ,
\end{eqnarray}
where we also have used that $\textbf{p}_{1}=-\textbf{p}_{2}$.

The integral from Eq.\ (\ref{eq:optical_simple}) becomes
\begin{eqnarray}
\Gamma^{\text{tree}}(s=M^{2}) & = & \frac{g^{2}}{(2\pi)^{2}M}\int\text{d}^{3}p_{1}\int\text{d}^{3}p_{2} \ \frac{F^{2}(|\textbf{p}_{1}|)}{2E_{\textbf{p}_{1}}2E_{\textbf{p}_{2}}}\underbrace{\delta^{(4)}(p-p_{1}-p_{2})}_{=\delta^{(3)}(\textbf{p}_{1}+\textbf{p}_{2})\delta(M-E_{\textbf{p}_{1}}-E_{\textbf{p}_{2}})} \\
& = & \frac{g^{2}}{(2\pi)^{2}M}\int\text{d}^{3}p_{1} \ \frac{F^{2}(|\textbf{p}_{1}|)}{(2E_{\textbf{p}_{1}})^{2}} \ \delta(M-2E_{\textbf{p}_{1}}) \nonumber \\
\nonumber \\
& = & \frac{4\pi g^{2}}{(2\pi)^{2}M}\int_{0}^{\infty}\text{d}u \ \frac{Mu^{2}F^{2}(u)}{16(u^{2}+m^{2})\sqrt{\frac{M^{2}}{4}-m^{2}}} \ \delta\big(u-\sqrt{M^{2}/4-m^{2}}\big) \ , \nonumber
\end{eqnarray}
with $u=|\textbf{p}_{1}|$.\ A real-valued expression requires a threshold value $2m$, so we add a Heaviside step function to the result:
\begin{equation}
\Gamma^{\text{tree}}(s=M^{2}) = \frac{k(s=M^{2})}{4\pi M^{2}}g^{2}F^{2}\left(\sqrt{\frac{M^{2}}{4}-m^{2}}\right)\Theta(M-2m) \ ,
\end{equation}
with the absolute value $|\textbf{p}_{1}|=|\textbf{p}_{2}|=k(s=M^{2})$ of the two outgoing $\phi$ particles' three-momenta.\ This is in fact the same expression we have already found in Eq.\ (\ref{eq:tree_simple}), if one evaluates the width for general values of the Mandelstam variable $s$.

\section{Derivative interaction of the form $\mathcal{L}_{\text{int}}=gS(\partial_{\mu}\phi)(\partial^{\mu}\phi)$}
\subsection{Canonical quantization}
We compute in this section the one-loop self-energy in the case of derivative interactions.
We first derive the interacting part of the Hamiltonian from the Lagrangian via
a Legendre transformation.\ We shall see that the derivative interactions give
rise to new interaction vertices.\ It will be demonstrated that, in a perturbative calculation of the 
one-loop self-energy, these terms are necessary to cancel additional terms arising from contractions of
gradients of fields. This proves that, at least at one-loop level, it is justified to apply
standard Feynman rules with the derivative interaction in ${\cal L}_{\text{int}}$.\
It will be also shown that a computation of the self-energy via the dispersion relation in Eq.\ (\ref{eq:Redispersionrel}) 
may require subtraction constants to agree with the perturbative calculation using Feynman rules.

Let us consider the same theory as before with the two scalar fields $S$ and $\phi$, but where the interaction Lagrangian now contains gradients of the $\phi$ fields:
\begin{equation}
\mathcal{L}_{\text{int}} = gS(\partial_{\mu}\phi)(\partial^{\mu}\phi) \ .
\label{eq:Lint}
\end{equation}
For perturbative calculations of $\hat{S}$-matrix elements or Green's functions, however, 
one needs the interaction part of the Hamilton operator in the interaction picture. 
We derive this operator via a Legendre transformation of ${\cal L}$ 
and subsequent canonical quantization in the interaction picture.\ As a byproduct of this calculation, we will explicitly show that the derivative interactions 
invalidate the commonly used relation 
$\mathcal{H}_{\text{int}}=-\mathcal{L}_{\text{int}}$\ \cite{reinhardt}. 

The canonically conjugate fields are 
\begin{eqnarray}
\pi_{S} & = & \frac{\partial\mathcal{L}}{\partial(\partial_{0}S)} \hspace{0.19cm} = \hspace{0.2cm} \partial^{0}S \ , \nonumber \\
\pi_{\phi} & = & \frac{\partial\mathcal{L}}{\partial(\partial_{0}\phi)} \hspace{0.22cm} = \hspace{0.2cm} \partial^{0}\phi+2gS\partial^{0}\phi \hspace{0.2cm} = \hspace{0.18cm} (1+2gS) \partial^0 \phi \ .
\end{eqnarray}
The Hamiltonian is defined via a Legendre transformation of ${\cal L}$,
\begin{eqnarray}
\mathcal{H} & = & \pi_{S}\partial^{0}S+\pi_{\phi}\partial^{0}\phi-\mathcal{L} \nonumber \\
& = & \frac{1}{2}\pi_{S}\pi_{S} +\frac{1}{2}\vec{\nabla}S\cdot\vec{\nabla}S
+\frac{1}{2}m_{0}^{2}S^{2}
+\frac{1}{2}\pi_{\phi}\pi_{\phi}(1+2gS)^{-1} \nonumber \\
& & + \ \frac{1}{2}\vec{\nabla}\phi\cdot\vec{\nabla}\phi
+\frac{1}{2}m^{2}\phi^{2}+gS\vec{\nabla}\phi\cdot\vec{\nabla}\phi \ .
\end{eqnarray}
For a perturbative calculation, we need to expand the denominator 
$(1+2gS)^{-1}$ and obtain the interaction part of the Hamiltonian as
\begin{equation}
\mathcal{H}_{\text{int}}=-gS\pi_{\phi}\pi_{\phi}+gS\vec{\nabla}\phi\cdot\vec{\nabla}\phi
+2g^{2}S^{2}\pi_{\phi}\pi_{\phi}+\mathcal{O}(g^{3}) \ .
\end{equation}
We may now quantize in the Heisenberg picture (indicated by a superscript $H$
at the respective operators). This is commonly done by
promoting fields to operators $S \rightarrow\hat S^{H}$, $\phi\rightarrow\hat\phi^{H}$, 
$\pi_{S}\rightarrow\hat\pi_{S}^{H}$, $\pi_{\phi}\rightarrow\hat\pi_{\phi}^{H}$,
and postulating certain commutation relations for them.
However, in perturbation theory we need the operators in the interaction picture.
The following relations hold\ \cite{reinhardt}:
\begin{align}
\hat S^{I} &= \hat U \hat S^{H} \hat U^\dagger \ , & \hat\phi^{I} &= \hat U \hat \phi^{H} \hat U^\dagger \ , & \hat \pi_S^{I} &= \hat U \hat \pi_S^{H} \hat U^\dagger \ , & \hat \pi_\phi^{I} &= \hat U \hat \pi_\phi^{H} \hat U^\dagger \ ,
\end{align}
where $\hat U=e^{i\hat H_{0}t} e^{-i\hat Ht}$ is the time-evolution
operator that relates operators in the Heisenberg picture with those in the
interaction picture.\
Finally, after replacing $\hat\pi^{I}_{S} = \partial^{0}\hat S^{I}$ and $\hat\pi^{I}_{\phi} = \partial^{0}\hat\phi^{I}$, this results in
\begin{equation}
\hat{\mathcal{H}}_{\text{int}}^{I}=-\hat{\mathcal{L}}_{\text{int}}^{I}
+2g^{2}\hat S^{I}\hat S^{I}(\partial_{0}\hat\phi^{I})(\partial^{0}\hat\phi^{I})+\mathcal{O}(g^{3}) \ .
\label{eq:Hint}
\end{equation}
As advertised, the second term shows that 
$\hat{\mathcal{H}}_{\text{int}}\neq-\hat{\mathcal{L}}_{\text{int}}$.\ This term
corresponds to a four-point vertex, so it will not appear in the tree-level decay width, which reads
\begin{equation}
\Gamma^{\text{tree}}(s=M^{2}) = \frac{k(s=M^{2})}{4\pi M^{2}}g^{2}\left(-\frac{M^{2}-2m^{2}}{2}\right)F^{2}\left(\sqrt{\frac{M^{2}}{4}-m^{2}}\right)\Theta(M-2m) \ .
\end{equation}
In contrast, in the one-loop self-energy, that term will give rise to an additional tadpole contribution.

\subsection{Perturbative calculation of the one-loop self-energy}
We now turn to the self-energy $\Pi(s)$ of the field $S$.\ At one-loop level,
the Feynman rules applied to $\hat{\cal H}_{\text{int}}$ tell us that 
we will have two contributions. The first contribution
comes from taking two three-point vertices of $\hat{\cal L}_{\text{int}}$, where the $\hat{\phi}$
legs are joined in a manner which gives a $1$PI diagram. A covariant derivative acts
on each $\hat{\phi}$ leg at each vertex. The second contribution is a tadpole term
arising from the four-point vertex in Eq.\ (\ref{eq:Hint}), which has two time derivatives 
on the internal leg. This can be graphically depicted as follows:
\vspace{0.3cm}
\begin{equation} \label{eq:Pigraph}
\includegraphics[scale=0.49]{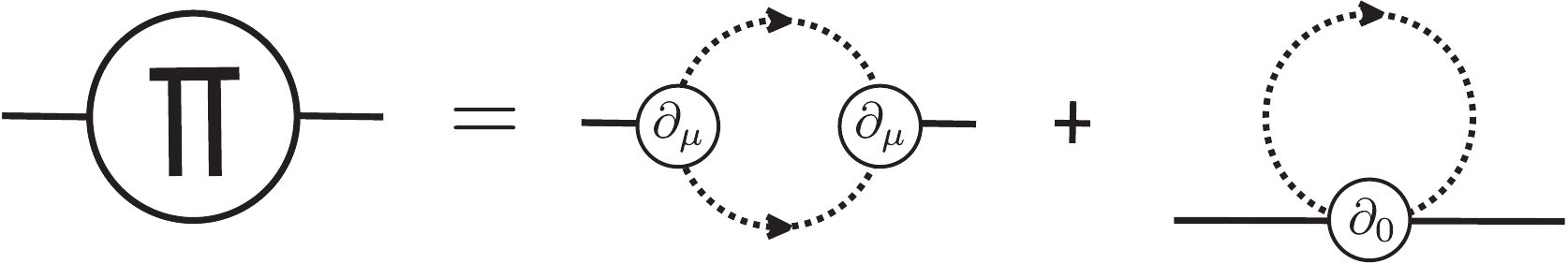}
\end{equation}
The usual Feynman propagator is defined as a contraction of two fields:
\begin{eqnarray} \label{eq:Feynmanprop}
\bcontraction{}{\hat{\phi}}{(x_{1})}{\hat{\phi}}\hat{\phi}(x_{1})\hat{\phi}(x_{2}) & = &
\langle0|\mathcal{T}\big{\{}\hat{\phi}(x_{1})\hat{\phi}(x_{2})\big{\}}|0\rangle \nonumber \\[0.1cm]
& = & \langle0|\hat{\phi}(x_{1})\hat{\phi}(x_{2})|0\rangle\Theta(x_{1}^{0}-x_{2}^{0})+\langle0|\hat{\phi}(x_{2})\hat{\phi}(x_{1})|0\rangle\Theta(x_{2}^{0}-x_{1}^{0}) \nonumber \\[0.1cm]
& = & i\Delta_{F}^{\phi}(x_{1}-x_{2}) \hspace{0.2cm} = \hspace{0.2cm} i\int\frac{\text{d}^{4}p}{(2\pi)^{4}}\frac{e^{-ip\cdot(x_{1}-x_{2})}}{p^{2}-m^{2}+i\epsilon} \ .
\end{eqnarray}
However, in the tadpole diagram, we have the contraction of two fields,
on each of which acts a time derivative. Because time-ordering has no
effect at the same space-time point, we obtain
\vspace{-0.19cm}
\begin{eqnarray}  \label{eq:tadpole1}
\langle0|\mathcal{T}\big{\{}\partial_{0}^{x}\hat{\phi}(x)\partial^{0,x}\hat{\phi}(x)\big{\}}|0\rangle
& = & \langle0|{}\partial_{0}^{x}\hat{\phi}(x)\partial^{0,x}\hat{\phi}(x)|0\rangle \nonumber \\
& = & i \int\frac{\text{d}^{4}p}{(2\pi)^{4}}\frac{E_{\textbf{p}}^{2}}{p^{2}-m^{2}+i\epsilon} \ .
\end{eqnarray}
To get this result, we inserted the standard Fourier decomposition of the
field operators,
\begin{equation}
\hat{\phi} (x) = \int\frac{\text{d}^{3}p}{(2\pi)^{3}}\frac{1}{\sqrt{2E_{\textbf{p}}}}
\Big(\hat{a}_{\textbf{p}}^{\dagger}e^{ip\cdot x}+\hat{a}_{\textbf{p}}e^{-ip\cdot x}\Big)\;,
\end{equation}
where $p^0 = E_{\bf p} = \sqrt{{\bf p}^2 +m^2}$ is the on-shell energy.
Thus, a time derivative acting
on a field operator brings down a factor of $\pm i$ times the on-shell energy in the 
corresponding Fourier representation.

One should realize that the result\ (\ref{eq:tadpole1}) is identical if we just act with
the time derivatives on the standard Feynman propagator\ (\ref{eq:Feynmanprop}):
\begin{eqnarray} \label{eq:tadpole2}
\partial_{0}^{x}\partial^{0,x} \langle0|{}\hat{\phi}(x)\hat{\phi}(x)|0\rangle
& = & \lim_{x_1\rightarrow x_2} \partial_{0}^{x_1}\partial^{0,x_2} 
\langle0|\mathcal{T}\big{\{}\hat{\phi}(x_1)\hat{\phi}(x_2)\big{\}}|0\rangle \nonumber \\
& = & i \int\frac{\text{d}^{4}p}{(2\pi)^{4}}\frac{E_{\textbf{p}}^{2}}{p^{2}-m^{2}+i\epsilon} \ .
\end{eqnarray}
In order to prove this, it is convenient to first perform the $p_0$ integration in
Eq.\ (\ref{eq:Feynmanprop}) and then take the time derivatives.
The equivalence of Eqs.\ (\ref{eq:tadpole1}) and (\ref{eq:tadpole2}) is graphically depicted as
\begin{table}[h]
\hspace{0.45cm}
\begin{tabular}{*{2}{m{0.48\textwidth}}}
\hspace{0.65cm} $\langle0|\mathcal{T}\big{\{}\partial_{0}^{x}\phi(x)\partial^{0,x}\phi(x)\big{\}}|0\rangle \ \sim$ & \vspace{0.27cm} \hspace{-1.75cm} \includegraphics[scale=0.5]{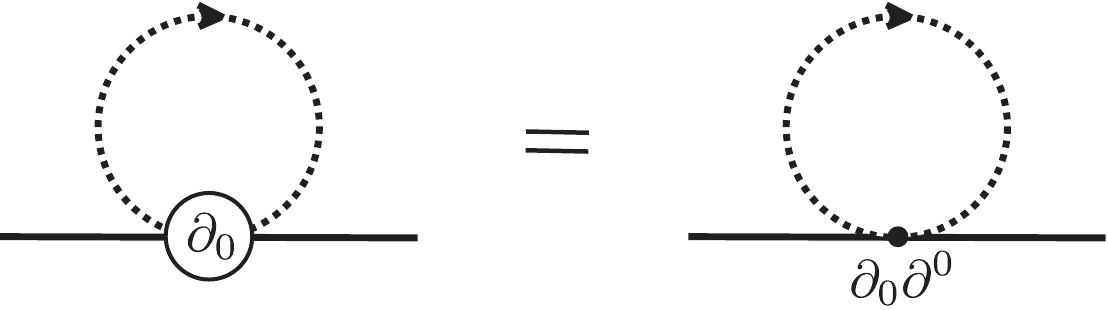} \ \ \ . \ \ \ \ \tagarray\label{eq:Pi12}
\end{tabular}
\end{table}
\\
In the perturbative series of the full propagator of the $S$ field, this tadpole contribution
appears in combination with two free $S$ field propagators (we omit the
superscript $S$):
\begin{align}
\frac{-2\hspace{0.01cm}i\hspace{0.01cm}g^2}{1!}\cdot2 \int\text{d}x^{\prime} \ i\Delta(x_{1}-x^{\prime})
\langle0|\mathcal{T}\big{\{}\partial_{0}^{x^{\prime}}\phi(x^{\prime})\partial^{0,x^{\prime}}
\phi(x^{\prime})\big{\}}|0\rangle\hspace{0.02cm}i\Delta(x^{\prime}-x_{2}) \nonumber \\[0.1cm]
= \ i\hspace{0.02cm}2g^{2}\cdot2\int\text{d}x^{\prime} \ \Delta(x_{1}-x^{\prime})
\Delta(x^{\prime}-x_{2})\langle0|\mathcal{T}\big{\{}\partial_{0}^{x^{\prime}}\phi(x^{\prime})
\partial^{0,x^{\prime}}\phi(x^{\prime})\big{\}}|0\rangle \ .
\label{eq:tadpole_pos}
\end{align}
The factor $-2i g^2$ is the factor accompanying the four-point vertex, 
see Eq.\ (\ref{eq:Hint}).\
A factor of $2$ arises because each $S$ propagator can be joined
with either one of the $S$ legs at the vertex.

We now compute the first diagram on the right side of Eq.\ (\ref{eq:Pigraph}). 
To this end, we need contractions of gradients of the $\phi$ fields.\ These
can be expressed in terms of gradients acting on the standard
Feynman propagator.
The gradient of the Feynman propagator\ (\ref{eq:Feynmanprop}) is
\begin{eqnarray}
i\hspace{0.01cm}\partial_{\nu}^{x_{2}}\Delta_{F}^{\phi}(x_{1}-x_{2}) & = & \partial_{\nu}^{x_{2}}
\langle0|\mathcal{T}\big{\{}\phi(x_{1})\phi(x_{2})\big{\}}|0\rangle \nonumber \\
& = & \langle0|\mathcal{T}\big{\{}\phi(x_{1})\partial_{\nu}^{x_{2}}\phi(x_{2})\big{\}}|0\rangle-\eta_{\nu0}\delta(x_{1}^{0}-x_{2}^{0}) 
\langle0|\underbrace{\big[\phi(x_{1}),\phi(x_{2})\big]}_{=0}|0\rangle \ , \nonumber \\
\end{eqnarray}
where we have used the explicit definition of the time-ordered product.\
The last term vanishes on account of the delta function, since it is an equal-time commutator
of two $\phi$ fields\ \cite{reinhardt}.\
Taking another gradient leads to
\begin{eqnarray}
i\hspace{0.02cm}\partial_{\mu}^{x_{1}}\partial_{\nu}^{x_{2}}\Delta_{F}^{\phi}(x_{1}-x_{2}) 
& = & \partial_{\mu}^{x_{1}}\langle0|\mathcal{T}\big{\{}\phi(x_{1})\partial_{\nu}^{x_{2}}
\phi(x_{2})\big{\}}|0\rangle \nonumber \\
& = & \langle0|\mathcal{T}\big{\{}\partial_{\mu}^{x_{1}}\phi(x_{1})\partial_{\nu}^{x_{2}}
\phi(x_{2})\big{\}}|0\rangle \nonumber \\
&  & + \ \eta_{\mu0}\delta(x_{1}^{0}-x_{2}^{0})\langle0|\underbrace{\big[\phi(x_{1}),
\partial_{\nu}^{x_{2}}\phi(x_{2})\big]}_{\neq0}|0\rangle \ .
\end{eqnarray}
Now, the second term does not vanish if $\nu=0$, because 
it involves a commutator of a field with its canonically conjugate field,
\begin{equation}
\eta_{\mu0}\delta(x_{1}^{0}-x_{2}^{0})\langle0|\big[\phi(x_{1}),\partial_{\nu}^{x_{2}}
\phi(x_{2})\big]|0\rangle = i\hspace{0.02cm}\eta_{\mu 0}\eta_{\nu 0}\delta^{(4)}(x_{1}-x_{2}) \ .
\end{equation}
An explicit calculation yields the same result:
\begin{eqnarray}
\eta_{\mu0}\delta(x_{1}^{0}-x_{2}^{0})\langle0|\big[\phi(x_{1}),\partial_{\nu}^{x_{2}}\phi(x_{2})\big]|0\rangle & = & \eta_{\mu0}\delta(x_{1}^{0}-x_{2}^{0})\int\frac{\text{d}^{3}p}{(2\pi)^{3}}\frac{1}{2E_{\textbf{p}}}ip_{\nu} \nonumber \\
& & \times\left[e^{i\textbf{p}\cdot(\textbf{x}_{1}-\textbf{x}_{2})}+e^{-i\textbf{p}\cdot(\textbf{x}_{1}-\textbf{x}_{2})}\right] \nonumber \\
& = & i\eta_{\mu0}\eta_{\nu0}\delta(x_{1}^{0}-x_{2}^{0})\int\frac{\text{d}^{3}p}{(2\pi)^{3}}\frac{2p_{0}}{2E_{\textbf{p}}}e^{i\textbf{p}\cdot(\textbf{x}_{1}-\textbf{x}_{2})} \nonumber \\
& = & i\eta_{\mu 0}\eta_{\nu 0}\delta(x_{1}^{0}-x_{2}^{0})\delta^{(3)}(\textbf{x}_{1}-\textbf{x}_{2}) \nonumber \\
& = & i\eta_{\mu 0}\eta_{\nu 0}\delta^{(4)}(x_{1}-x_{2}) \ .
\end{eqnarray}
We first have to use that $x_{1}^{0}=x_{2}^{0}$ because of the delta function in front of the commutator. Then we notice that the integral over the second exponential vanishes for spatial indices $\nu\ne0$, and that $p_{0}\equiv E_{\textbf{p}}$.

Collecting terms, we can express the contraction of two gradients of
the $\phi$ field as
\begin{equation} \label{eq:gradprop}
\langle0|\mathcal{T}\big{\{}\partial_{\mu}^{x_{1}}\phi(x_{1})\partial_{\nu}^{x_{2}}
\phi(x_{2})\big{\}}|0\rangle = i\hspace{0.02cm}\partial_{\mu}^{x_{1}}\partial_{\nu}^{x_{2}}\Delta_{F}^{\phi}(x_{1}-x_{2})-i\hspace{0.02cm}\eta_{\mu 0}\eta_{\nu 0}\delta^{(4)}(x_{1}-x_{2}) \ .
\end{equation}
In the perturbative series of the full propagator for the $S$ field,
the first diagram in Eq.\ (\ref{eq:Pigraph}) appears in a combination with such two-derivative $\phi$ field propagators:
\begin{eqnarray}
&  & \frac{(-ig)^{2}}{2!} \cdot 2\cdot 2
\int\text{d}x^{\prime}\int\text{d}x^{\prime\prime} \ i\Delta(x_{1}-x^{\prime})
i\Delta(x^{\prime\prime}-x_{2}) \nonumber \\[0.1cm]
&  & \times \ \langle0|\mathcal{T}\big{\{}\partial_{\mu}^{x^{\prime}}\phi(x^{\prime})
\partial_{\nu}^{x^{\prime\prime}}\phi(x^{\prime\prime})\big{\}}|0\rangle\langle0|
\mathcal{T}\big{\{}\partial^{\mu,x^{\prime}}\phi(x^{\prime})\partial^{\nu,x^{\prime\prime}}
\phi(x^{\prime\prime})\big{\}}|0\rangle \ .
\end{eqnarray}
Two factors of $-ig$ originate from the three-point vertices in $\hat{\cal L}_{\text{int}}$.\ 
The factor of $1/2!$ arises because the diagram is second order in perturbation theory. 
A factor of $2$ arises because each $S$ propagator can be joined
with one of the $S$ legs at the vertex.
Finally, another factor of $2$ comes from the fact that the two $\phi$ lines
at one vertex can be joined with corresponding lines at the other vertex in two
different ways. Successively inserting Eq.\ (\ref{eq:gradprop}), we compute
\begin{eqnarray}
&  & 2g^{2}\int\text{d}x^{\prime}\int\text{d}x^{\prime\prime} \ \Delta(x_{1}-x^{\prime})
\Delta(x^{\prime\prime}-x_{2}) 
\left[i\hspace{0.02cm}\partial_{\mu}^{x^{\prime}}\partial_{\nu}^{x^{\prime\prime}}
\Delta_{F}^{\phi}(x^{\prime}-x^{\prime\prime})-i\hspace{0.02cm}\eta_{\mu 0}\eta_{\nu 0}\delta^{(4)}(x^{\prime}
-x^{\prime\prime})\right] \nonumber \\[0.13cm]
&  & \times \ \langle0|\mathcal{T}\big{\{}\partial^{\mu,x^{\prime}}\phi(x^{\prime})
\partial^{\nu,x^{\prime\prime}}\phi(x^{\prime\prime})\big{\}}|0\rangle \nonumber \\
\nonumber \\
& = & 2g^{2}\int\text{d}x^{\prime}\int\text{d}x^{\prime\prime} \ \Delta(x_{1}-x^{\prime})
\Delta(x^{\prime\prime}-x_{2}) \ i\hspace{0.02cm}\partial_{\mu}^{x^{\prime}}
\partial_{\nu}^{x^{\prime\prime}}\Delta_{F}^{\phi}(x^{\prime}-x^{\prime\prime})
\langle0|\mathcal{T}\big{\{}\partial^{\mu,x^{\prime}}\phi(x^{\prime})
\partial^{\nu,x^{\prime\prime}}\phi(x^{\prime\prime})\big{\}}|0\rangle \nonumber \\
& & - \ i\hspace{0.02cm}2g^{2}\int\text{d}x^{\prime} \ \Delta(x_{1}-x^{\prime})\Delta(x^{\prime}
-x_{2})\langle0|\mathcal{T}\big{\{}\partial^{0,x^{\prime}}\phi(x^{\prime})\partial^{0,x^{\prime}}
\phi(x^{\prime})\big{\}}|0\rangle \nonumber \\
\nonumber \\
& = & - \ 2g^{2}\int\text{d}x^{\prime}\int\text{d}x^{\prime\prime} \ \Delta(x_{1}-x^{\prime})
\Delta(x^{\prime\prime}-x_{2}) \ \partial_{\mu}^{x^{\prime}}\partial_{\nu}^{x^{\prime\prime}}
\Delta_{F}^{\phi}
(x^{\prime}-x^{\prime\prime})\partial^{\mu,x^{\prime}}\partial^{\nu,x^{\prime\prime}}
\Delta_{F}^{\phi}(x^{\prime}-x^{\prime\prime}) \nonumber \\
& & - \ i\hspace{0.02cm}2g^{2}\int\text{d}x^{\prime}\int\text{d}x^{\prime\prime} \ 
\Delta(x_{1}-x^{\prime})\Delta(x^{\prime\prime}-x_{2})\hspace{0.02cm}i\hspace{0.02cm}
\partial_{0}^{x^{\prime}}\partial_{0}^{x^{\prime\prime}}\Delta_{F}^{\phi}(x^{\prime}
-x^{\prime\prime})\delta^{(4)}(x^{\prime}-x^{\prime\prime}) \nonumber \\
& & - \ i\hspace{0.02cm}2g^{2}\int\text{d}x^{\prime} \ \Delta(x_{1}-x^{\prime})
\Delta(x^{\prime}-x_{2})\langle0|\mathcal{T}\big{\{}\partial^{0,x^{\prime}}\phi(x^{\prime})
\partial^{0,x^{\prime}}\phi(x^{\prime})\big{\}}|0\rangle \ .
\end{eqnarray}
With Eq.\ (\ref{eq:Pi12}) one realizes that the last two terms are identical. The final
result is
\begin{eqnarray}
&  & - \ 2g^{2}\int\text{d}x^{\prime}\int\text{d}x^{\prime\prime} \ \Delta(x_{1}-x^{\prime})
\Delta(x^{\prime\prime}-x_{2})\partial_{\mu}^{x^{\prime}}\partial_{\nu}^{x^{\prime\prime}}
\Delta_{F}^{\phi}(x^{\prime}-x^{\prime\prime})\partial^{\mu,x^{\prime}}
\partial^{\nu,x^{\prime\prime}}\Delta_{F}^{\phi}(x^{\prime}-x^{\prime\prime}) \nonumber \\
&  & \ \ \ \ - \ i\hspace{0.02cm}2g^{2}\cdot 2\int\text{d}x^{\prime} \ \Delta(x_{1}-x^{\prime})
\Delta(x^{\prime}-x_{2})\langle0|\mathcal{T}\big{\{}\partial^{0,x^{\prime}}\phi(x^{\prime})
\partial^{0,x^{\prime}}\phi(x^{\prime})\big{\}}|0\rangle \ ,
\end{eqnarray}
which can be graphically depicted as
\vspace{0.25cm}
\begin{equation} \label{eq:Piregular}
\includegraphics[scale=0.5]{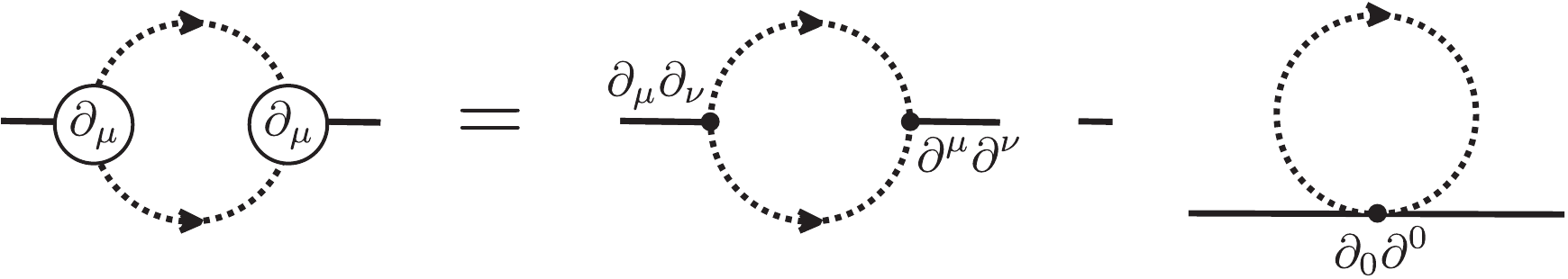} \ \ \ .
\end{equation}
Obviously, the second diagram on the right side cancels the tadpole contribution, Eq.\ (\ref{eq:Pi12}), 
in the one-loop self-energy from Eq.\ (\ref{eq:Pigraph}).

In summary, a derivative interaction in ${\cal L}_{\text{int}}$ produces an additional 
term in the interaction Hamiltonian and thus, after quantization, an additional vertex which
has to be taken into account in perturbative calculations via Feynman rules. This vertex leads to a tadpole diagram 
in the one-loop self-energy.\ Nevertheless, 
carefully computing contractions between gradients of the field operators we
demonstrated that these give a term which exactly cancels the tadpole diagram.
The remaining contribution is exactly equal to the self-energy when computed
with standard Feynman rules using $\hat{\cal L}_{\text{int}}$ and
derivatives acting on the usual Feynman propagators.

We have not delivered a rigorous proof of this tadpole cancellation to all orders in 
perturbation theory. However, since this seems to be just a demonstration of the validity 
of Matthews' theorem\ \cite{matthews}, which was investigated {\em e.g.}\ in 
Refs.\ \cite{duncan,barua,simon,knetter}, we also expect a similar cancellation to work
at higher-orders in perturbation theory.

\subsection{One-loop self-energy from a dispersion relation}
The second way to compute the self-energy is by applying the dispersion relation\ (\ref{eq:Redispersionrel}).
To this end, one needs the imaginary part of the self-energy in order to obtain
the real part; the imaginary part can be inferred from the decay width through the
optical theorem.\ For the one-loop self-energy, the cutting rules imply that the decay width
needs only to be known at tree-level:
\vspace{-2.2cm}
\begin{table}[h]
\hspace{1.2cm}
\begin{tabular}{*{4}{m{2.48\textwidth}}}
\hspace{0.31cm} $\text{d}\Gamma$ \hspace{3.10cm} $ = \ -\operatorname{Im}\bigg($ \hspace{1.29cm} $\bigg) \ = \ -\operatorname{Im}\bigg($ \hspace{1.34cm} $\bigg)$ \ \ . & \vspace{0.12cm} \hspace{-37.01cm} \includegraphics[scale=0.20]{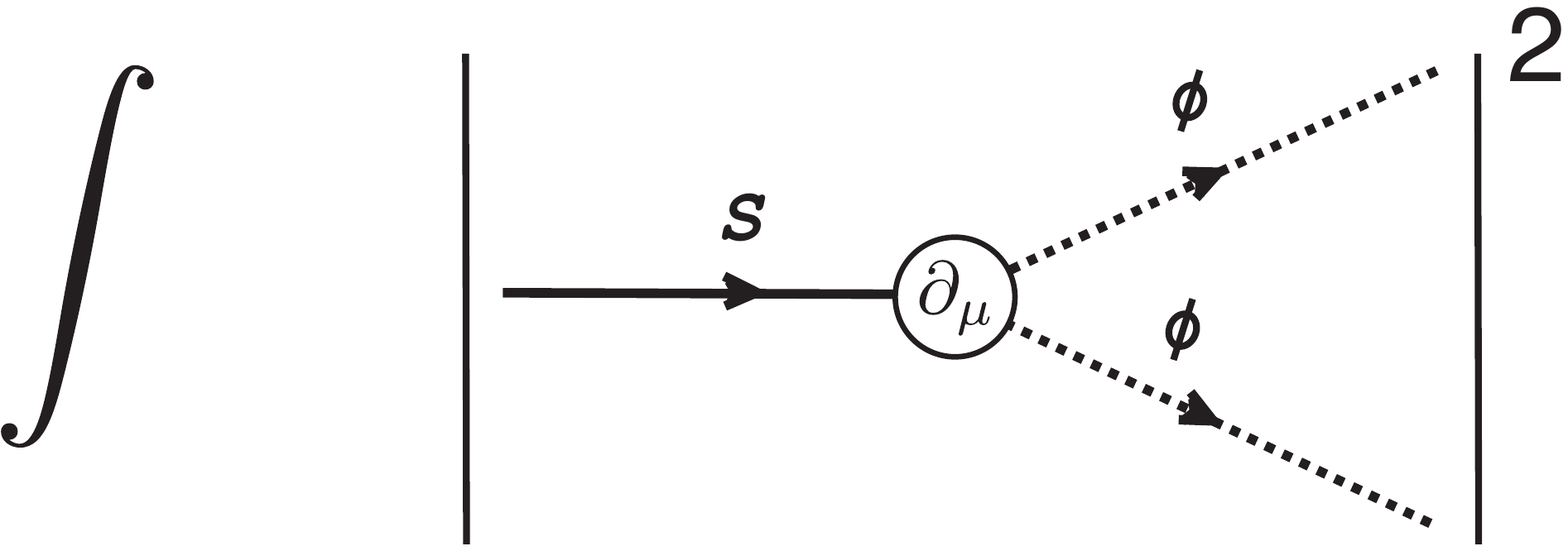} & \vspace{0.12cm} \hspace{-68.01cm} \includegraphics[scale=0.31]{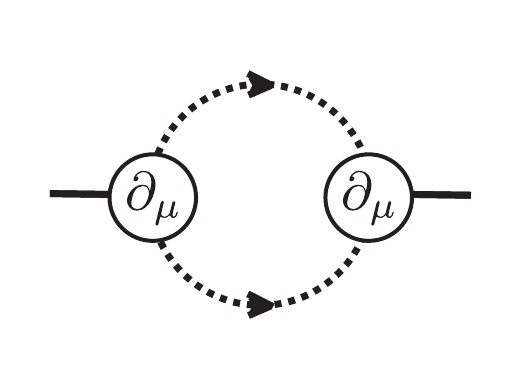} & \vspace{0.12cm} \hspace{-101.16cm} \includegraphics[scale=0.08]{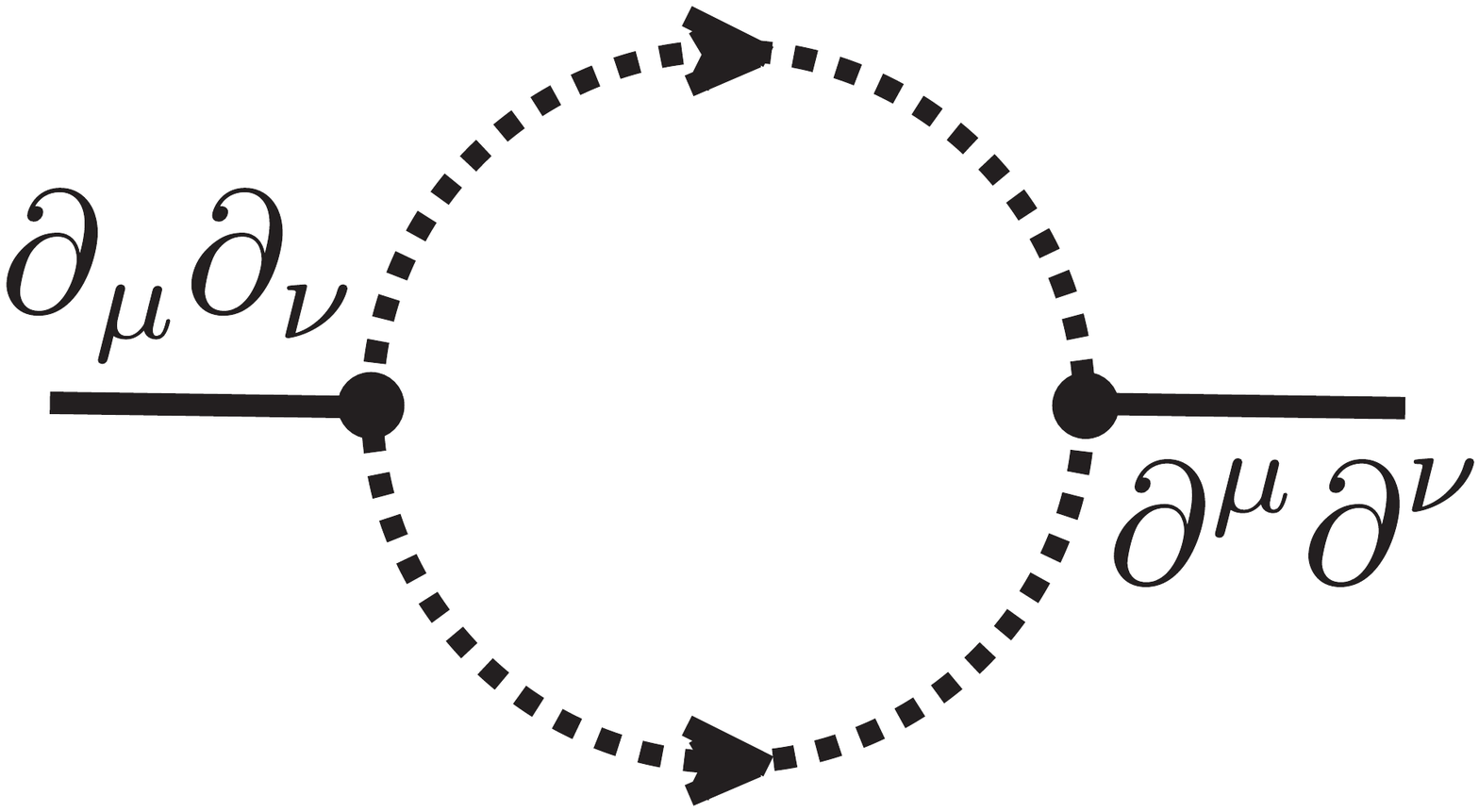} \ \ \ \ \tagarray\label{eq:impart}
\end{tabular}
\vspace{-2.0cm}
\end{table}
\\
The second equality is due to Eq.\ (\ref{eq:Piregular}) and the fact that the tadpole has no imaginary part, respectively.

The calculation of the tree-level decay width in momentum space
proceeds by replacing derivatives $\partial_{\mu}\rightarrow\pm ip_{\mu}$ (the lower/upper
sign stands for incoming/outgoing particles)
in the Lagrangian\ (\ref{eq:Lint}), {\em i.e.}, in our simple model the decay amplitude reads
\vspace{0.2cm}
\begin{equation}
\hspace{0.7cm}2ig\underbrace{\left(-\frac{s-2m^{2}}{2}\right)}_{=-p_{1}\cdot p_{2}} \ = \ \
\begin{minipage}{7cm}
\includegraphics[scale=0.39]{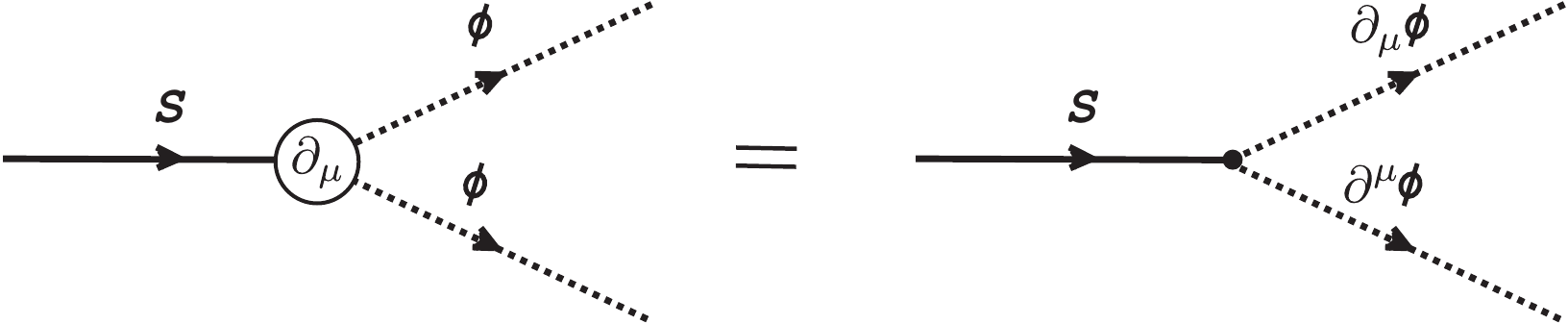}
\end{minipage} \ .
\end{equation}
The factor $2$ comes from the two identical particles in the outgoing channel.
The blob in the left diagram represents the vertex as given by Eq.\ (\ref{eq:Hint}), 
while in the right diagram the replacement $\partial_{\mu}\rightarrow\pm ip_{\mu}$ was performed in order to 
calculate the expression on the left-hand side.

Returning to the imaginary part\ (\ref{eq:impart}) of the self-energy, we 
observe that, on account of the fact that the tadpole does not contribute
to the imaginary part, with the dispersion
relation\ (\ref{eq:Redispersionrel}) one actually only computes the first diagram on the right-hand side of Eq.\ (\ref{eq:Pigraph}) and misses the tadpole contribution.\ In other words, as we have demonstrated
above, the
first diagram in Eq.\ (\ref{eq:Pigraph}) contains precisely the tadpole contribution, yet
with opposite sign, compare with Eq.\ (\ref{eq:Piregular}).\ Consequently, we need to add
this tadpole to the (real part of the) self-energy as computed via the dispersion relation -- in order to have the latter agree with the result obtained from the perturbative calculation.

Another possibility to demonstrate the emergence of a constant tadpole from the loop diagram in the self-energy $\Pi(s)$ is by writing down its analytic expression by using Feynman rules. After including a 3d form factor, one can rewrite the expression into a sum of the desired dispersion integral and a term that indeed is represented by a simple tadpole diagram. We first have to modify the result of Eq.\ (\ref{eq:loopSolution}) such that
\begin{equation}
\label{eq:Pi_Feynman_dd}
\Pi^{F}(s+i\epsilon) = -\frac{g^{2}}{\pi^{2}}\int_{0}^{\infty}\text{d}u \ \frac{\big[-3m^{4}-4m^{2}u^{2}+(m^{2}+u^{2})s\big]u^{2}F^{2}(u)}{\sqrt{u^{2}+m^{2}}\big[4(u^{2}+m^{2})-s-i\epsilon\big]} \ .
\end{equation}
Note that, for reasons of clarity, $\Pi^{F}(s)$ is the diagram where the derivatives have been replaced, clearly speaking it is the first diagram on the right-hand side of Eq.\ (\ref{eq:Piregular}).\ By using Ref.\ \cite{thomasthesis}, the numerator follows from the derivative interaction term and by replacing derivatives by momenta during all the steps performed there.\ What is left is to rewrite the above integral similarly as we did in Section\ \ref{seq:PolFunc}.\ Using the transformation $u\rightarrow\sqrt{t/4-m^{2}}$, a dispersion integral (with the correct imaginary part) appears:
\vspace{0.2cm}
\begin{eqnarray}
& = & \frac{g^{2}}{\pi^{2}}\int_{0}^{\infty}\text{d}u \ \frac{\big[-3m^{4}-4m^{2}u^{2}+(m^{2}+u^{2})s\big]u^{2}F^{2}(u)}{\sqrt{u^{2}+m^{2}}\big[s-4(u^{2}+m^{2})+i\epsilon\big]} \\[0.15cm]
& = & \frac{1}{\pi}\int_{4m^{2}}^{\infty}\text{d}t \ \frac{\frac{\sqrt{\frac{t}{4}-m^{2}}}{4\pi\sqrt{t}}g^{2}F^{2}\left(\sqrt{\frac{t}{4}-m^{2}}\right)\frac{1}{4}(4m^{4}-4m^{2}t+st)}{s-t+i\epsilon} \nonumber \\[0.15cm]
& = & \frac{1}{\pi}\int_{4m^{2}}^{\infty}\text{d}t \ \frac{\frac{\sqrt{\frac{t}{4}-m^{2}}}{4\pi\sqrt{t}}g^{2}F^{2}\left(\sqrt{\frac{t}{4}-m^{2}}\right)}{s-t+i\epsilon}\frac{1}{4}\big[(t^{2}-4m^{2}t+4m^{4})+t(s-t)\big] \nonumber \\[0.15cm]
& = & \frac{1}{\pi}\int_{4m^{2}}^{\infty}\text{d}t \ \frac{\frac{\sqrt{\frac{t}{4}-m^{2}}}{4\pi\sqrt{t}}g^{2}F^{2}\left(\sqrt{\frac{t}{4}-m^{2}}\right)\left(\frac{t}{2}-m^{2}\right)^{2}}{s-t+i\epsilon}+\frac{1}{\pi}\int_{4m^{2}}^{\infty}\text{d}t \ \frac{t}{4}\frac{\sqrt{\frac{t}{4}-m^{2}}}{4\pi\sqrt{t}}g^{2}F^{2}\left(\sqrt{\frac{t}{4}-m^{2}}\right) \nonumber \\[0.15cm]
& = & \frac{1}{\pi}\int_{4m^{2}}^{\infty}\text{d}t \ \frac{\frac{\sqrt{\frac{t}{4}-m^{2}}}{4\pi\sqrt{t}}g^{2}F^{2}\left(\sqrt{\frac{t}{4}-m^{2}}\right)\left(\frac{t}{2}-m^{2}\right)^{2}}{s-t+i\epsilon}+2g^{2}\cdot2\cdot\frac{1}{4\pi^{2}}\int_{0}^{\infty}\text{d}u \ u^{2}\sqrt{u^{2}+m^{2}}F^{2}(u) \ . \nonumber
\end{eqnarray}
The first term can be decomposed into the desired dispersion integral and the corresponding imaginary part. It is therefore the diagram on the left-hand side of Eq.\ (\ref{eq:Piregular}). The second term is nothing else than the tadpole, since
\begin{eqnarray}
\langle0|\mathcal{T}\big{\{}\partial_{0}^{x}\phi(x)\partial^{0,x}\phi(x)\big{\}}|0\rangle & = & \langle0|{}\partial_{0}^{x}\phi(x)\partial^{0,x}\phi(x)|0\rangle \nonumber \\[0.1cm]
& = & \int\frac{\text{d}^{4}p}{(2\pi)^{4}}\frac{iE_{\textbf{p}}^{2}}{p^{2}-m^{2}+i\epsilon}  \ , \\[0.15cm]
& \rightarrow & \frac{1}{4\pi^{2}}\int_{0}^{\infty}\text{d}k \ k^{2}\sqrt{m^{2}+k^{2}} \ \cdot F^{2}(k) \ ,
\end{eqnarray}
where in the last step the form factor was introduced.\ Again, this explicitly shows that the dispersion relation agrees with the result obtained from the perturbative calculation only up to a real constant, that is, some tadpole diagram.\ However, it is possible to correct for this term at the Lagrangian level; one only needs to absorb the constant in the bare mass $m_{0}\rightarrow\tilde{m}_{0}$ in Eq.\ (\ref{eq:LagSphiphi}) such that
\begin{equation}
\mathcal{L}_{S}\rightarrow\tilde{\mathcal{L}}_{S} = \frac{1}{2} \left( \partial_\mu S \partial^\mu S - \tilde{m}_{0}^2 S^2 \right) \ .
\end{equation}
The dispersion relation will then yield the correct result for the self-energy.

\section{Derivative interaction of the form $\mathcal{L}_{\text{int}}=g(\partial_{\mu}S)(\partial^{\mu}\phi_{1})\phi_{2}$}
The aim of this section is to demonstrate that derivative interaction terms of the form $\mathcal{L}_{\text{int}}=g(\partial_{\mu}S)(\partial^{\mu}\phi_{1})\phi_{2}$ not only spoil the dispersion relation in Eq.\ (\ref{eq:Redispersionrel}), but also the normalization of the spectral function.\ The first property will have the same origin as it was already discussed, {\em i.e.}, an additional vertex will appear during the quantization procedure and introduce a tadpole diagram. The second property is due to the nature of the interaction term: it will contribute to the kinetic term of the $S$ field, making a renormalization of the latter necessary.

\subsection{Quantization and Feynman result}
For pedagogical reasons, this time we start directly from the Feynman result and furthermore use distinguishable particles $\phi_{1}$ and $\phi_{2}$ with the same mass $m$.\footnote{Unequal masses would only make the formulas look more complicated, but would not change our general statement.} From
\begin{equation}
\mathcal{L}_{\text{int}} = g(\partial_{\mu}S)(\partial^{\mu}\phi_{1})\phi_{2} \ ,
\end{equation}
by using the conjugate momenta
\begin{eqnarray}
\pi_{S} & = & \frac{\partial\mathcal{L}}{\partial(\partial_{0}S)} \hspace{0.26cm} = \hspace{0.26cm} \partial^{0}S+g(\partial^{0}\phi_{1})\phi_{2} \ , \\
\pi_{\phi_{1}} & = & \frac{\partial\mathcal{L}}{\partial(\partial_{0}\phi_{1})} \hspace{0.14cm} = \hspace{0.14cm} \partial^{0}\phi_{1}+g(\partial^{0}S)\phi_{2} \ , \ \ \ \pi_{\phi_{2}} \hspace{0.2cm} = \hspace{0.22cm} \frac{\partial\mathcal{L}}{\partial(\partial_{0}\phi_{2})} \hspace{0.2cm}= \hspace{0.22cm} \partial^{0}\phi_{2} \ ,
\end{eqnarray}
we explicitly obtain to second order in the coupling constant $g$
\begin{equation}
\label{eq:delLag2}
\hat{\mathcal{H}}_{\text{int}}^{I}=-\hat{\mathcal{L}}_{\text{int}}^{I}+\frac{g^{2}}{2}\left[(\partial_{0}\hat \phi_{1}^{I})(\partial^{0}\hat \phi_{1}^{I})\hat\phi_{2}^{I}\hat\phi_{2}^{I}+(\partial_{0}\hat S^{I})(\partial^{0}\hat S^{I})\hat\phi_{2}^{I}\hat\phi_{2}^{I}\right]+\mathcal{O}(g^{3}) \ .
\end{equation}
One observes that the first term in brackets gives rise to vacuum fluctuations (bubble diagrams) which simply can be ignored provided that all graphs with disconnected vacuum bubble diagrams are left out. Then, just as for the Feynman result of the self-energy in Eq.\ (\ref{eq:Pi_Feynman_dd}), we start from
\begin{equation}
\Pi^{F}(s+i\epsilon) = -\frac{g^{2}}{\pi^{2}}\int_{0}^{\infty}\text{d}u \ \frac{\big[\frac{s(u^{2}+m^{2})}{2}-\frac{s^{2}}{4}\big]u^{2}F^{2}(u)}{\sqrt{u^{2}+m^{2}}\big[4(u^{2}+m^{2})-s-i\epsilon\big]} \ .
\end{equation}
The numerator follows as was described before.\ Applying again the transformation $u\rightarrow\sqrt{t/4-m^{2}}$, the expression for the Feynman result is
\vspace{0.2cm}
\begin{eqnarray}
& = & \frac{g^{2}}{\pi^{2}}\int_{0}^{\infty}\text{d}u \ \frac{\big[\frac{s(u^{2}+m^{2})}{2}-\frac{s^{2}}{4}\big]u^{2}F^{2}(u)}{\sqrt{u^{2}+m^{2}}\big[s-4(u^{2}+m^{2})+i\epsilon\big]} \nonumber \\[0.2cm]
& = & \frac{1}{\pi}\int_{4m^{2}}^{\infty}\text{d}t \ \frac{\frac{\sqrt{\frac{t}{4}-m^{2}}}{8\pi\sqrt{t}}g^{2}F^{2}\left(\sqrt{\frac{t}{4}-m^{2}}\right)\frac{s}{2}\big(s-\frac{t}{2}\big)}{s-t+i\epsilon} \nonumber \\[0.2cm]
& = & \frac{1}{\pi}\int_{4m^{2}}^{\infty}\text{d}t \ \frac{\frac{\sqrt{\frac{t}{4}-m^{2}}}{8\pi\sqrt{t}}g^{2}F^{2}\left(\sqrt{\frac{t}{4}-m^{2}}\right)\big[\frac{s^{2}}{4}+\frac{s}{4}(s-t)\big]}{s-t+i\epsilon} \nonumber \\[0.2cm]
& = & \frac{1}{\pi}\int_{4m^{2}}^{\infty}\text{d}t \ \frac{\frac{\sqrt{\frac{t}{4}-m^{2}}}{8\pi\sqrt{t}}g^{2}F^{2}\left(\sqrt{\frac{t}{4}-m^{2}}\right)\frac{s^{2}}{4}}{s-t+i\epsilon}+\frac{s}{4}\cdot\frac{1}{\pi}\int_{4m^{2}}^{\infty}\text{d}t \ \frac{\sqrt{\frac{t}{4}-m^{2}}}{8\pi\sqrt{t}}g^{2}F^{2}\left(\sqrt{\frac{t}{4}-m^{2}}\right) \nonumber \\[0.2cm]
& = & \frac{1}{\pi}\int_{4m^{2}}^{\infty}\text{d}t \ \frac{\frac{\sqrt{\frac{t}{4}-m^{2}}}{8\pi\sqrt{t}}g^{2}F^{2}\left(\sqrt{\frac{t}{4}-m^{2}}\right)\big(\frac{s^{2}}{4}-\frac{t^{2}}{4}+\frac{t^{2}}{4}\big)}{s-t+i\epsilon}+g^{2}\cdot\frac{s}{4}\cdot\frac{1}{2\pi^{2}}\int_{0}^{\infty}\text{d}t \ \frac{u^{2}F^{2}(u)}{\sqrt{u^{2}+m^{2}}} \ . \nonumber \\[0.4cm]
\end{eqnarray}
The first integral can be further decomposed:
\begin{eqnarray}
& = & \frac{1}{\pi}\int_{4m^{2}}^{\infty}\text{d}t \ \frac{\frac{\sqrt{\frac{t}{4}-m^{2}}}{8\pi\sqrt{t}}g^{2}F^{2}\left(\sqrt{\frac{t}{4}-m^{2}}\right)\frac{t^{2}}{4}}{s-t+i\epsilon}+\frac{1}{\pi}\int_{4m^{2}}^{\infty}\text{d}t \ \frac{\sqrt{\frac{t}{4}-m^{2}}}{8\pi\sqrt{t}}g^{2}F^{2}\left(\sqrt{\frac{t}{4}-m^{2}}\right)\frac{1}{4}(s+t) \nonumber \\[0.1cm]
& & + \ g^{2}\cdot\frac{s}{4}\cdot\frac{1}{2\pi^{2}}\int_{0}^{\infty}\text{d}t \ \frac{u^{2}F^{2}(u)}{\sqrt{u^{2}+m^{2}}} \nonumber \\[0.25cm]
& = & \frac{1}{\pi}\int_{4m^{2}}^{\infty}\text{d}t \ \frac{\frac{\sqrt{\frac{t}{4}-m^{2}}}{8\pi\sqrt{t}}g^{2}F^{2}\left(\sqrt{\frac{t}{4}-m^{2}}\right)\frac{t^{2}}{4}}{s-t+i\epsilon}+g^{2}\cdot2\cdot\frac{1}{4\pi^{2}}\int_{0}^{\infty}\text{d}u \ u^{2}\sqrt{u^{2}+m^{2}}F^{2}(u) \nonumber \\[0.1cm]
\label{eq:Feyn_adad}
& & + \ g^{2}\cdot\frac{s}{4\pi^{2}}\int_{0}^{\infty}\text{d}u \ \frac{u^{2}F^{2}(u)}{\sqrt{u^{2}+m^{2}}} \ .
\end{eqnarray}
We comment on this result: From what we know concerning the contraction of two gradients of field operators, it is obvious that the sum of the last two terms has to be a contribution emerging from the contraction of the gradients of the $\phi_{1}$ fields. In fact, in the perturbative series of the full propagator for the $S$ field, to $\mathcal{O}(g^{2})$ we find
\begin{eqnarray}
& & \frac{(-i)^{2}}{2!}\cdot g^{2}\cdot2\int\text{d}x^{\prime}\int\text{d}x^{\prime\prime} \ i\hspace{0.02cm}\partial^{\mu,x^{\prime}}\Delta(x_{1}-x^{\prime})\hspace{0.02cm}i\hspace{0.02cm}\partial^{\nu,x^{\prime\prime}}\Delta(x^{\prime\prime}-x_{2}) \nonumber \\
& & \times \ \langle0|\mathcal{T}\big{\{}\partial_{\mu}^{x^{\prime}}\phi_{1}(x^{\prime})\partial_{\nu}^{x^{\prime\prime}}\phi_{1}(x^{\prime\prime})\big{\}}|0\rangle\hspace{0.02cm}i\hspace{0.02cm}\Delta_{F}^{\phi_{2}}(x^{\prime}-x^{\prime\prime}) \nonumber \\
\nonumber \\[-0.14cm]
& = & g^{2}\int\text{d}x^{\prime}\int\text{d}x^{\prime\prime} \ \partial^{\mu,x^{\prime}}\Delta(x_{1}-x^{\prime})\hspace{0.02cm}\partial^{\nu,x^{\prime\prime}}\Delta(x^{\prime\prime}-x_{2}) \nonumber \\
& & \times\left[i\hspace{0.02cm}\partial_{\mu}^{x^{\prime}}\partial_{\nu}^{x^{\prime\prime}}\Delta_{F}^{\phi_{1}}(x^{\prime}-x^{\prime\prime})-i\eta_{\mu 0}\eta_{\nu 0}\delta^{(4)}(x^{\prime}-x^{\prime\prime})\right]i\hspace{0.02cm}\Delta_{F}^{\phi_{2}}(x^{\prime}-x^{\prime\prime})
\nonumber \\
\nonumber \\[-0.14cm]
& = & -g^{2}\int\text{d}x^{\prime}\int\text{d}x^{\prime\prime} \ \partial^{\mu,x^{\prime}}\Delta(x_{1}-x^{\prime})\hspace{0.02cm}\partial^{\nu,x^{\prime\prime}}\Delta(x^{\prime\prime}-x_{2})\hspace{0.02cm}\partial_{\mu}^{x^{\prime}}\partial_{\nu}^{x^{\prime\prime}}\Delta_{F}^{\phi_{1}}(x^{\prime}-x^{\prime\prime})\hspace{0.02cm}\Delta_{F}^{\phi_{2}}(x^{\prime}-x^{\prime\prime}) \nonumber \\
& & - \ ig^{2}\int\text{d}x^{\prime} \ \partial_{0}^{x^{\prime}}\Delta(x_{1}-x^{\prime})\hspace{0.02cm}\partial^{0,x^{\prime}}\Delta(x^{\prime}-x_{2})\hspace{0.02cm}\langle0|\mathcal{T}\big{\{}\phi_{2}(x^{\prime})\phi_{2}(x^{\prime})\big{\}}|0\rangle \ .
\end{eqnarray}
On the other hand, the first non-trivial term in the perturbative series is also to $\mathcal{O}(g^{2})$,
\begin{align}
\frac{-i}{1!}\cdot\frac{g^{2}}{2}\cdot2\cdot2\cdot\frac{1}{2}\int\text{d}x^{\prime} \ i\hspace{0.02cm}\partial_{\mu}^{x^{\prime}}\Delta(x_{1}-x^{\prime})\langle0|\mathcal{T}\big{\{}\phi_{2}(x^{\prime})\phi_{2}(x^{\prime})\big{\}}|0\rangle\hspace{0.02cm}i\hspace{0.02cm}\partial^{\mu,x^{\prime}}\Delta(x^{\prime}-x_{2}) \nonumber \\
= \ i\hspace{0.02cm}g^{2}\int\text{d}x^{\prime} \ \partial_{\mu}^{x^{\prime}}\Delta(x_{1}-x^{\prime})\partial^{\mu,x^{\prime}}\Delta(x^{\prime}-x_{2})\langle0|\mathcal{T}\big{\{}\phi_{2}(x^{\prime})\phi_{2}(x^{\prime})\big{\}}|0\rangle \ ,
\end{align}
and cancels the second term in the equation before.\ This cancellation just yields the Feynman result, as expected, and again demonstrates that the self-energy consists of more than the simple dispersion integral.

\subsection{Normalization}
Looking at Eq.\ (\ref{eq:Feyn_adad}), let us write this as $\Pi^{F}(s) = \Pi^{DR}(s)+\mathcal{C}_{1}+s\hspace{0.03cm}\mathcal{C}_{2}$.
The full inverse propagator of the field $S$ will then have the form
\begin{eqnarray}
G^{-1}(s) & = & s-m_{0}^{2}-\Pi^{F}(s) \nonumber \\[0.15cm]
& = & (1-\hspace{0.02cm}\mathcal{C}_{2})s-m_{0}^{2}-\Pi^{DR}(s)-\mathcal{C}_{1} \nonumber \\[0.15cm]
& = & (1-\hspace{0.02cm}\mathcal{C}_{2})\left[s-\frac{m_{0}^{2}}{1-\hspace{0.02cm}\mathcal{C}_{2}}-\frac{\Pi^{DR}(s)}{1-\hspace{0.02cm}\mathcal{C}_{2}}-\frac{\mathcal{C}_{1}}{1-\hspace{0.02cm}\mathcal{C}_{2}}\right] \ ,
\label{eq:normprop}
\end{eqnarray}
which causes the spectral function to be not normalized correctly. One needs to renormalize the field $S$ such that $S\rightarrow\sqrt{1-\mathcal{C}_{2}}\hspace{0.05cm}S$.\ This field transformation has of course no effect on the position of the propagator poles. By writing $m_{0}\rightarrow\tilde{m}_{0}$ and $g\rightarrow\tilde{g}$, we again bring the Lagrangian in a form so that the dispersion relation yields the correct result:
\begin{equation}
\mathcal{L}_{S}\rightarrow\tilde{\mathcal{L}}_{S} = \frac{1}{2} \left( \partial_\mu S \partial^\mu S - \tilde{m}_{0}^2 S^2 \right) \ , \ \ \ \mathcal{L}_{\text{int}}\rightarrow\tilde{\mathcal{L}}_{\text{int}} = \tilde{g}(\partial_{\mu}S)(\partial^{\mu}\phi_{1})\phi_{2} \ .
\end{equation}

\section{Non-derivative interaction of the form $\mathcal{L}_{\text{int}}=gS\phi\phi-\frac{\lambda}{4!}\phi^{4}$}
Although it will not be applied in this thesis, it may be useful for future work to investigate also contact interaction terms and their influence on the position of the resonance poles. The perturbative series of the full propagator for the $S$ field still can be resummed, however the self-energy $\Pi(s)$ is modified by chain diagrams:
\vspace{0.2cm}
\begin{eqnarray}
\includegraphics[scale=0.37]{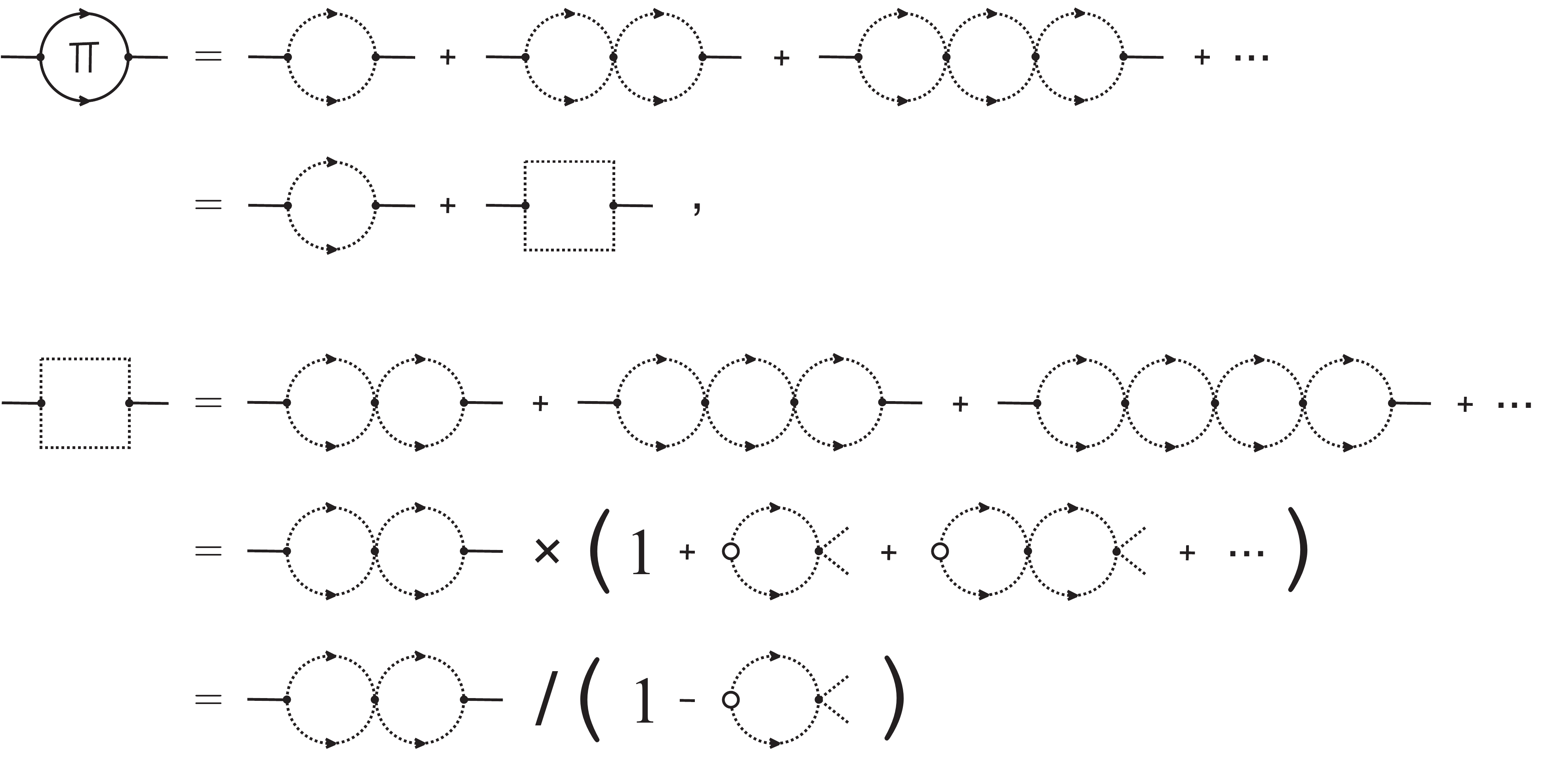} \label{eq:contact}
\end{eqnarray}
The box diagram represents the infinite sum of loops with intermediate contact vertices, starting from the number of vertices $n=3$.\ The analytic expression for the box term is
\begin{equation}
\sum_{n=3}^{\infty}\frac{(-i)^{n}}{n!}\cdot2\cdot2\cdot\left(\frac{1}{2}\right)^{n-1}\cdot\bigg(\begin{matrix}n \\ 2\end{matrix}\bigg)\cdot2(n-2)!\cdot4!^{n-2}\cdot\left(\frac{\lambda}{4!}\right)^{n-2}g^{2}\Sigma^{n-1}(s) \ .
\end{equation}
The function $g^{2}\Sigma(s)$ is the well-known loop integral from Eq.\ (\ref{eq:loopSolution}), but without accounting for any combinatorial aspects like vertex commutations or symmetry factors.\footnote{It is $2g^{2}\Sigma(s)=\Pi(s)$, with the latter from Eq.\ (\ref{eq:loopSolution}).} Let us discuss the origin of the upper factors:\ $(i)$ The first factor originates because the corresponding diagram is $n$-th order in perturbation theory.\
$(ii)$ The two factors of $2$ come from the fact that the two $\phi$ lines at each one of the two three-point vertices can be joined in two different ways.\ $(iii)$ The factor $(1/2)^{n-1}$ is a symmetry factor; one can interchange two $\phi$ lines in a loop with each other without altering the diagram.\ $(iv)$ The factorial $2(n-2)!$ is due to vertex commutations.\ $(v)$ The factor of $4!^{n-2}$ is present because the four $\phi$ lines at each one of the four-point vertices can be joined in $4!$ different ways.\ $(vi)$ The last factor arises from the four-point vertices. $(vii)$ Finally, what is left is the binomial coefficient.\ It is there since for each order $n$ in the perturbative series, we have in principle to include all the terms 
\begin{equation}
\left(S\phi\phi+\frac{-\lambda}{4!}\phi^{4}\right)^{n} = \sum_{k=0}^{n}\bigg(\begin{matrix}n \\ k\end{matrix}\bigg)(S\phi\phi)^{n-k}\left(\frac{-\lambda}{4!}\phi^{4}\right)^{k} \ .
\end{equation}
However, for each $n$ only that term with $k=n-2$ contributes.\ In the perturbative series, all those \(\binom{n}{n-2}\) = \(\binom{n}{2}\) possible diagrams are equivalent; in each diagram one can rename the integration variables.

The step in the last two lines of Eq.\ (\ref{eq:contact}) is simply the application of the geometric series formula:
\begin{equation}
i\lambda g^{2}\Sigma^{2}(s)\sum_{k=0}^{\infty}\frac{(-i\lambda)^{k}}{2^{k}}\Sigma^{k}(s) = \frac{i\lambda g^{2}\Sigma^{2}(s)}{1-\frac{(-i\lambda)}{2}\Sigma(s)} \ .
\end{equation}
The unfilled circle in both lines represents the three-point vertex without coupling constant $g$.\ It is $a\ priori$\ not clear if this equation holds in general and for all energies $s$.\ The validity will in fact depend on the choice of the form factor and the magnitude of the coupling constant $\lambda$. The explicit relation
\vspace{-0.2cm}
\begin{equation}
\lambda < \frac{2}{\abs{\Sigma(s)}}
\vspace{-0.2cm}
\end{equation}
has to be checked for a given value of $\lambda$ for the window of complex energies $s$ of interest.

\newpage
\clearpage

\

\newpage
\clearpage

\chapter{Dynamical generation: The $a_{0}(980)$}
\label{chap:chapter4}

\medskip

Among others, the study of T\"ornqvist and Roos\ \cite{tornqvist,tornqvist2} based on the Unitarized Quark Model (UQM) demonstrated that an isovector state above $1$ GeV may exist alongside a companion resonance pole below/near $1$ GeV. 
Indeed, as we shall show in this chapter for the heavy scalar--isovector seed state 
$a_{0}(1450)$, the coupling of this state to $\eta\pi$, $K\bar{K}$, and 
$\eta^{\prime}\pi$ dynamically generates the light $a_{0}(980)$ as a particular type of
four-quark meson. We will first illustrate the idea of dynamical generation 
introduced in the second chapter, {\em i.e.}, we will present the general formalism by discussing and extending 
previous calculations of T\"ornqvist and Roos and of Boglione and Pennington.\ After that, we will construct an effective 
Lagrangian where $a_{0}(1450)$ couples to pseudoscalar mesons by both non-derivative and derivative interactions.

\section{Some words on the scalar--isovector resonances}
The isotriplet state $a_{0}(1450)$ has been observed by the Crystal Barrel Collaboration in $p\bar{p}$ annihilation into the $\pi^{0}\pi^{0}\eta$ final state\ \cite{crysbar1}.\ There are also claims to have found this particle for the same annihilation into the final state $\pi^{+}\pi^{-}\omega$\ \cite{baker}.\ Early coupled-channel analysis on this (and previous) data required the $a_{0}(1450)$ to exist, in particular to provide a reasonable determination of the $a_{0}(980)$ parameters\ \cite{buggani}.\ The latter resonance is known for a longer time, see for example Refs.\ \cite{astier,morgan980,flatte}.\ It couples strongly to the $K\bar{K}$-channel\ \cite{crysbar2} and lies just below the threshold; this highly distorts its line shape and makes it difficult to determine the mass and width from a simple Breit--Wigner (for example, it was intensively discussed whether the width is large or not).\ For the same reason the determination of the relative coupling $K\bar{K}/\eta\pi$ and branching ratio, respectively, is difficult\ \cite{flatte,defoix,corden,barberis,crysbar1,crysbar2,janssen,bugg2008}.

The isovector sector was subject to a vast amount of studies using very different approaches, see for instance\ Refs.\ \cite{morgan980,flatte,jaffe,jaffe2,dullemond,janssen,tornqvist,tornqvist2,oller,oller2,oller3,oller4,boglione,pennington,fariborz,rodriguez,maiani,tqmix,hooft,zhiyong,denisphd,eLSM2,bugg2008,molecular,molecular2,molecular3,molecular4,molecular5}.\ Both states $a_{0}(980)$ and $a_{0}(1450)$ are nowadays established resonances included in the summary tables of the PDG\ \cite{olive}.\ Yet, as mentioned earlier, it is not possible to assign both as $q\bar{q}$ states. In particular, the positively charged $a_{0}^{+}$ state, if predominantly a quarkonium, may be realized as $a_{0}^{+}\sim u\bar{d}$, but it is not clear whether this quarkonium corresponds to $a_{0}(980)$ or $a_{0}(1450)$.

The $a_{0}(980)$ is therefore often interpreted as some type of four-quark state\ \cite{jaffe,alford,maiani,weinstein,hooft}, where this notion may contain tetraquarks and $K\bar{K}$ molecules (and mixtures of them). This is also supported from analyzing data on radiative $\phi\rightarrow\gamma a_{0}(980)$ decays, see Refs.\ \cite{achasov2,oller7,kalash,kalash2,achasov3} and references therein. In fact, there is a growing consensus that the scalar resonances $f_{0}(1370)$, $f_{0}(1500)$, 
$K_{0}^{\ast}(1430)$, and $a_{0}(1450)$
are predominantly quark-antiquark states, see for example Refs.\
\cite{close,close2,close3,close4,close5,close6,tqmix,eLSM1,eLSM1-2,eLSM1-3,eLSM2,oller4,oller5,oller6,zhiyong}.\ 
Then, the light scalar states $f_{0}(500)$, $f_{0}(980)$, 
$K_{0}^{\ast}(800)$, and $a_{0}(980)$ are (most likely) predominantly four-quark states 
(see \emph{e.g.}\ Refs.\ 
\cite{jaffe,jaffe2,maiani,hooft,fariborz,fariborz2,rodriguez,molecular,molecular2,molecular3,molecular4,molecular5,oller,oller2,oller3,oller4,oller5,oller6,tqmix,zhiyong} and references therein). The nature of these resonances can be also studied by looking at their pole trajectories in the large-$N_{c}$ limit\ \cite{pelaez,zhiyong}, with the conclusion that they behave very differently from ordinary quarkonia. Further insights into the inner structure of the $a_{0}(980)$ are expected from its mixing with $f_{0}(980)$\ \cite{achasov4,hanhart}.

A possible identification of the scalar states in the context of the eLSM was put forward in Refs.\ \cite{denisphd,eLSM2}.\ There, the isospin triplet was studied for $N_{f}=3$ and the corresponding resonance was assigned within two scenarios to be either the state below $1$ GeV or the one above $1$ GeV. An analogous investigation was performed simultaneously for the scalar kaon (see also next chapter).\ In the first scenario, that is, where the isotriplet was identified with the $a_{0}(980)$, the fit was aimed to bring the decay amplitudes for $a_{0}(980)\rightarrow\eta\pi$, $a_{0}(980)\rightarrow K\bar{K}$ together with the resonance mass in agreement with the experimental data\ \cite{pagliaraderivatives2}. In the second scenario, $a_{0}(1450)$ was fitted such that its decay width was in agreement with the PDG. The result of the fit clearly preferred the $a_{0}$ state to lie above $1$ GeV with $m_{a_{0}}=1.363\pm0.001$ GeV and $\Gamma=0.266\pm0.012$ GeV\ \cite{denisphd,eLSM2}.\footnote{In contrast to Ref.\ \cite{baker}, the decay channel $a_{0}(1450)\rightarrow\pi\pi\omega$ was found to be very small in the eLSM (compared for example to $a_{0}(1450)\rightarrow\eta\pi$). Therefore, this channel will not play any role in our further investigations.}

\section{The $a_{0}(980)$ revisited}
Following earlier work\ \cite{dullemond2,dullemond}, T\"ornqvist and Roos\ \cite{tornqvist,tornqvist2} 
(in the following denoted as TR) and later Boglione and Pennington\ \cite{boglione,pennington} (denoted as BP)
studied the mechanism of dynamical generation in the scalar sector. To this end, they investigated the influence of 
meson-loop contributions to the self-energy in the UQM. 
We now extend their studies and compare numerical 
results for the poles of the propagator
to the latest experimental data\ \cite{olive}.\ It turns out that (depending on the assignment 
of the poles to physical resonances) the widths of both the seed state 
$a_{0}(1450)$ and the dynamically generated state $a_{0}(980)$ are
by a factor of two larger than the experimental values.\
Moreover, the mass of $a_{0}(1450)$ is too large (by $\sim100$ MeV in TR and by $\sim400$ MeV 
in BP). It thus seems that, while qualitatively feasible, the dynamical generation of
resonances as companion poles in the propagator does not yield results
that are in quantitative agreement with experimental data.

\subsection{Approach of TR and BP}
The following two points are relevant in the mechanism of dynamical generation,
irrespective of the quantum numbers of the hadronic resonance considered: 
$(i)$\ The propagator of a quark-antiquark seed state is dressed by meson-loop
contribution to the self-energy. These contributions shift the mass of the state and 
change the form of its spectral function.\ When increasing the coupling, 
the corresponding pole moves away from the real axis and follows a certain trajectory 
in the complex plane. The mass and the width of the resonance are determined by the position 
of the complex pole of the dressed propagator on the appropriate 
Riemann sheet -- a procedure first proposed by Peierls a long time ago\ \cite{peierls}.\ 
$(ii)$\ If the interaction exceeds a critical value, a(t least one) companion pole can appear 
in the complex plane. If this pole is sufficiently close to the real axis, it can manifest itself
in the spectral function as an \emph{additional} resonance with the same quantum numbers
as the seed state\ \cite{morgan,tornqvist,tornqvist2,polosa}.\ 
Since the coupling of scalars to pseudoscalars is large,
the scalar sector is particularly affected by such distortions of the spectral function. 

We now recapitulate the seminal works TR\ \cite{tornqvist,tornqvist2} and 
BP\ \cite{pennington}, where the latter uses the same model as the former but with a slightly different set of 
parameters. The main goal is the determination of the inverse propagator of a 
resonance after applying a Dyson resummation of loop contributions to the self-energy:
\begin{equation}
G^{-1}(s)=s-m_{0}^{2}-\Pi(s)\ ,
\end{equation}
where $s$ is the first Mandelstam variable, $m_{0}$ is the
bare mass of the seed state, and $\Pi(s)=\sum_{i}\Pi_{i}(s)$ is the self-energy.\footnote{Note that we use a different sign convention 
for the propagator $G(s)$ and the self-energy $\Pi(s)$ than Refs.\
\cite{tornqvist,tornqvist2,pennington}.\ In the first reference, the pole-dominated scattering amplitude is studied, 
but only its denominator is important here, which has the same form as the inverse propagator.}\ 
Here, the sum runs over the loops emerging from the coupling of the resonance
to various decay channels. The imaginary part of $\Pi_i(s)$ corresponds to the partial
decay width of the resonance into mesons in channel $i$.\
The real part of $\Pi(s)$ on the real axis is related to the imaginary part by the 
dispersion relation
\begin{equation}
\operatorname{Re}\Pi(s)=\frac{1}{\pi}-\hspace{-0.435cm}\int\text{d}s^{\prime} \ 
\frac{-\operatorname{Im}\Pi(s^{\prime})}{s-s^{\prime}} \ ,
\label{eq:disp}
\end{equation}
compare with Eq.\ (\ref{eq:Redispersionrel}).
TR and BP now assume a simple model for the imaginary part of each $\Pi_i(s)$, 
see Refs.\ \cite{tornqvist,tornqvist2,boglione,pennington,isgurcomment,tornreply1,harada,harada2} for details:\footnote{It is actually not clear how a Lagrangian formulation of such a model would look like; at least we think that it may not be simple to construct it.}
\begin{equation}
\operatorname{Im}\Pi_i(s) 
=-g_{i}^{2}\frac{k_{i}(s)}{\sqrt{s}}(s-s_{A,i})F_{i}^{2}(s)\Theta(s-s_{th,i}) \ .
\label{eq:ImPi}
\end{equation}
In the scalar--isovector sector the Adler zeros $s_{A,i}$ are set to zero for simplicity\ 
\cite{tornqvist,tornqvist2,boglione,pennington}.\ The form factor is chosen to be a simple exponential,
\begin{equation}
F_{i}(s)=\exp\left[-k_{i}^{2}(s)/(2k_{0}^{2})\right] \ ,
\label{eq:formfactor}
\end{equation}
where $k_{0}$ is a cutoff parameter and $k_{i}(s)$ is the absolute value of the 
three-momentum of the decay particles in the rest frame of the resonance,
\begin{equation}
k_{i}(s) = \frac{1}{2\sqrt{s}}\sqrt{s^{2}+(m_{i1}^{2}-m_{i2}^{2})^{2}-2(m_{i1}^{2}+m_{i2}^{2})s} \ .
\label{eq:modk}
\end{equation}
An explicit derivation of this formula can be found in Appendix\ \ref{app:appendixC}. 
Here, $m_{i1}$ and $m_{i2}$ are the masses of the decay particles, {\em i.e.}, in our case
the pseudoscalar mesons.\footnote{TR and BP did not quote values for the $m_{i1}$, $m_{i2}$.\ 
We used in this section the isospin-averaged numerical values 
given in the PDG from 2002, the year of publication of BP. Note that these values 
differ from the ones used in TR, so that our results for the pole positions 
slightly differ numerically from theirs.} 
The form factor $F_{i}(s)$ guarantees that the imaginary part of $\Pi(s)$ vanishes 
sufficiently fast for $s \rightarrow \infty$ (the inverse cutoff $k_0$ corresponds 
to the non-vanishing size of a typical hadron and is taken to be equal in all channels\ \cite{tornqvist,giacosapagliara}). The step function in Eq.\ (\ref{eq:ImPi}) 
ensures that the decay channel $i$ contributes only when the squared energy of the
resonance exceeds the threshold value $s_{th,i}$.\ Finally, the coupling constants 
$g_{i}$ are related by $SU(3)$ flavor symmetry.

Note that one may also define the so-called Breit--Wigner mass of a resonance
as the real-valued root of the real part of the inverse propagator, $\operatorname{Re}G^{-1}(s)=0$.\ These roots can be
found by identifying the intersections of the so-called 'running mass\grq
\begin{equation}
m^{2}(s)=m_{0}^{2}+\operatorname{Re}\Pi(s) 
\label{eq:runningmass}
\end{equation}
with the straight line $f(s)=s$,
where $s$ is purely real.\ This definition of the mass of the resonance is also used in
TR and BP. However, it was already emphasized that the Breit--Wigner 
mass does not necessarily correspond to a pole in the complex energy plane or to a
peak in the spectral function.

For the scalar--isovector sector, the main results of TR and BP can be summarized
as follows:
\begin{enumerate}
\item TR found a pole on the second Riemann sheet with coordinates\footnote{In agreement with TR and 
in contrast to the definition in 
Eq.\ (\ref{eq:polecoor}), we apply the convenient parameterization $s_{\text{pole}}=m_{\text{pole}}^{2}-i\hspace{0.02cm}m_{\text{pole}}\Gamma_{\text{pole}}$ \hspace{0.02cm}for propagator poles.} 
$m_{\text{pole}}=1.084$ GeV and $\Gamma_{\text{pole}}=0.270$ GeV, which is a 
companion pole corresponding to the resonance $a_{0}(980)$.\ A reanalysis\ \cite{tornqvist2} where
the complex plane was investigated more carefully revealed another pole 
with $m_{\text{pole}}=1.566$ GeV and $\Gamma_{\text{pole}}=0.578$ GeV on the third 
sheet.\ This pole is indeed the original seed state and describes the resonance 
$a_{0}(1450)$.\ It was suggested that, although the numerical agreement was not yet 
satisfactory, an improved model could in principle be capable of describing the whole 
scalar--isovector sector up to $1.6$ GeV. TR also reported one (but not more) intersection 
point(s) of the running mass from Eq.\ (\ref{eq:runningmass}).
\item BP used the same approach but did not look for poles of the propagator.
Instead, the authors considered the Breit--Wigner mass. Compared to TR, also the 
values of the bare mass parameter $m_{0}$ as well as the overall strength of the couplings
$g_i$ in Eq.\ (\ref{eq:ImPi}) were changed. 
BP found two intersection points for the running mass from Eq.\ 
(\ref{eq:runningmass}), one in the region around $1$ GeV 
corresponding to $a_{0}(980)$ (like TR), and another one at about $1.4$ GeV 
(absent in TR). This latter intersection was interpreted as the state $a_{0}(1450)$.\
Note that, although BP did not investigate the poles of the propagator, a pole and an
 intersection were reported in an earlier work\ \cite{boglione}.
\end{enumerate}

Apparently, the situation is not yet conclusive regarding the number and location
of poles of the propagator and/or intersection points of the running mass. 
Therefore, we decided to repeat the study of TR and BP and to investigate
the propagator in the complex plane including all Riemann sheets nearest to the
first (physical) sheet in order to clarify this problem \cite{a0}.
%\begin{table}[t]
%\center
%\begin{tabular}
%[c]{lcccc}
%\toprule Thresholds & Sheet & Sheet Number &  & \\
%\midrule $\eta\pi$, $K\bar{K}$, $\eta^{\prime}\pi$ & $-,+,+$ & II &  & \\
%& $-,-,+$ & III &  & \\
%& $-,-,-$ & VI &  & \\
%\toprule &  &  &  &
%\end{tabular}
%\caption{Sheet numbering.}
%\label{tab:tab1}
%\end{table}
The self-energy on the unphysical sheet(s) is obtained by analytic continuation.
To this end, one first computes the discontinuity of the self-energy across the
real $s$-axis,
\begin{equation}
\disc\Pi(s)=2i\lim_{\epsilon\rightarrow0^{+}}\sum_{i}
\operatorname{Im}\Pi_{i}(s+i\epsilon) \ , \ \ \ s \in \mathbb{R} \ .
\end{equation}
Then, the appropriately continued self-energy $\Pi^{c}(s)$ on the next Riemann sheet is 
obtained via
\begin{equation}
\Pi^{c}(s)=\Pi(s)+\disc\Pi(s) \ .
\end{equation}
This expression is valid on the whole Riemann sheet, {\em i.e.}, the energy $s$ is complex-valued.
Note that in our case there are three thresholds, in successive order corresponding 
to the decays of $a_0$ into $\eta\pi$, $K\bar{K}$, and $\eta^{\prime}\pi$.\ These channels will
be numbered $i=1,2,3$ in the following.\
Thus, crossing the real $s$-axis at values of $s$ in the interval $(s_{th,1},s_{th,2}]$, we
move from the first to the second Riemann sheet, in the following denoted by roman
numeral II. Analogously, crossing the real
$s$-axis in the interval $(s_{th,2},s_{th,3}]$, we move from the first to the third (III) sheet.
Finally, crossing the real $s$-axis at $s > s_{th,3}$, we move from the first
to the sixth (VI) sheet (in the standard notation, see also Table\ \ref{tab:sheetNum}). Since we will also show plots of 
the spectral function $d(x)$, we recall its definition,
\begin{equation}
d(x)=-\frac{2x}{\pi}\lim_{\epsilon\rightarrow0^{+}}\operatorname{Im}
G(x^{2}+i\epsilon) \ , 
\end{equation}
where $x=\sqrt{s}$ and $G(x^{2})$ is the full propagator of the resonance.
\begin{table}[t]
\center
\begin{tabular}[c]{lcc}
\toprule Passed thresholds & Signs & Sheet number\\
\midrule\\[-0.25cm]
$\eta\pi$ & $-,+,+$ & II\\
$\eta\pi$, $K\bar{K}$ & $-,-,+$ & III\\
$\eta\pi$, $K\bar{K}$, $\eta^{\prime}\pi$ & $-,-,-$ & VI\\
\\[-0.1cm]
\toprule
\end{tabular}
\caption{Sheet numbering. The column in the middle indicates the signs of the momenta in Eq.\ (\ref{eq:modk}), after analytic continuation when passing the thresholds.}
\label{tab:sheetNum}
\end{table}

\subsection{Spectral functions and poles}
We also introduce a dimensionless parameter 
$\lambda \in [0,1]$ and replace the coupling constants in Eq.\ (\ref{eq:ImPi}) by
$g_{i}^{2}\rightarrow \lambda g_{i}^{2}$.\ This is completely equivalent to a large-$N_{c}$ study upon setting
\begin{equation}
\lambda = \frac{3}{N_{c}} \ .
\label{eq:lambdaFac}
\end{equation}
In consequence, for $\lambda =0$ the self-energy
vanishes and we just obtain the spectral function of the non-interacting
seed state, that is, a delta function.\ The corresponding pole lies on the real 
$\sqrt{s}$-axis.
Increasing $\lambda$ from zero to $1$, the interaction is successively increased and
we can monitor in a controlled manner how the spectral function changes.\ In the
following figures, we will show the spectral function for the physical value $\lambda=1.0$ 
and for the intermediate value $\lambda=0.4$.\
Changing $\lambda$ from zero to $1$, we will also see how the pole of the seed state 
moves off the real axis and other poles emerge, which correspond to the 
dynamically generated resonances.\ A continuous
change of $\lambda$ will trace out pole trajectories in the complex $\sqrt{s}$-plane.\ 
The final and physical locations of the poles are reached
when $\lambda=1.0$ and are indicated by a dot in the figures.\ 
We consider the three Riemann sheets nearest to the physical region ({\em i.e.}, 
the first sheet) 
in one figure (a list of the poles corresponding to the resonances of interest 
can be found at the end of this chapter in Table\ \ref{tab:tab2}). 
For TR, we use the values $g_{1}=1.2952$ GeV, $g_{2}=0.8094$ GeV, $g_{3}=0.9461$ GeV, and 
$k_{0}=0.56$ GeV, and for BP $g_{1}=1.7271$ GeV, $g_{2}=1.0975$ GeV, $g_{3}=1.4478$ 
GeV, and $k_{0}=0.56$ GeV. The number of digits does not correspond to the numerical precision.

\begin{figure}[ht!]
\vspace{-0.14cm}
\hspace*{-0.9cm} \begin{minipage}[hbt]{8cm}
\centering
\includegraphics[scale=0.8]{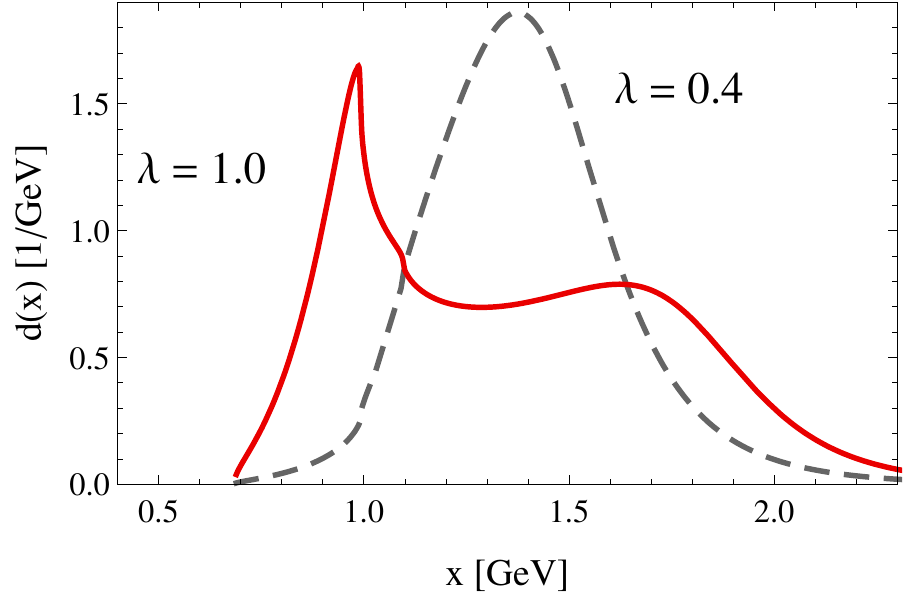}
\end{minipage}
\begin{minipage}[hbt]{8cm}
\centering
\vspace{0.2cm}
\includegraphics[scale=0.869]{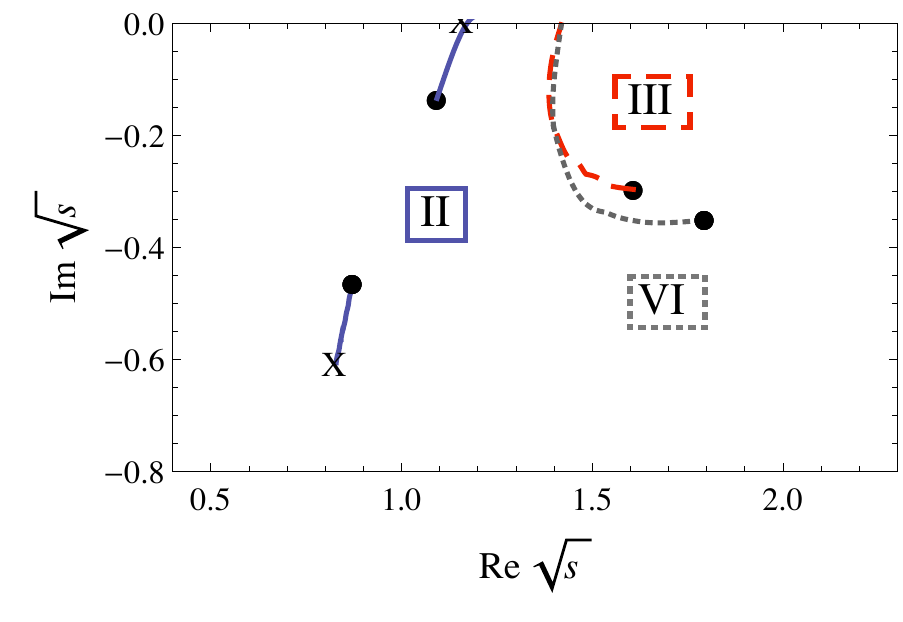}
\end{minipage}
\hspace*{-0.9cm} \begin{minipage}[hbt]{8cm}
\centering
\includegraphics[scale=0.8]{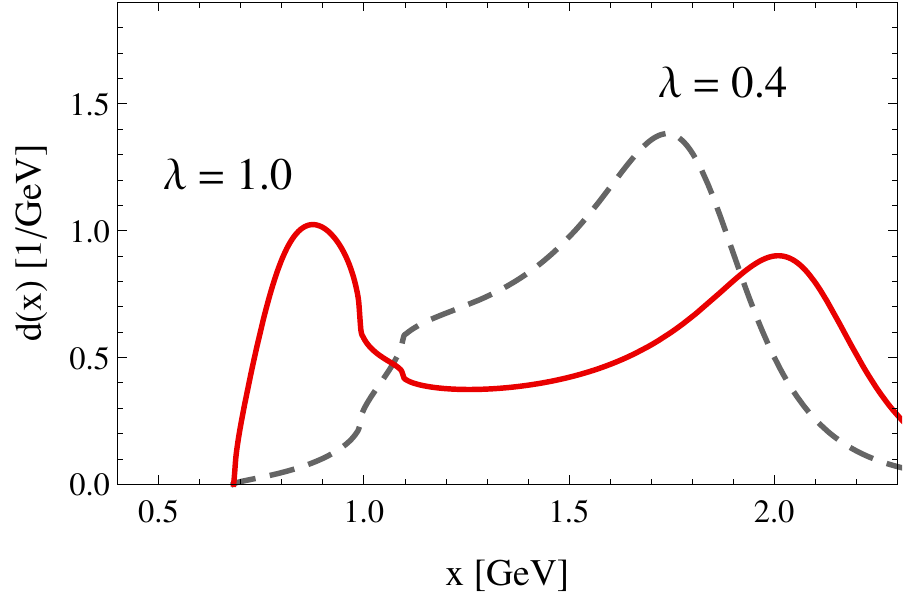}
\end{minipage}
\begin{minipage}[hbt]{8cm}
\centering
\vspace{0.2cm}
\includegraphics[scale=0.869]{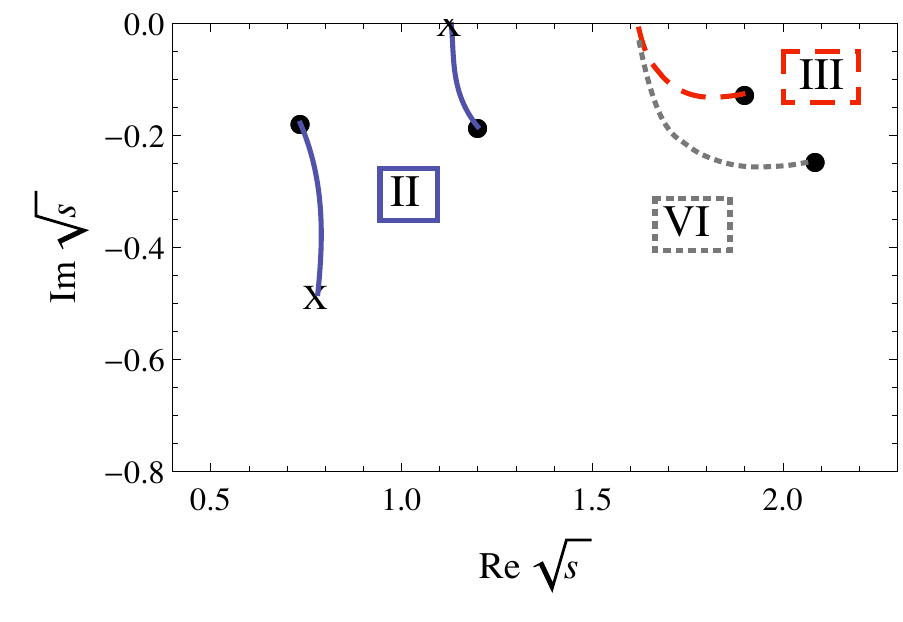}
\end{minipage}
\caption{Spectral functions (left panels) and position of poles in the complex 
$\sqrt{s}$-plane (right panels) for the parameter sets of
TR (upper row) and BP (lower row). Spectral functions are shown for $\lambda =0.4$ (dashed
gray lines) and $\lambda=1.0$ (solid red lines). The pole trajectories of
the seed state are indicated by gray dotted or red dashed lines (for details, see text), 
the one for the dynamically generated
resonance by solid blue lines.\ The roman numerals 
indicate the Riemann sheets where the respective poles can be found.
Final pole positions $(\lambda =1.0)$ are indicated by solid black dots, pole positions at
$\lambda_{c,i}$, {\em i.e.}, where the pole $i$ first emerges, are indicated by X.}
\label{fig:spec_and_poles}
\end{figure}

The results are shown in Figure\ \ref{fig:spec_and_poles}.\ We first discuss the results for
the TR parameterization, and then those for BP.
\begin{enumerate}
\item The two panels in the upper row of Figure\ \ref{fig:spec_and_poles} show the 
results for TR. In the case of $\lambda=1.0$ the spectral function exhibits a narrow peak in the region 
around $1$ GeV that was interpreted by TR as the $a_{0}(980)$ resonance.\ 
We furthermore observe a broad structure above $1.5$ GeV. For decreasing coupling 
strength the narrow peak around $1$ GeV vanishes, while the broad structure becomes more 
pronounced. It is located around $1.4$ GeV, which is the location of the
seed state. The width of the peak decreases with $\lambda$, such that we obtain
a delta function for $\lambda$ equal zero, as expected (not displayed here).

The behavior described above can also be understood considering the pole structure
in the complex $\sqrt{s}$-plane. 
The narrow peak around $1$ GeV for $\lambda=1.0$ corresponds 
to a pole at $s\approx(1.084^{2}-i\hspace{0.02cm}1.084\cdot0.270)$ GeV$^2$, 
which TR has found on the 
second sheet. This pole is indeed present only if $\lambda$ exceeds the critical value 
$\lambda_{c,1}^{\text{TR}} \approx 0.75$.\ The pole emerges close to (but not on) 
the real axis for $\lambda_{c,1}^{\text{TR}}$
and descends down into the complex plane on the second sheet for increasing coupling 
strength. One can interpret this appearance and motion of a pole as a feature typical
for the kind of dynamical generation we are interested in.

However, we also find another pole on the second sheet emerging at a
large imaginary value of $\sqrt{s}$ and moving up toward the real axis.\ 
It first appears for $\lambda_{c,2}^{\text{TR}} \approx0.84$.\ Its effect on the spectral
function is hard to discern, since (the absolute value of) its imaginary part ({\em i.e.}, its 
decay width) is still too large at $\lambda=1.0$.\ This pole was not reported in TR, 
yet, in Ref.\ \cite{comm_beveren}, a similar situation was described where 
the $a_{0}(980)$ was taken to have such a behavior, strictly speaking, its pole was coming 
from the region of large negative imaginary parts of $\sqrt{s}$ and heading toward the 
real axis. This is, however, not the case for the pole of TR, which is dynamically
generated near the real axis and then shifts toward larger (negative) imaginary
values of $\sqrt{s}$.

On the third sheet, TR reports another pole.\ One could think that this
pole corresponds to the seed state, since for $\lambda=0$ the pole trajectory starts 
on the real axis at the mass of the seed state. However, the pole lies on the third sheet,
so prior to crossing the $\eta^{\prime}\pi$ threshold, but its location is above
that threshold. Therefore, this pole does not induce the broad bump above $1$ GeV 
in the spectral function. However, there is also a pole on the sixth sheet which also 
starts at the mass of the seed state. From its position this pole can also be considered 
to generate the broad resonance shape in the spectrum above $1.5$ GeV. 
It is interesting that the pole on the 
third sheet was suggested in TR to correspond to the $a_{0}(1450)$ resonance. 
From our point of view, because of the above arguments it is more natural to 
take the pole on the sixth sheet. A close inspection of the peak position of the
broad bump in the spectral function reveals that it
corresponds more closely to the real part of the pole on the sixth sheet than that on
the third sheet, which corroborates our interpretation.

\item In the lower row of Figure\ \ref{fig:spec_and_poles} we present the results for the 
parameter choice in BP. We find that the qualitative behavior is very similar to the one in TR. 
Quantitatively, we find that the bump in the spectral function corresponding to the 
$a_{0}(980)$ resonance is now somewhat wider. The broad structure at large
$\sqrt{s}$ is more pronounced and now lies around $2$ GeV. For decreasing coupling 
strength the peak becomes narrower and moves toward $1.6$ GeV (because the 
seed state is located there).

We find again two poles on the second sheet.\ 
The right pole appears first for $\lambda_{c,1}^{\text{BP}}\approx0.69$ and the
left one for $\lambda_{c,2}^{\text{BP}}\approx0.66$.\ The parameter set of BP does 
not yield a pole structure from which one can infer which pole corresponds to the 
$a_{0}(980)$.\ Both poles give too large widths, and the left one is too light, while the right 
one is too heavy. It seems that both of them are relevant in the generation of the 
bump at $1$ GeV in the spectrum. Moreover, it does not seem to be appropriate
to assign the poles on the other two sheets to $a_{0}(1450)$.\ At least within this model 
and with the chosen parameters, the pole masses are definitely too high.\footnote{From the discussion of the poles, we can furthermore
conclude that crossings of the running mass are not really indicative of poles
in the propagator. BP reports three crossings, the first at a mass value close to $1$ GeV,
the second one around $1.4$ GeV, and a third one located around $1.8$ GeV.
The latter was discarded in BP as unphysical 
(see Ref.\ \cite{boglione} for more details). From our point of view it is not possible
to unambiguously assign poles to these crossings.}
\end{enumerate}

\section{Effective model with derivative interactions}
\label{seq:simpleEff}
In the previous section, we have re-examined the approach of
TR and BP in order to dynamically generate resonances in the scalar--isovector sector.\ 
We now apply the above mechanism of dynamical generation of resonances 
using a formulation based on an interaction Lagrangian. In contrast to what was discussed before, 
we show that the mechanism of dynamical 
generation in fact produces results which are in quantitative agreement with the data. 
To this end, we introduce a Lagrangian inspired by the eLSM\ \cite{eLSM1,eLSM1-2,eLSM1-3,denisphd,eLSM2}, 
where the mesons interact via derivative and non-derivative couplings. In our case, the
Lagrangian contains a single isotriplet seed state $a_{0}$ that corresponds to the 
resonance $a_{0}(1450)$.\ A careful analysis of the pole structure of the corresponding
propagator shows that it is indeed possible to obtain a narrow resonance with 
mass around $1$ GeV, the 
pole coordinates of which fit quite well with those of the physical $a_{0}(980)$ resonance, 
and \emph{simultaneously} obtain a pole for the seed in agreement with that
for the $a_{0}(1450)$\ \cite{olive}.

\vspace{0.15cm}
\subsection{Interactions with derivatives: A lesson from the eLSM}
The way a scalar field couples to pseudoscalar states depends on the effective 
approach used. Let us, for instance, consider the coupling of $a_{0}$ to kaons.
In chiral perturbation theory (chPT)\ \cite{chpt,chpt2}, 
which is based on the nonlinear realization of chiral symmetry, 
only derivative couplings of the type $a_{0}^{0}\partial_{\mu}K^{0}\partial^{\mu}\bar{K}^{0}$
can appear in the chiral limit\ \cite{Ecker}.\ Away
from the chiral limit, a non-derivative coupling $a_{0}^{0}K^{0}\bar{K}^{0}$ appears, too, 
but its strength is proportional to $m_{K}^{2}$, {\em i.e.}, via the Gell-Mann--Oakes--Renner
relation proportional to the explicit breaking of chiral
symmetry by nonzero quark masses.\ On the other hand, if the standard
linear sigma model (without vector degrees of freedom) is considered, the coupling is 
only of the non-derivative type $a_{0}^{0}K^{0}\bar{K}^{0}$.\ 
At tree-level both chPT and the sigma model can coincide, but when loops are included 
differences arise due to the different $s$-dependence in the amplitudes.

Studying the spectral function of 
$\phi\rightarrow a_{0}(980)\gamma\rightarrow\eta\pi^{0}\gamma$ measured by the 
KLOE Collaboration\ \cite{kloe}, it was shown in Refs.\ \cite{pagliaraderivatives,pagliaraderivatives2} 
that a derivative coupling of the type 
$a_{0}^{0}\partial_{\mu}K^{0}\partial^{\mu}\bar{K}^{0}$ seems to be necessary.\ 
As we shall demonstrate here, we come to the same conclusion: 
a derivative coupling is necessary for the simultaneous description of both 
resonances $a_{0}(980)$ and $a_{0}(1450)$.\ Interestingly, the eLSM naturally contains both non-derivative and derivative 
coupling terms -- as mentioned in Chapter \ref{chap:chapter2}, this feature is due to the inclusion of (axial-)vector degrees
of freedom in the model, for more details see Refs.\ \cite{eLSM1,eLSM1-2,eLSM1-3,denisphd,eLSM2}.\ In this approach the resonance $a_{0}(1450)$ turns out to be predominantly 
a quark-antiquark (seed) state with a (bare) mass of $m_{a_{0}}=1.363$ GeV.

The Lagrangian for the scalar--isovector sector emerging from the eLSM has the 
following form for the neutral state:
\begin{eqnarray}
\mathcal{L}_{a_{0}\eta\pi}^{\text{eLSM}} & = & A_{1}^{\text{eLSM}}a_{0}^{0}\eta\pi^{0}
+B_{1}^{\text{eLSM}}a_{0}^{0}\partial_{\mu}\eta\partial^{\mu}\pi^{0}
+C_{1}^{\text{eLSM}}\partial_{\mu}a_{0}^{0}(\pi^{0}\partial^{\mu}\eta
+\eta\partial^{\mu}\pi^{0}) \ , \nonumber \\[0.15cm]
\mathcal{L}_{a_{0}\eta'\pi}^{\text{eLSM}} & = & A_{2}^{\text{eLSM}}a_{0}^{0}\eta'\pi^{0}
+B_{2}^{\text{eLSM}}a_{0}^{0}\partial_{\mu}\eta'\partial^{\mu}\pi^{0}
+C_{2}^{\text{eLSM}}\partial_{\mu}a_{0}^{0}(\pi^{0}\partial^{\mu}\eta'
+\eta'\partial^{\mu}\pi^{0}) \ , \nonumber \\[0.15cm]
\mathcal{L}_{a_{0}K\bar{K}}^{\text{eLSM}} & = & A_{3}^{\text{eLSM}}
a_{0}^{0}(K^{0}\bar{K}^{0}-K^{-}K^{+})+B_{3}^{\text{eLSM}}a_{0}^{0}
(\partial_{\mu}K^{0}\partial^{\mu}\bar{K}^{0}-\partial_{\mu}K^{-}\partial^{\mu}K^{+}) 
\nonumber \\[0.15cm]
&  & + \ C_{3}^{\text{eLSM}}\partial_{\mu}a_{0}^{0}(K^{0}\partial^{\mu}\bar{K}^{0}
+\bar{K}^{0}\partial^{\mu}K^{0}-K^{-}\partial^{\mu}K^{+}-K^{+}\partial^{\mu}K^{-}) \ ,
\label{eq:Lagrangian}
\end{eqnarray}
where $A_{i}^{\text{eLSM}}$, $B_{i}^{\text{eLSM}}$, and $C_{i}^{\text{eLSM}}$ are coupling 
constants that are functions of the parameters of the model\ \cite{denisphd,eLSM2}.\ 
Note that both non-derivative 
and derivative interactions appear.\ The derivatives in front of the fields produce 
an $s$-dependence in the decay amplitudes, $-i\mathcal{M}_{i}^{\text{eLSM}}(s)$, 
which enter the tree-level expressions of the decay widths,
\begin{equation}
\Gamma_{i}^{\text{eLSM}}(s) = \frac{k_{i}(s)}{8\pi s}|\hspace{0.03cm}\text{--}\hspace{0.05cm}
i\mathcal{M}_{i}^{\text{eLSM}}(s)|^{2}\hspace{0.02cm}\Theta(s-s_{th,i}) \ ,
\end{equation}
which have to be evaluated for $s = m_{a_0}^2$.\ The amplitudes read
\begin{equation}
\mathcal{M}_{i}^{\text{eLSM}}(s) = A_{i}^{\text{eLSM}}
-\frac{1}{2} B_{i}^{\text{eLSM}}\left(s-m_{i1}^{2}-m_{i2}^{2} \right)+C_{i}^{\text{eLSM}}s \ ,
\label{eq:eLSMamp}
\end{equation}
where the masses $m_{i1}$ and $m_{i2}$ are the pseudoscalar masses in the relevant channels.
The parameters of the eLSM were determined from a $\chi^2$ fit to tree-level 
masses and decay widths.\ So far, no loop corrections were considered.\
For a consistent loop calculation one would have to perform a new fit of the parameters, 
which is an 
interesting project for future work.

A first attempt to incorporate loop corrections in a scheme inspired by the eLSM was
presented in Refs.\ \cite{procEEF70,proceqcd}.\ There, the $s$-dependence of the 
amplitudes (\ref{eq:eLSMamp}) was completely neglected and the form factor from Eq.\ (\ref{eq:formfactor}) was introduced,
\begin{equation}
-\hspace{-0.08cm}i\mathcal{M}_{i}^{\text{eLSM}}(s) \ \rightarrow \ -i\mathcal{M}_{i}(s) 
= -i\mathcal{M}_{i}^{\text{eLSM}}(m_{a_{0}}^{2})F_{i}(s) \ .
\end{equation}
After that, the imaginary part of the self-energy was computed using the optical theorem from Eq.\ (\ref{eq:optical_simple}),
\begin{equation}
\operatorname{Im}\Pi_{i}(s) = - \sqrt{s}\, \Gamma_{i}^{\text{tree}}(s) = - \frac{k_{i}(s)}{8\pi 
\sqrt{s}}|\hspace{0.03cm}\text{--}\hspace{0.05cm}i\mathcal{M}_{i}(s)|^{2}\hspace{0.02cm}\Theta(s-s_{th,i}) \ ,
\label{eq:optical}
\end{equation}
and the real part from the dispersion relation\ (\ref{eq:disp}). This effectively reduced the interaction part of the eLSM Lagrangian to a pure non-derivative form,
\begin{equation}
\mathcal{L}_{\text{red}}^{\text{eLSM}} = A_{\text{red}}^{\text{eLSM}}a_{0}^{0}\eta\pi^{0}+B_{\text{red}}^{\text{eLSM}}a_{0}^{0}\eta^{\prime}\pi^{0}+C_{\text{red}}^{\text{eLSM}}a_{0}^{0}(K^{0}\bar{K}^{0}-K^{-}K^{+}) \ ,
\end{equation}
where $A_{\text{red}}^{\text{eLSM}}$, $B_{\text{red}}^{\text{eLSM}}$, and $C_{\text{red}}^{\text{eLSM}}$ are combinations of the coupling constants and masses\footnote{Note that in the mentioned work we had used the pseudoscalar masses as they were obtained from the eLSM.} (values on-shell) deduced from Eq.\ (\ref{eq:eLSMamp}).

It turns out that the complex propagator pole for the seed state $a_{0}(1450)$ on the sheet nearest to the physical region is too close to the real axis, hence yields a too small decay width, if the cutoff parameter $\Lambda=\sqrt{2}\hspace{0.02cm}k_{0}$ is too small. Only if $\Lambda\simeq1.4$ GeV it is possible to obtain values for the width in terms of pole coordinates that are in agreement with the experiment. However, in any case the resulting mass is too small. Increasing the value of the cutoff further makes the $a_{0}(1450)$ even lighter and broader.\ On the other hand, we do not find any companion pole of $a_{0}(1450)$ and this result does not change upon variations of $\Lambda$.\ This simply means that no additional pole corresponding to the $a_{0}(980)$ is dynamically generated. Obviously, neglecting the $s$-dependence of the amplitudes is an oversimplification. One has to take into account the derivatives in some way; at the same time care is needed when derivative interactions appear in a Lagrangian, as we have demonstrated in Chapter \ref{chap:chapter3}.

Note that since the values of the parameters of the eLSM were determined at tree-level only\ \cite{denisphd,eLSM2}, one should not use them here in the expressions for $A_i^{\text{eLSM}}$, $B_i^{\text{eLSM}}$, and $C_i^{\text{eLSM}}$.\ Therefore, we shall treat the latter as free parameters
in the following and will compare them to the ones of the eLSM later on.

\subsection{Effective model with both non-derivative and derivative interactions}
We now consider an effective model for the isotriplet, containing the same 
decay channels as the eLSM and including also non-derivative and derivative interactions.\footnote{From now on we consistently use physical masses for the pseudoscalar mesons as given by the PDG\ \cite{olive} throughout this chapter.}\ 
The Lagrangian (for the neutral state) is given by the sum of the following terms:
\begin{eqnarray}
\mathcal{L}_{a_{0}\eta\pi} & = & A_{1}a_{0}^{0}\eta\pi^{0}
+B_{1}a_{0}^{0}\partial_{\mu}\eta\partial^{\mu}\pi^{0} \ , \label{eq:Lag_eff} \\[0.1cm]
\mathcal{L}_{a_{0}\eta^{\prime}\pi} & = & A_{2}a_{0}^{0}\eta^{\prime}\pi^{0}
+B_{2}a_{0}^{0}\partial_{\mu}\eta^{\prime}\partial^{\mu}\pi^{0} \ , \nonumber \\[0.1cm]
\mathcal{L}_{a_{0}K\bar{K}} & = & A_{3}a_{0}^{0}(K^{0}\bar{K}^{0}-K^{-}K^{+})
+B_{3}a_{0}^{0}(\partial_{\mu}K^{0}\partial^{\mu}\bar{K}^{0}-\partial_{\mu}K^{-}\partial^{\mu}K^{+}) \nonumber \ .
\end{eqnarray}
Formally it can be obtained by rewriting the terms proportional to $C_{i}^{\text{eLSM}}$ 
in Eq.\ (\ref{eq:Lagrangian}) by an integration by parts to get rid of
the derivatives in front of the $a_0$ fields:
\begin{eqnarray}
C_{i}^{\text{eLSM}}a_{0}^{0}\partial^{\mu}\eta\partial_{\mu}\pi^{0} & = & C_{i}^{\text{eLSM}}\partial^{\mu}\left(a_{0}^{0}\partial_{\mu}\eta\pi^{0}\right)-C_{i}^{\text{eLSM}}\partial^{\mu}a_{0}^{0}\partial_{\mu}\eta\pi^{0}-C_{i}^{\text{eLSM}}a_{0}^{0}\square\eta\pi^{0} \nonumber \\[0.15cm]
\Rightarrow \ \ \ C_{i}^{\text{eLSM}}\partial^{\mu}a_{0}^{0}\partial_{\mu}\eta\pi^{0} & = & -C_{i}^{\text{eLSM}}a_{0}^{0}(\square\eta\pi^{0}+\partial^{\mu}\eta\partial_{\mu}\pi^{0}) \ .
\label{eq:LagDiff}
\end{eqnarray}
Subsequently, one replaces the emerging second derivatives with the help of the Klein--Gordon equation, {\em e.g.}\ $\square\eta=-m_{\eta}^{2}\eta$.\ One should realize that this last step is actually a simplification. It can be 
easily shown that the tree-level decay amplitudes coming from both types of interaction Lagrangians in Eq.\ (\ref{eq:LagDiff}) are the same. However, it is $a\ priori$ not clear if this holds true also for the loop integrals, since a d'Alembert operator is involved. For what concerns its structure we will therefore treat our Lagrangian (\ref{eq:Lag_eff}) only as being inspired by the eLSM. Then, it gives rise to the following $s$-dependent amplitudes,
\begin{equation}
\mathcal{M}_{i}^{\text{eff}}(s) = \left[A_{i}-\frac{1}{2} B_{i}\left(s-m_{i1}^{2}-m_{i2}^{2}\right)
\right] F_{i}(s) \ ,
\end{equation}
where we have already included the form factor $F_{i}(s)$ as defined in Eq.\ (\ref{eq:formfactor}) with $\Lambda=\sqrt{2}\hspace{0.02cm}k_{0}$.\ 
We stress once more that the constants $A_{i}$ and $B_{i}$ will not
be computed from the numerically determined parameters of the eLSM, but
will be determined independently in order to produce the masses and decay widths of the resonances
under study. Note that in chPT the parameters
$A_{i}$ are proportional to the masses of the pseudo-Nambu--Goldstone bosons as
$A_{1}\propto m_{\pi}^{2}+m_{\eta}^{2}$, $A_{2}\propto m_{\pi}^{2}
+m_{\eta^{\prime}}^{2}$, $A_{3}\propto2m_{K}^{2}$ and thus vanish in the chiral limit. Thus, also from this
consideration, we expect that the derivative terms are sizable and crucial for
the determination of the resonance poles. We computed the real and imaginary part of the self-energy as 
was described in the third chapter, {\em i.e.}, we computed the tree-level decay widths and used the optical 
theorem from Eq.\ (\ref{eq:optical}) to obtain the imaginary part of the self-energy. We then applied the dispersion relation\ (\ref{eq:disp}) in order to calculate the corresponding real part (including the tadpole diagrams).

There are eight parameters in our approach:\ $m_{0}$, $\Lambda$, and the
six coupling constants $A_i$,\ $B_i$ $(i=1,2,3)$.\ We vary the numerical values
of $m_{0}$ and $\Lambda$ within reasonable intervals $m_{0} \in (0.8,1.5)$ GeV
and $\Lambda \in (0.4,1.5)$ GeV. Each time we perform a fit of the six
coupling constants to six experimental quantities:\ one pole in the PDG range
for $a_{0}(980) \ (\text{in our case }\sqrt{s}=(0.980-i\hspace{0.02cm}0.038) \ \text{GeV})$ and one for 
$a_{0}(1450) \ (\text{in our case }\sqrt{s}=(1.474-i\hspace{0.02cm}0.133) \ \text{GeV})$, and 
the central values of the branching ratios of $a_{0}(1450)$ evaluated at the physical 
mass, see Eq.\ (\ref{eq:branching}). 
By this, all six free parameters can be fixed.

It turns out that there is only a \textit{narrow} range of suitable values of
the parameters $m_{0}$ and $\Lambda$ for which the fit of the six coupling
constants is possible:\ approximately $m_{0} \in (0.9,1.2)$ GeV and $\Lambda \in (0.4,0.9)$ GeV. 
Here, 'approximately\grq\ refers to 
the fact that, due to the interdependence of the parameters, the window is not rectangular. 
However, a small change in $m_{0}$ and/or $\Lambda$ by $\sim50$ MeV near the borders of the quoted interval
does not allow one to reproduce the data anymore.\ Thus, although we have eight parameters, 
we are severely constrained in their choice in order to describe the $I=1$ resonances. 
As we will see below, the present parameters also explain why $a_{0}(980)$ couples strongly to kaons.
In order to have similar values for the cutoff and the bare mass as in the previous section, we chose as final values for
the parameters and coupling constants
\vspace{-0.2cm}
\begin{align}
m_{0}&=1.2 \ \text{GeV} \ , & \Lambda &= 0.6 \ \text{GeV} \ , \\
A_{1} &= 2.69 \ \text{GeV} \ , & B_{1} &= -7.95 \ \text{GeV}^{-1} \ , \nonumber \\
A_{2} &= 8.42 \ \text{GeV} \ , & B_{2} &= 7.24 \ \text{GeV}^{-1} \ , 
\nonumber \\
A_{3} &= -6.89 \ \text{GeV} \ , & B_{3} &= -1.66 \ \text{GeV}^{-1} \ . \label{eq:AiBi}
\end{align}
Just as in the discussion of TR and BP, we rescale the above coupling constants by a 
common factor $\sqrt{\lambda}$, see Eq.\ (\ref{eq:lambdaFac}), and compute the corresponding spectral functions.\ 
The result is shown in the left panel of Figure\ \ref{fig:toy}.\ We also compute the
pole trajectories in the complex $\sqrt{s}$-plane by varying $\lambda$ from zero to $1$.\ 
The following comments are in order:
\begin{figure}[t]
\hspace*{-1.4cm}
\begin{minipage}[hbt]{8cm}
\centering
\includegraphics[scale=0.805]{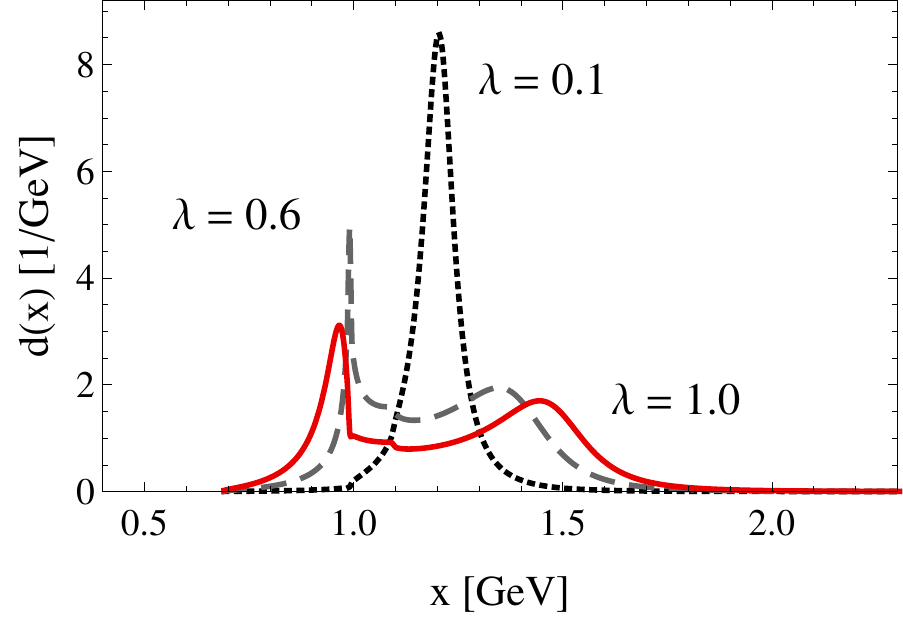}
\end{minipage}
\hspace*{0.0cm}
\begin{minipage}[hbt]{8cm}
\centering
\vspace*{0.16cm}
\includegraphics[scale=0.9]{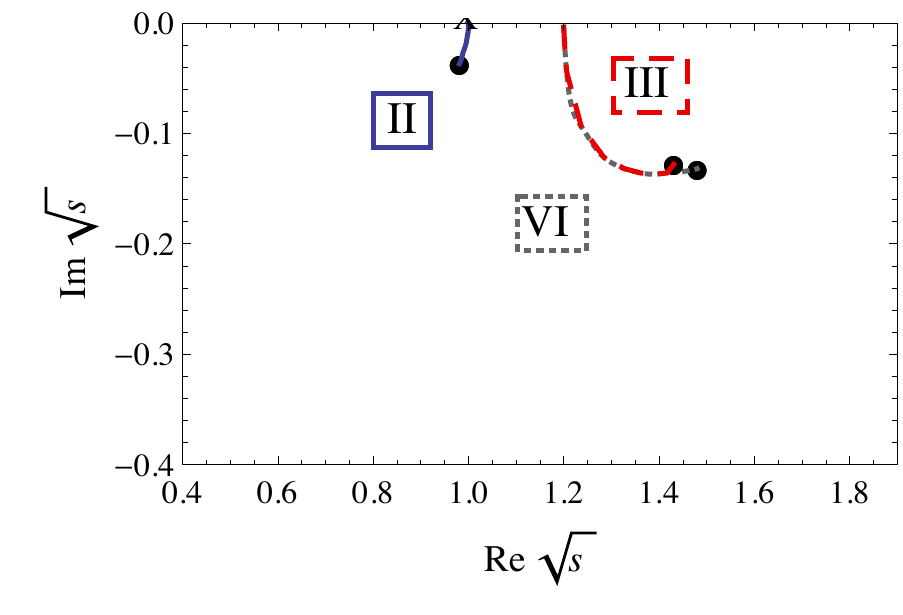}
\end{minipage}
\caption{In the left panel we
show the spectral functions for three different values of $\lambda$.\ In the right panel
we display pole trajectories obtained by varying $\lambda$ from zero to $1$.\ 
Black dots indicate the position of the poles for $\lambda=1.0$.\ The X indicates the
pole position for $\lambda_c$, {\em i.e.}, when the pole first emerges.
The roman numeral indicates on which sheet the respective pole can be found.}
\label{fig:toy}
\end{figure}
\begin{enumerate}
\item The spectral function shows a narrow peak for $\lambda=1.0$ at a value of 
$x= \sqrt{s}$ slightly smaller than $1$ GeV, which can be interpreted as the $a_{0}(980)$.\ 
The form is distorted by the nearby $K\bar{K}$ threshold and resembles the Flatt\'{e} 
distribution\ \cite{flatte,flatte2}, see also Refs.\ \cite{giacosapagliara,baru} and references therein.\ 
The pole corresponding to this peak lies on the second sheet and has coordinates
\begin{equation}
\sqrt{s} = (0.981-i\hspace{0.02cm}0.037) \ \text{GeV} \ ,
\end{equation}
meaning that we find the $a_{0}(980)$ to have a mass of $m_{\text{pole}}=0.980$ GeV 
and a width of $\Gamma_{\text{pole}}=0.075$ GeV. 
This pole appears only if $\lambda$ exceeds $\lambda_c\approx0.61$ (note that the pole trajectory is very different 
to the one reported in Ref.\ \cite{zhiyong}).\ The corresponding
position is indicated by an X in the right panel of Figure\ \ref{fig:toy}.\ The important thing here 
is that, in contrast to what we have found for the TR and BP parameterizations, there is 
only one pole for the $a_{0}(980)$ and thus there is no ambiguity on the identification of the resonances.\footnote{There are two additional poles in the 
relevant part of the complex plane which will not be displayed and discussed here:\ 
$(i)$ a pole deep in the imaginary region on the second sheet, and $(ii)$ a pole close
to the imaginary axis on the sixth sheet.\ Both have no physical impact.\ 
Quite interestingly, we do not observe any virtual bound states. 
Such poles were described in models without derivative interactions {\em e.g.}\ in Refs.\ \cite{e38,thomasthesis}.}

\item There is also a broad structure around $1.5$ GeV that corresponds to the resonance 
$a_{0}(1450)$.\ For decreasing $\lambda$, both peaks merge and settle around 
$1.2$ GeV, where the seed state is located.

\item As expected, there is (only) one pole present on the third sheet with 
coordinates $\sqrt{s}=(1.431-i\hspace{0.02cm}0.128)$ GeV. However, as in 
TR and BP, we find a pole on the sixth sheet, too. Its coordinates are
\begin{equation}
\sqrt{s} = (1.480-i\hspace{0.02cm}0.132)  \ \text{GeV} \ ,
\end{equation}
or $m_{\text{pole}}=1.474$ GeV and $\Gamma_{\text{pole}}=0.265$ GeV. 
This is the pole which is responsible for the peak around $1.5$ GeV in the spectrum, and thus
we assign it to the $a_{0}(1450)$.\ Nevertheless, since the pole on the third sheet also
reproduces the mass and width of $a_{0}(1450)$ to reasonable accuracy, 
it is in principle possible to regard this one 
as the pole corresponding to $a_{0}(1450)$, too.
\end{enumerate}
\clearpage

The present study demonstrates that, by starting with a unique seed state, it is indeed 
possible to find two poles for the isovector states, both of which reproduce the
masses and widths of $a_0(980)$ and $a_0(1450)$ reasonably well. A summary of our results and, 
for comparison, those of T\"ornqvist and Roos, and Boglione and Pennington, can be found in Table\
\ref{tab:tab2}.
\begin{table}[t]
\center
\begin{tabular}[c]{lcccc}
\toprule & \multicolumn{2}{c}{$a_{0}(980)$} & \multicolumn{2}{c}{$a_{0}(1450)$}\\[0.1cm]
& $m_{\text{pole}}$ [GeV] & $\Gamma_{\text{pole}}$ [GeV] & $m_{\text{pole}}$ [GeV] 
& $\Gamma_{\text{pole}}$ [GeV]\\[0.05cm]
\midrule\\[-0.25cm]
TR\ \cite{tornqvist,tornqvist2} & $1.084$ & $0.270$ & $1.566$ & $0.578$\\
BP\ \cite{pennington} & $\hspace{0.175cm} 1.186^{*}$ & $\hspace{0.165cm} 0.373^{*}$ & $1.896$ & $0.250$\\
Our results \ & $0.980$ & $0.075$ & $1.474$ & $0.265$\\
&  &  &  &\\[-0.15cm]
\vspace{0.3cm}PDG\ \cite{olive} & $ \ 0.980\pm0.020 \ $ & $ \ 0.050$ to $0.100$ \ & $ \ 
1.474\pm0.019 \ $ 
& $ \ 0.265\pm0.013$\\
\multicolumn{5}{l}{${}^{*}$\footnotesize{In order to compare to TR, the right pole on the second 
sheet was chosen.}}\\[0.05cm]
\toprule\\
\end{tabular}
\caption{Numerical results for the pole coordinates in the scalar--isovector sector in 
TR, BP, and our effective model, compared to the PDG values. In the case of the 
$a_{0}(1450)$, the poles listed for TR and BP are located on the third sheet, while our pole 
lies on the sixth sheet.\ All poles for the $a_{0}(980)$ are found on the second sheet. 
Note that all poles listed for BP were obtained performing the analytic 
continuation of the propagator given by BP.}
\label{tab:tab2}
\end{table}

\subsection{Branching ratios and coupling constants for $a_{0}(980)$}
For completeness, we report the branching ratios of our effective model by using 
the tree-level decay widths obtained from the optical theorem\ (\ref{eq:optical}). 
The partial widths are evaluated at the peak value of the spectral function above $1$ GeV, 
$m_{\text{peak}}=1.444$ GeV. For the resonance $a_{0}(1450)$ this leads to
\begin{equation}
\frac{\Gamma_{a_{0}(1450)\rightarrow\eta^{\prime}\pi}^{\text{tree}}}{
\Gamma_{a_{0}(1450)\rightarrow\eta\pi}^{\text{tree}}}\simeq0.41 \ , \ \ \ 
\frac{\Gamma_{a_{0}(1450)\rightarrow K\bar{K}}^{\text{tree}}}{
\Gamma_{a_{0}(1450)\rightarrow\eta\pi}^{\text{tree}}}\simeq0.96 \ ,
\label{eq:ourbranching}
\end{equation}
which can be compared to the experimental values\ \cite{olive}:
\begin{equation}
\frac{\Gamma_{a_{0}(1450)\rightarrow\eta^{\prime}\pi}}{\Gamma_{a_{0}(1450)\rightarrow\eta\pi}}
=0.35\pm0.16 \ , \ \ \ \frac{\Gamma_{a_{0}(1450)\rightarrow K\bar{K}}}{\Gamma_{a_{0}(1450)\rightarrow\eta\pi}}
=0.88\pm0.23 \ .
\label{eq:branching}
\end{equation}

Concerning the resonance $a_{0}(980)$, we give the following estimates for the coupling 
constants in the $\eta\pi$- and $K\bar{K}$-channels: We calculate the partial widths 
$\Gamma_i^{\text{tree}}(s)$, this time with $\sqrt{s}$ equal to 
the peak mass of the spectral function 
below $1$ GeV, $m_{\text{peak}}=0.966$ GeV. Then, Eq.\
(\ref{eq:optical}) is used to solve for the absolute values of the amplitudes, $-i\mathcal{M}_{i}(s)$.\ The result
is multiplied with the root of the wave function renormalization factor, $\sqrt{Z}=0.588$, 
which is its value at the Breit--Wigner mass of the $a_{0}(980)$.\ Thus, we obtain
the coupling constants in the $\eta\pi$- and $K\bar{K}$-channels as
\begin{equation}
g_{\eta\pi}=2.320 \ \text{GeV} \ , \ \ \ g_{K\bar{K}}=5.611 \ \text{GeV} \ .
\end{equation}
It is remarkable that the coupling of $a_{0}(980)$ to kaons turns out to be sizably larger 
than the coupling to $\eta\pi$.\ This is in agreement with various other works on this topic\
\cite{molecular,molecular2,molecular3,molecular4,molecular5,pagliaraderivatives,pagliaraderivatives2,maiani}: 
virtual kaon-kaon pairs near the corresponding threshold are important for the dynamical generation 
of the resonance $a_{0}(980)$.

\subsection{Phase shifts and inelasticity}
We now compute the phase shifts $\delta_{1}=\delta_{\eta\pi}(s)$, $\delta_{2}=\delta_{K\bar{K}}(s)$, and the inelasticity parameter $\eta=\eta(s)$ of our effective model by including all the three channels $\eta\pi$, $K\bar{K}$, and $\eta^{\prime}\pi$.\ However, we will restrict ourselves to energies $s\leq s_{th,3}$, that is, the $\eta^{\prime}\pi$ threshold. This allows us to use only expressions for the two-channel case; the $\hat S$-matrix is then parameterized as usually by
\begin{equation}
\hat S = \begin{pmatrix}\eta e^{2i\delta_{1}} & i\sqrt{1-\eta^{2}}e^{i(\delta_{1}+\delta_{2})} & \dots
 \\ i\sqrt{1-\eta^{2}}e^{i(\delta_{1}+\delta_{2})} & \eta e^{2i\delta_{2}} & \dots \\ \dots & \dots & \dots \end{pmatrix} \ .
\end{equation}
The relevant $\hat S$-matrix elements are shown above and are related to the pole-dominated amplitudes\ \cite{ishida1,ishida2,tornqvist,pennington}
\vspace{-0.1cm}
\begin{eqnarray}
a_{ij} = \frac{-\sqrt{\sqrt{s}\, \Gamma_{i}^{\text{tree}}(s)\sqrt{s}\, \Gamma_{j}^{\text{tree}}(s)}}{s-m_{0}^{2}-\Pi(s)} \ .
\label{eq:deltaAmp}
\end{eqnarray}
This means that
\vspace{-0.3cm}
\begin{eqnarray}
\hat S_{11} & = & \eta e^{2i\delta_{1}} \ \ = \ \ 1+2ia_{11} \ , \nonumber \\
\hat S_{12} & = & i\sqrt{1-\eta^{2}}e^{i(\delta_{1}+\delta_{2})} \ \ = \ \ 2ia_{12} \ , \nonumber \\
\hat S_{22} & = & \eta e^{2i\delta_{2}} \ \ = \ \ 1+2ia_{22} \ . \label{eq:deltaSmatrix}
\end{eqnarray}
By imposing unitarity and restricting to real values only, the phase shifts and the inelasticity can be extracted numerically.

\begin{figure}[t!]
\hspace*{-0.95cm} \begin{minipage}[hbt]{8cm}
\centering
\hspace{0.1cm}
\includegraphics[scale=0.55]{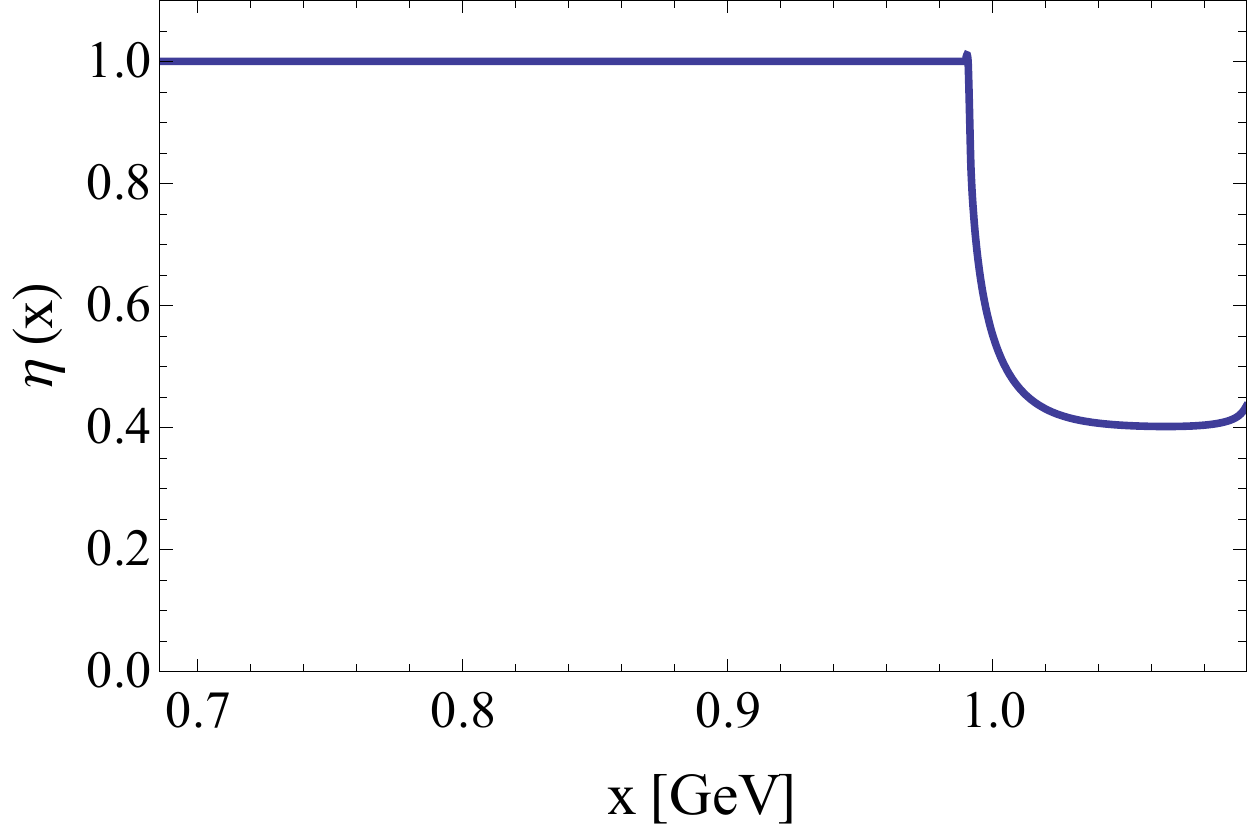}
\end{minipage}
\vspace{0.5cm}
\begin{minipage}[hbt]{8cm}
\centering
\includegraphics[scale=0.55]{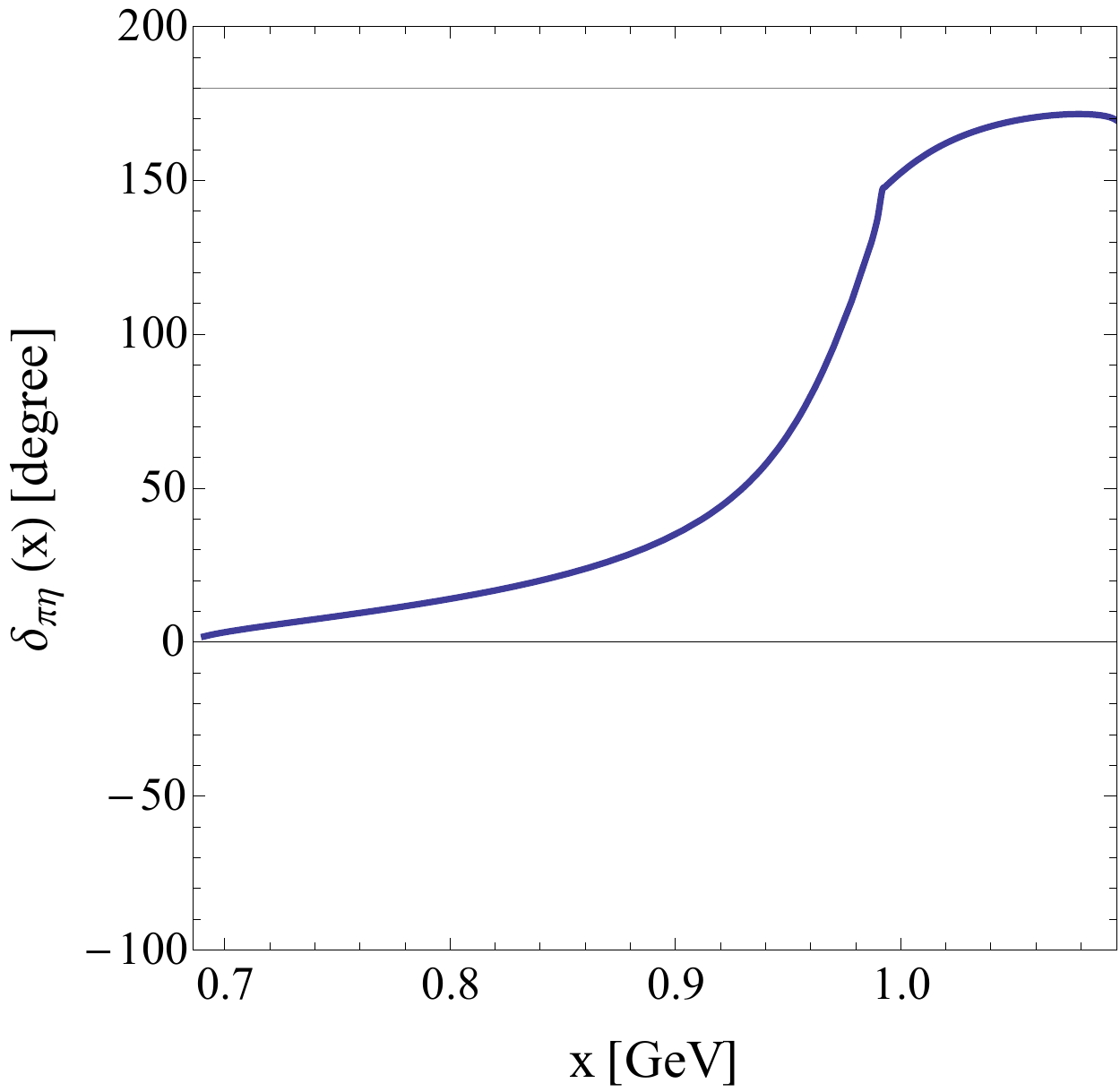}
\end{minipage}
\hspace*{-0.95cm} \begin{minipage}[hbt]{8cm}
\centering
\includegraphics[scale=0.55]{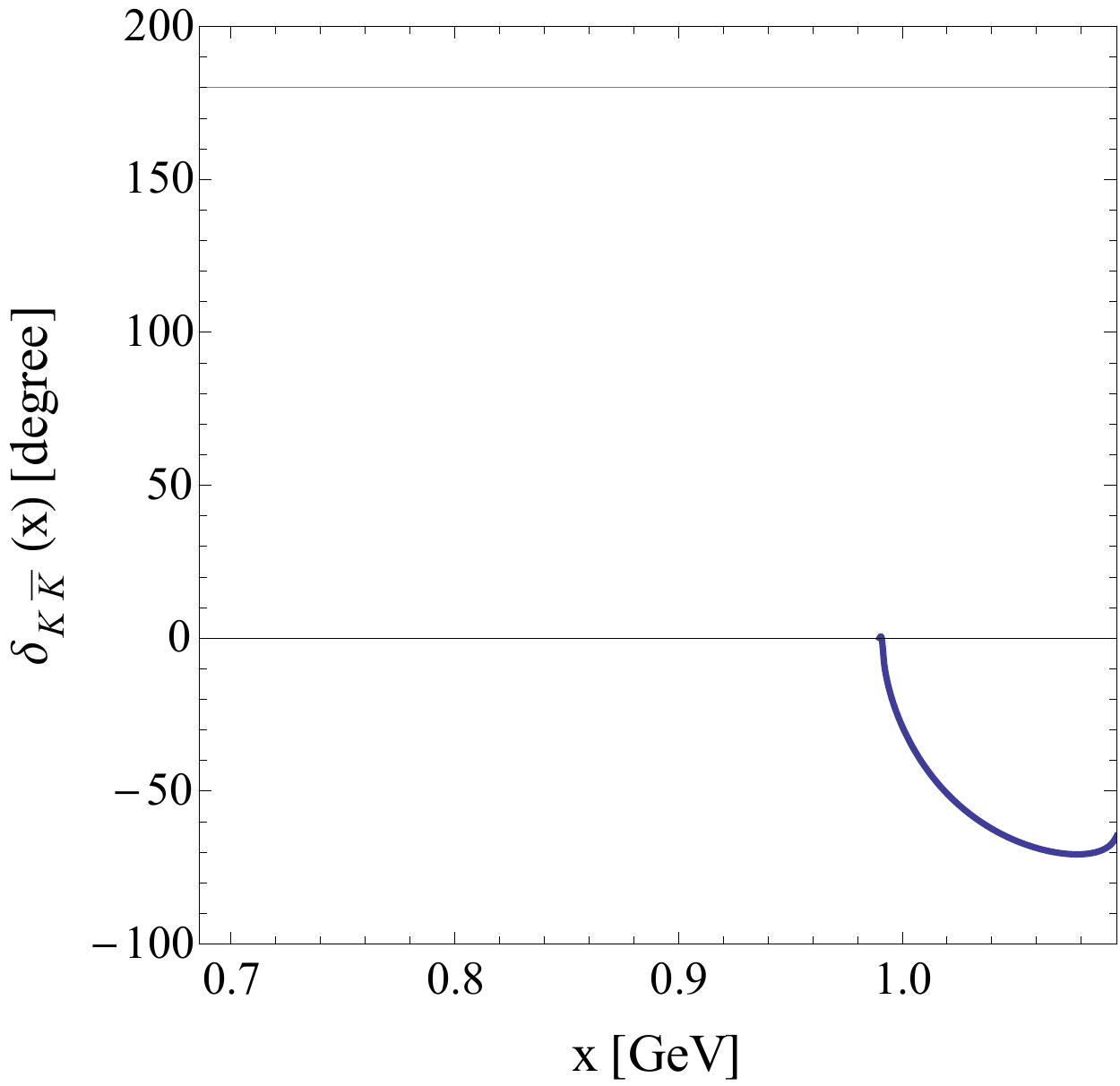}
\end{minipage}
\begin{minipage}[hbt]{8cm}
\centering
\includegraphics[scale=0.55]{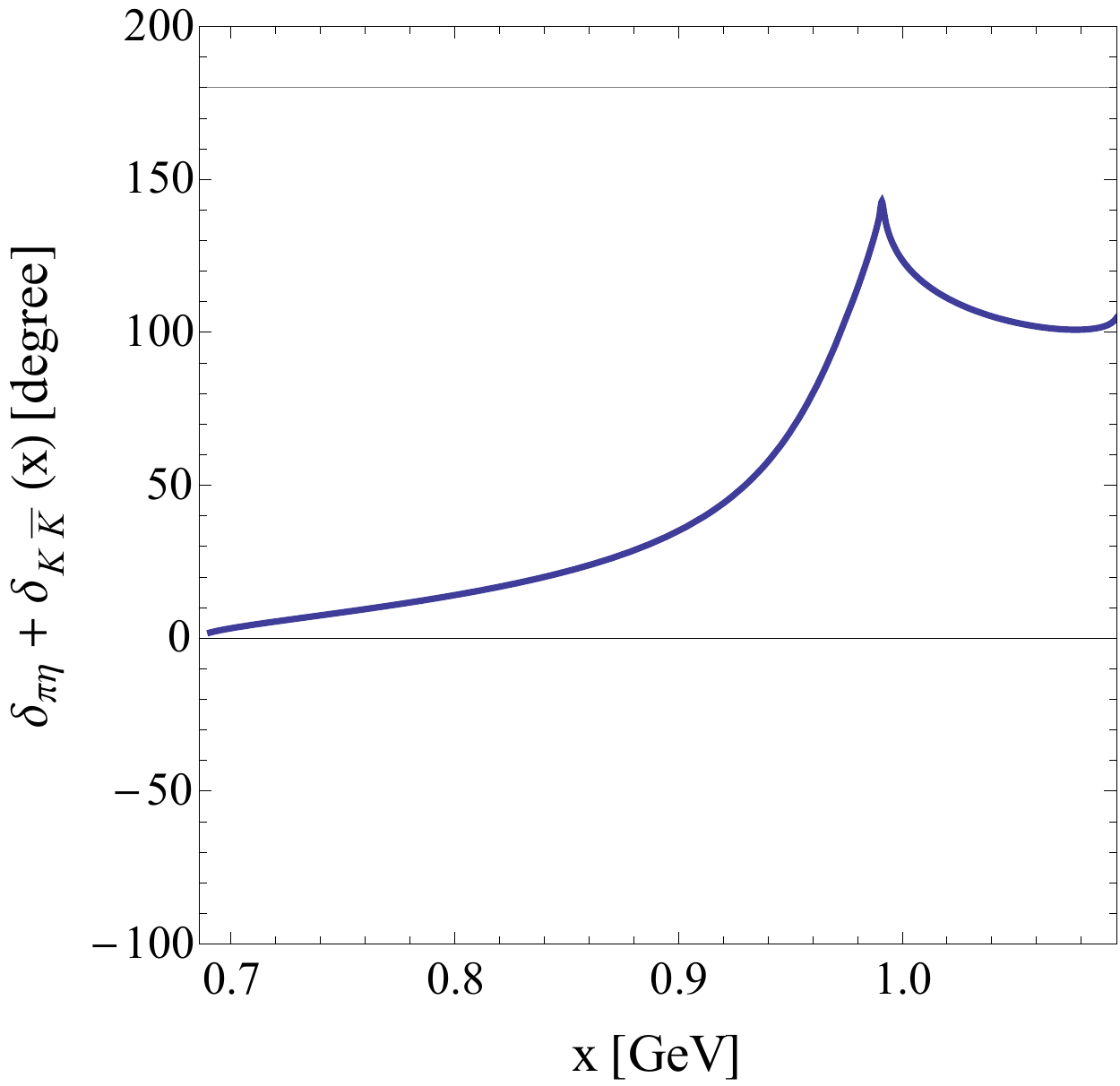}
\end{minipage}
\caption{Inelasticity parameter $\eta=\eta(s)$ and phase shifts $\delta_{1}=\delta_{\eta\pi}(s)$, $\delta_{2}=\delta_{K\bar{K}}(s)$, and the combination $\delta_{\eta\pi}(s)+\delta_{K\bar{K}}(s)$ with respect to the energy $x=\sqrt{s}$.}
\label{fig:shifts}
\end{figure}

The results are shown in Figure\ \ref{fig:shifts} and are in agreement with previous works on the subject\ \cite{oller3,abaphase}:\ $(i)$\ We observe a rapidly falling inelasticity when the $K\bar{K}$ threshold is reached. This is natural because it is known that this channel is very important for the isovector states and the low-lying $a_{0}(980)$, respectively\ \cite{olive}.\ $(ii)$\ The inelasticity has also a local minimum.\ $(iii)$\ Besides that, the phase shift $\delta_{1}$ increases with rising curvature until the $K\bar{K}$ threshold. After that it reaches a local maximum and starts to decrease. $(iv)$\ On the other hand, the phase shift $\delta_{2}$ is negative starting from the $K\bar{K}$ threshold. Note that so far no reliable data exist to which we could compare.

Our formalism is based on the assumption that the $s$-channel propagation
dominates, see also Ref.\ \cite{olive}.\ The validity of this assumption (and thus
neglecting the contributions from the $t$- and $u$-channel exchange diagrams) was
extensively discussed in the literature\ \cite{isgurcomment,tornreply1,harada,harada2,ruppcomment,beverenV}.\ 
It was especially demonstrated that this approximation alters only slightly the position of (some) resonance
poles and is therefore very suitable for our purpose.

\subsection{Some considerations: The importance of the derivative interactions}
In the following, we investigate the role of the derivative interaction terms in the Lagrangian of our model (see also Ref.\ \cite{procPisa}).\ We introduce a dimensionless 
parameter $\delta \in [0,1]$ by replacing the derivative coupling constants in 
Eq.\ (\ref{eq:Lag_eff}) such that $B_{i}^{2}\rightarrow \delta B_{i}^{2}$.\ As a consequence, for $\delta =0$ the self-energy 
contains only non-derivative interactions, while for $\delta=1.0$ we reproduce the poles found for our effective model. 
Increasing $\delta$ from zero to $1$, the derivative interaction is successively increased and we can monitor 
in a controlled manner how the pole structure changes.
\begin{figure}[t]
\hspace{2.2cm}
\includegraphics[scale=1.0]{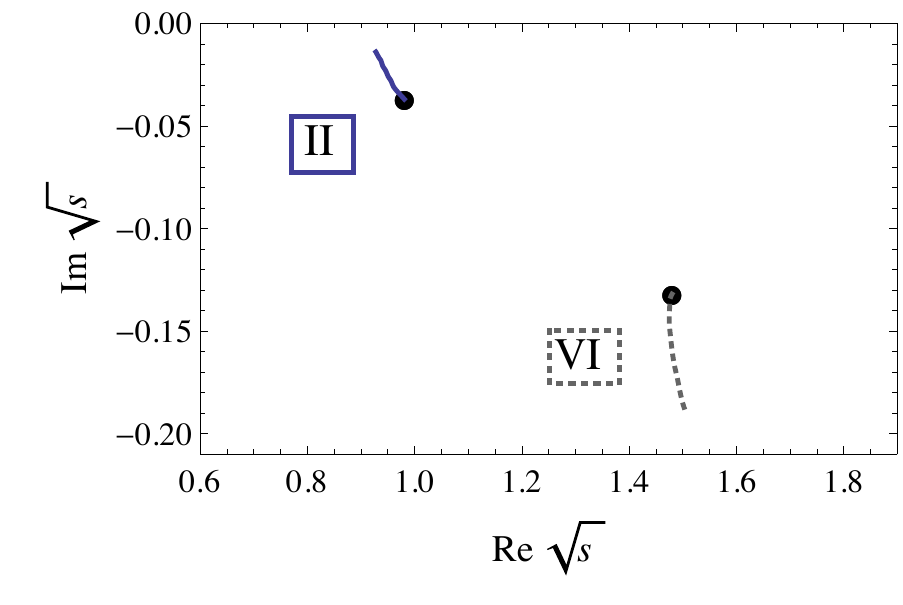}
\caption{Pole structure of our effective model in dependence of $\delta$.\ 
Black dots indicate the position of the poles for $\delta=1.0$.\ 
The roman numerals indicate on which sheet the respective poles can be found.}
\label{fig:delta}
\end{figure}

The result can be seen in Figure \ref{fig:delta}. It turns out that it is 
possible to obtain two poles even for vanishing $\delta$ where the derivative interactions 
give no contribution. In this case the real part of the corresponding pole for $a_{0}(980)$ (second sheet) is somewhat too small, 
but the imaginary part is definitely too small. 
On the other hand, the imaginary part of the pole for $a_{0}(1450)$ (sixth sheet) is obviously 
too large. In the limit $\delta\rightarrow1.0$, both poles reach their final positions in different ways: For the pole 
on the second sheet both real and imaginary parts increase. Concerning the pole on the sixth 
sheet, the real and imaginary parts decrease, but the latter is more strongly affected by a change of $\delta$.\ From the variation 
of the overall coupling strength $\delta$ one can see that both types of interaction terms (derivative and non-derivative) seem to be equally important.\footnote{It will be demonstrated in the next chapter for the case of isospin $I=1/2$ that in fact the derivative terms are responsible for the dynamical generation of the light $K_{0}^{\ast}(800)$ alias $\kappa$.}

\section{Comparison to the eLSM}
By applying Eq.\ (\ref{eq:LagDiff}) to the eLSM Lagrangian\ (\ref{eq:Lagrangian}) for example in the $\eta\pi$-channel, we end up with
\begin{equation}
\mathcal{L}_{a_{0}\eta\pi}^{\text{eLSM}} = \underbrace{\big[A_{1}^{\text{eLSM}}+C_{1}^{\text{eLSM}}(m_{\pi}^{2}+m_{\eta}^{2})\big]}_{=A_{1}}a_{0}^{0}\eta\pi^{0}
+\underbrace{\big(B_{1}^{\text{eLSM}}-2C_{1}^{\text{eLSM}}\big)}_{=B_{1}}a_{0}^{0}\partial_{\mu}\eta\partial^{\mu}\pi^{0} \ .
\end{equation}
Corresponding transformations yield similar expressions for the other two channels. If we use the parameters and the coupling constants of the eLSM\ \cite{denisphd,eLSM2}, respectively, the couplings of our effective model should be
\begin{align}
A_{1} &= -20.027 \ \text{GeV} \ , & B_{1} &= -21.510 \ \text{GeV}^{-1} \ , \nonumber \\
A_{2} &= -11.439 \ \text{GeV} \ , & B_{2} &= -21.211 \ \text{GeV}^{-1} \ , 
\nonumber \\
A_{3} &= 11.089 \ \text{GeV} \ , & B_{3} &= 12.379 \ \text{GeV}^{-1} \ , \label{eq:eLSMcompPar}
\end{align}
which is different from Eq.\ (\ref{eq:AiBi}).\ We demonstrate in the next chapter that these values are indeed not reliable at one-loop level (for example, the tree-level calculation does not make use of any form factor). Besides that, our bare mass is $m_{0}=1.2$ GeV, while the eLSM tree-level mass is $m_{a_{0}}=1.363$ GeV. Furthermore, the branching ratios of the eLSM are\ \cite{denisphd,eLSM2}
\begin{equation}
\frac{\Gamma_{a_{0}(1450)\rightarrow\eta^{\prime}\pi}^{\text{eLSM}}}{
\Gamma_{a_{0}(1450)\rightarrow\eta\pi}^{\text{eLSM}}}=0.19\pm0.02 \ , \ \ \ 
\frac{\Gamma_{a_{0}(1450)\rightarrow K\bar{K}}^{\text{eLSM}}}{
\Gamma_{a_{0}(1450)\rightarrow\eta\pi}^{\text{eLSM}}}=1.12\pm0.07 \ ,
\end{equation}
which differ from ours, see Eq.\ (\ref{eq:ourbranching}).

\section{Summary and conclusions}
We have repeated previous calculations of T\"ornqvist and Roos\ \cite{tornqvist,tornqvist2} and Boglione and 
Pennington\ \cite{pennington}\ concerning the scalar isotriplet channel.\ These studies have been extended by us to the 
complex plane on all Riemann sheets nearest to the physical one.\ As a result, some issues concerning the dynamical 
generation in the isovector sector have been clarified. Moreover, by using a hadronic model 
based on an effective Lagrangian approach, it was demonstrated that it is in fact possible to correctly describe the 
$a_{0}(980)$ and $a_{0}(1450)$ states as propagator poles in a unique framework, starting from one single seed state only. 
Our model contains both derivative and 
non-derivative interaction terms inspired by the eLSM\ \cite{eLSM1,eLSM1-2,eLSM1-3,denisphd,eLSM2}. Both terms were 
found to be equally important. A summary of our results and, for comparison, those of TR and BP, can be found in 
Table\ \ref{tab:tab2}.

It became clear that $a_{0}(980)$ can be regarded as a four-quark object, obtained as a dynamically generated pole 
which is not present in the original formulation of our hadronic model.\ Its existence requires hadronic loop contributions to be 
included in the formalism; a large-$N_{c}$ study confirmed this statement. Also, our estimates for the coupling constants 
in the $\eta\pi$- and $K\bar{K}$-channels indicate the very importance of the latter for the dynamical generation of the 
resonance below $1$ GeV.

Besides that, we computed phase shifts and inelasticity predictions which turned out to be in qualitative agreement with other studies on the subject.\ Further work on a comparison with recent lattice QCD results \cite{dudek} are ongoing \cite{privPenn}.

\newpage
\clearpage
\

\newpage
\clearpage

\chapter{Dynamical generation: The $K_{0}^{\ast}(800)$}
\label{chap:chapter5}

\medskip

The aim of this chapter is to apply the same quantum field theoretical approach as before in order
to investigate the existence and nature of the light $\kappa$ resonance.\ A single (quark-antiquark) seed state, roughly 
corresponding to the well-known resonance $K_{0}^{\ast}(1430)$, is described by an effective Lagrangian containing both 
derivative and non-derivative interaction terms.\ We will demonstrate that the simultaneous presence of poles for both states 
ensures a good description of scattering data: After computing the full one-loop resummed propagator, we perform a
fit to experimental $\pi K$ phase shift data from Ref.\ \cite{astonpion}.\ We find that, 
besides the expected resonance pole of
$K_{0}^{\ast}(1430)$, a pole corresponding to the light $\kappa$ naturally
emerges on the unphysical Riemann sheet.

\section{Some words on the scalar isodoublet resonances}
The lightest scalar resonance with isospin $I=1/2$ is the state $K_{0}^{\ast
}(800)$, also denoted as $\kappa$.\ This state is not yet listed in the summary
tables of the PDG\ \cite{olive} although it is listed in the meson tables.\ The confirmation of
$\kappa$ is important, since it would complete the nonet of light scalar
states below $1$ GeV. As mentioned before, the light scalar
mesons are excellent candidates to be non-conventional states, {\em i.e.},
four-quark objects realized as diquark-antidiquark states
\cite{jaffe,jaffe2,jaffe3,maiani,tqmix2,tqmix,pagliaraderivatives2,fariborz2,fariborz,rodriguez,fariborz3}
and/or as dynamically generated molecular-like states
\cite{tornclose,pelaez,pelaez2,oller,oller2,oller3,oller4,oller5,oller6,2006beveren,
morgan,dullemond,tornqvist,tornqvist2,boglione,pennington}
(for review, see also Refs.\ \cite{amslerrev,amslerrev2}). For various determinations of the
pole position of $\kappa$ see {\em e.g.}\ Refs.\
\cite{ishida2,descotes,magalhaes,pelaez2,zheng,zheng2,black,fariborzk,oller5}.\ Recently, the low-mass 
$\kappa$ was also reported by the BES Collaboration in 
$J/\psi$ decays to $\bar{K}^{\ast}(892)^{0}K^{+}\pi^{-}$\ \cite{beskappa,beskappa3}, see Ref.\ \cite{beskappa2} 
and references therein for other experimental evidence. A first scattering study on the lattice for $\pi K$ can be 
found in Ref.\ \cite{dudek2}.

The second state with $I=1/2$, the heavier $K_{0}^{\ast}(1430)$, is well-known from $K^{-}\pi^{+}$ elastic scattering from 
$Kp$ production at SLAC, widely known as LASS data\ \cite{astonpion}.\ 
After performing an energy-independent partial wave analysis of the 
$K^{-}\pi^{+}$ system, the $K^{-}\pi^{+}$ scattering amplitude was obtained as the sum of the $I=1/2$ 
and $I=3/2$ components. The $S$-wave amplitude for the former channel was then determined by subtracting the $I=3/2$ contribution.\ This data will be used in the present chapter.\ Various studies agree that a resonance pole for the $K_{0}^{\ast}(1430)$ with a width of about $270$ MeV is present in the data\ \cite{astonpion,tornqvist,tornqvist2,cherry,bugg2010,zheng,anisovichK0,ishida2,magalhaes}.

Concerning the nature of the two resonances, it basically holds true what was said about the isovector states: there is 
a growing consensus that the state above $1$ GeV is a member of the nonet of scalar quark-antiquark states. The important difference between $a_{0}(980)$ and $K_{0}^{\ast}(800)$ is the huge width of the latter, making the situation quite similar to the case of the famous $f_{0}(500)$ resonance (for a recent review see Ref.\ \cite{pelaez3}).

In Refs.\ \cite{denisphd,eLSM2} it was found within the eLSM that the scenario in which the $K_{0}^{\ast}(800)$ is a quarkonium is not favorable. In particular, the result for the mass was too high. In a second scenario, in which the assignment to the $K_{0}^{\ast}(1430)$ resonance was studied, the same fit reproduced both the mass and the width very well. This clearly favored the $K_{0}^{\ast}$ quarkonium state to lie above $1$ GeV. The final values for the resonance above $1$ GeV have been given as $m_{K_{0}^{\ast}}=1.450\pm0.001$ GeV and $\Gamma=0.285\pm0.012$ GeV\ \cite{denisphd,eLSM2}.

\section{Fitting phase shift data: Different model approaches}
\subsection{Effective model with both non-derivative and derivative interactions}
As for the isovector case, see Eq.\ (\ref{eq:Lag_eff}), our model for $I=1/2$ consists of an interaction 
Lagrangian describing the decay of a single scalar kaonic seed state, denoted as
$K_{0}^{\ast}$, into one pion and one kaon\ \cite{K0Milena}.\ We therefore again have two types of terms, {\em i.e.}, one without and one with derivatives:
\begin{eqnarray}
\mathcal{L}_{K_{0}^{\ast}\pi K} & = & AK_{0}^{\ast-}\big(\pi^{0}K^{+}+\sqrt{2}\pi^{+}K^{0}\big)+BK_{0}^{\ast-}\big(\partial_{\mu}\pi^{0}\partial^{\mu}K^{+}+\sqrt{2}\partial_{\mu}\pi^{+}\partial^{\mu}K^{0}\big) \nonumber \\
& & + \ \dots \ , \label{eq:lag}
\end{eqnarray}
where dots represent analogous interaction terms for the other three members of the isospin multiplets of scalar kaons. The energy-dependent decay 
width of $K_{0}^{\ast}$ reads
\begin{equation}
\Gamma_{K_{0}^{\ast}}^{\text{tree}}(s)=3\hspace{0.02cm}\frac{k(s)}{8\pi s}\left(
A-B\hspace{0.02cm}\frac{s-m_{\pi}^{2}-m_{K}^{2}}{2}\right)^{2}%
F^{2}(s)\hspace{0.02cm}\Theta\left[s-(m_{\pi}+m_{K})^{2}\right] \ ,\label{eq:width}%
\vspace{0.15cm}
\end{equation}
where the factor of $3$ comes from summing over isospin (of the decay products).\footnote{It should be clear that $\Gamma_{K_{0}^{\ast-}\rightarrow\pi^{-}\bar{K}^{0}}=2\hspace{0.02cm}\Gamma_{K_{0}^{\ast-}\rightarrow\pi^{0}K^{-}}$\hspace{0.03cm}.}\ The form factor $F(s)$
and the modulus of the three-momentum $k(s)$ of the outgoing particles are the same as in 
Eqs.\ (\ref{eq:formfactor}) and\ (\ref{eq:modk}). When the form factor is set to unity in Eq.\ (\ref{eq:width}) and\ $s\simeq(1.43)^{2}$
GeV$^{2}$, we obtain a constant tree-level decay width. It can be identified with
the physical width of the $K_{0}^{\ast}(1430)$ in (some) phenomenological
models in which this resonance is interpreted as a quarkonium, in particular in the eLSM\ \cite{denisphd,eLSM2}.\
As we shall see, the bare seed state $K_{0}^{\ast}$ in our Lagrangian\
(\ref{eq:lag}) in fact corresponds roughly to the resonance
$K_{0}^{\ast}(1430)$ -- this is in agreement with various phenomenological
studies of the scalar sector\ \cite{fariborz2,fariborz,rodriguez,fariborz3,eLSM2,close01,close3,close7,close5,isgur}.

Following closely the formalism we presented in the previous chapter, the inverse propagator of the
scalar kaonic field is given by the well-known expression
\begin{equation}
G^{-1}(s) = s-m_{0}^{2}-\Pi(s) \ ,
\label{eq:propagator2}
\end{equation}
where $m_{0}$ is the bare mass of the scalar kaon and $\Pi(s)$ is the sum of
all one-loop contributions with one pion and one kaon circulating in the loop.\ We will once more use the spectral function
\vspace{0.15cm}
\begin{equation}
d(x)=-\frac{2x}{\pi}\lim_{\epsilon\rightarrow0^{+}}\operatorname{Im}
G(x^{2}+i\epsilon) \ , 
\label{eq:spectral}
\vspace{0.15cm}
\end{equation}
with $x=\sqrt{s}$ and $d(x)$ having the correct normalization $\int_{0}^{\infty}d_{K_{0}^{\ast}}(x)\mathrm{dx}=1$, and that according to the optical theorem
$\operatorname{Im}\Pi(s)=- \sqrt{s}\,\Gamma_{K_{0}^{\ast}}^{\text{tree}}(s)$.

Just as in our study of the isovector sector, we assume the $I=1/2$, $J=0$ phase shift for 
$\pi K$ scattering up to $1.8$ GeV to be dominated by the scalar kaonic resonances(s). When applying Eqs.\ (\ref{eq:deltaAmp}) and\ (\ref{eq:deltaSmatrix}) to find the imaginary part
\vspace{0.15cm}
\begin{equation}
\operatorname{Im}\frac{e^{2i\delta_{\pi K}(x)}-1}{2i} \ = \ \operatorname{Im}\frac{-x\hspace{0.02cm}\Gamma_{K_{0}^{\ast}}^{\text{tree}}(x)}{s-m_{0}^{2}-\Pi(s)} \ ,
\end{equation}
we arrive at
\begin{eqnarray}
\frac{1}{2}\big[1-\cos{2\delta_{\pi K}(x)}\big] & = & \frac{-x\hspace{0.02cm}\Gamma_{K_{0}^{\ast}}^{\text{tree}}(x)\operatorname{Im}\Pi(s)}{\big[s-m_{0}^{2}-\operatorname{Re}\Pi(s)\big]^{2}+\big[\operatorname{Im}\Pi(s)\big]^{2}} \nonumber \\[0.2cm]
\frac{1}{2}\big[1-\cos{2\delta_{\pi K}(x)}\big] & = & \frac{\pi}{2}\Gamma_{K_{0}^{\ast}}^{\text{tree}}(x)d(x) \ ,
\end{eqnarray}
such that the phase shift can be expressed as
\begin{equation}
\delta_{\pi K}(x) = \frac{1}{2}\arccos\left[1-\pi\Gamma_{K_{0}^{\ast}}^{\text{tree}}(x)\hspace{0.02cm}d(x)\right] \ . \label{eq:phaseshift}
\end{equation}
However, one should be very careful with this formula. The inverse of the cosine function, $\arccos(x)$, is a multi-valued function and defined only on the domain $-1\leq x\leq1$.\ In particular, the branch cut is usually set on the negative real axis for $x<-1$.\ This property was not considered during the above derivation, thus restricting the phase shift formula to a value below $90^{\circ}$; or putting it another way, until the argument of the inverse of the cosine reaches a value of $-1$.\ It is easy to show\footnote{Another way to see this is by taking the logarithm of Eq.\ (\ref{eq:deltaSmatrix}) and solving for the phase shift. The logarithm will pick up a factor of $2\pi i$ when its argument passes the negative imaginary axis.\ This expression gives the same numbers as our equivalent formula\ (\ref{eq:equivPhase}).} that the analytically continued phase shift is
\begin{equation}
\delta_{\pi K}^{c}(x) = -\delta_{\pi K}(x)+\pi \ .
\label{eq:equivPhase}
\end{equation}

In contrast to the isovector case, we will apply in the following both phase shift formulas in order to fit $\pi K$ phase shift data. Whenever we will refer to the phase shift in Eq.\ (\ref{eq:phaseshift}), this shall automatically include the continuation to the second sheet of the inverse cosine function. The fit to data will determine our free model parameters and we will be able to study the pole structure of the scattering amplitude and the propagator\ (\ref{eq:propagator2}), respectively, on the second Riemann sheet. Before doing so, some comments are in order:
\begin{enumerate}
\item Similarly to the isovector states, Eq.\ (\ref{eq:phaseshift}) is based on the assumption 
that the $s$-channel propagation dominates, see also Ref.\ \cite{olive}.\ The validity of this assumption (and thus
neglecting the contributions from the $u$-channel exchange diagrams) was
extensively discussed in the literature\ \cite{isgurcomment,tornreply1,harada,harada2,ruppcomment,beverenV}.\ 
It was especially demonstrated that this approximation alters only slightly the position of (some) resonance
poles: it is therefore very suitable for our purpose.
\item Note also that we do not use any constant background term in our
model.\ This is different from many previous works on the subject (see for example 
Ref.\ \cite{ishida2} or, more recently, Ref.\ \cite{fariborzk}); instead, we utilize derivative
interactions. In order to illustrate this, we rely on an analogy with
the old linear sigma model which contains a non-derivative interaction as
well as a background term. The potential of the model has the usual Mexican
hat form, $V=\frac{\lambda}{4}(\vec{\pi}^{2}+\sigma^{2}-F^{2})^{2}%
-\varepsilon\sigma.$ The field $\sigma$ has a non-vanishing vacuum expectation
value $\phi;$ as a consequence (after performing the shift $\sigma
\rightarrow\sigma+\phi$) the mass of $\sigma$ reads $M_{\sigma}^{2}%
=\lambda\phi^{2}$, while the pion mass reads $M_{\pi}^{2}=\varepsilon/\phi$
and vanishes in the chiral limit (where $\varepsilon\propto m_{q}$ vanishes).

Retaining only the interaction terms relevant for $\pi\pi$ scattering, we have
$V=$ $\frac{\lambda}{4}\vec{\pi}^{4}+\lambda\sigma\vec{\pi}^{2}+\dots$, thus
one is left with a non-derivative interaction through $\sigma$ exchange, as
well as a four-leg repulsion term. After transforming the fields into a polar
form by $(\sigma,\vec{\pi})\rightarrow\sigma e^{i\vec{t}\cdot\vec{\pi}}$ (an
intermediate step toward chiral perturbation theory), we obtain $V=\frac
{1}{\phi}\sigma(\partial_{\mu}\vec{\pi})^{2}-\frac{M_{\pi}^{2}}{2\phi}%
\sigma\vec{\pi}^{2}+\dots$, that is, no background term of type $\vec{\pi}^{4}$
is present, instead a dominant derivative interaction has emerged.

The non-derivative interaction is subdominant and vanishes in the chiral
limit: this is in agreement with low-energy chiral theorems. The interchange
of one pion field with one kaon field allows to pass from the case of the
$\sigma$ to that of the kaonic sector studied here (formally, it is a simple
rotation in flavor space) -- but the very same intuitive arguments show why the
use of derivative interactions is important for scalar mesons in general.
Moreover, the contemporary presence of derivative and non-derivative
interactions implies that the structure giving rise to Adler's zero is
automatically fulfilled (we thus do not have to add the Adler's zero
separately, as done for example in TR, BP, and Ref.\ \cite{zhiyong}).
\vspace{0.1cm}
\item Our model is designed to study the scattering in the $I=1/2$ channel
only, in which the $s$-channel exchange of a scalar kaon can be considered as
dominant.\ Indeed, the scalar kaon contributes also through $u$-channel
exchange diagrams to the cross-section.\ Experimentally, the $I=3/2$ phase
shift is negative (meaning that there is a repulsion in this channel) but is at
least a factor of four smaller than for $I=1/2$.\ This shows also that the enhanced
intensity in the $I=1/2$ channel can be ascribed to the $s$-channel exchange of a
scalar kaon.
\end{enumerate}
\vspace{0.1cm}

\begin{figure}[t]
\centering
\vspace{0.3cm}
\hspace{-0.5cm}
\includegraphics[scale=0.65]{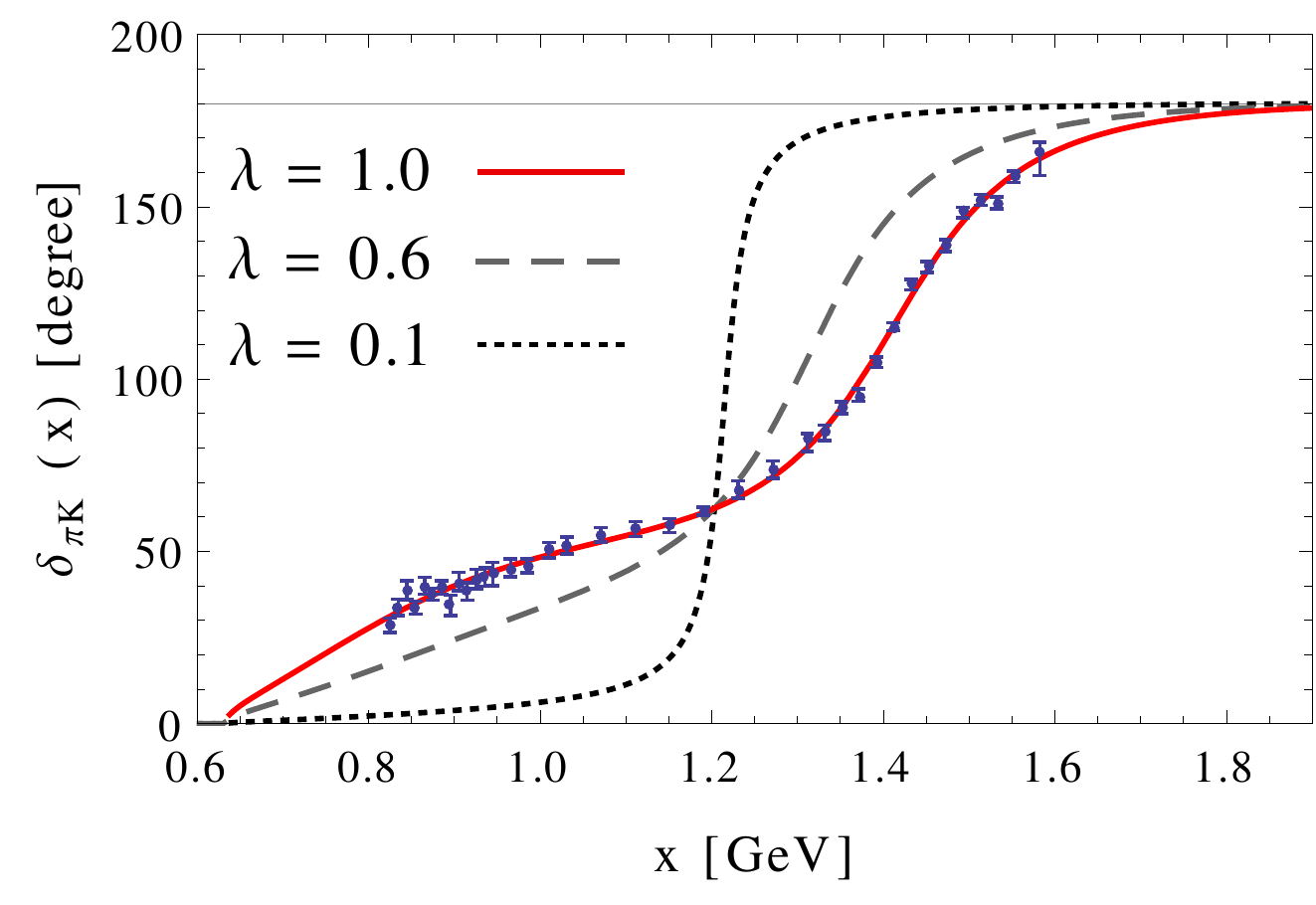}\caption{\label{fig:fit}The solid (red) curve shows our fit
result for the phase shift from Eq.\ (\ref{eq:phaseshift}) with respect to the
four model parameters $A$, $B$, $\Lambda$, and $m_{0}$ (see Table\ \ref{tab:parameters}).
The blue points are the data of Ref.\ \cite{astonpion}.\ The rescaling parameter\ 
$\lambda$ from Eq.\ (\ref{eq:lambdaFac}) is set to $1.0$.\ The other two curves correspond to $\lambda=0.6$ (long-dashed) and $\lambda=0.1$ (short-dashed).}
\end{figure}

The data for our least-square fit are taken from Ref.\ \cite{astonpion}.\ The model parameters are the two coupling constants $A$ and $B$, the familiar cutoff $\Lambda$, and the bare seed mass $m_{0}$.\ Their errors are calculated as the square roots of the diagonal elements of the inverted Hessian matrix obtained from the $\chi^{2}$ function. The result is shown as solid (red) curve in Figure\ \ref{fig:fit} and the values of the parameters, together with their errors, are reported in Table\ \ref{tab:parameters}.\ The value of $\chi^{2}$ is: $\chi_{0}
^{2}/d.o.f.=1.25$, explaining the very good agreement of our model result with
data. By comparing the coupling constants it turns out that the derivative
coupling is dominant, which is expected by chPT\ \cite{chpt,Ecker,chpt2} and
by other studies\ \cite{blackphi,pagliaraderivatives}.

Using the parameters listed in Table\ \ref{tab:parameters}, we continue the
propagator from Eq.\ (\ref{eq:propagator2}) into the second Riemann sheet and
scan the complex plane for poles.
\begin{table}[t]
\center
\vspace{0.45cm}
\begin{tabular}[c]{cc}
\toprule Parameter & Value\\
\midrule\\[-0.25cm]
$A$ & $1.60\pm0.22$ GeV\\
$B$ & \hspace{-0.09cm}$-11.16\pm0.82$ GeV$^{-1}$\\
$\Lambda$ & $0.496\pm0.008$ GeV\\
$m_{0}$ & $1.204\pm0.008$ GeV\\
\\[-0.1cm]
\toprule
\end{tabular}
\caption{Results of the fit; $\chi_{0}^{2}/d.o.f.=1.25$.}
\label{tab:parameters}
\end{table}
We find two poles which we assign in the following way:
\begin{align}
K_{0}^{\ast}(1430)  & :\ (1.413\pm0.002)-i\hspace{0.02cm}(0.127\pm0.003)\ \text{GeV} \ ,\\
K_{0}^{\ast}(800)\   & :\ (0.746\pm0.019)-i\hspace{0.02cm}(0.262\pm0.014)\ \text{GeV} \ .\label{eq:poles}%
\end{align}
This means for the masses and widths:
\begin{align}
K_{0}^{\ast}(1430)  & :\ m_{\text{pole}} = (1.407\pm0.002) \ \text{GeV} \ , \ \ \Gamma_{\text{pole}} = (0.255\pm0.007) \ \text{GeV} \ ,\\
K_{0}^{\ast}(800)\   & :\ m_{\text{pole}} = (0.698\pm0.025) \ \text{GeV} \ , \ \ \Gamma_{\text{pole}} = (0.600\pm0.036) \ \text{GeV} \ .\label{eq:masswidths}%
\end{align}
The PDG\ \cite{olive} reports for $K_{0}^{\ast}(1430)$ a mass of
$(1.425\pm0.050)$ GeV and a width of $(0.270\pm0.080)$ GeV. All our values fit very
well into these windows. In particular, the pole width obtained by doubling the negative 
imaginary part reads $(0.254\pm0.006)$ GeV and is thus determined with a small
error. Remarkably, a pole corresponding to the light $\kappa$ emerges very naturally in our
calculation. For $K_{0}^{\ast}(800)$ the PDG reports a mass of $(0.682\pm0.029)$ GeV and
a width of $(0.547\pm0.024)$ GeV, which are also in agreement with our values
(although our result for the pole mass points to a somewhat larger value).\ The 
mass $(0.746\pm0.019)$ GeV and width $(0.524\pm0.028)$ GeV for the $\kappa$ are, however, more precise
than most of the results listed in Ref.\ \cite{olive}.
\begin{figure}[t]
\hspace*{-1.5cm}
\begin{minipage}[hbt]{8cm}
\centering
\includegraphics[scale=0.6]{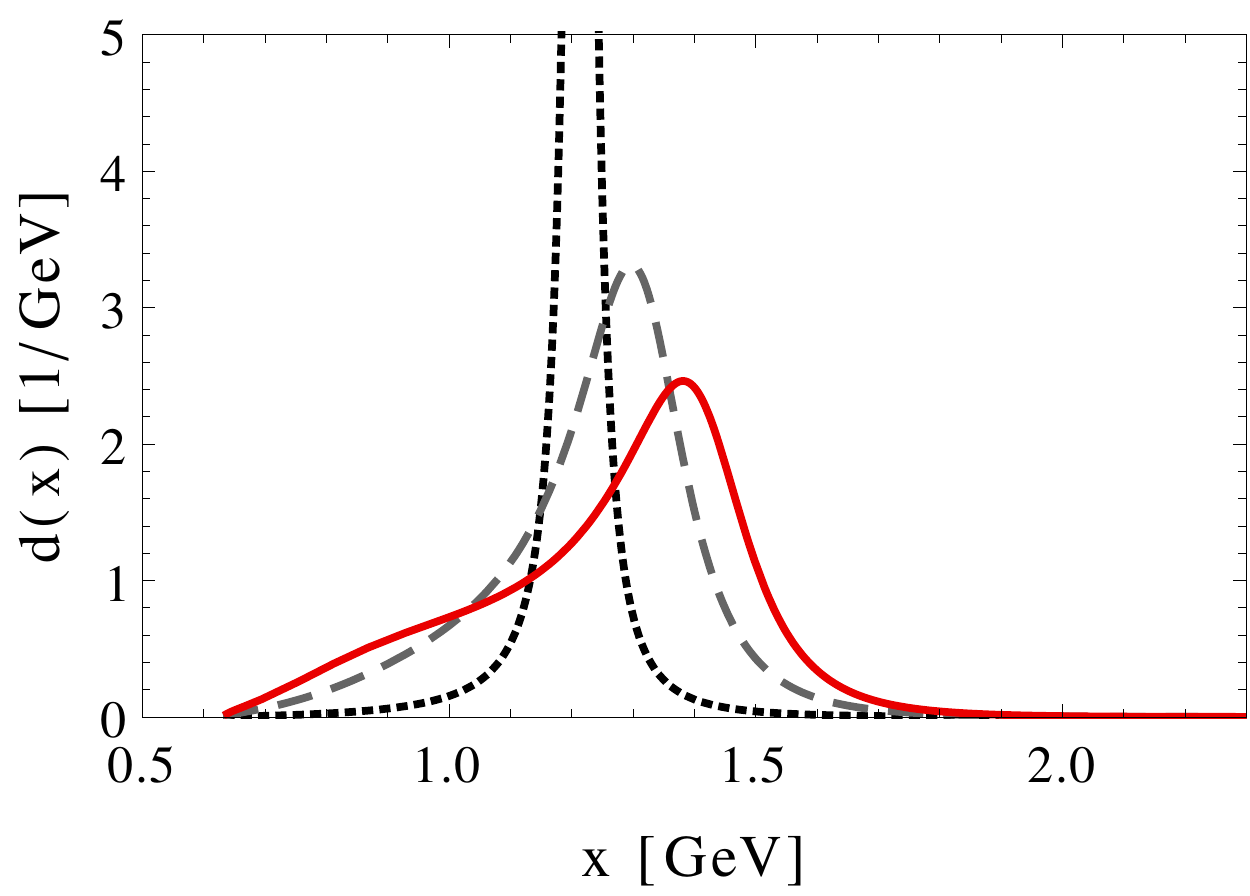}
\end{minipage}
\hspace*{0.3cm}
\begin{minipage}[hbt]{8cm}
\centering
\vspace*{0.32cm}
\includegraphics[scale=0.691]{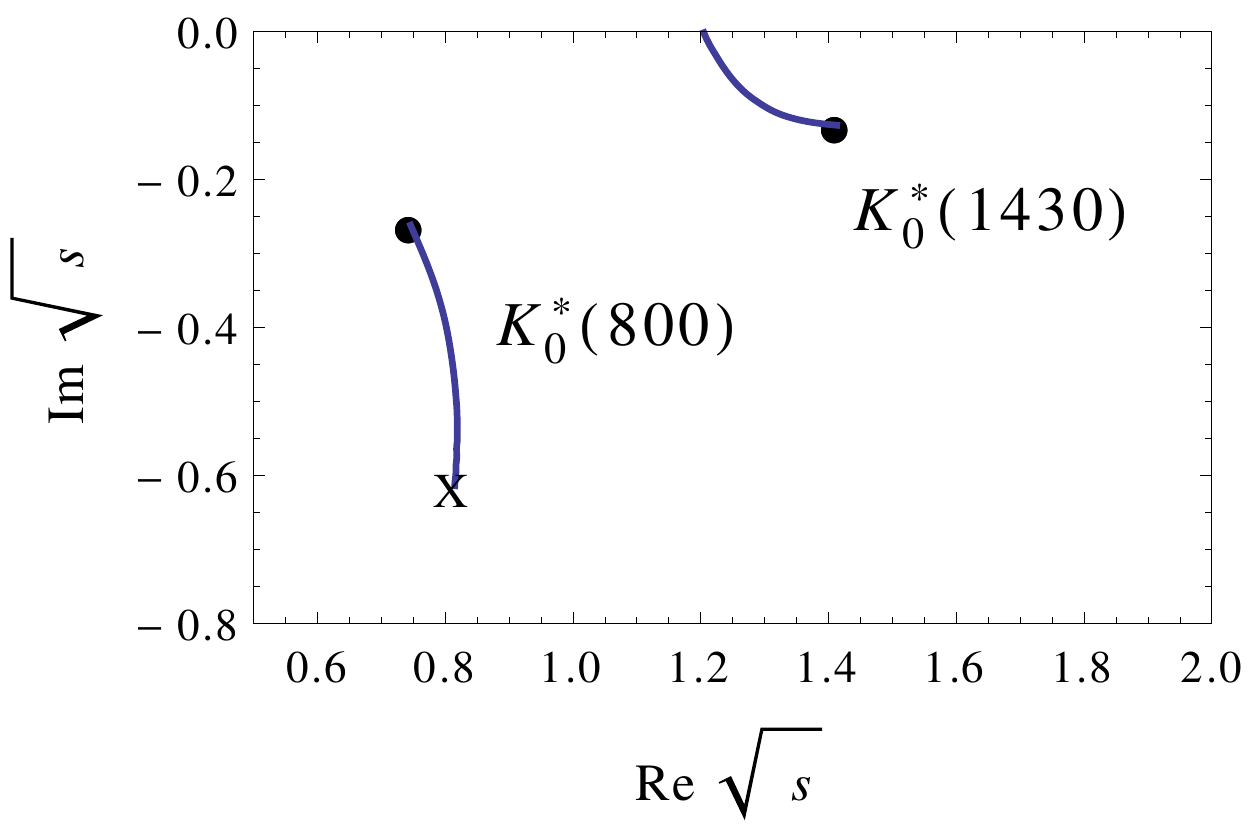}
\end{minipage}
\caption{\label{fig:sf}In the left panel we
show the spectral functions for the three different values of $\lambda$ indicated in Figure\ \ref{fig:fit}.\ In the right panel
we display pole trajectories obtained by varying $\lambda$ from zero to $1$.\ 
Black dots indicate the position of the poles for $\lambda=1.0$.\ The X indicates the
pole position for $\lambda_c\approx0.24$, {\em i.e.}, when the pole first emerges.
Both poles are on the second sheet.}
\end{figure}

In the left panel of Figure\ \ref{fig:sf} we show the spectral functions for the
parameters of Table\ \ref{tab:parameters} for three different values of the rescaling parameter $\lambda$, 
compare with Eq.\ (\ref{eq:lambdaFac}). A low-energy enhancement is present for $\lambda=1.0$,
but no peak.\ Obviously, the low-energy enhancement becomes
smaller for decreasing $\lambda$, {\em i.e.}, for increasing $N_{c}$.\ The absence of a peak is one of the 
reasons why the acceptance of the $\kappa$ might be considered to be controversial. However, if resonance
poles on unphysical Riemann sheets are the relevant quantities, it turns out
that the existence of the broad $\kappa$ is a consequence of our model. 
Similar statements can be made concerning the broad isoscalar state
$f_{0}(500)$: its pole is widely accepted, while a clear peak in the spectral
function is not present.\ On the contrary, the two scalar states $a_{0}(980)$
and $f_{0}(980)$ are pretty narrow: although their couplings are large, these
resonances sit just at the kaon-kaon threshold, making their decays into kaons
to be kinematically suppressed. As advertised, all those states together with
$\kappa$ seem to have their common origin in the presence of quantum fluctuations.

\begin{figure}[t]
\centering
\vspace{0.3cm}
\hspace{-0.5cm}
\includegraphics[scale=0.65]{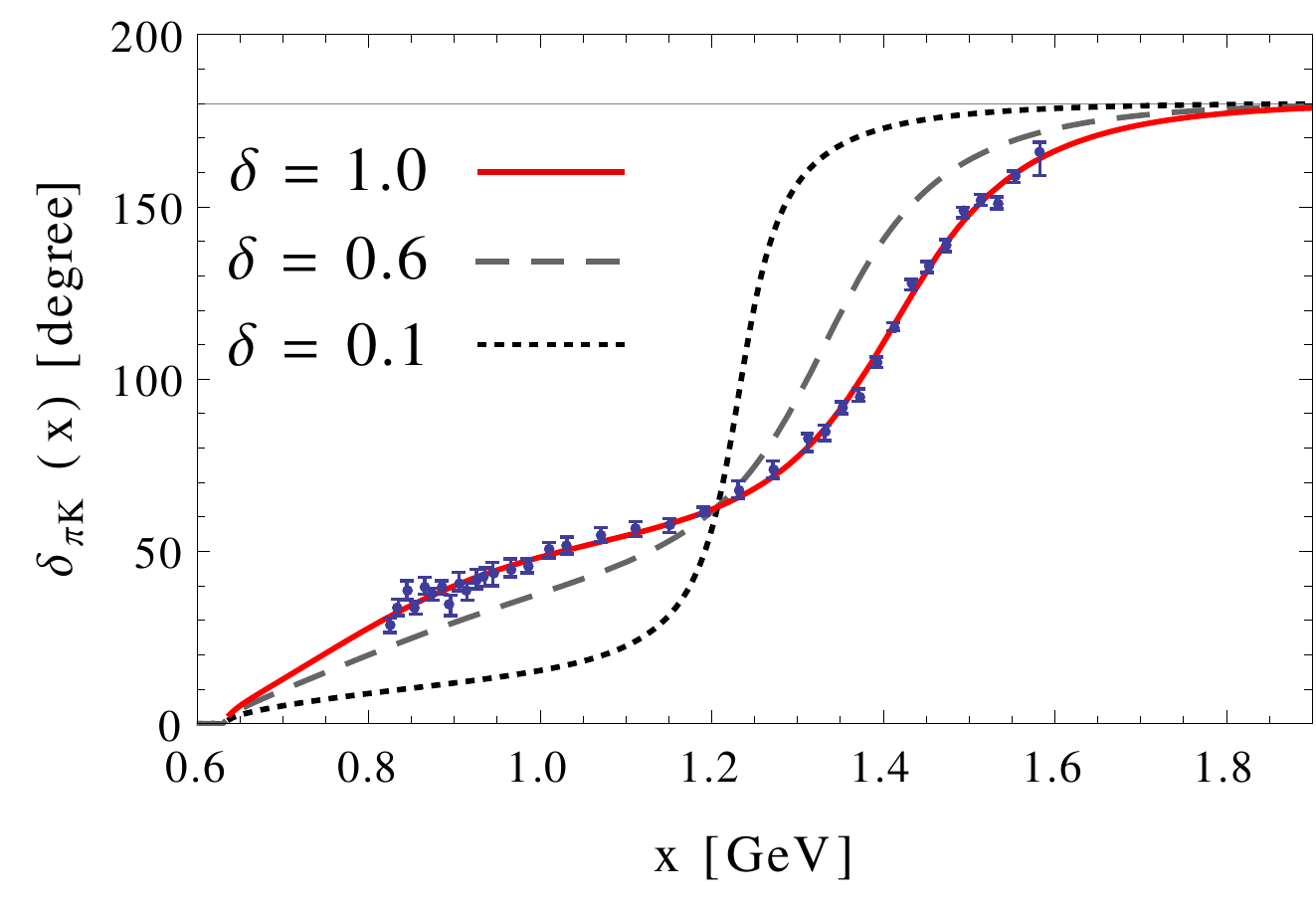}\caption{\label{fig:phase2}The solid (red) curve shows our fit
result for the phase shift from Eq.\ (\ref{eq:phaseshift}) with respect to the
four model parameters $A$, $B$, $\Lambda$, and $m_{0}$ (see Table\ \ref{tab:parameters}). 
The rescaling parameter $\delta$ is set to $1.0$.\ The other two curves correspond to $\delta=0.6$ (long-dashed) and $\delta=0.1$ (short-dashed).}
\end{figure}
We present the pole trajectories within our model as function of $\lambda$ in the right panel of Figure\
\ref{fig:sf}.\ One observes that the pole of $K_{0}^{\ast}(1430)$ moves
toward the real axis, a behavior expected for a quarkonium state and completely equivalent to what we have found for 
the $a_{0}(1450)$.\ The pole of $K_{0}^{\ast}(800)$, however, 
moves away from the real axis and disappears for
$\lambda_{c}\approx0.24$ (or $N_{c}\approx13$) deep in the complex plane.\ This explains why the low-energy 
enhancement in the spectral function vanishes and the phase shift gains a typical 
single-resonance shape for decreasing $\lambda$.\ From this it also follows that the pole of
$K_{0}^{\ast}(800)$ belongs to a dynamically generated state not surviving the
large-$N_{c}$ limit.\ We have found a similar behavior for the $a_{0}(980)$, too, but it was also reported for instance in Refs.\ \cite{zhiyong,ollerNC}.

\begin{figure}[t]
\hspace*{-1.5cm}
\begin{minipage}[hbt]{8cm}
\centering
\includegraphics[scale=0.6]{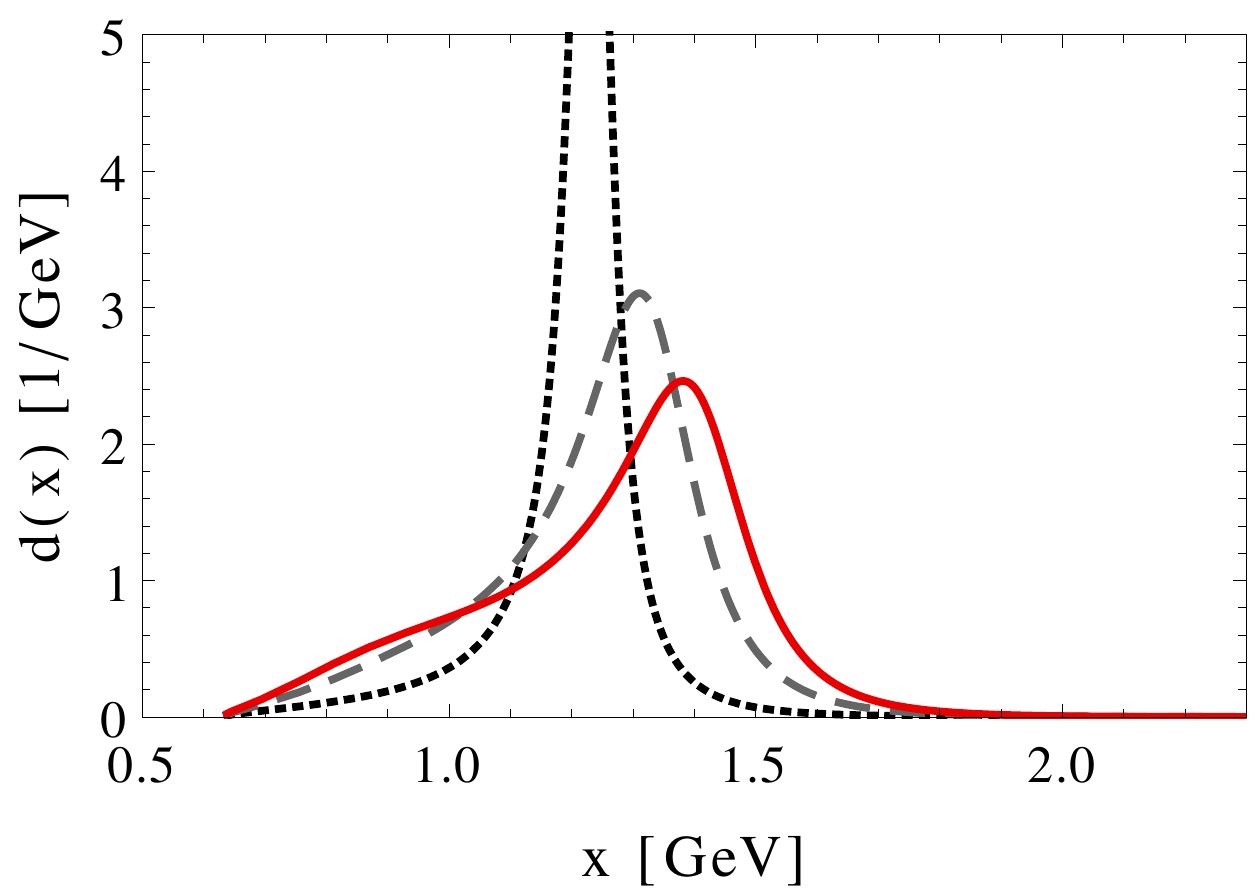}
\end{minipage}
\hspace*{0.3cm}
\begin{minipage}[h]{8cm}
\centering
\vspace*{0.32cm}
\includegraphics[scale=0.691]{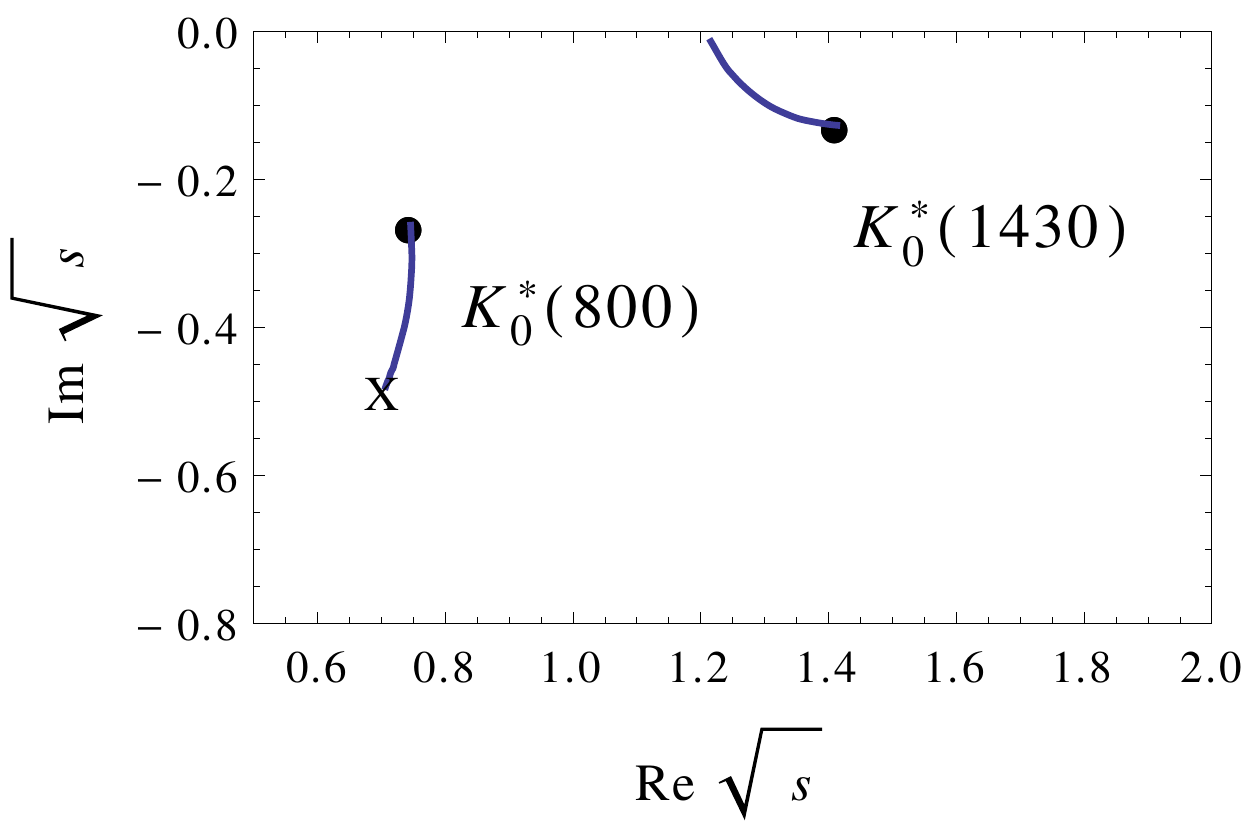}
\end{minipage}
\caption{\label{fig:delta2}In the left panel we
show the spectral functions for the three different values of $\delta$ indicated in Figure\ \ref{fig:phase2}.\ In the right panel
we display pole trajectories obtained by varying $\delta$ from zero to $1$.\ 
Black dots indicate the position of the poles for $\delta=1.0$.\ The X indicates the
pole position for $\delta_c\approx0.35$, {\em i.e.}, when the pole first emerges.
Both poles are on the second sheet.}
\end{figure}

Finally, we investigate the role of the derivative interaction terms in the Lagrangian of our model by again introducing the dimensionless 
parameter $\delta \in [0,1]$, and replacing the derivative coupling constant $B$ according to $B^{2}\rightarrow \delta B^{2}$.\ This is the same procedure as in the isovector case: When increasing $\delta$ from zero to $1$, the derivative interaction is successively increased and we can monitor how the pole structure changes. The results can be seen in Figure\ \ref{fig:phase2} and\ \ref{fig:delta2}.
In contrast to the isotriplet states, it is not possible to obtain two poles for vanishing $\delta$; the $\kappa$ pole only 
appears for $\delta_{c}\approx0.35$.\ From this one can conclude that in fact the derivative interaction term is {\em essential} for the existence and dynamical generation of the $K_{0}^{\ast}(800)$ on the second sheet.

At this point it should be stressed that the choice of the form factor\ (\ref{eq:formfactor}) is
model-dependent. A Gaussian form as implement here is a standard choice when
investigating mesonic resonances and the position of their poles, respectively,
see also the discussion in Refs.\ \cite{isgurcomment,tornreply1,harada,harada2,ruppcomment}.\ Yet, in 
the next subsection we
investigate a possible variation of the form factor -- and indeed find that this modification
is not capable of reproducing the phase shift data correctly. At the same
time, we will vary our model by discarding either one of the two types of interaction terms and we will also 
investigate the statistical significance of the various fits.

\subsection{Variations of the model}
We study different scenarios in order to better understand the situation discussed above. To this end, we first perform
two fits to the phase shift data: one in which we consider only the
non-derivative terms in Eq.\ (\ref{eq:lag}) (we set $B=0$), and one in which we consider
only the derivative terms (we set $A=0$).\ The results are presented in Figure\ \ref{fig:varmodels1} and Table\ \ref{tab:variations}.\ The first entry of Table\ \ref{tab:variations} 
summarizes what was found in the previous subsection.\ The second and third
entries represent the two cases $B=0$ and $A=0$, respectively.

\begin{enumerate}
\item As can be seen from the third column, 
in both cases the $\chi^{2}$ has increased, hinting at a worse agreement than with our first fit.
\begin{figure}[t]
\hspace*{-1.1cm}
\begin{minipage}[hbt]{8cm}
\centering
\includegraphics[scale=0.6]{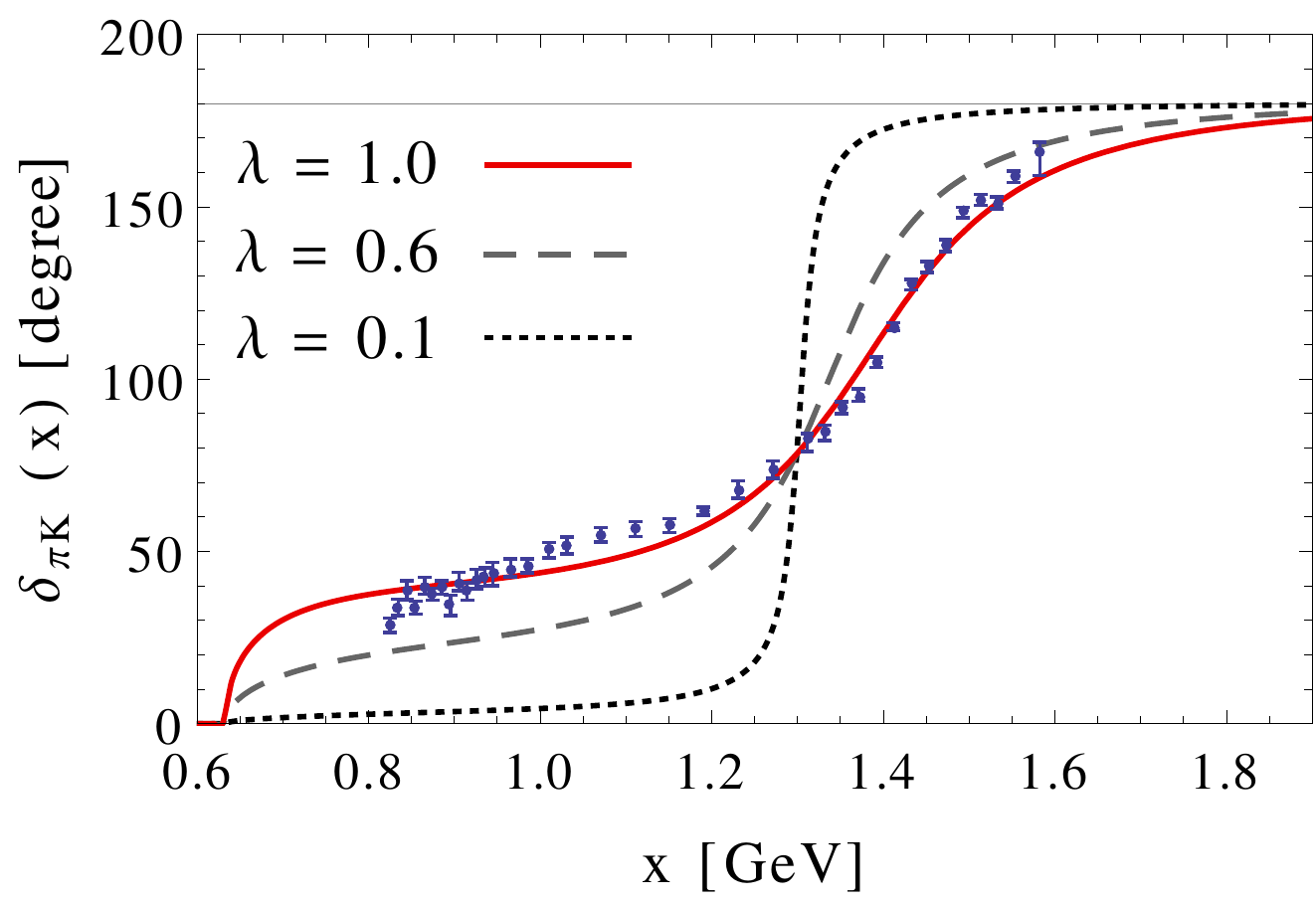}
\end{minipage}
\hspace*{0.4cm}
\begin{minipage}[hbt]{8cm}
\centering
\includegraphics[scale=0.6]{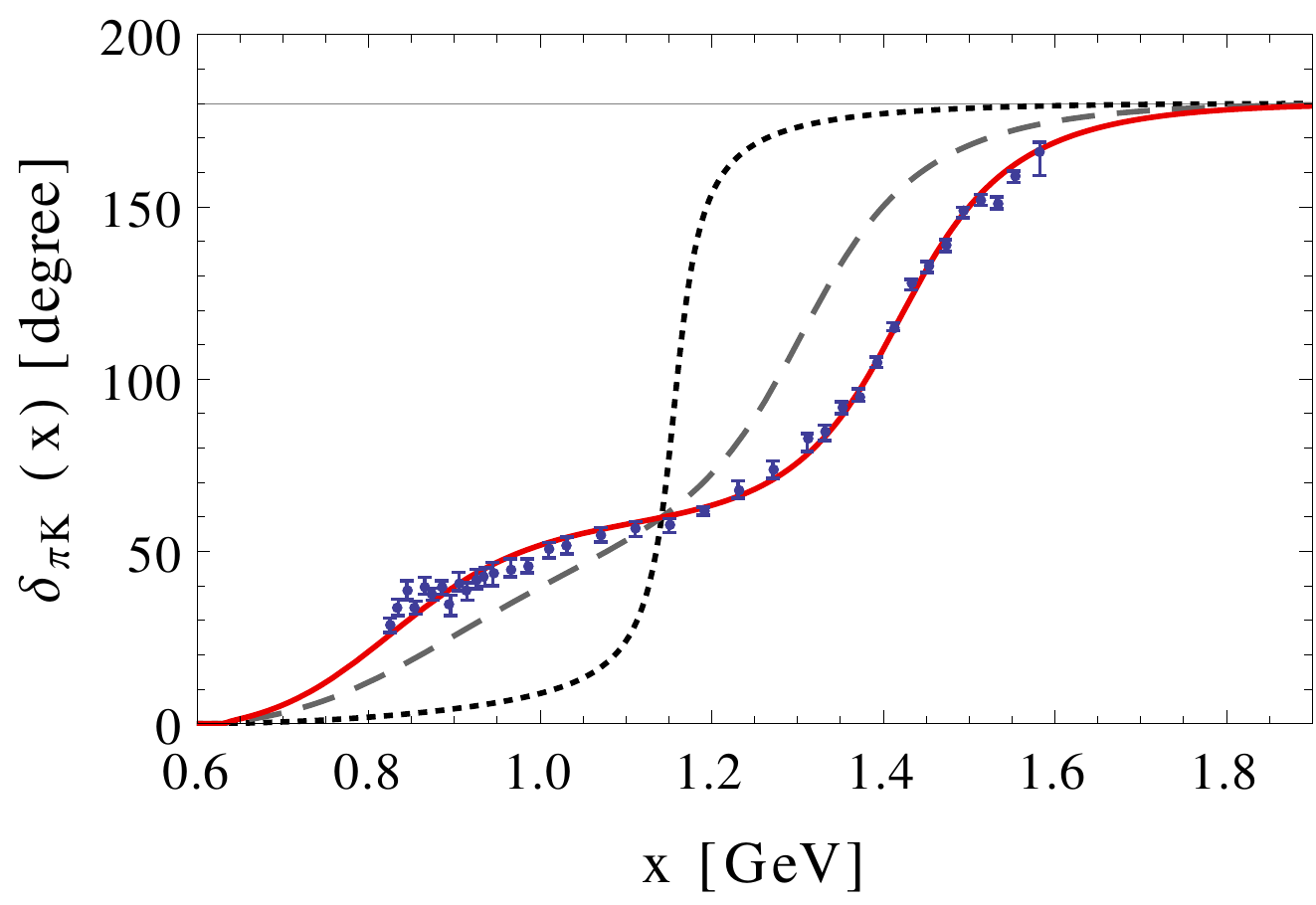}
\end{minipage}
\caption{\label{fig:varmodels1}The left panel shows the case in which we consider only the
non-derivative term in Eq.\ (\ref{eq:lag}) ($B=0$), while in the right panel the case in which we consider
only the derivative term ($A=0$) is displayed. The solid (red) curves represent the fit
results for the phase shift from Eq.\ (\ref{eq:phaseshift}) with respect to the
three model parameters $A$ or $B$, $\Lambda$, and $m_{0}$ (see Table\ \ref{tab:variations}). 
The rescaling parameter $\lambda$ is set to $1.0$.\ The other two curves correspond to $\lambda=0.6$ (long-dahed) and $\lambda=0.1$ (short-dashed).}
\end{figure}
However, in order to be more quantitative, we report in the fourth column a statistical test of the goodness of the fit: The quantity
\begin{equation}
p\left(\chi^{2}>\chi_{0}^{2}\right)=\frac{1}{2^{d/2}\Gamma(d/2)}\int_{\chi_{0}^{2}}^{\infty}\text{d}x \ x^{\frac{d}{2}-1}e^{-x/2}
\end{equation}
(with $d=d.o.f.$) is the probability to obtain a larger value of the $\chi
^{2}$ than $\chi_{0}^{2}$, if a new experiment shall be performed (by using, of course, the same
theoretical function in the fit). When this probability is very small, one may
conclude that $(i)$ the theoretical model is not correct (a reasonable
conclusion) or that $(ii)$ the theoretical model is correct, but the experimental
results show -- accidentally -- a statistical fluctuation.\ For instance, a probability smaller than $5\%$ excludes 
the theoretical model at the $95\%$ confidence level.\ In our case, the preferred solution from the
previous subsection gives $p\left(\chi^{2}>\chi_{0}^{2}\right)=15.3\%$, which implies that the
theoretical model {\em cannot} be rejected (here, $d.o.f.=37-4=33$). On
the contrary, the models with either only non-derivative or derivative interactions 
can be rejected with a very high level of accuracy.\
While this result is expected for the non-derivative terms, because the shape of
the theoretical function does not match the data (see left panel of Figure\ \ref{fig:varmodels1}), the situation
is more subtle in the case of only derivative terms.
Here, the form is qualitatively correct, but the statistical test reveals that it is not in
agreement with the experiment (with $d.o.f.=37-3=34$).
\item A very interesting observation is the fact that for all cases the bare mass $m_{0}$ varies only little ($\sim50-150$ MeV), whereas the cutoff is nearly doubled for $B=0$. A much higher cutoff is necessary to get the phase shift curve near the data, as can be verified by driving this parameter away from the minimum. Most remarkably is, however, the following: In the fifth and sixth columns we report the pole positions
\begin{sidewaystable}[H]
\begin{tabular}[c]{clcccc}
\toprule Scenario & \hspace{0.9cm}Parameters & $\chi_{0}^{2}/d.o.f.$ & $p\left(\chi^{2}>\chi_{0}^{2}\right)$ & Pole for $K_{0}^{\ast}(800)$ & Pole for $K_{0}^{\ast}(1430)$\vspace{0.05cm}\\\hline\vspace{-0.25cm}\\
$A,B\neq0$, Gaussian & {$\!\begin{aligned}A &= 1.60\pm0.22 \text{ GeV} \\B &= -11.16\pm0.82 \text{ GeV}^{-1} \\ \Lambda &= 0.496\pm0.008 \text{ GeV} \\m_{0} &=1.204\pm0.008 \text{ GeV} \end{aligned}$} & 1.25 & $0.15$ & {$\!\begin{aligned}(0.746\pm0.019) \\ -i\hspace{0.02cm}(0.262\pm0.014) \end{aligned}$} & {$\!\begin{aligned}(1.413\pm0.002) \\ -i\hspace{0.02cm}(0.127\pm0.003) \end{aligned}$}\vspace{0.05cm}\\\hline\vspace{-0.25cm}\\
$B=0$, Gaussian & {$\!\begin{aligned}A &= 4.06\pm0.04 \text{ GeV} \\ \Lambda &= 0.902\pm0.015 \text{ GeV} \\m_{0} &=1.299\pm0.002 \text{ GeV} \end{aligned}$} & 5.41 & $1.72\cdot10^{-22}$ & - & {$\!\begin{aligned}(1.385\pm0.002) \\ -i\hspace{0.02cm}(0.146\pm0.003) \end{aligned}$}\vspace{0.05cm}\\\hline\vspace{-0.25cm}\\
$A=0$, Gaussian & {$\!\begin{aligned}B &= -17.10\pm0.17 \text{ GeV}^{-1} \\ \Lambda &= 0.453\pm0.002 \text{ GeV} \\m_{0} &=1.142\pm0.002 \text{ GeV} \end{aligned}$} & 2.54 & $1.92\cdot10^{-6}$ & {$\!\begin{aligned}(0.820\pm0.003) \\ -i\hspace{0.02cm}(0.187\pm0.002) \end{aligned}$} & {$\!\begin{aligned}(1.419\pm0.001) \\ -i\hspace{0.02cm}(0.112\pm0.002) \end{aligned}$}\vspace{0.05cm}\\\hline\vspace{-0.25cm}\\
$A,B\neq0$, $F(s)=e^{-k^{4}(s)/\Lambda^{4}}$ & {$\!\begin{aligned}A &= 2.32\pm0.09 \text{ GeV} \\ B &= -3.40\pm0.26 \text{ GeV}^{-1}\\ \Lambda &= 0.652\pm0.006 \text{ GeV} \\m_{0} &=1.248\pm0.003 \text{GeV} \end{aligned}$} & 2.86 & $7.98\cdot10^{-8}$ & {$\!\begin{aligned}(0.863\pm0.008) \\ -i\hspace{0.02cm}(0.339\pm0.017) \end{aligned}$} & {$\!\begin{aligned}(1.433\pm0.002) \\ -i\hspace{0.02cm}(0.112\pm0.003) \end{aligned}$} \\ \\[-0.3cm]
\toprule
\end{tabular}
\caption{Fitting results for our effective model (first entry) and its variations. Poles are given in GeV.}
\label{tab:variations}
\end{sidewaystable}
for the various models.\ In view of the statistical analysis, only our original fit in the first row can be regarded as reliable.\ Nevertheless, we see that for $B=0$ it is not possible to find a good fit where a $\kappa$ pole is present (or even a second pole somewhere nearby).\ The low-energy enhancement right at threshold originates due to a virtual bound state approaching the branch point from below the second sheet (not shown in Figure\ \ref{fig:varmodels2}, and similar to what we described in Ref.\ \cite{e38}). It appears in the spectrum if the coupling constant $A$ is further increased.
\vspace{-0.05cm}
\item In the left panels of Figure\ \ref{fig:varmodels2} we show the spectral functions for the
parameters of Table\ \ref{tab:variations} for the three different values of $\lambda$ indicated in Figure\ \ref{fig:varmodels1}, 
and in the right panels the corresponding pole trajectories as function of $\lambda$. The first row displays the case in which 
we consider only the non-derivative terms in Eq.\ (\ref{eq:lag}) ($B=0$), and the second row is for the case where only the 
derivative terms are considered ($A=0$). Note the very different trajectory for the $K_{0}^{\ast}(1430)$ pole in the second case: 
both the pole mass and width appear to be highly influenced by derivative coupling terms. 
This is also visible in the movement of the 
spectral function peaks. Besides that, the low-energy enhancement because of the dynamically generated pole is nicely 
pronounced, since the pole moves much nearer to the real axis than in our original fit.\ The very importance of the derivative 
interactions is also deduced from $\lambda_{c}\approx0.08$: an additional pole on the second Riemann sheet is already 
generated for very small derivative coupling constants.
\vspace{-0.05cm}
\item As a next step, we investigate a different choice for the form factor $F(s)$.\ As shown throughout all this thesis, 
the Gaussian form factor is rather standard in various works on the
subject -- and it is also easy to use. Especially in presence of derivative
interactions it is very practical because it cuts off the integrand in the loop
integrals sufficiently fast.\footnote{When using a form like
$\left(1+k^{n}/\Lambda^{n}\right)^{-2}$, one usually observes non-physical
bumps at high energies\ \cite{tqmix2,tqmix}.}\ However, there is no fundamental reason why the
Gaussian should be the best one to apply. It is therefore important to check
variations of it. We test the following simple modification:
\vspace{-0.25cm}
\begin{equation}
F(s)=\exp\left[-k^{4}(s)/\Lambda^{4}\right] \ . \\[-0.25cm]
\label{eq:modform}
\end{equation}
The result of the fit for this choice is reported in Figure\ \ref{fig:phases}, as well as in the last entry of
Table\ \ref{tab:variations}.\ Also in this case the theoretical curve shows a qualitative agreement between the model
and data.\ Yet, the statistical test excludes this version at a very high level
of accuracy. From this perspective it is not surprising to find a putative pole for the 
$\kappa$ to be not in agreement with our result from the previous
\clearpage
\begin{figure}[h]
\hspace*{-1.5cm} \begin{minipage}[hbt]{8cm}
\centering
\includegraphics[scale=0.6]{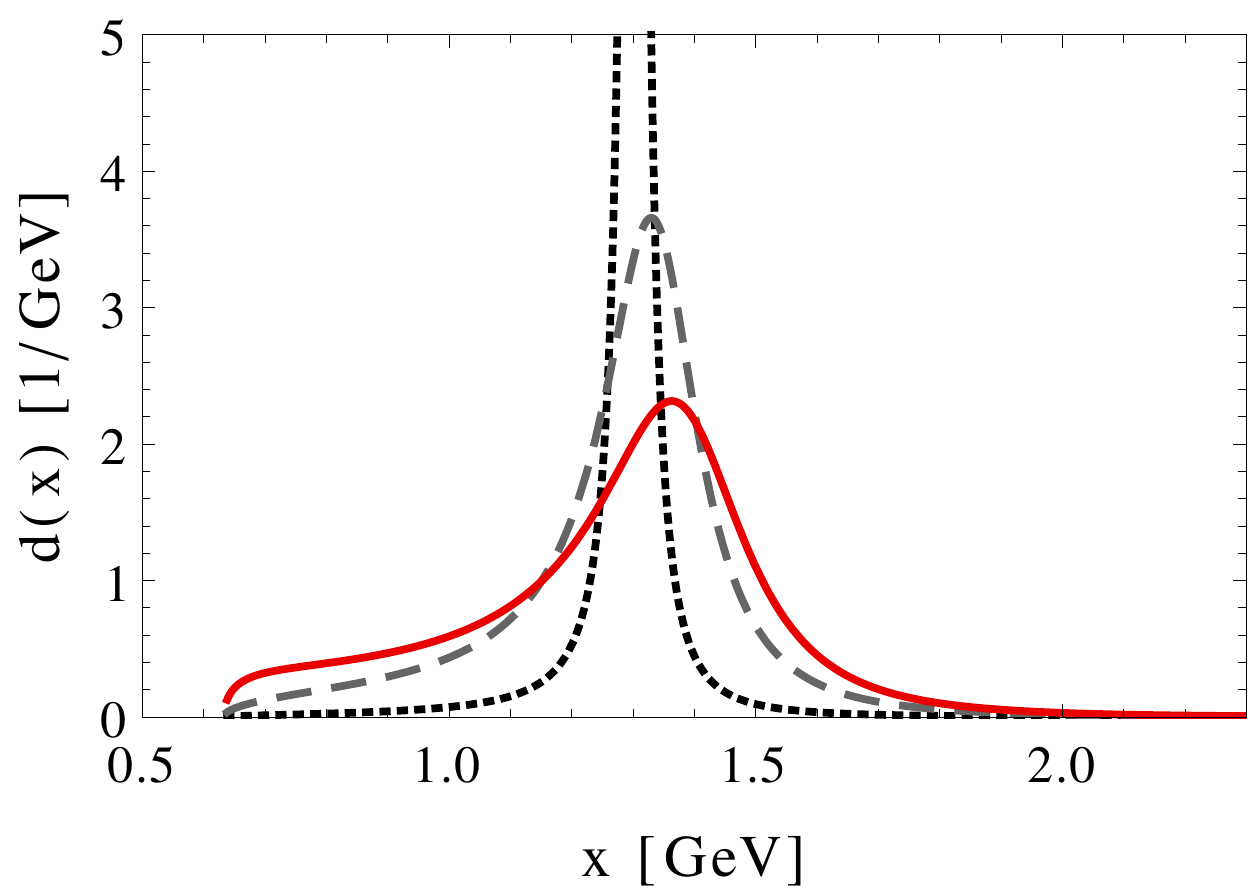}
\end{minipage}
\hspace*{0.3cm}
\begin{minipage}[hbt]{8cm}
\centering
\vspace*{0.32cm}
\includegraphics[scale=0.691]{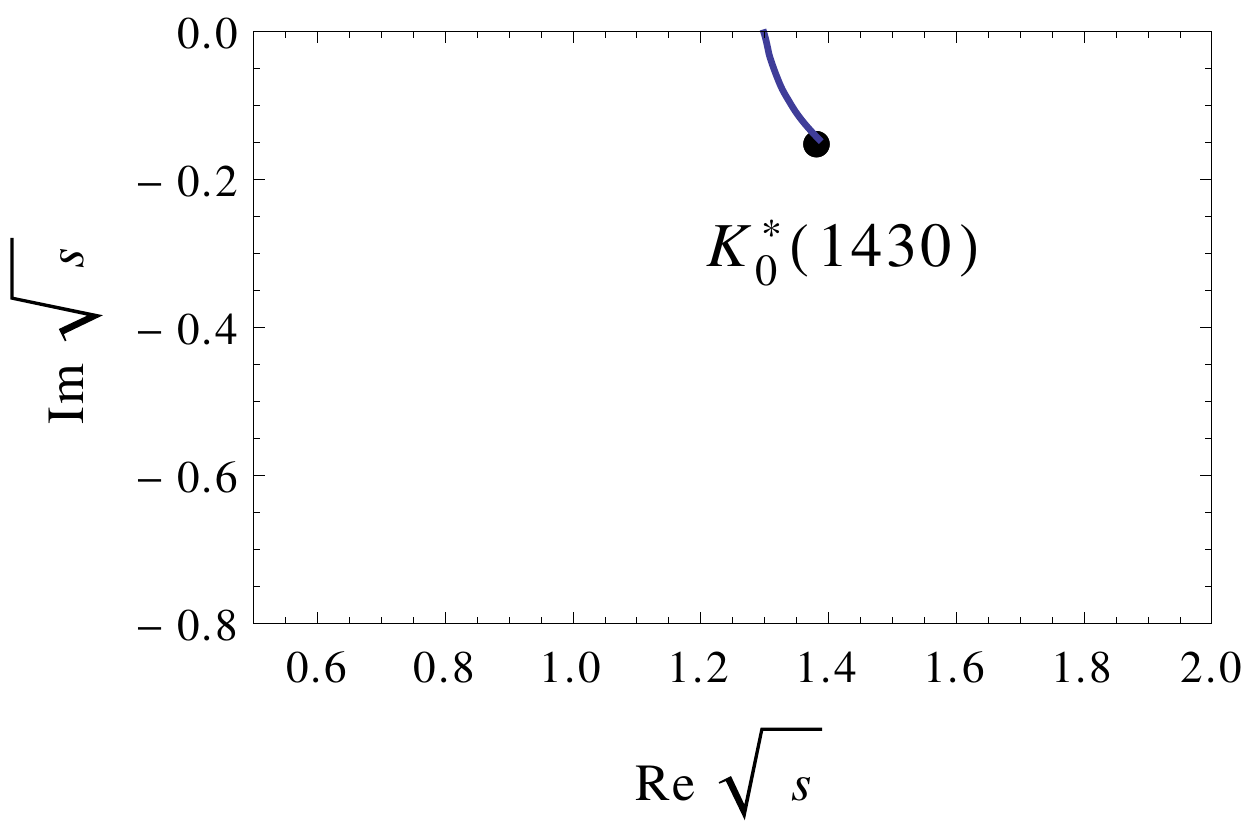}
\end{minipage}
\hspace*{-1.5cm} \begin{minipage}[hbt]{8cm}
\centering
\includegraphics[scale=0.6]{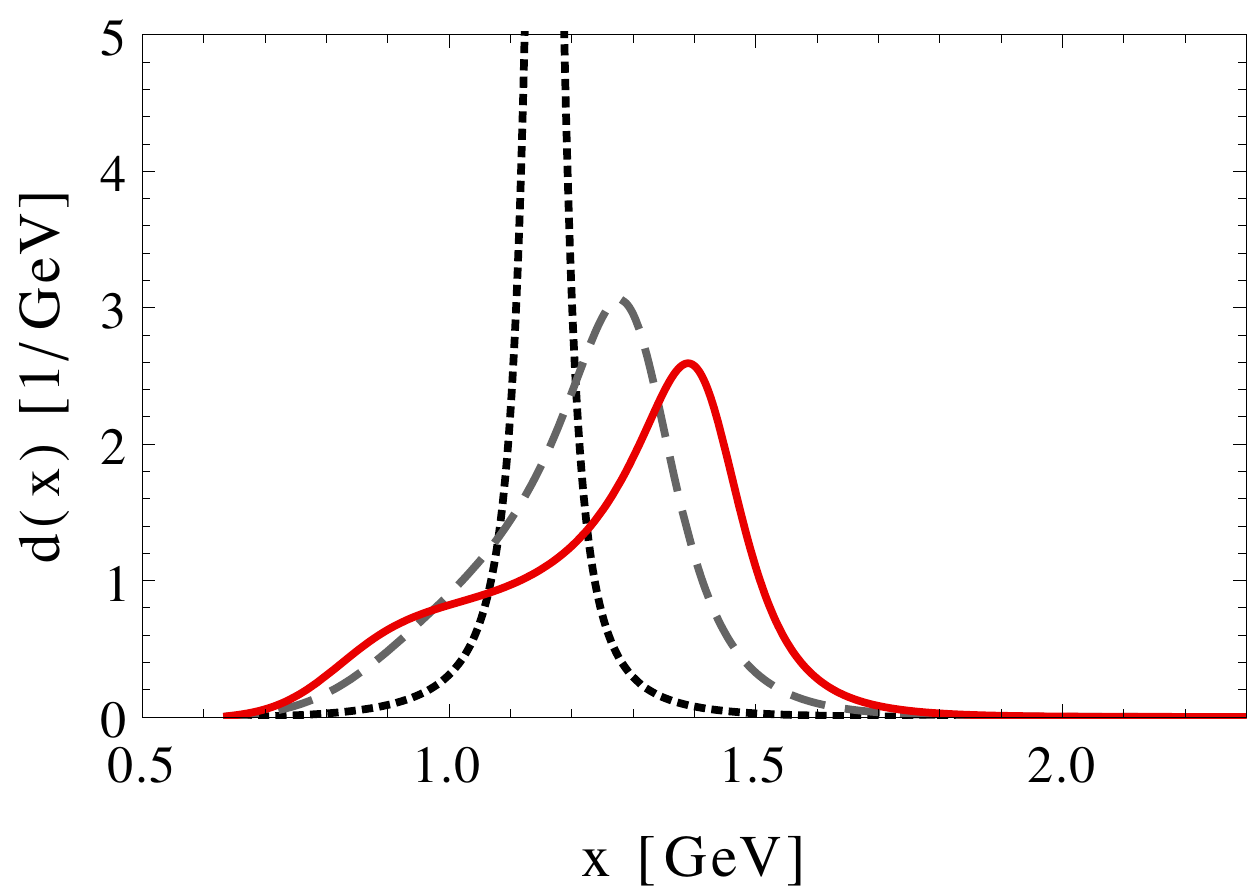}
\end{minipage}
\hspace*{0.3cm}
\begin{minipage}[hbt]{8cm}
\centering
\vspace*{0.32cm}
\includegraphics[scale=0.691]{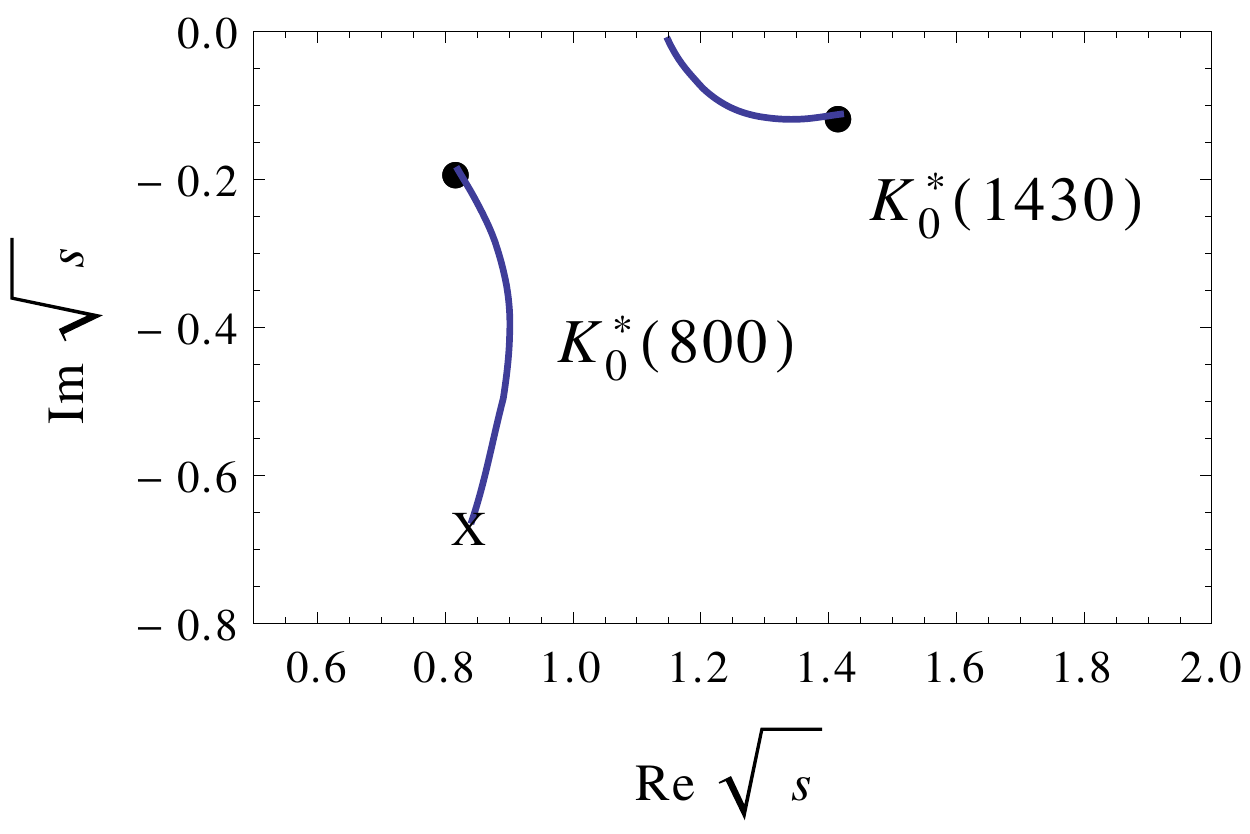}
\end{minipage}
\caption{\label{fig:varmodels2}The first row is for the case in which we consider only the
non-derivative terms in Eq.\ (\ref{eq:lag}) ($B=0$), and the second row shows the case in which we consider
only the derivative terms ($A=0$).\ In the left panels we
show the spectral functions for the three different values of $\lambda$ indicated in Figure\ \ref{fig:varmodels1}.\ 
In the right panels we display pole trajectories obtained by varying $\lambda$ from zero to $1$.\ 
Black dots indicate the position of the poles for $\lambda=1.0$.\ The X indicates the
pole position for $\lambda_c\approx0.08$, {\em i.e.}, when the pole first emerges.
All poles are on the second sheet.}
\end{figure}
\begin{figure}[t]
\centering
\vspace{0.3cm}
\hspace{-0.5cm}
\includegraphics[scale=0.65]{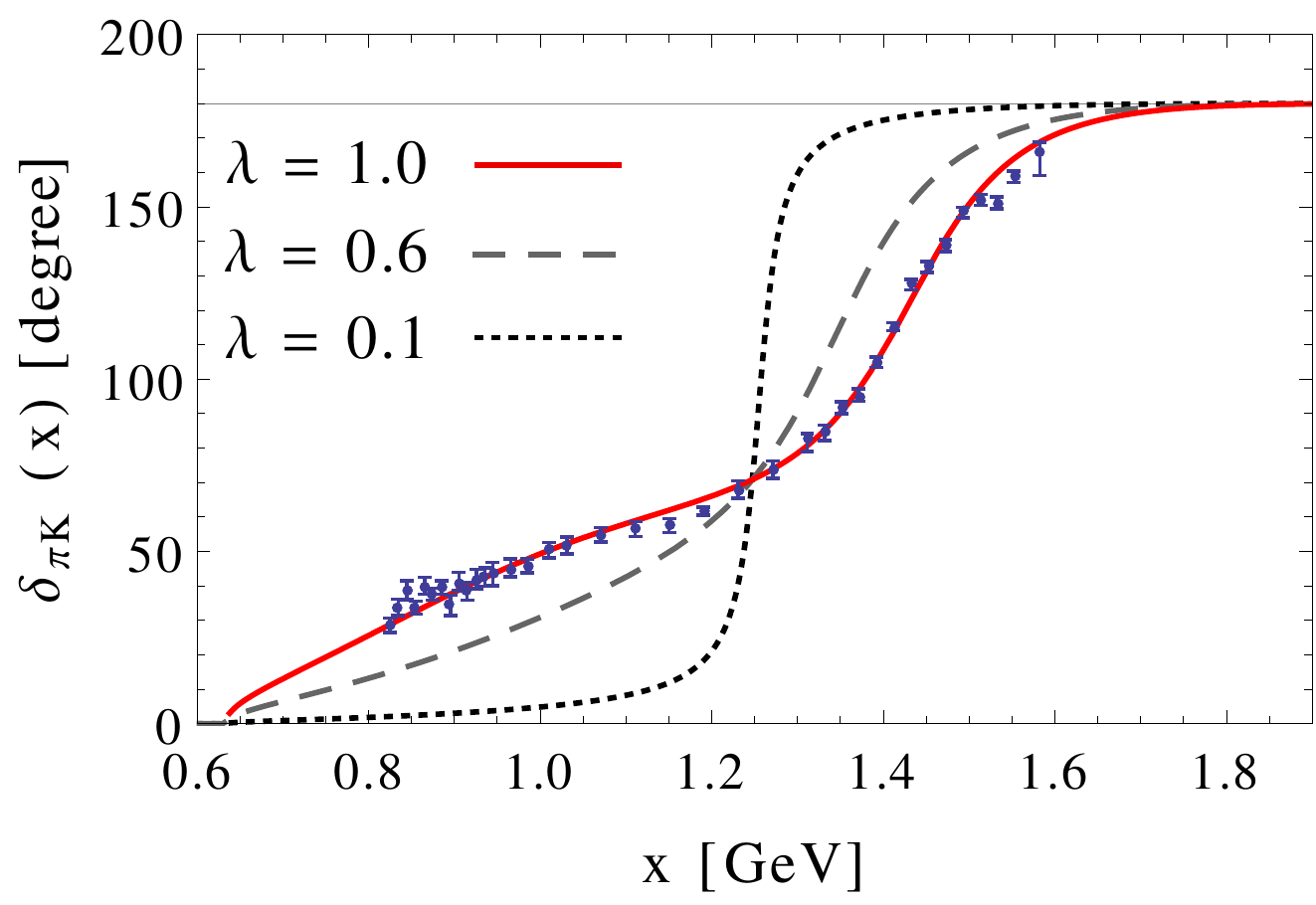}\caption{\label{fig:phases}The solid (red) curve 
shows our fit for the modified form factor in Eq.\ (\ref{eq:modform}) with respect to the
four model parameters $A$, $B$, $\Lambda$, and $m_{0}$ (see Table\ \ref{tab:parameters}).
The blue points are the data of Ref.\ \cite{astonpion}.\ The rescaling parameter 
$\lambda$ from Eq.\ (\ref{eq:lambdaFac}) is set to $1.0$.\ The other two curves correspond to $\lambda=0.6$ (long-dashed) and $\lambda=0.1$ (short-dashed).}
\end{figure}
subsection
and with other listings in the PDG\ \cite{olive}.\ Thus, changing the form factor does not
guarantee a good description of data, especially for what concerns the $\kappa$.\ For the sake of completeness we also 
show the spectral functions and the pole trajectories in Figure\ \ref{fig:phasesPT}.\ The spectral functions are not altered very much compared to our original fit, but the pole movement of the $\kappa$ has considerably changed. In particular, it was not possible to determine a value for $\lambda_{c}$; if it is finite, then we can only give the upper bound $\lambda_{c}<0.005$.
\end{enumerate}

\begin{figure}[t]
\hspace*{-1.5cm}
\begin{minipage}[hbt]{8cm}
\centering
\includegraphics[scale=0.6]{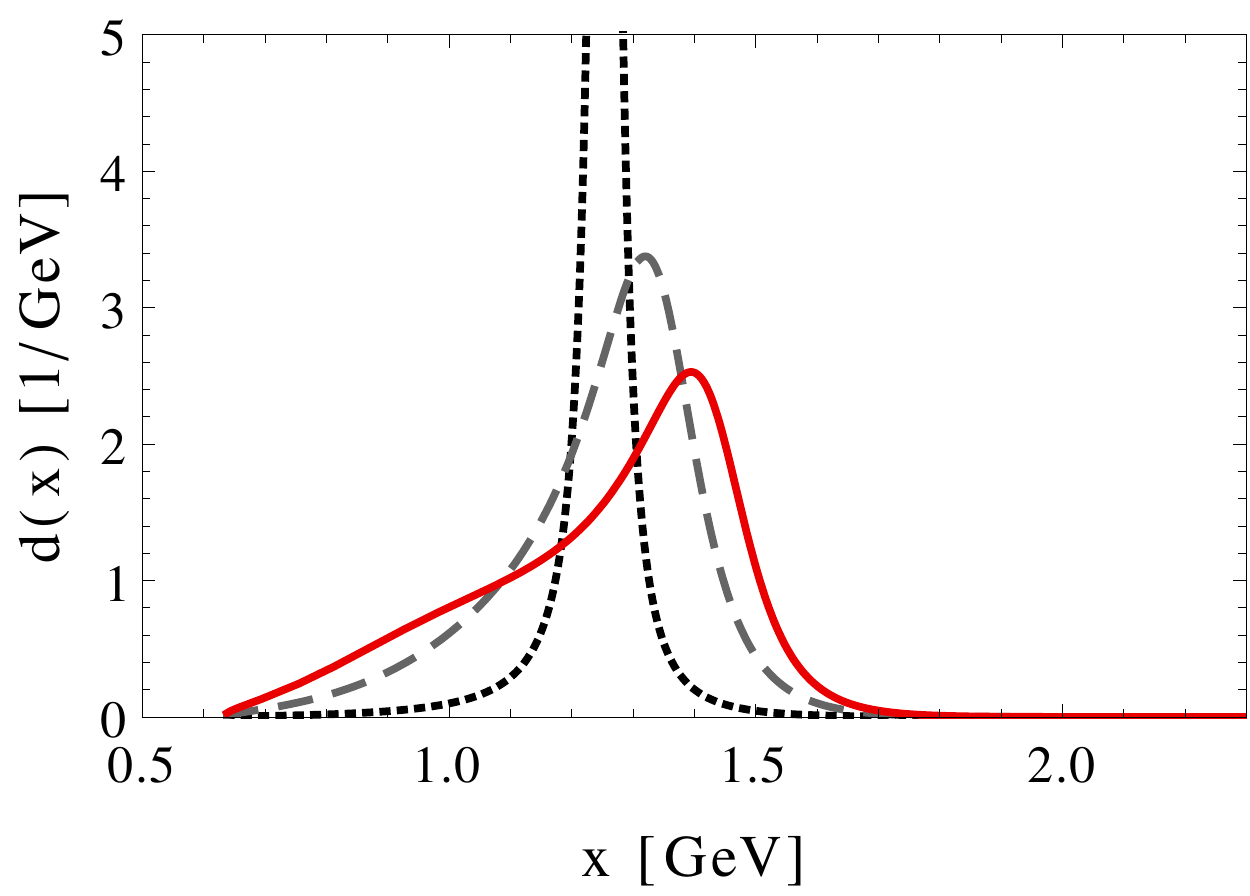}
\end{minipage}
\hspace*{0.3cm}
\begin{minipage}[hbt]{8cm}
\centering
\vspace*{0.32cm}
\includegraphics[scale=0.691]{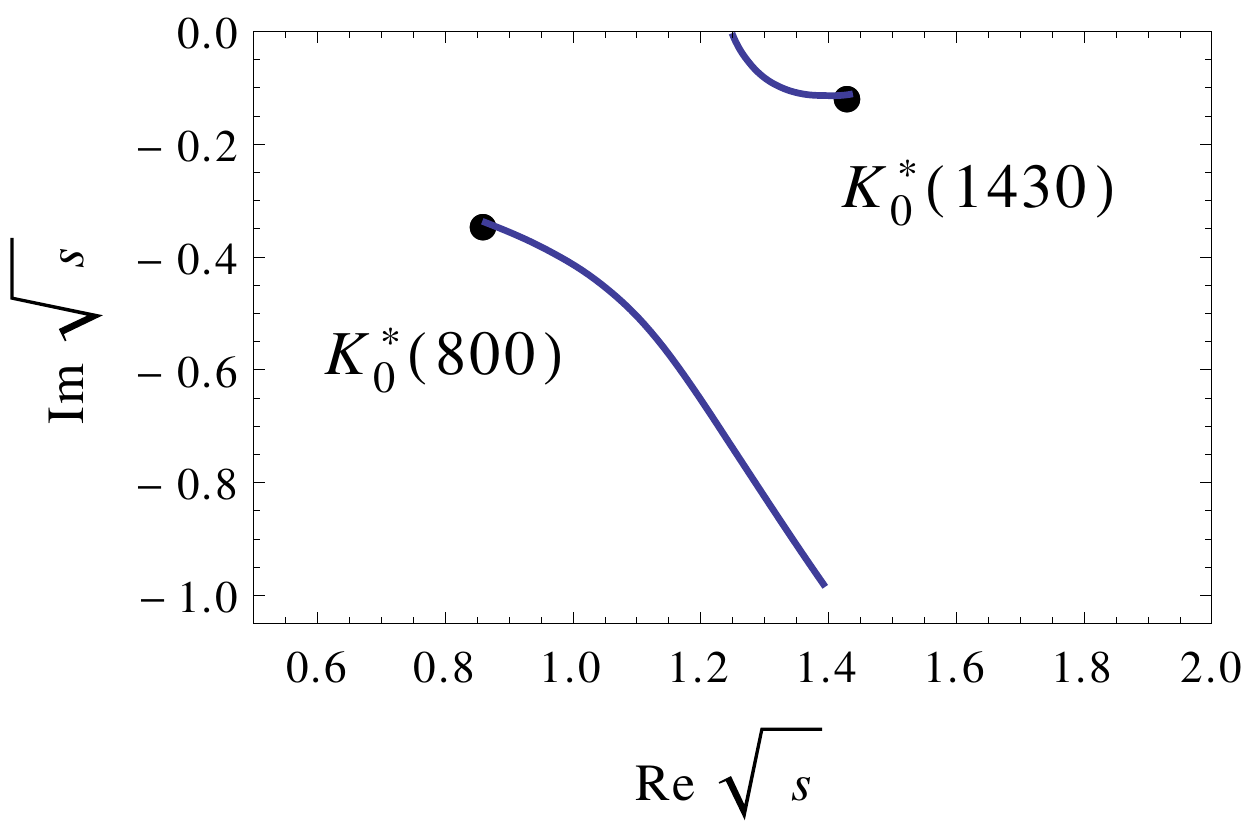}
\end{minipage}
\caption{\label{fig:phasesPT}In the left panel we
show the spectral functions for the three different values of $\lambda$ indicated in Figure\ \ref{fig:phases}.\ In the right panel
we display pole trajectories obtained by varying $\lambda$ from zero to $1$.\ 
Black dots indicate the position of the poles for $\lambda=1.0$.\ Both poles are on the second sheet.}
\end{figure}

In conclusion, our study confirms that the Gaussian form factor is an adequate
choice for mesonic interactions, leading to results that are in a good
agreement with the data up to $\sim1.8$ GeV, when {\em both} types of interaction terms, a (dominant)
derivative and a (subdominant) non-derivative one, are {\em simultaneously} taken into account.

\section{Comparison to the eLSM}
\subsection{Introductory remarks about the eLSM}
So far, we have exploited our effective model with derivative and non-derivative interaction terms for the scalar isotriplet 
and isodoublet cases. For the second one, we also performed a fit to experimental phase shift data on $\pi K$ scattering and extracted pole parameters for the two resonances with $I=1/2$.\ The final step would now be to fit the same data by applying the full eLSM Lagrangian. This requires to first also study interactions terms with a derivative in front of the decaying particle. In the third chapter we have explained that such a coupling apparently spoils the normalization of the spectral function, see Eq.\ (\ref{eq:normprop}).\ At first glance, this would influence our phase shift formula\ (\ref{eq:phaseshift}) and we risk to loose the practicality of our formalism, because each iteration step in the fit would require to renormalize the bare mass $m_{0}$ as well as all the coupling constants. Nevertheless, by carefully comparing the two equations under investigation, one realizes that the (squared) field renormalization factor, which modifies the spectral function as a multiplicative constant, also renormalizes the bare mass parameter $m_{0}$ and all the coupling constants as an inverse multiplicative constant. Since the couplings enter the expression for the tree-level decay width as coefficients, the field renormalization factor is automatically canceled -- and we are left with the same phase shift formula\ (\ref{eq:phaseshift}). This holds true as long as unrenormalized parameters are studied. It should be stressed again that all this has no effect on pole coordinates.

\begin{figure}[t]
\centering
\vspace{0.3cm}
\hspace{-0.5cm}
\includegraphics[scale=0.65]{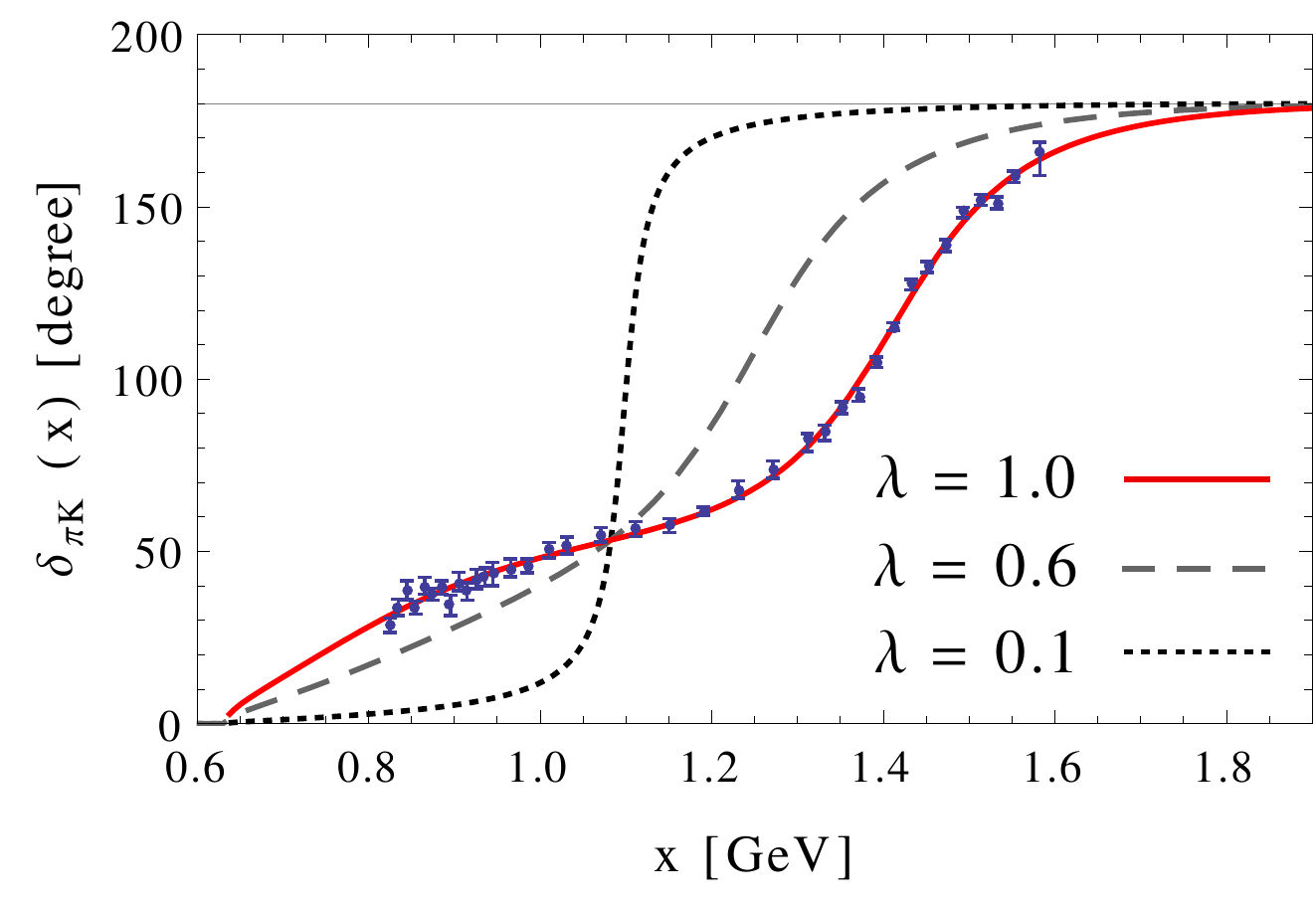}\caption{\label{fig:fitader}The solid (red) curve shows our fit
result when the model in Eq.\ (\ref{eq:adermodel}) is used with respect to the
four model parameters $C_{1}$, $C_{2}$, $\Lambda$, and $m_{0}$ (see Table\ \ref{tab:parametersader}).
The blue points are the data of Ref.\ \cite{astonpion}.\ A very good agreement is obtained. The rescaling parameter 
$\lambda$ is set to $1.0$.\ The other two curves correspond to $\lambda=0.6$ (long-dashed) and $\lambda=0.1$ (short-dashed).}
\end{figure}

\begin{figure}[t]
\hspace*{-1.5cm}
\begin{minipage}[hbt]{8cm}
\centering
\includegraphics[scale=0.6]{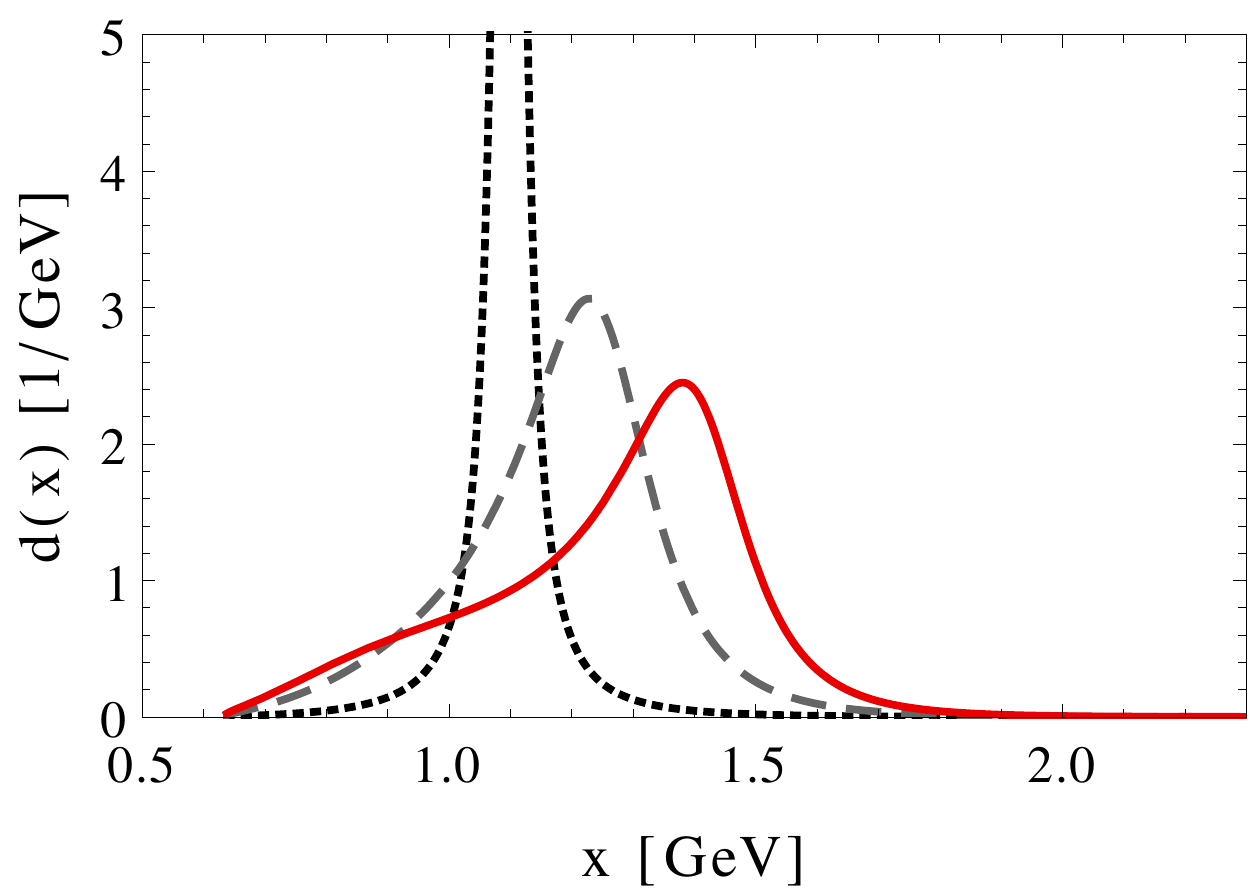}
\end{minipage}
\hspace*{0.3cm}
\begin{minipage}[hbt]{8cm}
\centering
\vspace*{0.32cm}
\includegraphics[scale=0.691]{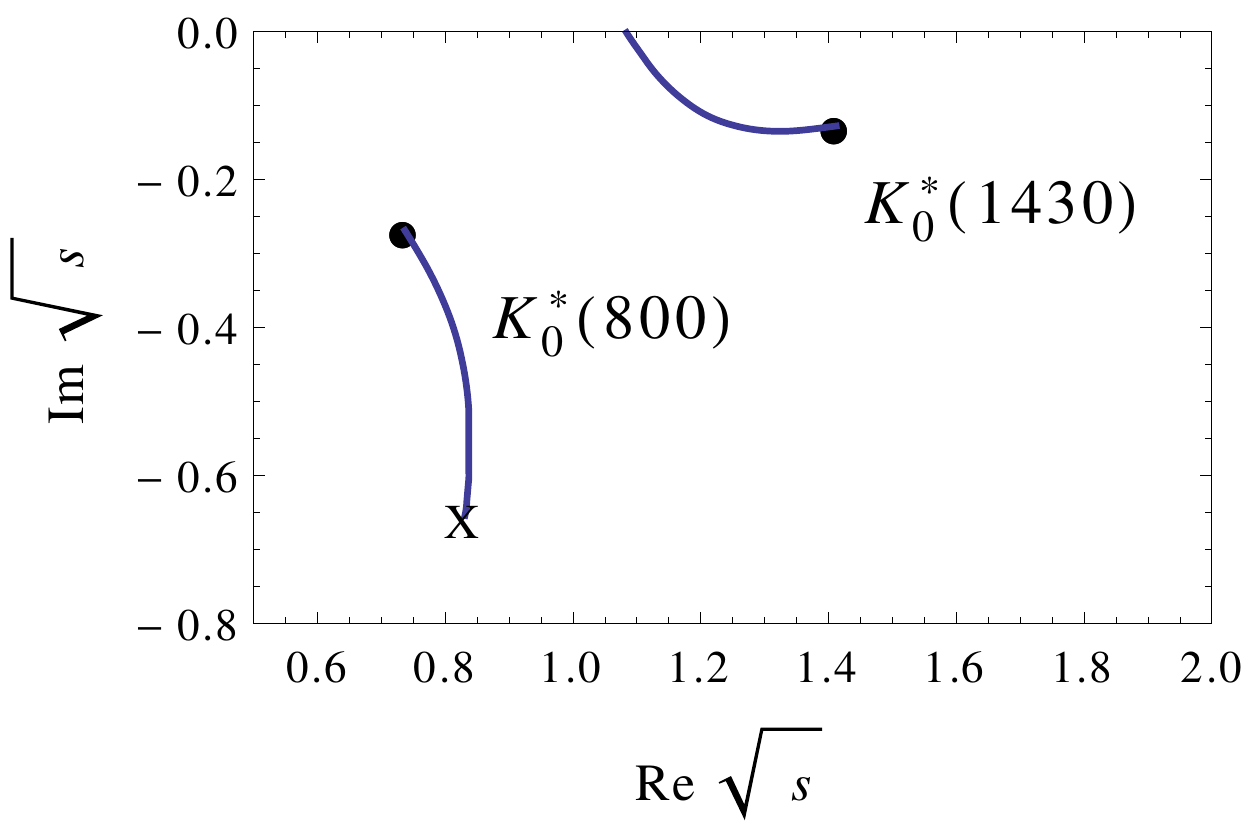}
\end{minipage}
\caption{\label{fig:aderPS}In the left panel we
show the (normalized) spectral functions for the three different values of $\lambda$ indicated in Figure\ \ref{fig:fitader}.\ In the right panel
we display pole trajectories obtained by varying $\lambda$ from zero to $1$.\ 
Black dots indicate the position of the poles for $\lambda=1.0$.\ The X indicates the
pole position for $\lambda_c\approx0.24$, {\em i.e.}, when the pole first emerges.
Both poles are on the second sheet.}
\end{figure}

A model with derivatives in front of the $K_{0}^{\ast}$ state has the form
\begin{eqnarray}
\mathcal{L}_{K_{0}^{\ast}\pi K} & = & C_{1}\partial_{\mu}K_{0}^{\ast-}\big(\partial^{\mu}\pi^{0}K^{+}+\sqrt{2}\partial^{\mu}\pi^{+}K^{0}\big)+C_{2}\partial_{\mu}K_{0}^{\ast-}\big(\pi^{0}\partial^{\mu}K^{+}+\sqrt{2}\pi^{+}\partial^{\mu}K^{0}\big) \nonumber \\
& & + \ \dots \ . \label{eq:adermodel}
\end{eqnarray}
From the fit to data, see Figures\ \ref{fig:fitader} and\ \ref{fig:aderPS}, and Table\ \ref{tab:parametersader}, the poles are found to be
\begin{align}
K_{0}^{\ast}(1430)  & :\ (1.412\pm0.002)-i\hspace{0.02cm}(0.128\pm0.003)\ \text{GeV} \ ,\\
K_{0}^{\ast}(800)\   & :\ (0.737\pm0.023)-i\hspace{0.02cm}(0.269\pm0.017)\ \text{GeV} \ .\label{eq:aderpoles}
\end{align}
Clearly, the fit is very good and delivers a value for $\chi^{2}$ close to our best fit from the previous section. It is remarkable that within this different approach the pole positions are similar, even if the values of the fitting parameters are different -- especially the bare mass is lowered. This supports our conclusion from the previous subsection: in order to fit the experimental data accurately, one has to introduce interaction terms which introduce a $\kappa$ pole dynamically. However, there is something puzzling here, namely the error of the coupling constant $C_{1}$, which is quite large.\ We suspect that this issue is caused by the presence of the interaction type with a derivative in front of the decaying particle.
\begin{table}[t]
\center
\vspace{0.45cm}
\begin{tabular}[c]{cc}
\toprule Parameter & Value\\
\midrule\\[-0.25cm]
$C_{1}$ & \hspace{0.4cm}$3.73\pm1.93$ GeV$^{-1}$\\
$C_{2}$ & \hspace{0.4cm}$6.03\pm0.97$ GeV$^{-1}$\\
$\Lambda$ & $0.500\pm0.009$ GeV\\
$m_{0}$ & $1.083\pm0.015$ GeV\\
\\[-0.1cm]
\toprule
\end{tabular}
\caption{Results of the fit; $\chi_{0}^{2}/d.o.f.=1.27$.}
\label{tab:parametersader}
\end{table}

Before proceeding, we apply similar transformations as in Eq.\ (\ref{eq:LagDiff}) to compare the parameters found here with the ones obtained by our original fit. Obviously,
\begin{eqnarray}
C_{1}\partial^{\mu}K_{0}^{\ast-}\partial_{\mu}\pi^{0}K^{+} & = & -C_{1}K_{0}^{\ast-}(\square\pi^{0}K^{+}+\partial^{\mu}\pi^{0}\partial_{\mu}K^{+}) \ , \\
C_{2}\partial^{\mu}K_{0}^{\ast-}\pi^{0}\partial_{\mu}K^{+} & = & -C_{2}K_{0}^{\ast-}(\pi^{0}\square K^{+}+\partial^{\mu}K^{+}\partial_{\mu}\pi^{0}) \ ,
\end{eqnarray}
leading to
\vspace{-0.2cm}
\begin{equation}
\mathcal{L}_{K_{0}^{\ast}\pi K} = \underbrace{\big(C_{1}m_{\pi}^{2}+C_{2}m_{K}^{2}\big)}_{=A}K_{0}^{\ast-}\pi^{0}K^{+}
+\underbrace{\big(-C_{1}-C_{2}\big)}_{=B}K_{0}^{\ast-}\partial_{\mu}\pi^{0}\partial^{\mu}K^{+}+\dots \ . \\[-0.1cm]
\end{equation}
We find that the coupling constants of our effective model read
\vspace{-0.2cm}
\begin{align}
A &= 1.541 \ \text{GeV} \ , & B &= -9.758 \ \text{GeV}^{-1} \ , \\[-0.9cm]
\nonumber
\end{align}
which are slightly different from those of Table\ \ref{tab:parameters}, although their order or magnitude is reproduced.\ With this we conclude that the comparisons shown in Eq.\ (\ref{eq:eLSMcompPar}) are not reliable on the one-loop level, because one cannot replace the d'Alembert operators with the pseudoscalar masses.\ Our original effective model therefore {\em is} different to the one here and consequently different to the eLSM.

\subsection{Full eLSM case}
\begin{figure}[t]
\centering
\vspace{0.3cm}
\hspace{-0.5cm}
\includegraphics[scale=0.65]{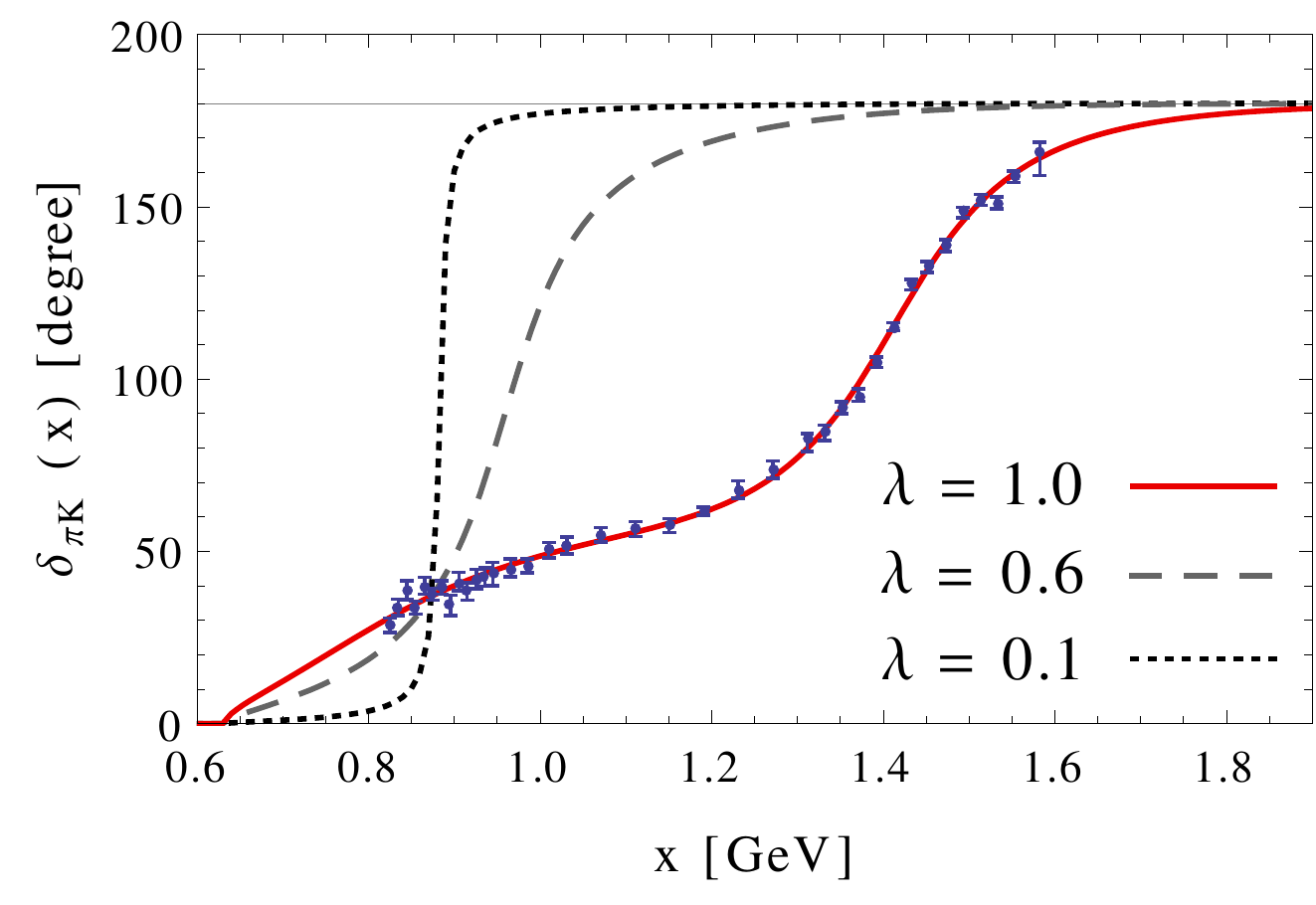}\caption{\label{fig:fiteLSM}The solid (red) curve shows our fit
result when the model in Eq.\ (\ref{eq:eLSMLagrangian}) is used with respect to the
six model parameters $A$, $B$, $C_{1}$, $C_{2}$, $\Lambda$, and $m_{0}$ (see Table\ \ref{tab:parametereLSMcase}).\
The blue points are the data of Ref.\ \cite{astonpion}.\ A very good agreement is obtained. The rescaling parameter 
$\lambda$ is set to $1.0$.\ The other two curves correspond to $\lambda=0.6$ (long-dashed) and $\lambda=0.1$ (short-dashed).}
\end{figure}

\begin{figure}[t]
\hspace*{-1.45cm}
\begin{minipage}[hbt]{8cm}
\centering
\includegraphics[scale=0.6]{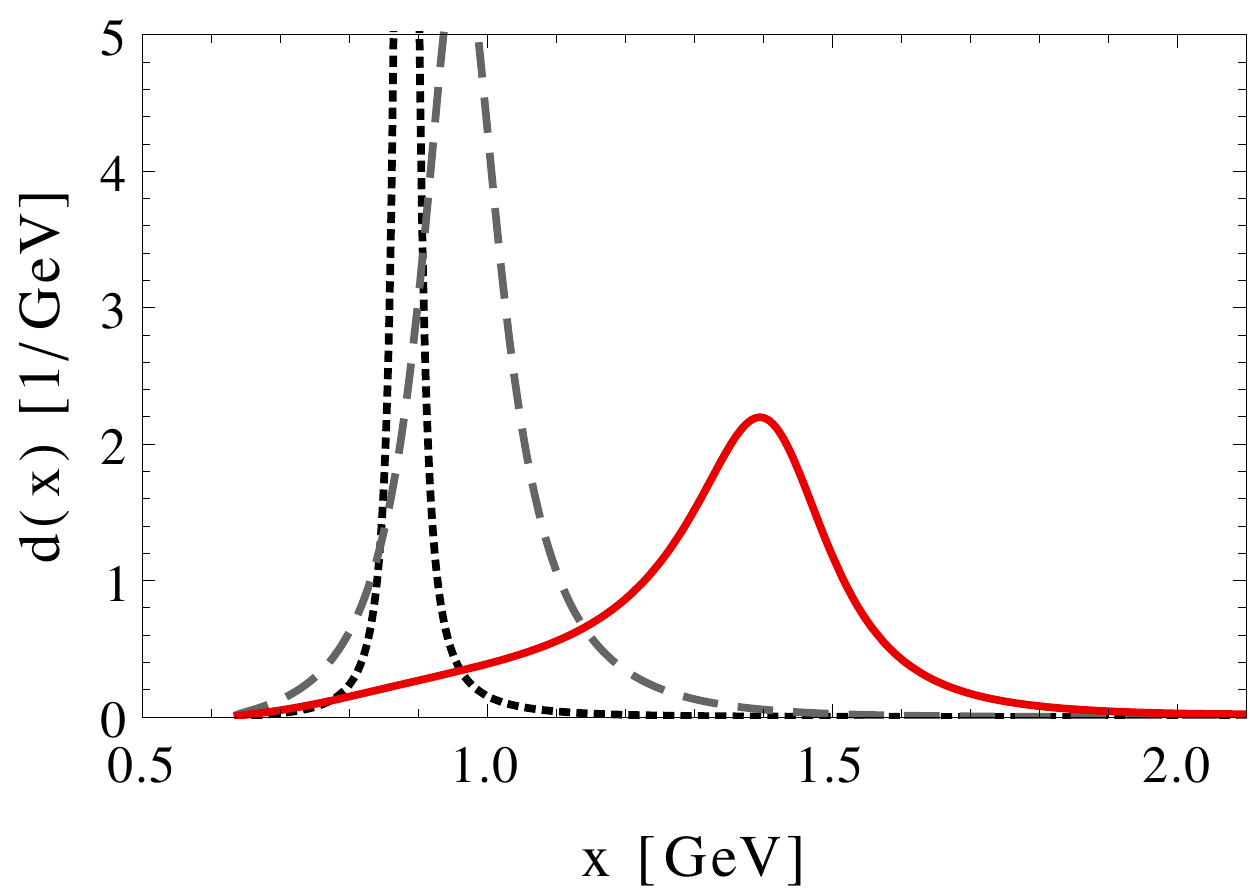}
\end{minipage}
\hspace*{0.3cm}
\begin{minipage}[hbt]{8cm}
\centering
\vspace*{0.32cm}
\includegraphics[scale=0.691]{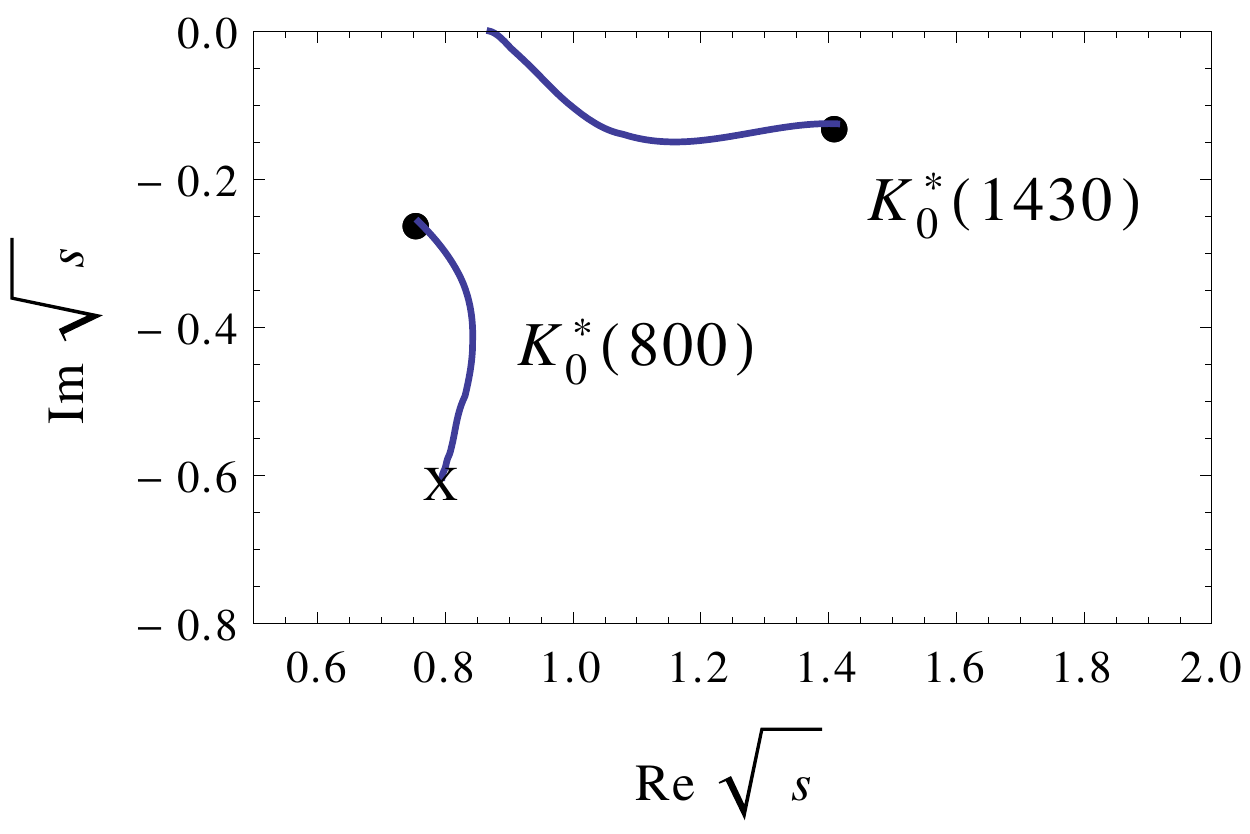}
\end{minipage}
\caption{\label{fig:eLSMcasePS}In the left panel we
show the (normalized) spectral functions for the three different values of $\lambda$ indicated in Figure\ \ref{fig:fiteLSM}.\ 
In the right panel we display pole trajectories obtained by varying $\lambda$ from zero to $1$.\ 
Black dots indicate the position of the poles for $\lambda=1.0$.\ The X indicates the
pole position for $\lambda_c\approx0.43$, {\em i.e.}, when the pole first emerges.\
Both poles are on the second sheet.}
\end{figure}

We are finally able to perform the fit to phase shift data by using the full eLSM interaction Lagrangian:
\vspace{-0.15cm}
\begin{eqnarray}
\mathcal{L}_{K_{0}^{\ast}\pi K} & = & AK_{0}^{\ast-}\big(\pi^{0}K^{+}+\sqrt{2}\pi^{+}K^{0}\big)+BK_{0}^{\ast-}\big(\partial_{\mu}\pi^{0}\partial^{\mu}K^{+}+\sqrt{2}\partial_{\mu}\pi^{+}\partial^{\mu}K^{0}\big) \nonumber \\
&  & + \ C_{1}\partial_{\mu}K_{0}^{\ast-}\big(\partial^{\mu}\pi^{0}K^{+}+\sqrt{2}\partial^{\mu}\pi^{+}K^{0}\big)+C_{2}\partial_{\mu}K_{0}^{\ast-}\big(\pi^{0}\partial^{\mu}K^{+}+\sqrt{2}\pi^{+}\partial^{\mu}K^{0}\big) \nonumber \\
& & + \ \dots \ .
\label{eq:eLSMLagrangian} \\[-0.9cm]
\nonumber
\end{eqnarray}
The results are presented in Figures\ \ref{fig:fiteLSM} and\ \ref{fig:eLSMcasePS}, and Table\ \ref{tab:parametereLSMcase}.\ 
We find two poles which we assign in the following way:
\begin{align}
K_{0}^{\ast}(1430)  & :\ (1.413\pm0.007)-i\hspace{0.02cm}(0.125\pm0.007)\ \text{GeV} \ ,\\
K_{0}^{\ast}(800)\   & :\ (0.758\pm0.015)-i\hspace{0.02cm}(0.257\pm0.014)\ \text{GeV} \ .\label{eq:eLSMpoles}%
\end{align}
This means for the masses and widths:
\begin{align}
K_{0}^{\ast}(1430)  & :\ m_{\text{pole}} = (1.408\pm0.008) \ \text{GeV} \ , \ \ \Gamma_{\text{pole}} = (0.251\pm0.015) \ \text{GeV} \ ,\\
K_{0}^{\ast}(800)\   & :\ m_{\text{pole}} = (0.713\pm0.020) \ \text{GeV} \ , \ \ \Gamma_{\text{pole}} = (0.546\pm0.034) \ \text{GeV} \ .\label{eq:eLSMmasswidths}
\end{align}
The most important observation here is the fact that the poles lie exactly within the error ranges obtained from our original 
effective model. As expected, this seems to be due to the quality of our fit: the $\chi^{2}$ 
has a value of $\chi_{0}^{2}/d.o.f.=1.31$.\ One may suggest that by adopting the type of models we have investigated, poles 
are automatically induced such that the experimental results are well described.

Again it turns out that the cutoff $\Lambda$ needs to be around $\sim0.5$ GeV, hence rather small compared 
to the scale we are interested in of $\sim1$ GeV. The single exception where 
this was not the case is the model with only non-derivative interactions -- and in fact this was the single one where no 
pole for the $K_{0}^{\ast}(800)$ was generated. We therefore suggest that a low value for the cutoff parameter 
is a universal feature in the type of models investigated.\footnote{The case of Eq.\ (\ref{eq:modform}) gave a little higher cutoff.}\ Whereas this seems to be a common property, the values for the bare seed mass $m_{0}$ undergo high variations. One observes that derivative interactions decrease the value of $m_{0}$.\ The strongest effect occurs when all eLSM interactions are considered: the bare mass falls deep below $1$ GeV, although the large-$N_{c}$ study proves that the intrinsic pole associated with this state is a quarkonium.\ This is even more puzzling because other investigations of the $I=1/2$ sector usually add the (bare) mass of the strange quark to the mass parameter $m_{0}$, automatically increasing its value\ \cite{tornqvist,tornqvist2,boglione,pennington}.\ As a last comment, we would like to remark that the errors for $m_{0}$ and $\Lambda$ stated in Table\ \ref{tab:parametereLSMcase} are only upper bounds. Our computation reported even smaller ones. In fact, a closer look at all the errors reveals that the ones for the coupling constants $A$ and especially for $C_{2}$ are quite large. The inclusion of interactions of the type where a derivative is in front of the decaying particle seems to produce this kind of behavior -- either from a conceptual or numerical point of view.
\begin{table}[t!]
\center
\vspace{0.45cm}
\begin{tabular}[c]{cc}
\toprule Parameter & Value\\
\midrule\\[-0.25cm]
$A$ & $0.39\pm0.19$ GeV\\
$B$ & \hspace{0.22cm}$22.80\pm1.20$ GeV$^{-1}$\\
$C_{1}$ & \hspace{0.22cm}$28.68\pm0.34$ GeV$^{-1}$\\
$C_{2}$ & \hspace{0.11cm}$-0.06\pm0.70$ GeV$^{-1}$\\
$\Lambda$ & $0.440\pm0.001$ GeV\\
$m_{0}$ & $0.874\pm0.001$ GeV\\
\\[-0.1cm]
\toprule
\end{tabular}
\caption{Results of the fit; $\chi_{0}^{2}/d.o.f.=1.31$.}
\label{tab:parametereLSMcase}
\end{table}

Besides that, we observe at least one non-physical bump at an energy about $2.3$ GeV (not visible in the left panel of Figure\ \ref{fig:eLSMcasePS}), similar to those reported in Refs.\ \cite{tqmix2,tqmix}.\ Neither the eLSM nor all our presented models preserve their validity for energies higher than $\sim1.8$ GeV, thus we ignore these anomalies.

\section{Summary and conclusions}
We have concentrated in this chapter on the channel $I=1/2$, $J=0$.\ 
Our starting model contained non-derivative and derivative interactions in agreement with low-energy 
effective approaches to QCD. It was
demonstrated that, by using a single kaonic seed state, both scalar resonances
$K_{0}^{\ast}(1430)$ and $K_{0}^{\ast}(800)$ (known as $\kappa$) can be
described as complex propagator poles -- the two poles are required in order to
correctly reproduce phase shift data of $\pi K$ scattering.\ The spectral
function of our model turned out to be not of the ordinary Breit--Wigner type, due to strong distortions in the low-energy regime as a direct
consequence of the $\kappa$ pole.

In the large-$N_{c}$ limit, this pole finally disappeared; the corresponding
state is therefore not a conventional quarkonium.\ On the contrary, the pole
corresponding to $K_{0}^{\ast}(1430)$ approached the real energy axis for
large values of $N_{c}$, hence became very narrow, which is a general feature of a
quark-antiquark state. It must be stressed that the presence of derivative interactions turned out to be crucial
for our results.\ They are the dominant contributions toward the
description of the $\pi K$ phase shift, however, they need to be accompanied by non-derivative terms.\ Otherwise, 
no reliable fit to the data can be achieved. Moreover, a good fit cannot be obtained when a simple variation of 
the form factor is performed.

Additionally, we have performed a fit by using a model with the same interaction terms as are present in the eLSM. 
Although the results were almost as good as for our starting model, large errors emerged for some of the model parameters. 
We suspect that this problem originates because of the interaction terms with a derivative in front of the 
decaying particle.

Based on our results, we suggest to include the light $\kappa$ in the summary tables of the PDG. For the future, one should use more complete models than the one presented at the beginning of 
this chapter.\ In particular, a model is desired which allows to study
simultaneously the $I=1/2$ and the $I=3/2$ channels.\ One should stress that the full version of the eLSM can be applied
for this purpose.

\newpage
\clearpage

\

\newpage
\clearpage

\chapter{Summary and conclusions}
\label{chap:conclusions}

\medskip

Experimental data exhibit several puzzling facts about the light scalar mesons:\ 
$f_{0}(500)$ (or $\sigma$) and $K_{0}^{\ast}(800)$ (or $\kappa$) have 
large decay widths, while $f_{0}(980)$ and $a_{0}(980)$ are narrow, 
but their line shapes show threshold distortions due to the nearby $K\bar{K}$ decay channel opening. 
It is nowadays recognized that these states do not fit into the ordinary $q\bar{q}$ picture 
based on a simple representation of $SU(3)$ flavor symmetry.\ In fact, there is a growing consensus that the 
scalar resonances $f_{0}(1370)$,\ $f_{0}(1500)$,\ $K_{0}^{\ast}(1430)$,\ and $a_{0}(1450)$\ are predominantly 
quark-antiquark states, while the light resonances below $1$ GeV, namely $f_{0}(500)$, $f_{0}(980)$, $K_{0}^{\ast}(800)$, 
and $a_{0}(980)$ should be predominantly some sort of four-quark objects.\ However, there is no agreement on the precise 
mechanism creating them.
\\
\\
We have shown that one can regard those light particles as dynamically generated states: they\ $(i)$\ are not present in the original 
formulation of a hadronic model, and $(ii)$ emerge as companion poles after incorporating hadronic loop corrections. This is 
often called unitarization.\ It establishes access to the four-quark contributions in the Fock space of an unstable particle
 -- while those contributions usually shift the resonance pole from the real energy axis into the complex plane, in the case a 
scalar state they may create new poles in the propagator and scattering matrix, respectively.
\\
\\
We have concentrated in this work on the isotriplet $I=1$ and isodoublet $I=1/2$ resonances.\ The effective models 
we applied to study those two scalar sectors contain derivative and non-derivative interaction vertices.\ It was therefore 
necessary to work out in Chapter\ \ref{chap:chapter3} how such interactions appear in a unitarization scheme. In particular, 
we investigated the use of Feynman rules in the context of quantum field theories with derivative interactions, and 
demonstrated how to solve apparent discrepancies between ordinary Feynman rules and dispersion relations.
\\
\\
In Chapter\ \ref{chap:chapter4}, we then have repeated and extended previous calculations of T\"ornqvist and 
Roos\ \cite{tornqvist,tornqvist2} and Boglione and Pennington\ \cite{pennington}\ in order to clarify some open issues concerning the dynamical generation in the isovector sector ($I=1$).\ From this basis, we introduced a 
Lagrangian inspired 
by the extended Linear Sigma Model (eLSM)\ \cite{denisphd,eLSM2}, where the mesons indeed interact via 
derivative and non-derivative couplings. In our case, the Lagrangian contained a single scalar isotriplet $a_{0}$ seed state, 
corresponding to the resonance $a_{0}(1450)$.\ A careful analysis of its propagator pole structure revealed an additional pole 
of a narrow state with mass around $1$ GeV. We identified this pole with the physical $a_{0}(980)$ particle.

\vspace{0.1cm}
\noindent Our fit 
reproduced not only the pole position of both states, but also the experimental branching ratios of $a_{0}(1450)$, 
previous theoretical estimates for the couplings of $a_{0}(980)$ to its decay channels, and delivered predictions for 
the phase shifts and the inelasticity. The latter two turned out to be in qualitative agreement with former studies on 
the subject. A large-$N_{c}$ study of the pole trajectories finally confirmed that the state below $1$ GeV is not a 
quark-antiquark state but rather some sort of four-quark object, while the resonance above $1$ GeV is predominantly 
a quarkonium.
\\
\\
Following this, the same type of model and formalism were used in Chapter\ \ref{chap:chapter5} for the isodoublet case ($I=1/2$).\ The 
single seed state was assigned to the $q\bar{q}$ state $K_{0}^{\ast}(1430)$.\ In contrast to the previous investigation, we 
applied our model to fit experimental $\pi K$ phase shift data from Ref.\ \cite{astonpion}.\ Along this line, several variations 
of the model have been tested on the data set:\ it was found that both derivative and non-derivative interactions are 
needed for a satisfactory fit and are thus highly important.\ The model with either only derivatives or only non-derivatives 
turned out to diagree with the experiment, the same as the full model with form factor different from the 
usual Gaussian.

\vspace{0.1cm}
\noindent Besides the expected resonance pole of $K_{0}^{\ast}(1430)$, a pole corresponding to the light $\kappa$ naturally
emerged on the unphysical second Riemann sheet. We determined the position of the poles for both states with surprisingly 
small errors. A large-$N_{c}$ study demonstrated that the $\kappa$ pole finally disappears; as a consequence, this very 
broad state is predominantly not a quarkonium, but rather a dynamically generated meson. Just as the famous $f_{0}(500)$, we think that one therefore should include the light $\kappa$ in the summary tables of the PDG. The state above $1$ GeV, however, turned out to be predominantly a quark-antiquark state.

\vspace{0.1cm}
\noindent Finally, we proceeded to fit the data with a Lagrangian that includes all the relevant interaction vertices as the eLSM, in particular, vertices with a derivative in front of the decaying particle. From a first study of the model with the latter interaction(s) only, we observed that although such a theory can describe the data very well, some model parameters are badly determined. In the end, the full eLSM also showed this problem, but we furthermore found an unphysical peak above $2$ GeV in the spectral function. This needs to be investigated in more detail.

\begin{center}
$\ast$
\end{center}

\noindent The mechanism of dynamical generation is applicable to the other light and some of the heavy scalar states. 
It therefore seems promising to extend the present study in the low-energy regime in order to include the isoscalars, where the resonances $f_{0}(500)$ and $f_{0}(980)$ should be dynamically generated. In this case, $f_{0}(1370),$ $f_{0}(1500)$, and $f_{0}(1710)$ would be predominantly a non-strange quarkonium, a strange quarkonium, and a scalar glueball, respectively.\ The main task of such work would be a simultaneous fit to at least $I=0$ $\pi\pi$ and $I=1/2$ $\pi K$ scattering data, which are available in very good quality.\ This requires to extend our formalism to a situation with two $q\bar{q}$ seed states for the isoscalars (here, without a glueball). This again would just reflect a small part of the eLSM and hence would be a putative proof of concept. It is desirable to apply the eLSM Lagrangian as shown in Eq.\ (\ref{eq:eLSMLag}), strictly speaking, one has to unitarize the eLSM from the very beginning and to perform an exhaustive simultaneous fit to more data. The foundations were laid in this work.
\\
\\
Another interesting subject is the study of dynamical generation in the framework of 
'puzzling\grq\ resonances in the charmonium sector\ \cite{PenningtonCharm}, see for example Ref.\ \cite{brambilla} 
and references therein.\ Namely, a whole class of mesons, called $X$, $Y$, and $Z$ states, 
has been experimentally discovered but is so far not fully understood \cite{braaten,braaten2,maianix}.\
As demonstrated in Ref.\ \cite{coitox} for the case of $X(3872)$, some of the $X$ and $Y$ states 
could emerge as companion poles of quark-antiquark states.

\vspace{0.1cm}
\noindent This may also be true for the scalar strange-charmed state $D_{S0}^{\ast}(2317)$: it decays to $D$ mesons and kaons and was described in an extension of the eLSM with a mass above the $DK$ threshold\ \cite{eshraim}.\ This led to a very large decay width such that it could not be seen in experiment. However, the PDG gives a mass below the $DK$ threshold and a width smaller than $5$ MeV\ \cite{olive}.\ The solution could be that the $D_{S0}^{\ast}(2317)$ is actually not a quarkonium (as assumed in the eLSM), but a dynamically generated four-quark object.

\newpage
\clearpage

\

\appendix

\chapter{Mathematical formulas}
\label{app:appendixA}

\medskip

\begin{itemize}
\item Fourier transformation:
\begin{eqnarray}
\text{three dimensions:} \ \ \ f(\textbf{x}) & = & \int\frac{\text{d}^{3}k}{(2\pi)^{3}} \ e^{i\textbf{k}\cdot\textbf{x}}\tilde{f}(\textbf{k}) \ , \\
\tilde{f}(\textbf{k}) & = & \int\text{d}^{3}x \ e^{-i\textbf{k}\cdot\textbf{x}}f(\textbf{x}) \ , \\
\Rightarrow \ \ \  \delta^{(3)}(\textbf{k}-\textbf{q}) & = & \int\frac{\text{d}^{3}x}{(2\pi)^{3}} \ e^{-i(\textbf{k}-\textbf{q})\cdot\textbf{x}} \\ \nonumber \\
\text{four dimensions:} \ \ \ \ f(x) & = & \int\frac{\text{d}^{4}k}{(2\pi)^{4}} \ e^{-ik\cdot x}\tilde{f}(k) \ , \\
\tilde{f}(k) & = & \int\text{d}^{4}x \ e^{ik\cdot x}f(x) \ , \\
\Rightarrow \ \ \  \delta^{(4)}(k-q) & = & \int\frac{\text{d}^{4}x}{(2\pi)^{4}} \ e^{i(k-q)\cdot x}
\end{eqnarray}
\item Useful representation of the delta function:
\begin{equation}
\delta(a-b) = \lim_{\epsilon \to 0^{+}}\frac{1}{\pi}\frac{\epsilon}{(a-b)^{2}+\epsilon^{2}}
\end{equation}
\item Sokhotski--Plemelj theorem:
\begin{eqnarray}
\lim_{\epsilon \to 0^{+}}\int_{a}^{b}\text{d}x \ \frac{f(x)}{x\pm i\epsilon} & = & -\hspace{-0.415cm}\int_{a}^{b}\text{d}x \ \frac{f(x)}{x}\mp i\pi f(0) \ , \\
\Rightarrow \ \ \ \frac{1}{x\pm i\epsilon} & = & \mathcal{P}\left(\frac{1}{x}\right)\mp i\pi\delta(x) \ ,
\end{eqnarray}
with the Cauchy principal value
\begin{equation}
-\hspace{-0.435cm}\int_{a}^{c}\text{d}x \ f(x) = \mathcal{P}\int_{a}^{c}\text{d}x \ f(x) = \lim_{\eta \to 0^{+}}\left[\int_{a}^{b-\eta}\text{d}x \ f(x)+\int_{b+\eta}^{c}\text{d}x \ f(x)\right]
\end{equation}
\item Polar form of a complex number $z=x+iy=\rho e^{i\phi}$ for $\phi\in(-\pi,\pi]$:
\begin{eqnarray}
\rho & = & \sqrt{x^{2}+y^{2}} \ , \\[0.3cm]
\phi & = & \arg z \ \ = \ \ \begin{cases} \arctan\frac{y}{x} & x>0 \\
\arctan\frac{y}{x}+\pi & x<0, \ y\ge0 \\
\arctan\frac{y}{x}-\pi & x<0, \ y<0 \\
\frac{\pi}{2} & x=0, \ y>0 \\
-\frac{\pi}{2} & x=0, \ y<0 \\
\text{undefined} & x=0, \ y=0
\end{cases}
\end{eqnarray}

\end{itemize}

\clearpage

\chapter{Conventions}
\label{app:appendixB}

\medskip

\begin{itemize}
\item Natural units:
\begin{equation}
c = \hbar = 1
\end{equation}
\item Minkowski metric:
\begin{equation}
\text{d}s^{2} = \text{d}t^{2}-\text{d}\textbf{r}^{2} \ , \ \ \ \ \ (\eta_{\mu\nu}) = \diag(1,-1,-1,-1)
\end{equation}
\item Normalization of states and commutator relations:
\begin{eqnarray}
|\textbf{p}\rangle & = & \sqrt{2E_{\textbf{p}}}a_{\textbf{p}}^{\dagger}|0\rangle \ , \\[0.15cm]
\langle \textbf{p}|\textbf{q}\rangle & = & 2E_{\textbf{p}}(2\pi)^{3}\delta^{(3)}(\textbf{p}-\textbf{q}) \ , \\[0.15cm]
\phi(x) & = & \int\frac{\text{d}^{3}p}{(2\pi)^{3}}\frac{1}{\sqrt{2E_{\textbf{p}}}}\Big(a_{\textbf{p}}^{\dagger}e^{ip\cdot x}+a_{\textbf{p}}e^{-ip\cdot x}\Big) \ , \\[0.15cm]
{[}a_{\textbf{p}},a_{\textbf{q}}^{\dagger}{]} & = & (2\pi)^{3}\delta^{(3)}(\textbf{p}-\textbf{q})
\end{eqnarray}
\item Spectral representation of Green's functions (with $\epsilon \to 0^{+}$):
\begin{eqnarray}
G(p^{2}) & = & \frac{1}{\pi}\int_{0}^{\infty}\text{d}x^{2} \ \frac{\rho(x^{2})}{p^{2}-x^{2}+i\epsilon} \nonumber \\
& = & \frac{1}{\pi}\int_{0}^{\infty}\text{d}x \ \frac{2x\rho(x^{2})}{p^{2}-x^{2}+i\epsilon} \nonumber \\
& = & \int_{0}^{\infty}\text{d}x \ \frac{d(x)}{p^{2}-x^{2}+i\epsilon} \ , \\[0.5cm]
\rho(s) & = & -\operatorname{Im}G(s+i\epsilon) \ , \\
d(x=\sqrt{s}) & = & -\frac{2x}{\pi}\operatorname{Im}G(x^{2}+i\epsilon)
\end{eqnarray}

\end{itemize}

\clearpage

\chapter{Kinematics of two-body decays}
\label{app:appendixC}

\medskip

Consider a particle $S$ in its rest frame, decaying into two particles $\phi_{1}$ and $\phi_{2}$.\ Kinematic relations fix the expression for $k=|\textbf{k}|$, {\em i.e.}, the absolute value of the three-momentum of the outgoing particles:
\\
\begin{equation}
(p_{S}^{\mu}) = \begin{pmatrix}m_{S} \\ 0 \\ 0 \\ 0\end{pmatrix} = \begin{pmatrix}\sqrt{k^{2}+m_{1}^{2}} \\ k \\ 0 \\ 0\end{pmatrix}+\begin{pmatrix}\sqrt{k^{2}+m_{2}^{2}} \\ -k \\ 0 \\ 0\end{pmatrix} \ ,
\end{equation}
\begin{eqnarray}
m_{S}^{2} & = & m_{1}^{2}+m_{2}^{2}+2k^{2}+2\sqrt{(m_{1}^{2}+k^{2})(m_{2}^{2}+k^{2})} \ , \\
\nonumber \\
\Rightarrow \ \ \ k & = & \frac{1}{2m_{S}}\sqrt{m_{S}^{4}+(m_{1}^{2}-m_{2}^{2})^{2}-2(m_{1}^{2}+m_{2}^{2})m_{S}^{2}} \ .
\end{eqnarray}
\begin{figure}[h]
\centering
\includegraphics[scale=0.5]{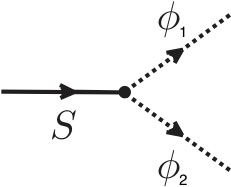}
\caption{Schematic decay process $S\rightarrow \phi_{1}\phi_{2}$.}
\label{figure_SABdecay}
\end{figure}
\\
For $m_{1}=m_{2}$ this simplifies to
\begin{eqnarray}
k & = & \frac{1}{2m_{S}}\sqrt{m_{S}^{4}-4m_{1}^{2}m_{S}^{2}} \nonumber \\[0.15cm]
& = & \sqrt{\frac{m_{S}^{2}}{4}-m_{1}^{2}} \ .
\end{eqnarray}

\clearpage

\chapter{Multi-valued complex functions and Riemann sheets}
\label{app:appendixD}

\medskip

\section{Introductory example}
\begin{figure}[t]
\begin{center}
\includegraphics[scale=0.5]{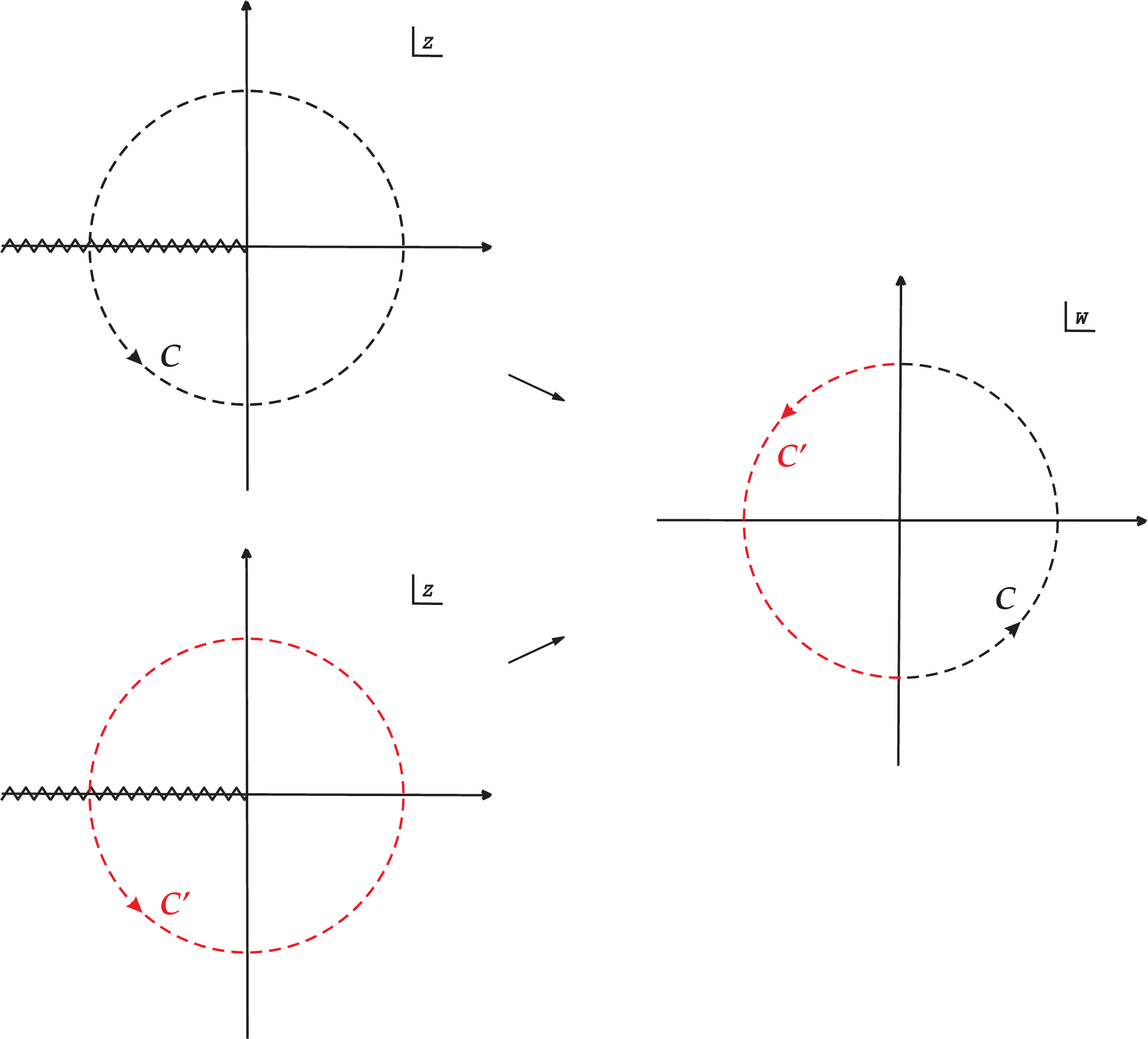}
\caption{Multi-valued character of $f(z)=\sqrt{z}$ with paths $\mathcal{C}$ (dashed black) and $\mathcal{C}^{\prime}$ (dashed red) in the complex $z$- and $w$-planes.}
\label{figure_rootzwplane}
\end{center}
\end{figure}

Consider the complex root function:\footnote{A more detailed version of this appendix can be found in Ref.\ \cite{thomasthesis}.}
\begin{equation}
f:\mathbb{C}\rightarrow\mathbb{C}, \ z\mapsto+\sqrt{z}=\sqrt{z}=w \ . \label{equation_complexrootdefinition}
\end{equation}
Using polar coordinates, $z=\rho e^{i\phi}$, the root function can be written as
\begin{equation}
f(z) = \sqrt{z} = \sqrt{\rho} e^{i\frac{\phi}{2}} \ , \ \ \ \phi\in(-\pi,\pi] \ .
\end{equation}
This function is not single-valued, meaning that it is not well-defined for all $z\in\mathbb{C}$.\ We can convince ourselves of this fact by looking at the complex $z$- and $w$-planes, see Figure\ \ref{figure_rootzwplane}.\ Taking a path $\mathcal{C}$ in the $z$-plane starting from the point $z=-\rho$ and walking counterclockwise back to that point, this corresponds to a path starting from $-i\sqrt{\rho}$ and ending at $i\sqrt{\rho}$ in the $w$-plane, which is a semicircle on the right side of the imaginary axis. So, coming back to the starting point in the $z$-plane does not give us the same value for $\sqrt{z}=w$.\ In fact, we must take the same path as before in the $z$-plane, denoted as $\mathcal{C}^{\prime}$, in order to arrive at the same point in the $w$-plane.

One can investigate the behavior of $f(z)$ by approaching the negative real axis from the upper and lower side of the complex $z$-plane:
\begin{eqnarray}
\lim_{\epsilon \to 0^{+}}f(-\rho+i\epsilon) & = & \sqrt{\rho}e^{i\frac{\pi}{2}} \nonumber \\
& = & i\sqrt{\rho} \ , \label{equation_rootdisc1} \\
\lim_{\epsilon \to 0^{+}}f(-\rho-i\epsilon) & = & \sqrt{\rho}e^{-i\frac{\pi}{2}} \nonumber \\
& = & -i\sqrt{\rho} \ . \label{equation_rootdisc2}
\end{eqnarray}
As stated before, coming from different directions yields different limits of the root function on the negative real axis. One can change the definition of $f(z)$ to get rid of its multi-valued character by introducing a branch cut:
\begin{equation}
f:\mathbb{C}\rightarrow\mathbb{C}\backslash\{w\in\mathbb{C}:\operatorname{Re}w<0\vee(\operatorname{Re}w=0\wedge \operatorname{Im}w<0)\}, \ z\mapsto+\sqrt{z}=\sqrt{z}=w \ .
\end{equation}
This is called the {\em principal branch} of the complex root function.

\section{Riemann sheets}
By cutting the complex $z$-plane along a line (or line segment), a multi-valued complex function can be made well-defined, because the branch cut effectively behaves like an edge for every path taken from one side to the other. Within this framework it is not possible to cross it. Yet, a more general interpretation of multi-valued complex functions is obtained by looking at Figure\ \ref{figure_rootzwplane}: We notice that the two sketched $z$-planes are actually the same, although the value of $w$ in both planes is not. It turns out that the problem of obtaining a single-valued function is only a question of choosing one of those $z$-planes. Exploiting this idea, the complex function $f(z)$ is then a mapping from those two planes onto a single $w$-plane, and the distinction between the two planes is accomplished by introducing a $k$-value, which is referring to either the first ($k=1$) or the second ($k=2$) one. It is fully natural to take this value as an additional coordinate; giving all three coordinates, namely the real- and imaginary part of a given point in the complex $z$-plane and the $k$-value of that plane, we can assign exactly one value $w$ to every given point $z$.\ The function appears to be single-valued now.

\begin{figure}[t]
\begin{center}
\includegraphics[scale=0.49]{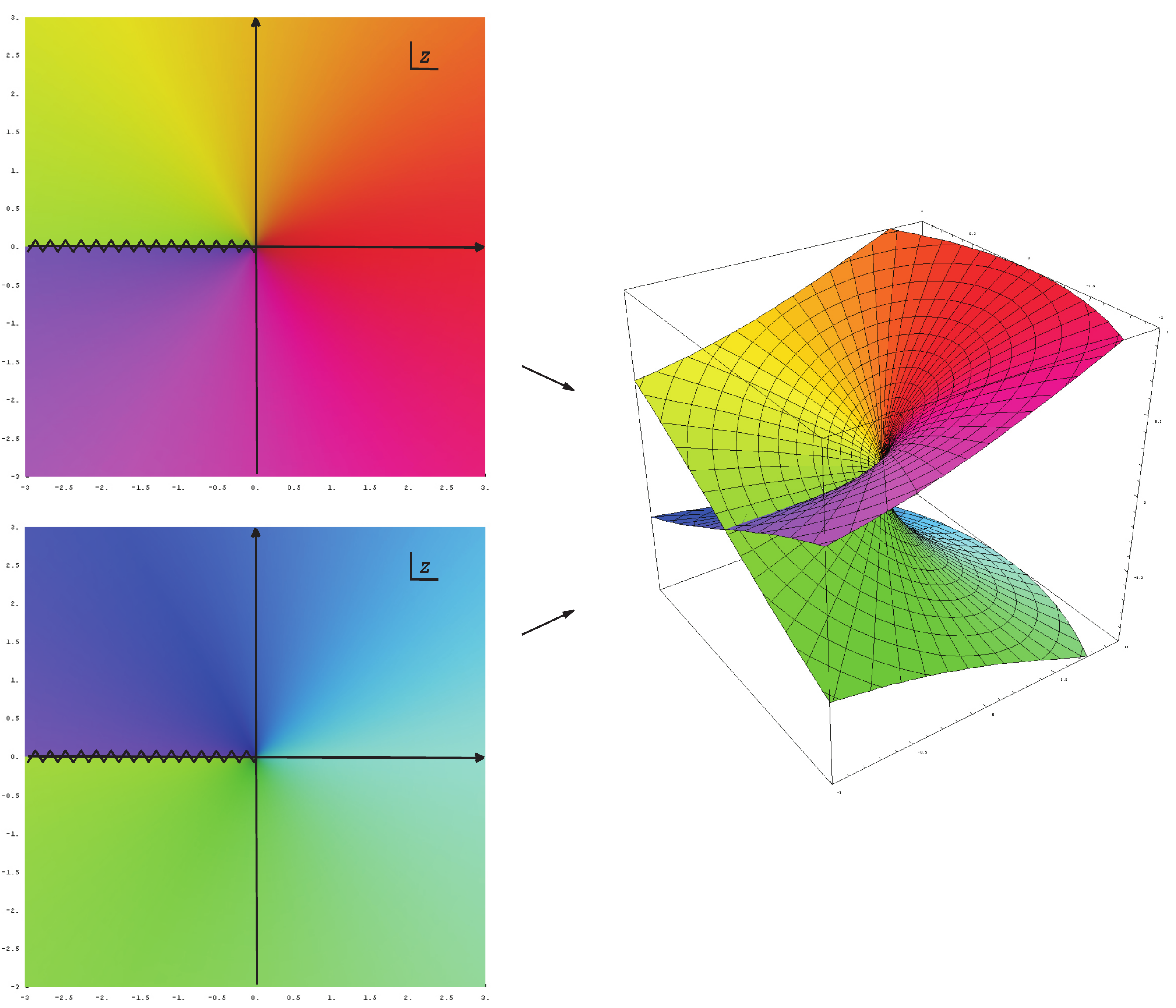}
\caption{Riemann surface of the complex root function. Each complex value $w$ is represented as a particular color: the $\arg$ of the complex number is encoded as the hue of the color, the modulus as its saturation (the colored background graphics on the left as well as the figure on the right were created by \href{http://livingmatter.physics.upenn.edu/node/38}{Jan Homann} from the University of Pennsylvania).}
\label{figure_rootsurface}
\end{center}
\end{figure}

Since there is no discontinuity at all when slipping from one plane to the other, the branch cut can be understood as the connection between the two different $z$-planes. Both planes can be attached along the cut such that circling around the origin will make us leave the first plane after a full polar angle of $2\pi$, walking onto the second plane, and letting us arrive at the starting point in the first plane after a (global) turn of $4\pi$.\ The structure of the two combined $z$-planes appears as a very simple closed spiral stairway, see Figure\ \ref{figure_rootsurface}.\ We call every individual $z$-plane a {\em Riemann sheet} of the {\em Riemann surface}.\ As a result, our final definition of the single-valued complex root function is
\begin{equation}
f:X\rightarrow\mathbb{C}, \ f(z) = \sqrt{z} = w \ ,
\end{equation}
where $X$ is the Riemann surface sketched in Figure\ \ref{figure_rootsurface}, a one-dimensional complex manifold. Note that the very important {\em identity theorem} holds true on a Riemann surface. This is why we can apply analytic continuation not only from the real to the complex, but also from one Riemann sheet to the other. For a deeper mathematical introduction on Riemann surfaces see for example the standard textbook by Forster\ \cite{forster}. A very good presentation covering a wide range of the practical aspects for calculations can be found in Ref.\ \cite{lepage}. For just a quick but adequate look we highly recommend Ref.\ \cite{nearing}.

\section{Analytic continuation}
If we would like to pass through the branch cut, we need to apply analytic continuation down into the second sheet. In general, suppose that we have two holomorphic functions $f_{1}(z)$ and $f_{2}(z)$, defined each on domains $\Omega_{1},\Omega_{2}\subset\mathbb{C}$ such that both domains have a non-empty intersection $\Omega=\Omega_{1}\cap\Omega_{2}$ with $f_{1}(z)=f_{2}(z)$ for every $z\in\Omega$.\ Then, both functions are analytic in their domains and can be expressed in terms of a power series,
\begin{equation}
f_{1}(z) = \sum_{n=0}^{\infty}a_{n}(z-z_{1})^{n} \ , \ \ \ \ \ f_{2}(z) = \sum_{n=0}^{\infty}b_{n}(z-z_{2})^{n} \ ,
\end{equation}
where $z_{1}\in\Omega_{1}$ and $z_{2}\in\Omega_{2}$.\ Either of the two expressions is valid in the intersection region $\Omega$ and is consequently the series representation of one and the same analytic function $f(z)$ around the points $z_{1}$ and $z_{2}$, respectively\ \cite{bronstein}.\ Thus, the function $f_{2}(z)$ is called the analytic continuation of $f_{1}(z)$ onto $\Omega_{2}$ and vice versa.

\begin{figure}[h]
\begin{center}
\includegraphics[scale=0.55]{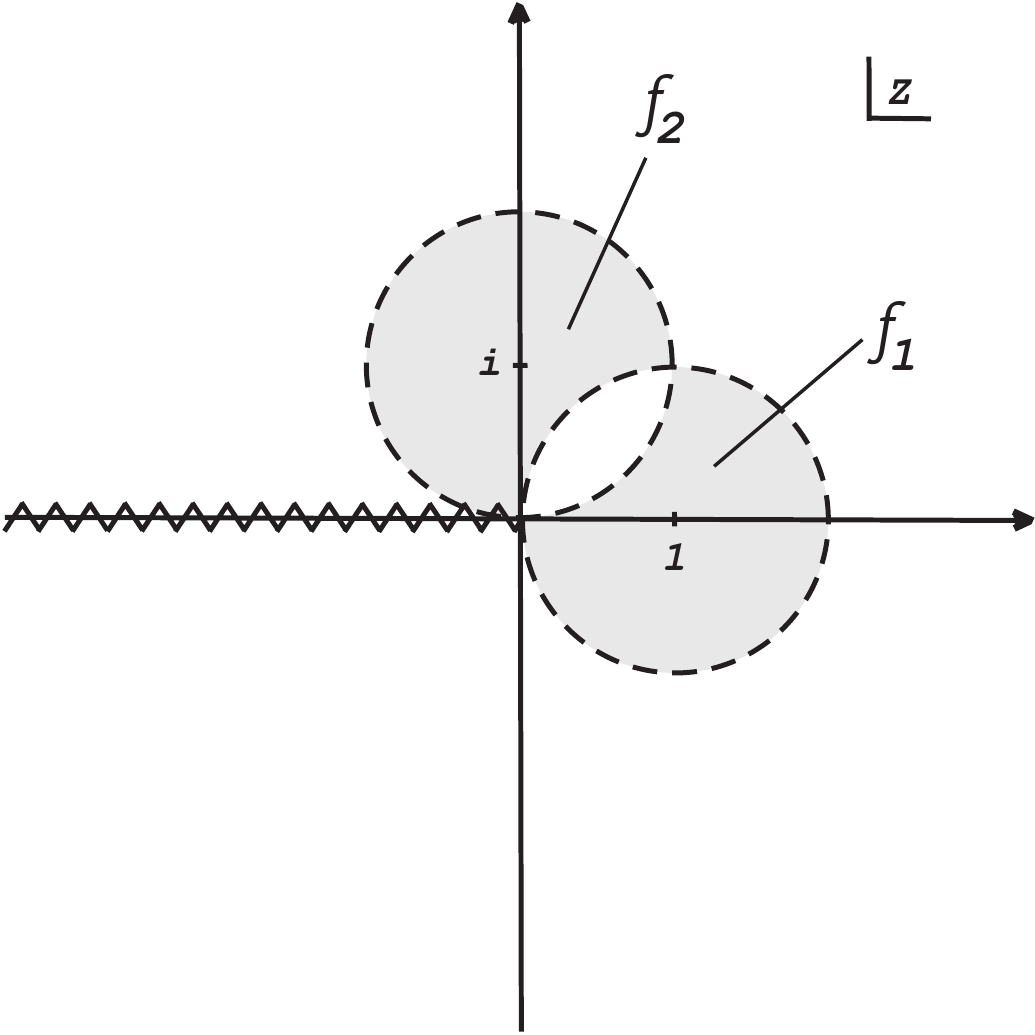}
\caption{Analytic continuation of the complex root function by expanding it in a power series in two different discs (gray) and realizing that both representations equal each other for every $z$ in the intersection region (white).}
\label{figure_rootanalytic}
\end{center}
\end{figure}

We can immediately apply this procedure to the complex root function $f(z)=\sqrt{z}$ around the point $z_{1}=1$ on an open disc with radius one. The latter limitation reflects the radius of convergence of the power series of the complex root, hence it seems to be only analytic on the disc. However, we now expand $f(z)$ around the point $z_{2}=i$ in another open disc, but with the same radius as before. This creates an overlapping region, see Figure\ \ref{figure_rootanalytic}.\ The expanded function is again holomorphic in the second disc and, moreover, it is still the root function! In fact, we could take the point $z_{2}$ in such a way that it lies in the first disc and where a point $z=i$ is included in the new disc resulting from the series expansion around that $z_{2}$.\ Either way, we realize that both resulting functions agree for every $z$ in the intersection region and are consequently the same in both discs.
\vspace{0.5cm}

It is possible to extend the natural domain ({\em i.e.}, a disc with finite radius of convergence) of a single series representation of a function without passing through any singularity by adopting the outlined method. For instance, if we add a third disc with center at $z_{3}=-1$ in Figure\ \ref{figure_rootanalytic}, the root function is extended from the second disc to negative real numbers. Note that there is no branch cut preventing us to get on and go beyond the negative real axis. Instead we would clearly rediscover its necessity when continuing further back to our starting point at $z=1$, because we would end up with the negative root function, the analytic continuation in the second Riemann sheet:
\begin{eqnarray}
f(z) & = & \sqrt{z} \ , \\
f_{\text{\Romannum{2}}}(z) & = & -f(z) \ .
\end{eqnarray}
We have denoted the second Riemann sheet with a roman numeral. Formally, we get the above result by approaching the branch cut on the negative real axis from two different directions, compare with Eqs.\ (\ref{equation_rootdisc1}) and (\ref{equation_rootdisc2}). The observed discontinuity is of course only the difference between two sheets and not an intrinsic property of the root function. At $z=-\rho$ it simply reads
\begin{eqnarray}
\disc f(-\rho) & = & \lim_{\epsilon \to 0^{+}}\Big[f(-\rho+i\epsilon)-f(-\rho-i\epsilon)\Big] \nonumber \\
& = & i\sqrt{\rho}-(-i\sqrt{\rho}) \nonumber \\
& = & 2i\sqrt{\rho} \ .
\end{eqnarray}
The analytic continuation of the root function down into the second sheet is then performed by accepting the requirement that the value $f_{\text{\Romannum{2}}}(-\rho-i\epsilon)$ ({\em i.e.}, the function just below the cut in the second sheet) equals the value $f(-\rho+i\epsilon)$ ({\em i.e.}, the function just above the cut in the first sheet) along the whole negative real axis:
\begin{eqnarray}
\lim_{\epsilon \to 0^{+}}f_{\text{\Romannum{2}}}(-\rho-i\epsilon) & = & \lim_{\epsilon \to 0^{+}}f(-\rho+i\epsilon) \nonumber \\
& = & \lim_{\epsilon \to 0^{+}}f(-\rho-i\epsilon)+2i\sqrt{\rho} \nonumber \\
& = & -i\sqrt{\rho}+2i\sqrt{\rho} \nonumber \\
& = & i\sqrt{\rho} \ . \label{equation_rootsecondinfinitesimal}
\end{eqnarray}
The analytic extension of either this result or the second last line into the lower half plane is what we were looking for,
\begin{equation}
f_{\text{\Romannum{2}}}(z) = f(z)+\disc f(z) = -\sqrt{z} \ , \label{equation_rootsecondsheet}
\end{equation}
and exactly the second branch of the complex root in the context of our first interpretation of multi-valued functions. Note the very useful identity between the imaginary part of $f(z)$, taken right above the real axis, and its discontinuity:
\begin{equation}
\disc f(-\rho) = 2i\lim_{\epsilon \to 0^{+}}\operatorname{Im}f(-\rho+i\epsilon) \ . \label{equation_usefullidentity}
\end{equation}

One may ask why the extension (\ref{equation_rootsecondsheet}) of the purely imaginary result from Eq.\ (\ref{equation_rootsecondinfinitesimal}) is so simple. In fact, we should be more precise in this point: The identity theorem states that if two given functions $f(z)$ and $g(z)$, both holomorphic on a domain $\Omega$, equal each other on some line segment $U$ lying in $\Omega$, then they equal each other on the whole domain $\Omega$\ \cite{fischerkaul}.\ Here, we see at first glance what is really the difference between real and complex analysis: a holomorphic function defined on $\Omega$ is completely determined by its values on a line segment $U\subset\Omega$.\ So, there is no problem in continuing Eq.\ (\ref{equation_rootsecondinfinitesimal}) into the complex plane.

\newpage
\clearpage

\thispagestyle{empty}
\

\newpage
\clearpage

\pagenumbering{roman}
\setcounter{page}{\value{bibpage}}

\bibliographystyle{apsrev4-1}
\bibliography{thesis}

\newpage
\clearpage

\thispagestyle{empty}
\

\newpage
\clearpage

\thispagestyle{empty}
\section*{Acknowledgements}

\medskip

I want to thank my supervisor Dr.\ Francesco Giacosa for being my mentor and friend during the past couple of years. He gave me the necessary freedom to deliver this work. I am grateful for plenty of fruitful discussions about physics, mathematics, and philosophy -- it is amazing how much Francesco has taught me about particle physics ever since when I started performing my first calculations.\ I also thank him and his family, his students, Prof.\ Dr.\ Wojciech Broniowski, and the staff of the physics department in Kielce for the wonderful time in Poland.
\\[0.3cm]
Moreover, I would like to thank Prof.\ Dr.\ Dirk H.\ Rischke who gave me the opportunity to do my PhD studies in his group. I am thankful to him for carefully reading the thesis and for his important remarks.\ I also thank him and Prof.\ Dr.\ Jochen Wambach for their useful assistance in the PhD committee meetings.
\\[0.3cm]
I also thank Dr.\ Dennis D.\ Dietrich and Dr.\ Robert Kaminski for their helpful comments on technical aspects of derivative interactions. I am very grateful to Prof.\ Dr.\ Michael R.\ Pennington from Jefferson Lab for providing me with further information about his publications on the scalar--isovector sector.
\\[0.3cm]
I want to thank all the members of the chiral group in Frankfurt -- new and former ones -- for years of pleasant time and funny stories.\ In particular, I owe special thanks to my colleagues Tim and J\"urgen with whom I shared a common office. I think, the past years would have never been so enjoyable without them.
\\[0.3cm]
I also would like to thank Dr.\ Gerhard Burau and the whole team of HGS-HIRe for making my studies possible.\ I am grateful for financing my work through a fellowship (especially during the last six months) and for organizing all the fantastic accompanying courses. I thank Emma Ford for being my mentor during the soft-skill trainings -- this I regard as highly valuable for my self-awareness.
\\[0.3cm]
Time has passed so quickly$\dots$
\\[0.3cm]
\begin{otherlanguage}{german}
Vielen lieben Dank an Kira und Micha f\"ur ihre moralische Unterst\"utzung, aufmunternde Art und f\"ur viele Gespr\"ache. Ich bedauere, dass wir so wenig Zeit zusammen verbringen konnten.
\\[0.3cm]
Ich danke Jessica f\"ur ihr ungeahntes Ma{\ss} an Verst\"andnis.
\\[0.3cm]
Ich danke Helene und Matilda f\"ur jeden Tag, an dem sie so sind, wie sie sind.
\end{otherlanguage}

\newpage
\clearpage

\thispagestyle{empty}
\

\newpage
\clearpage

\pagenumbering{gobble}

%\begin{wrapfigure}{r}{0.23\textwidth}
%\begin{center}
%\framebox{\includegraphics[width=0.26\textwidth]{wolkanowski}}
%\end{center}
%\end{wrapfigure}
\section*{Curriculum vitae}
\begin{table}[!h]
\begin{tabular}{l l}
 &  \\
{\em Personal data:} &  \\
 &  \\
Name: & Wolkanowski-Gans \\
Prename: & Thomas \\
Date of birth: & 6th August 1986 \\
Place of birth: & Opole (Poland) \\
Nationality: & German \\
Marital status: & Married, 2 children \\
 &  \\
 &  \\
{\em Education:} &  \\
 &  \\
02/2013 - 08/2016 & Ph.D.\ in Physics, Goethe-Universit\"at Frankfurt am Main \\
 & ``Dynamical generation of hadronic resonances in effective models \\
 & with derivative interactions'' \\
 & Supervisor: PD Dr. Francesco Giacosa \\[0.2cm]
 & Fellowship of the ``Helmholtz Graduate School for Hadron and \\
 & Ion Research'', funded by F\&E GSI/GU and HIC for FAIR Frankfurt \\
 &  \\
10/2010 - 01/2013 & M.Sc.\ in Physics, Goethe-Universit\"at Frankfurt am Main \\
 & ``Resonances and poles in the second Riemann sheet'' \\
 & Supervisor: Prof. Dr. Dirk H. Rischke \\
 &  \\
10/2007 - 09/2010 & B.Sc.\ in Physics, Goethe-Universit\"at Frankfurt am Main \\
 & ``Einfluss des NN-Wirkungsquerschnitts auf die Dissipation in \\
 & Schwerionenkollisionen'' \\
 & Supervisor: Prof. Dr. Joachim A. Maruhn \\
 &  \\
06/2006 & Allgemeine Hochschulreife, Goetheschule, Wetzlar \\
\end{tabular}
\end{table}

\begin{table}[!h]
\begin{tabular}{l l}
{\em Experience:} &  \\
 &  \\
10/2009 - 09/2013 & Teaching assistant \\
 & Group of Prof. Dr. Hartmut G. Roskos, Prof. Dr. Dirk H. Rischke, \\
 & and Prof. Dr. Owe Philipsen \\
 &  \\
11/2008 - 04/2009 & Research assistant \\
 & Group of Prof. Dr. Hartmut G. Roskos \\
 & Field of research: femtosecond spectroscopy of molecules \\
\end{tabular}
\end{table}

\clearpage

\begin{table}[!h]
\begin{tabular}{l l}
 &  \\
 &  \\
 &  \\
{\em Publications:} &  \\
 &  \\
06/2016 & ``$K_{0}^{\ast}(800)$ as a companion pole of $K_{0}^{\ast}(1430)$'', T. Wolkanowski, M. So\l tysiak,\\
 & and F. Giacosa, Nucl.\ Phys.\ B \textbf{909}, 418\\
 & \\
01/2016 & ``$a_{0}(980)$ revisited'', T. Wolkanowski, F. Giacosa, and D. H. Rischke, \\
 & Phys. Rev. D \textbf{93}, 014002 \\
 & \\
10/2014 & ``The role of the next-to-leading order triangle-shaped diagram in two-body \\
 & hadronic decays'', J. Schneitzer, T. Wolkanowski, and F. Giacosa, \\
 & Nucl. Phys. B \textbf{888}, 287 \\
 & \\
12/2012 & ``Propagator poles and an emergent stable state below threshold: \\
 & general discussion and the $E(38)$ state, F. Giacosa and T. Wolkanowski, \\
 & Mod. Phys. Lett. A \textbf{27}, 1250229 \\
 &  \\
 &  \\
 {\em Conferences:} &  \\
 &  \\
03/2016 & DPG-Fr\"uhjahrstagung der Fachverb\"ande Physik der Hadronen und Kerne, \\
 & Darmstadt, with talk \\
 &  \\
07/2015 & ``The 8th International Workshop on Chiral Dynamics 2015'', Pisa, \\
 & with poster and proceedings \\
 &  \\
09/2014 & ``EEF70 -- Workshop on Unquenched Hadron Spectroscopy'', Coimbra, \\
 & with talk and proceedings \\
 &  \\
03/2014 & DPG-Fr\"uhjahrstagung der Fachverb\"ande Physik der Hadronen und Kerne, \\
 & Didaktik der Physik, Frankfurt am Main, with talk \\
 &  \\
02/2014 & ``Excited QCD 2014'', Bjelasnica Mountain, Sarajevo, \\
 & with talk and proceedings \\
 &  \\
03/2013 & 77. Jahrestagung der DPG und DPG-Fr\"uhjahrstagung, Dresden, \\
 & with talk \\
 &  \\
02/2013 & ``Excited QCD 2013'', Bjelasnica Mountain, Sarajevo, \\
 & with talk and proceedings \\
\end{tabular}
\end{table}

\clearpage

\begin{table}[!h]
\begin{tabular}{l l}
 &  \\
 &  \\
 &  \\
{\em Training:} &  \\
 &  \\
06/2015 & Soft skill course ``Basic Course III: Leadership and Career \\
 & Development'' as part of the graduate school program \\
 &  \\
 05/2015 & Postgraduate training ``Lecture Week on Hadron Physics: \\
 & Hadron physics at the Belle and BES experiments'' as part of \\
 & the graduate school program \\
 &  \\
06/2014 & Soft skill course ``Basic Course II: Leading Teams in a Research \\
 & Environment'' as part of the graduate school program \\
 &  \\
05/2014 & Postgraduate training ``Lecture Week on Hadron Physics: \\
 & In-medium properties of hadrons'' as part of the graduate school \\
 & program \\
 &  \\
09/2013 & Soft skill course ``Basic Course I: Making an Impact \\
 & as an Effective Researcher'' as part of the graduate school program \\
 &  \\
07/2013 & Postgraduate training ``Lecture Week on Hadron Physics: \\
 & Measuring properties of hadrons in experiments and on the lattice''  \\
 & as part of the graduate school program \\
 &  \\
 &  \\
{\em Computer skills:} \\
 &  \\
 & Operating systems: Mac OS X, Linux (Ubuntu), Microsoft Windows \\
 & Software: LaTeX, Microsoft Office \\
 & Programming languages: Mathematica, Fortran \\
 &  \\
 &  \\
{\em Languages:} &  \\
 &  \\
German & mother tongue \\
English & fluent \\
Polish & good \\
French & basic \\
\end{tabular}
\end{table}

\newpage
\clearpage

\thispagestyle{empty}
\

\newpage
\clearpage

\thispagestyle{empty}
\

\end{document}